\documentclass[a4paper]{aa}

\usepackage{graphicx}
\usepackage{color}
\usepackage{txfonts}
\definecolor{myblue}{rgb}{0.00, 0.0, 0.9}

\usepackage[breaklinks,  citecolor=myblue, linkcolor=myblue, urlcolor=myblue, colorlinks=true, debug, baseurl=' ']{hyperref}
\begin{document}

   \title{Galactic magnetic field reconstruction using the polarized diffuse Galactic emission: Formalism and application to {\it Planck} data}

   \author{V. Pelgrims
          \inst{1,2,3}
          \and J.F. Mac{\'i}as-P{\'e}rez
          \inst{1}
          \and F. Ruppin
          \inst{1,4}
          }

   \institute{Univ. Grenoble Alpes, CNRS, Grenoble INP, LPSC-IN2P3, 38000 Grenoble, France
   \and Institute of Astrophysics, Foundation for Research and Technology-Hellas, GR-71110 Heraklion, Greece
   \and Department of Physics, and Institute for Theoretical and Computational Physics, University of Crete, GR-70013 Heraklion, Greece
   \and Kavli Institute for Astrophysics and Space Research, Massachusetts Institute of Technology, Cambridge, MA 02139, USA
   \\
              \email{pelgrims@physics.uoc.gr ; macias@lpsc.in2p3.fr}
             }

   \date{Received July 26, 2018; accepted May 11, 2021}
        \titlerunning{GMF Reconstruction}

  \abstract
   {

   The polarized Galactic synchrotron and thermal dust emission
   constitutes a major tool in the study of the Galactic magnetic
   field (GMF) and in constraining its strength and geometry for the
   regular and turbulent components.
   In this paper, we review the modeling of these two components
   of the polarized Galactic emission and present our strategy for
   optimally exploiting the currently existing data sets.
   We investigate a Markov Chain Monte Carlo (MCMC) method to
   constrain the model parameter space through maximum-likelihood
   analysis, focusing mainly on dust polarized emission.
   Relying on simulations,
   we demonstrate that our methodology can be used to constrain the
   regular GMF geometry. Fitting for the reduced Stokes parameters, this
   reconstruction is only marginally dependent of the accuracy of the reconstruction
   of the Galactic dust grain density distribution.
   However, the reconstruction degrades, apart from the pitch angle, when including a turbulent component on
   the order of the regular one as suggested by current observational constraints.
    Finally, we applied this methodology to a set of  \textit{Planck} polarization
    maps at 353~GHz to obtain the first MCMC based constrains on
        the large-scale regular-component of the GMF from the polarized diffuse
        Galactic thermal dust emission.
        By testing various models of the dust density distribution and of the GMF
        geometry, we prove that it is possible to infer the large-scale geometrical
        properties of the GMF.
        We obtain coherent three-dimensional (3D) views of the GMF, from which we
        infer a mean pitch angle of 27 degrees with 14 \% scatter, which is in agreement
        with results obtained in the literature from synchrotron emission.
   }


   \maketitle
%

\section{Introduction}
\label{sec:intro}
From a cosmological perspective, the characterization of the polarized diffuse Galactic emission from synchrotron and thermal dust is of prime importance. This emission dominates the signal in the frequency range of interest for the observation of the  cosmic microwave background (CMB) polarization anisotropies (e.g., see \citealt{PlanckX2016}).

In particular, possible contamination from the polarized diffuse
Galactic emission has been shown to be one of the major limitations for
the detection of primordial CMB polarization B-modes related to the
inflationary era in the early universe
\citep{2015PhRvL.114j1301B,2016PhRvL.116c1302B}.
Therefore, providing an accurate characterization and modeling of this
polarized Galactic emission is of great importance as it would strengthen the level of confidence
with regard to our understanding of its physical scope and would allow for accurate testing of 
the results obtained from elaborated component-separation techniques 
\citep[e.g.,][]{PlanckIX2016,PlanckX2016} which are
used to extract the cosmological CMB signal. \\

At microwave frequencies that are typically relevant for CMB experiments, the polarized sky is dominated by synchrotron emission below  $\sim$ 80 GHz and by thermal dust emission above that frequency.
Both emission components result from a line-of-sight integration of local emission and these offer the possibility to infer the properties of the Galactic magnetic field (GMF), an important constituent and actor in the ecosystem of our Galaxy.
The diffuse polarized Galactic synchrotron emission is produced by relativistic electrons that spiral along the GMF  lines \citep[see][for a review]{1966SvPhU...8..674G}.
Equivalently, the polarized thermal dust emission is produced by rotating aspherical dust grains that are totally or partially aligned with the GMF lines (\citealt{1951ApJ...114..206D}; \citealt{1973IAUS...52..197K}; \citealt{1996ASPC...97...72O}; \citealt{1996ApJ...472..240L}; \citealt{2000ApJ...533..298O}; \citealt{2002ApJ...575..886E}; \citealt{2009MNRAS.400..536J}; \citealt{2013lcdu.confE...3V}; \citealt{2015ARA&A..53..501A}; \citealt{2017arXiv170401721H}).

Over the last two decades, the polarized diffuse Galactic emission has been measured up to a relatively high level of accuracy and high angular resolution by the \textit{WMAP}\footnote{\url{https://map.gsfc.nasa.gov/}}
and \textit{Planck}\footnote{\url{http://www.esa.int.Planck}} satellite experiments (\citealt{Pag2007}; \citealt{2013ApJS..208...20B}; \citealt{PlanckX2016}), which have observed the sky in polarization in a large range of frequency going from 23 to 353~GHz.
The wealth of information present in these full-sky observations is extremely valuable in modeling the various components of the Galaxy.
In light of these polarization data, there have been attempts to constrain the geometry of the GMF at the largest Galactic scales
and the relative amplitude of their main components (regular, turbulent,...) and of the Galactic matter content
(\citealt{Pag2007}; \citealt{Rui2010}; \citealt{Jaff2010}; \citealt{Fau2011}; \citealt{Jan2012a}; \citealt{Jan2012b}; \citealt{2012A&A...540A.122F}; \citealt{Jaff2013}).

Because full-sky polarization data first came  at low frequencies, these
investigations were driven by the study of the full-sky synchrotron
emission.
\cite{Pag2007} used the three-year full-sky maps from the \textit{WMAP}
satellite at 22 GHz (the K-band) and fitted a parametric model using the
polarization position angles of the emission. 
\cite{Rui2010} used the five-year \textit{WMAP} polarization data at the
same frequency and searched for the best fits of several parametric models
on a grid-based exploration of the parameter space,
thus providing the first systematic comparison of GMF models.
\cite{Sun2008}, \cite{Sun2010}, \cite{Jan2012a},
\cite{Jan2012b}, \cite{Jaff2010}, and \cite{Jaff2013}
built more sophisticated GMF models and used the same \textit{WMAP} data to constrain them, complementing (for the most part) the synchrotron data with Faraday rotation or dispersion measures on Galactic or extragalactic sources.
Recently, \cite{PlanckXLII2016} used the synchrotron data from the \textit{Planck} satellite to update
GMF models that were previously constrained from \textit{WMAP} data and rotation measure data. The latter study has shown the limitations of the models at reproducing the current data sets. See also \citep{Ste2018} for a recent work.

Similar studies have been carried out
using the  polarized thermal dust emission. \cite{Pag2007} showed that the 94 GHz band of the \textit{WMAP} satellite measured the thermal dust emission and used it to constrain the GMF. \cite{Fau2011} used the 353-GHz data from the
\textit{ARCHEOPS} balloon experiment \citep{Ben2004} in addition to the \textit{WMAP} satellite 22-GHz channel data (tracing the polarized diffuse Galactic synchrotron emission) to constrain GMF models.
\cite{Jaff2013} used the full-sky \textit{WMAP} 94-GHz polarization maps and showed that the diffuse Galactic emission observed at this frequency is not compatible with GMF configuration that fits best the polarized synchrotron emission as traced by the \textit{WMAP} low frequency data. More recently, \cite{PlanckXLII2016} showed that the full-sky 353-GHz data from \textit{Planck} is in conflict with predictions from the reconstructed large-scale GMF obtained from the synchrotron emission data.
Other studies have considered the thermal dust emission only from the polar caps to adjust models of the magnetic field in the neighborhood of the Sun (\citealt{PlanckXLIV2016}; \citealt{Alv2018}; \citealt{Pel2020}).

\smallskip

From these different works it has been established that the GMF is made of a regular (or coherent) component of an amplitude of a few micro Gauss and to which one or two turbulent components, random and ordered-random, are added \citep{Jan2012a,Jan2012b,Jaff2013,PlanckXLII2016}.
However, GMF models turn to be complex and highly dimensional (large number of parameters).
As a consequence, they are highly degenerated and difficult to constrain, particularly as a result of the fact that according to some authors, the turbulent components (random or ordered-random) have amplitudes that might exceed that of the regular part and also because the density distributions of the Galactic matter content are poorly known.

\medskip

In principle, a combined study of the diffuse synchrotron and thermal dust emission should allow us to constrain the GMF characteristics in three dimensions and to mitigate the loss of information introduced by the integration along the lines of sight that induces degeneracies. However, an additional complexity occurs. Three-dimensional (3D) density distributions of relativistic electron and of dust grains differ. Therefore, the GMF is not sampled in the exact same way along the lines of sight by the two species of the interstellar matter and the respective emission components may often reflect properties of the magnetized ISM at different locations.
This characteristic, which could be used to our advantage to model the GMF, may likely be source of confusion and of apparent discrepancies if it is not properly accounted for.

To face this complexity, one approach would be to constrain models of the GMF and of the matter density distributions through a joint modeling of both emission components.
Such an approach, followed by the \textsc{Imagine} Consortium (\citealt{Bou2018}), requires us to simultaneously constrain several models, with each residing in potentially high-dimensional space.
Performing such an ambitious fit through, for instance, a Bayesian method should ideally be the aim in this instance.

In this paper we motivate and explore a somewhat different approach to infer the main characteristics of the large-scale GMF.
In Sect.~\ref{sec:ModAndImp}, we review the modeling of the polarized diffuse Galactic emission from the 3D models of the distribution of matter in the Galaxy and of the GMF.
In Sect.~\ref{sec:Methodo}, we discuss the main methodology used as well as the data and simulations used. 
Section~\ref{sec:GMFrecons} discusses the fitting procedure in intensity and polarization for the dust thermal and its application in simulating the data.
In Sect.~\ref{sec:Iqu_fit}, we discuss the quality of our reconstruction of the regular part of the large-scale GMF, along with possible limitations and biases. 
Sect.~\ref{sec:synchrotronproof} evaluates the impact of those limitations when independently tackling Galactic synchrotron emission.
We apply our methodology to the full-sky polarization 353~GHz Planck maps in Sect.~\ref{sec:fit2Planck}. 
Finally, we summarize our findings and present our conclusions in Sect.~\ref{sec:conclu}.

\section{Modeling \& implementation}
\label{sec:ModAndImp}

\subsection{Thermal dust polarized emission}
\begin{figure}
\includegraphics[width=.9\columnwidth]{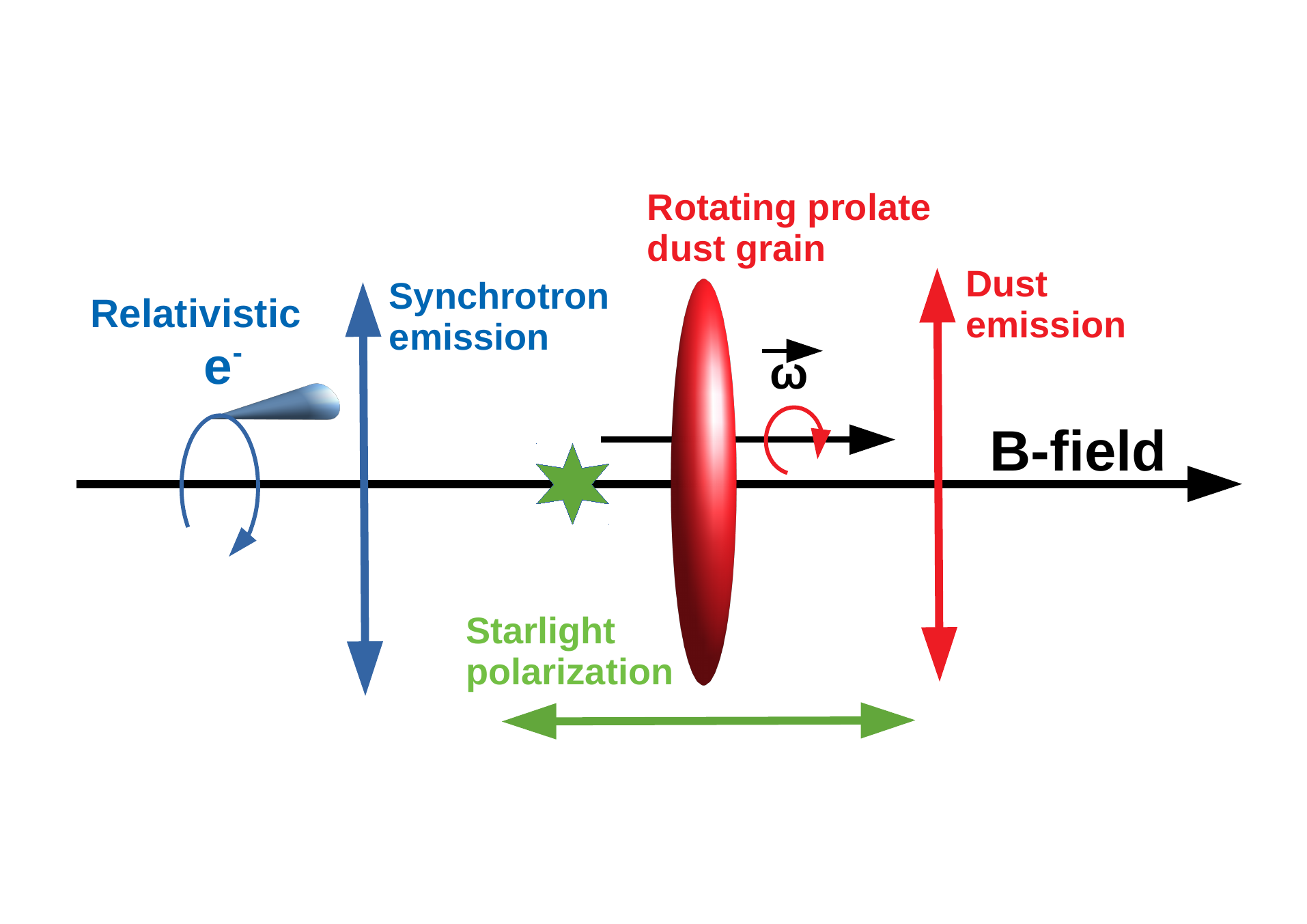}
\caption{Schematic view of the polarization direction of the Galactic
synchrotron and dust thermal emission as a function of the Galactic
magnetic field (GMF). For completeness, we also show the direction
of the starlight polarization.}
\label{fig:sketchdustsyncpol}
\end{figure}

The thermal dust emission dominates the polarized signal above
80 GHz and is the most significant foreground of the CMB at those
frequencies \citep[e.g.,][]{PlanckXXX2014}.
If the GMF is coherent in a given region of the sky, dielectric aspherical dust grains tend to align with  their major axis perpendicular to the field lines in the region and rotate with their spin axis parallel to the field lines \citep[e.g.,][]{Mar2007, Fau2011}.
As a consequence, the thermal dust emission arising at sub-millimeter and millimeter wavelengths is expected to be linearly polarized perpendicular to the GMF lines, as sketched in Fig.~\ref{fig:sketchdustsyncpol}.
Tracing the same dust, the optical polarized emission of stars is expected
to be parallel to the GMF lines as this is due to anisotropic absorption in the plane of polarization.

To model the diffuse thermal emission from optically thin Galactic dust, we adopted a parameterization close to that of \cite{Fau2011}, but which follows the physically motivated modeling of the emission introduced by \cite{Lee1985} and reviewed by \cite{Pel2020}.
We assume that the dust emission comes from a single population of dust grains heated at the same temperature from the interstellar radiation field, which implies a constant dust emissivity and a spatially constant intrinsic degree of polarization.
We make the additional assumption that the degree of misalignment of the dust grains with respect to the GMF lines is spatially uniform. \\

Specifically, we model the intensity and the linear polarization Stokes parameters as
\begin{small}
\begin{align}
I_{\rm{d}}(\mathbf{n}) = & \, \epsilon_{\nu}^{\rm{d}} \,
\int_{0}^{+\infty}{dr \, n_{\rm{d}}(r,\mathbf{n}) \, \left\lbrace
1 + p^{\rm{d}}\, f_{\rm{ma}} \left(\frac{2}{3} - \sin^2 \alpha(r,\mathbf{n})\right)
\right\rbrace }
\nonumber \\ 
Q_{\rm{d}}(\mathbf{n}) = & \, \epsilon_{\nu}^{\rm{d}} \, p^{\rm{d}}\, f_{\rm{ma}}
\int_{0}^{+\infty}{dr \, n_{\rm{d}}(r,\mathbf{n}) \, 
\sin^2 \alpha(r,\mathbf{n}) \, \cos[2\, \gamma(r,\mathbf{n})]}
\nonumber \\
U_{\rm{d}}(\mathbf{n}) = & \, \epsilon_{\nu}^{\rm{d}} \, p^{\rm{d}}\, f_{\rm{ma}}
\int_{0}^{+\infty}{dr \, n_{\rm{d}}(r,\mathbf{n}) \, 
\sin^2 \alpha(r,\mathbf{n}) \, \sin[2\, \gamma(r,\mathbf{n})]} \, ,
\label{eq:DUSTEMISSION}
\end{align}
\end{small}
where $r$ is the radial distance from the observer along the line-of-sight at sky position, $\mathbf{n}$.
The different terms in the equation are:
\begin{itemize}
\item
$\epsilon_{\nu}^{\rm{d}}$, the dust emissivity at observational frequency $\nu$, which is linked to the dust temperature through a grey-body's law (e.g.,\citealt{PlanckXX2015}, Appendix B). 
\item
$p^{\rm{d}}$, the so-called intrinsic degree of linear polarization
of the dust that depends on the properties of the dust grains. It represents the maximum value of the degree of linear polarization of the radiation emitted by an hypothetical ensemble of perfectly aligned dust grains from a small volume; it only depends on the geometry of the dust grains.
\item
$f_{\rm{ma}}$, the misalignment term which generally should depend on the dust population. It characterizes the average tendency of the dust grains to align with the magnetic field line.
\item
$n_{\rm{d}}(r,\,\mathbf{n})$, the 3D Galactic dust grain density.
\item
$\alpha(r,\,\mathbf{n})$, the inclination angle of the GMF line with the line of sight at $(r,\,\mathbf{n}).$
\item
$\gamma(r,\,\mathbf{n})$, the so-called local polarization angle.
\end{itemize}
The local polarization angle is defined in the plane orthogonal to the line of sight as the angle between the polarization vector direction and the local meridian. Expressed in terms of the vector component of the ambient GMF, this angle is given as:\\ 
\begin{equation}
\gamma(r, \, \mathbf{n}) = \frac{1}{2} \arctan
\left( \frac{-2 \, B_\theta(r, \, \mathbf{n}) \, B_\phi(r, \, \mathbf{n})}
{B_\phi(r, \, \mathbf{n})^2 - B_\theta(r, \, \mathbf{n})^2} \right) \, ,
\label{eq:Gamma_def}
\end{equation}
with $B_\theta$ and $B_\phi$ the local transverse components of the magnetic field in the local spherical coordinate basis ($\mathbf{e}_r,\,\mathbf{e}_\theta,\,\mathbf{e}_\phi$) with $\mathbf{e}_\theta$ pointing towards the South pole. Thus, Eq.~\ref{eq:Gamma_def}  gives the polarization position angle of the polarization vectors stemming from the small space volume in the HEALPix (or COSMO) convention \citep{Gor2005}, which differs from the more commonly used IAU one. This angle is defined in the range $\left[ 0 , \; 180 \right[$ degrees.
We highlight that none of the values of $I_{\rm{d}}$, $Q_{\rm{d}}$ and $U_{\rm{d}}$ depend on the amplitude of the magnetic field, but only on its geometrical structure through the angles $\alpha$ and $\gamma$.

\medskip

Finally, in the following, we assume that the second term in the parenthesis of the top equation of Eqs.~\ref{eq:DUSTEMISSION} is negligible compared to the first term:
\begin{equation}
I_{\rm{d}}(\mathbf{n}) = \, \epsilon_\nu^{\rm{d}} \,
\int_{0}^{+\infty} dr \, n_d(r,\mathbf{n}) \;.
\label{eq:Idust}
\end{equation}
This simplifying assumption has been made in all previous works in the field (\cite{Fau2011,Jaff2013,PlanckXLII2016}). Relying on simulations, we found that in pixel space, the relative difference is at most of 10 percent (see also Sect.~\ref{sec:Methodo}).

\subsection{Synchrotron polarized emission}
\label{sec:syncmodeling}
The diffuse Galactic synchrotron emission is produced by relativistic electrons that spiral around the GMF lines \citep{1966SvPhU...8..674G,Ryb1979}.
This emission is to be polarized perpendicularly to the  GMF lines as sketched in Fig.~\ref{fig:sketchdustsyncpol}. For the contribution of synchrotron emission to CMB frequencies ($\gtrsim$ 20 GHz) we follow the modeling of the emission presented by \cite{Pag2007} and \cite{Fau2011} that relies on \cite{Ryb1979}. We adopt the notation of \cite{Fau2011}.
Assuming a power law for the relativistic electron energy the linear polarization Stokes parameters for the diffuse Galactic synchrotron emission is given as:
\begin{small}
\begin{align}
I_{\mathrm{s}}(\mathbf{n}) = & \, \epsilon_\nu^{\rm{s}} \, \int_{0}^{+\infty} dr\, n_{\rm{e}}(r,\mathbf{n})\,
\left(\mathbf{B}_\perp(r,\mathbf{n}) ^2 \right)^{(s+1)/4}                                               \nonumber \\
Q_{\mathrm{s}}(\mathbf{n}) = & \, \epsilon_\nu^{\rm{s}} \, p_{\rm{s}} \,\int_{0}^{+\infty} dr\, n_{\rm{e}}(r,\mathbf{n})\,
\left(\mathbf{B}_\perp(r,\mathbf{n}) ^2 \right)^{(s+1)/4} \, \cos[2\gamma(r,\mathbf{n})] \nonumber \\
U_{\mathrm{s}}(\mathbf{n}) = & \, \epsilon_\nu^{\rm{s}} \, p_{\rm{s}} \, \int_{0}^{+\infty} dr\, n_{\rm{e}}(r,\mathbf{n})\,
\left(\mathbf{B}_\perp(r,\mathbf{n}) ^2 \right)^{(s+1)/4}  \, \sin[2\gamma(r,\mathbf{n})],
\label{eq:SYNCEMISSION}
\end{align}
\end{small}
where $\epsilon_\nu^{\rm{s}}$ is the synchrotron emissivity, $n_{\rm{e}}(r,\mathbf{n})$ is the local density of relativistic electrons, $p_{\rm{s}}$ is the intrinsic synchrotron polarization fraction which is related to the relativistic electron energy spectral index ($s$) as follows:
\begin{equation}
p_{\rm{s}} = \frac{s+1}{s+7/3} \;.
\end{equation}

The angle $\gamma$ found in the expression of $Q_{\rm{s}}$ and
$U_{\rm{s}}$ is the same as in Eq. \ref{eq:Gamma_def}.
It marks the orientation perpendicular to the plane-of-sky component
of the magnetic field ($\mathbf{B}_\perp$).
We note that $\mathbf{B}_\perp(r,\mathbf{n}) ^2 = \mathbf{B}(r,\mathbf{n})^2 \sin^2[\alpha(r,\mathbf{n})]$, where the angle $\alpha(r,\mathbf{n})$ is the same as in Eq.~\ref{eq:DUSTEMISSION}; namely, it is the inclination angle of the GMF vector with respect to the line of sight.

\begin{figure*}
\centering
\begin{tabular}{ccc}
\includegraphics[width=.3\linewidth]{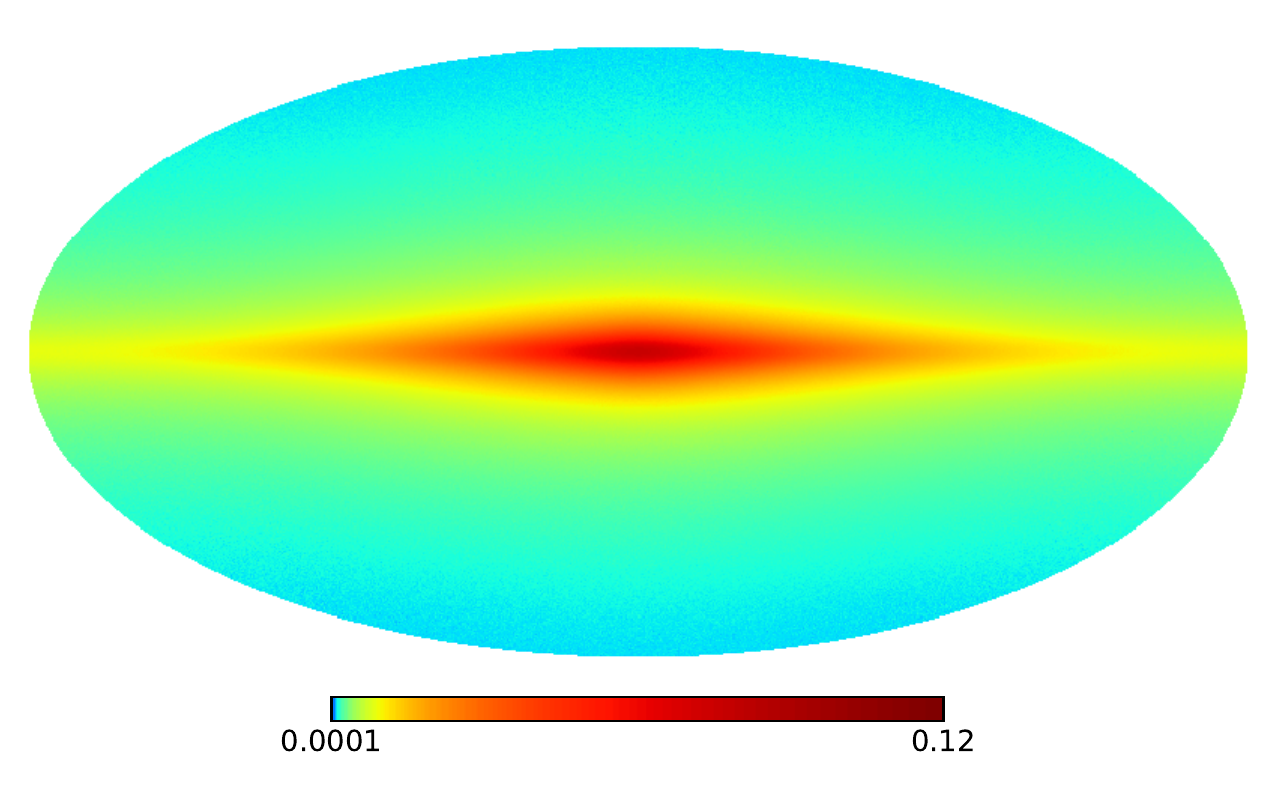}
        &       \includegraphics[width=.3\linewidth]{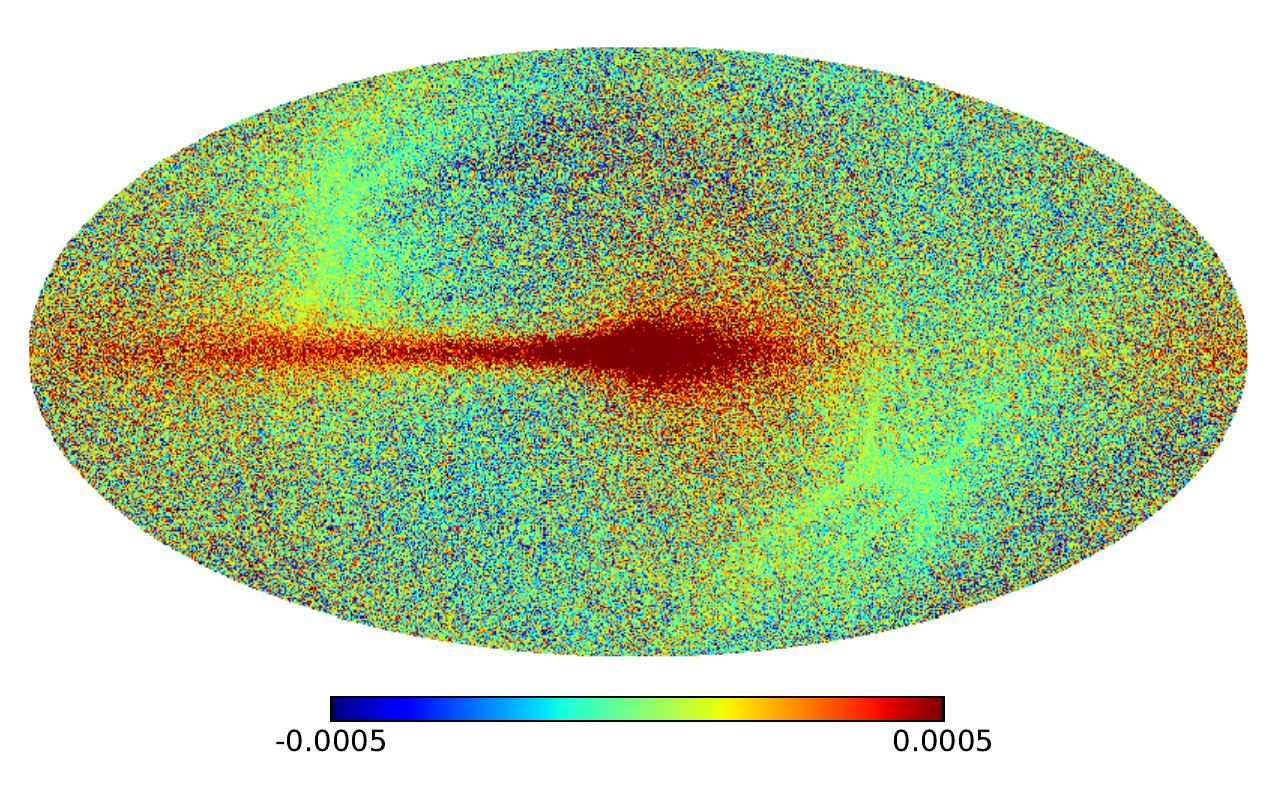}
                &       \includegraphics[width=.3\linewidth]{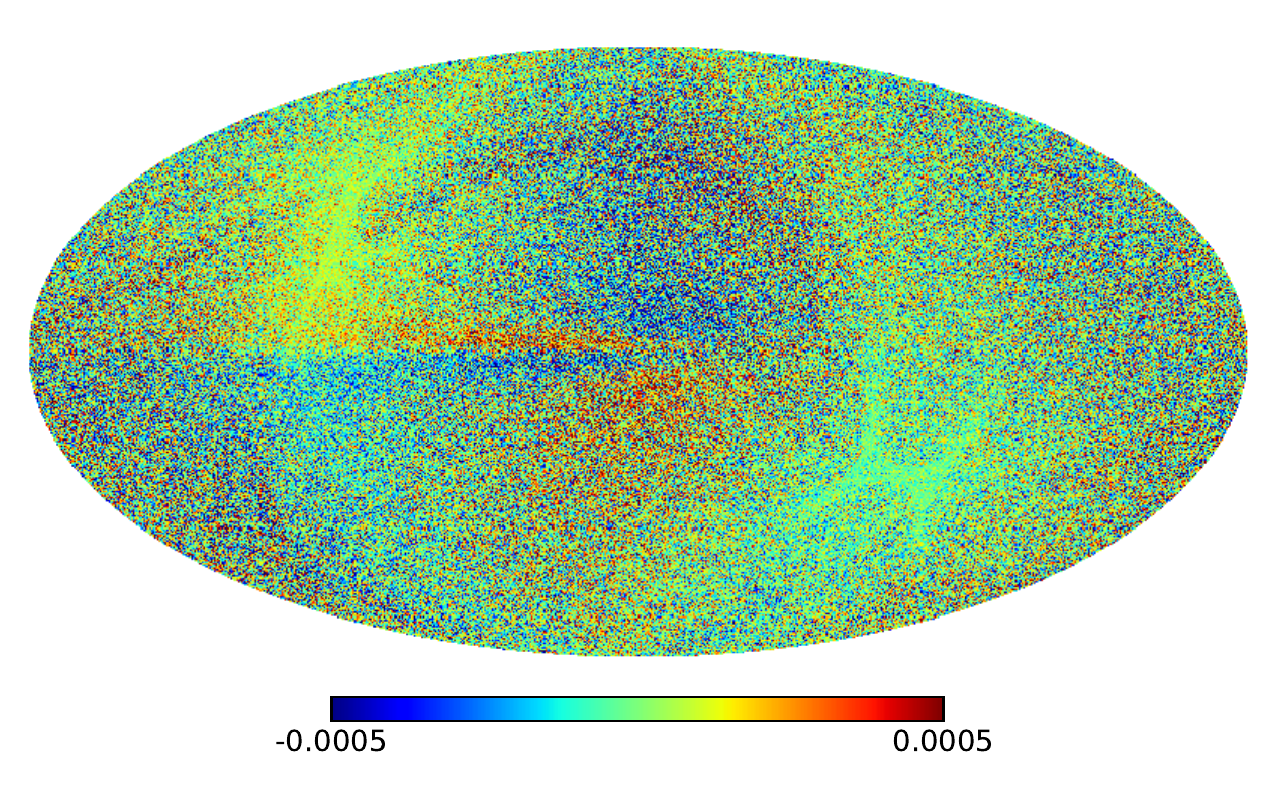} \\
        
\includegraphics[width=.3\linewidth]{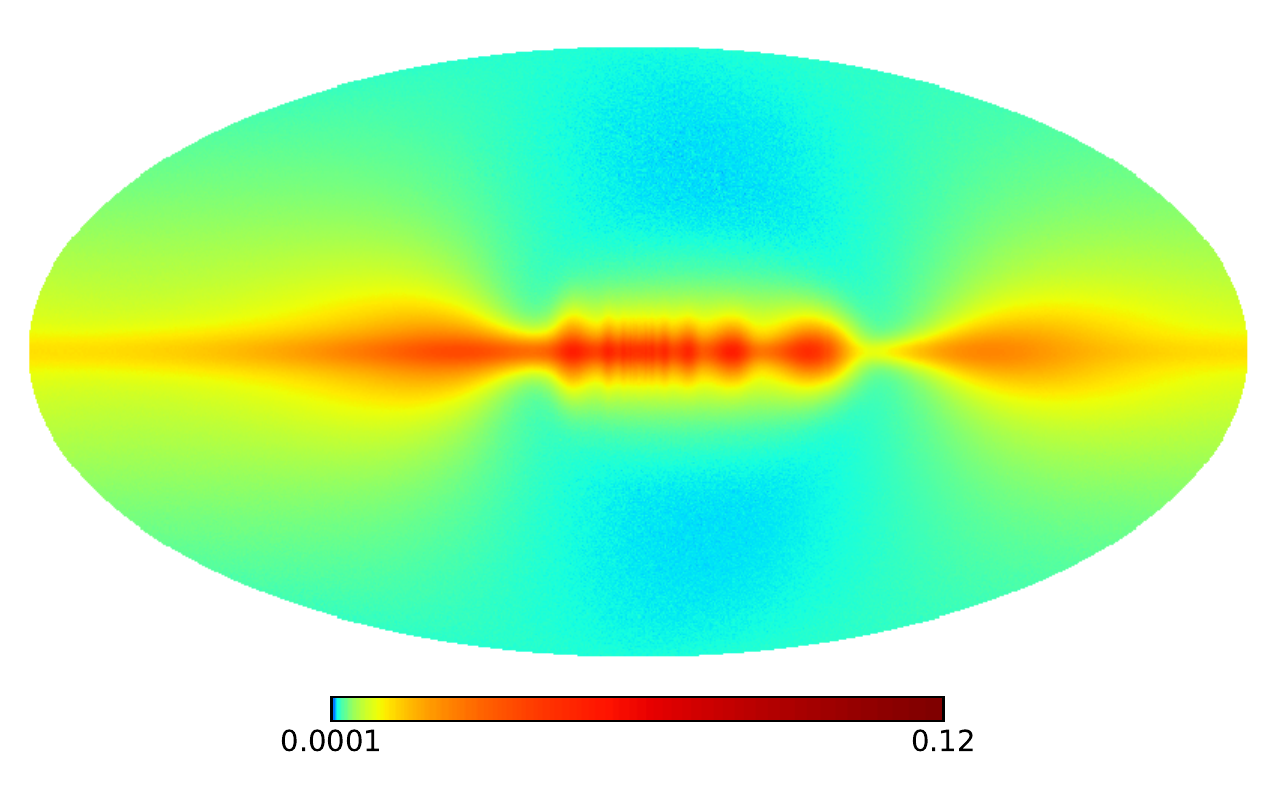}
        &       \includegraphics[width=.3\linewidth]{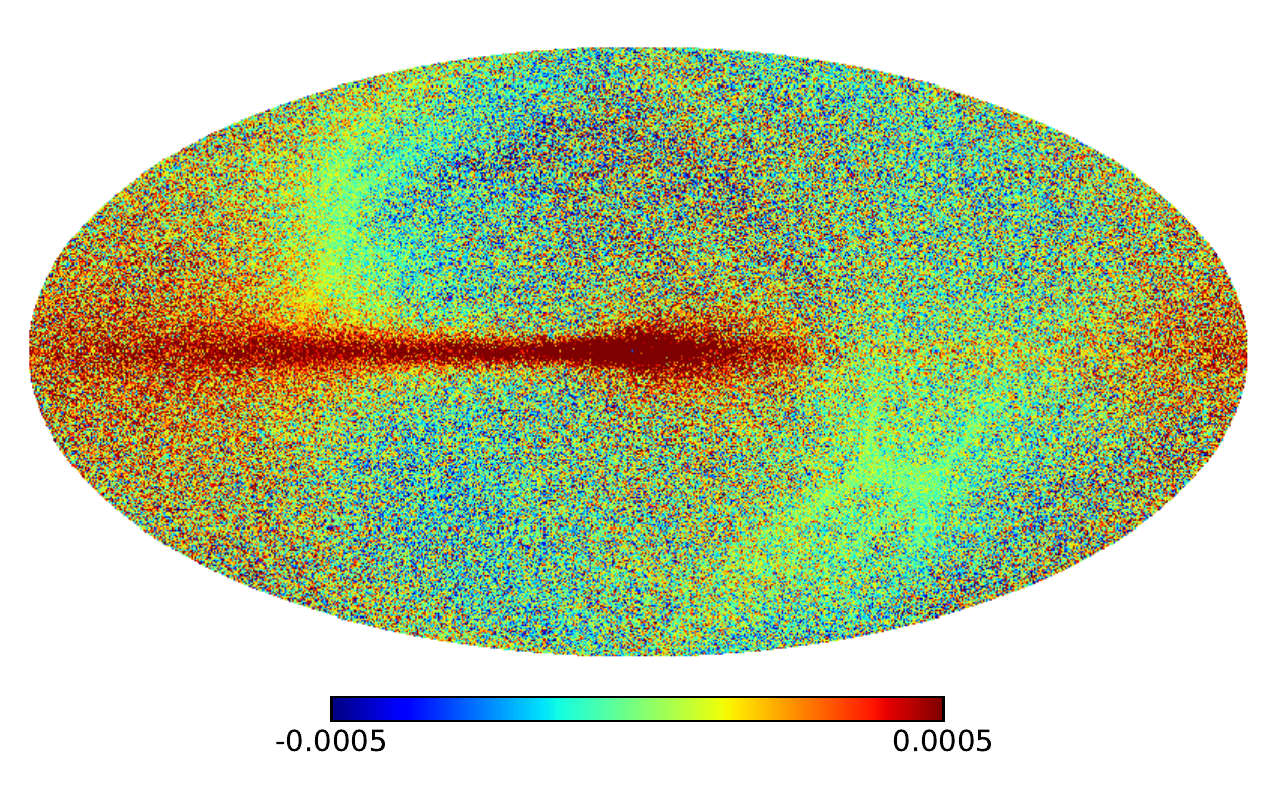}
                &       \includegraphics[width=.3\linewidth]{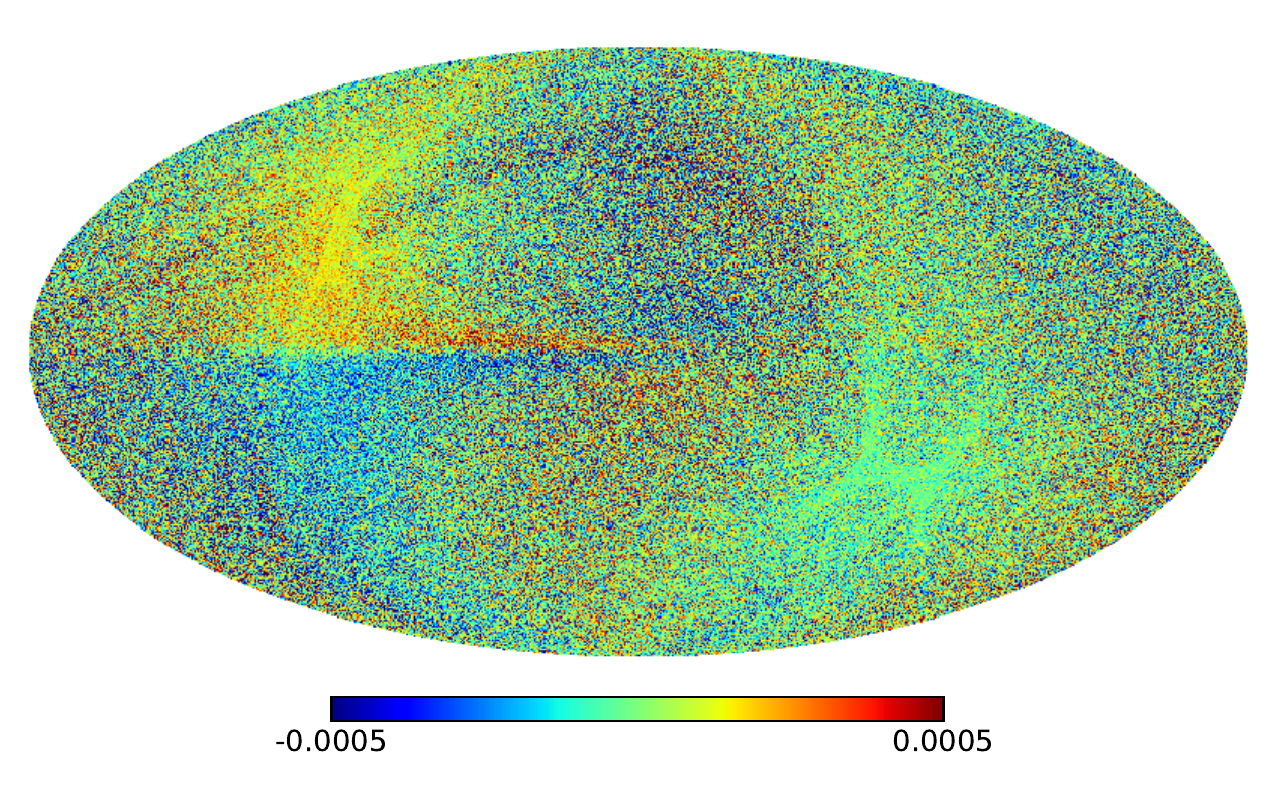} 
\end{tabular}
\caption{High-resolution simulations of the $I$, $Q$, $U$ Stokes parameters
(from left to right) obtained for \texttt{S1} (top) and \texttt{S2} (bottom).
Units are supposedly K$_{\rm{CMB}}$. Intensity map is given in colored-logarithmic scale and polarization maps in linear scale. Full-sky maps are in Galactic coordinates, centered on the Galactic center. Galactic longitude decrease towards the right-hand side of the plots.}
\label{fig:HR_models}
\end{figure*}

\subsection{Implementation}

To simulate the Stokes parameters $I$, $Q,$ and $U$ of the polarized diffuse thermal dust and synchrotron Galactic emission, we follow Eqs.~\ref{eq:DUSTEMISSION} and~\ref{eq:SYNCEMISSION}.
We chose to sample the Galactic space according to a spherical coordinate system centered on the Sun. The radial distance to the observer is linearly sampled and we adopt the equal-area HEALPix tessellation \citep{Gor2005} to  uniformly sample the angular coordinates as this is the most commonly used format when dealing with CMB data.
We thus consider as many spherical shells as radial bins. Then, the matter density distribution and the GMF models are evaluated at each 3D location along the lines of sight. 

In this work, we consider parametric models both for the matter density distribution and for the regular and stochastic (turbulent) GMF components.
We assume that these quantities are constant within the elemental volume and we integrate them according to the midpoint rule.
Our choice of the sampling of the Galactic space and the overall implementation are thus similar to the one adopted in the \texttt{hammurabi} code implementation \citep{Wae2009}, except that we do not consider refining the angular grid as the radial distance increases. At this stage, we do not consider this simplification as a substantial issue. \\

Within the framework of the {\sc Radioforegrounds} project, we have developed \texttt{gpempy}, a self-consistent suite a \textit{Python} codes that follows the above implementations.
It is publicly available\footnote{\url{http://www.radioforegrounds.eu/pages/software/gmf-reconstruction.php}}
where we describe the architecture that is optimized for quick and user-friendly simulations of the polarized sky.
These codes are not optimized for an MCMC exploration of parameter spaces of models.

\subsection{Parametric models}
In this paper, we consider gentle parametric models for the matter density distribution and for the large-scale regular GMF. The models are reviewed in details in Appendix~\ref{sec:AppendixA}.

\begin{table*}[t]
\centering
\caption{Free parameters of the parametric models are given for the two dust density distribution models (ED and ARM4) and for the GMF model (LSA). Column three gives the input values to create the \texttt{S1} and \texttt{S2} realistic simulations, Column four gives the ranges that are explored during the MCMC analysis.
Column five to seven give the best-fit parameter values and the corresponding 1$\sigma$ MCMC uncertainties obtained for the cases A, B, and C, described in Sect.~\ref{sec:Iqu_fit}.}
\label{tab:paramVal}
{\footnotesize{
\begin{tabular}{clll | ccc}
\hline
\hline
\\[-.5ex]
models & parameters     &       input values & explored ranges  & case A        & case B  & case C                \\
\hline
\\[-1.ex]
ED      & $\rho_0$ (kpc)        & $5.0$         & $]0,\, 100]$          & $5.0048 \pm 10^{-4}$    &       $100.000000 \pm 10^{-6}$ & -- -- -- \\
                & $z_0$  (kpc)          & $0.75$        & $]0,\, 10]$           & $0.75135 \pm 10^{-5}$   &       $1.83333 \pm 2\,10^{-5}$ & -- -- -- \\
\\[-1.5ex]
ARM4
                & $p$ ($^\circ$)                                                & $24$                    & $]0,\, 45]$           & -- -- --      & -- -- --      &       $24.0075 \pm 0.0003$                     \\
                & $\phi_{\rm{00}}$ ($^\circ$)   & $-76.868$             & $[-90,\, 360]$  & -- -- --      & -- -- --      &       $-76.934 \pm 0.004$             \\
                & $\rho_c$ (kpc)                                                & $0.7$                   & $]0,\, 20]$           & -- -- --      & -- -- --      &       $0.721 \pm 0.002$                      \\
                & $\sigma_\rho$ (kpc)                           & $8.0$                 & $]1,\, 45]$             & -- -- --      & -- -- --      &       $7.994 \pm 0.001$                          \\
                & $\sigma_z$ (kpc)                                      & $0.7$                   & $]0,\, 8]$                    & -- -- --      & -- -- --        &       $0.701330 \pm 1.5\, 10^{-5}$    \\
                & $\phi_0$ ($^\circ$)                           & $15$                  & $]0,\, 45]$             & -- -- --      & -- -- --      &       $15.0175 \pm 0.0002$                     \\
                & $A_{i = 2, 3, 4}$             & $[1.0,\,1.0\,1.0]$    & $[10^{-3},\, 10^3]$& -- -- -- & -- -- -- &      $[1.0020,\,1.0000,\,0.9990] \pm 10^{-4}$    \\
\\[-1.5ex]
LSA
                & $\psi_0$ ($^\circ$)   &       $ 27.0$ & $]0,\, 55]    $                       &       $26.98 \pm 0.01$       & $25.491 \pm 0.009$            &       $26.980 \pm 0.009$  \\
                & $\psi_1$ ($^\circ$)   &       $ 0.9$  & $]-180,\, 360]        $       &       $0.82 \pm 0.03$       & $1.60 \pm 0.04$               &       $0.84 \pm 0.04$  \\
                & $\chi_0$ ($^\circ$)   &       $ 25.0$ & $]-180,\, 180]        $       &       $24.9 \pm 0.1$        & $27.1 \pm 0.2$                &       $25.1 \pm 0.1$  \\
                & $z_0$ (kpc)                   &       $1.0$   & $]0,\, 20]$                    &       $1.001 \pm 0.006$       & $1.24 \pm 0.01$                       &       $1.007 \pm 0.007$  \\
\\[-.5ex]
\hline
\end{tabular}
}}
\end{table*}

\section{Methodology, data \& simulations}
\label{sec:Methodo}

\subsection{Methodology}
\label{sec:meth}
To efficiently constrain, from the existing data, the large-scale GMF features relying on parametric models, it  would be best to search for a combination of observables that allows us to tackle the lowest levels of complexity at a time.
Based on the modeling of the emission presented
in Eqs.~\ref{eq:DUSTEMISSION} and Eqs.~\ref{eq:SYNCEMISSION},
we observe that the Stokes parameters of the linear polarization result from a non-trivial mixing of the matter density distribution and the geometrical structure of the GMF.
However, we also notice that the thermal dust intensity is to first-order independent from the GMF (see Eq.~\ref{eq:Idust}).
Furthermore, the dust polarization emission is only affected by the geometry of the GMF and not by its intensity. These two facts open up the possibility to constrain the dust density distribution separately from the GMF and to considerably reduce  the number of free parameter in the fitting procedure.
Ideally armed with a good model for the dust density distribution, we can use it to constrain GMF models using polarization data. Here, we consider using the intensity normalized (or reduced) Stokes parameters $q_{\rm{d}} = Q_{\rm{d}}/I_{\rm{d}}$ and $u_{\rm{d}} = U_{\rm{d}}/I_{\rm{d}}$, rather than the Stokes $Q_{\rm{d}}$ and $U_{\rm{d}}$.
This choice is motivated by the fact that the reduced Stokes parameters
$q_{\rm{d}}$ and $u_{\rm{d}}$ may be regarded as the intensity-weighted
mean of the GMF geometry.
Therefore, we argue that the integrated GMF geometry dominates these quantities more than the dust density. For the same reason, we expect that possible biases or mis-modeling of the dust density distribution have less of an effect on the final reconstruction of the GMF geometry (see also our Sect.~\ref{sec:reconstructionGMF}).

\medskip

In the case of the synchrotron emission, it is more difficult to find an approach that separates the matter density distribution from the GMF. Furthermore, unlike the case of thermal dust emission, it is risky to consider the reduced Stokes parameters at low frequency as other Galactic emission components (e.g., from anomalous microwave emission and free-free) are also important in intensity at CMB frequencies.
Thus, a careful separation of these Galactic components is needed to safely compute the reduced synchrotron Stokes parameters
$q_{\rm{s}} = Q_{\rm{s}}/I_{\rm{s}}$ and $u_{\rm{s}} = U_{\rm{s}}/I_{\rm{s}}$.
In contrast, assuming that a
good model of the geometrical structure of the GMF can be obtained from thermal dust emission, this model could then be used to constrain
the relativistic electron density distribution and of the GMF strength through a fit of the synchrotron data.

\subsection{Data}
\label{sec:data}
In this paper, we mainly concentrate on the diffuse polarized thermal dust emission.
We use the \textit{Planck} single-frequency intensity and polarization maps at 353 GHz from the second release (PR2)\footnote{These maps correspond to the dataset selected for the {\sc Radioforegrounds} project. We note that more recent sets of polarization maps have been delivered since the work and analyses presented in this paper were conducted (see, e.g., the third release of Planck data (PR3) \citealt{PlanckIII2018}).} that are available on the Planck Legacy Archive\footnote{http://pla.esac.esa.int/pla/\#maps}.
We refer the reader to \citet{PlanckXIX2015, PlanckXXI2015} for details and discussions regarding these data.
The Planck HFI 353 GHz maps have a native resolution\footnote{https://wiki.cosmos.esa.int/planckpla2015/index.php/HFI\_ \\ performance\_summary} of about 4.94' \citep{PlanckXIX2015} and are given in a HEALPix\footnote{See http://healpix.sourceforge.net}
grid tessellation corresponding to $N_{\rm{side}} = 2048$ \citep{Gor2005}.
At the instrument resolution, the 353-GHz polarization Stokes $Q$ and $U$ maps are noise-dominated \citep{PlanckXIX2015}.
This is particularly true at high Galactic latitudes.
The dispersion arising from CMB polarization anisotropies is much lower than the instrumental noise for $Q$ and $U$ \citep{PlanckVI2014}. Thus, its impact on our analysis is expected to be negligible. The cosmic infrared background (CIB) is considered to be unpolarized \citep{PlanckXXX2014}.
At this frequency, subdominant contributions are expected in the intensity map either from the CMB, Galactic, and extragalactic point sources, the CIB and the zodiacal light (e.g., \citealt{PlanckXXI2015}). We note that we  carry out our work at low resolution so that contributions from point sources and the CIB are not expected to be significant. The CMB temperature anisotropies and zodiacal light contributions are expected to be subdominant with respect to the thermal dust emission in intensity. To consolidate the results of our analysis for the intensity part, we also make use of the full-sky dust extinction map derived in \cite{PlanckXI2014}, specifically, the map of the dust optical depth ($\tau_{\rm{353}}$). This quantity is related to the dust column density integrated along the line of sight. It has been shown to be a good tracer of the dust column density at high Galactic latitudes.
Deviation occurs in denser molecular region and, thus, towards regions of the Galactic plane  \citep{PlanckXI2014}.

\subsection{Simulated maps}
\label{sec:simu}
For the simulated data used in the following, we consider two dust density distribution models: the regular {\it Logarithmic Spiral Arm} (LSA) GMF model and the (random) turbulent GMF model, both presented in Appendix~\ref{sec:AppendixA}.
Using these parametric models and realistic {\it Planck} noise maps, we produced different sets of high-resolution maps that simulate the \textit{Planck} 353-GHz linear polarization Stokes parameter maps. These sets, hereafter called \texttt{S1}, \texttt{S2,} and \texttt{S2-turb} are defined as follows.

First, \texttt{S1} is built by combining the exponential-disk (ED) dust density distribution model with the LSA GMF model, with no turbulent component.
Second, \texttt{S2} is built by combining the four-spiral-arms dust density model (ARM4) with the LSA GMF model; no turbulent component. 
And finally, \texttt{S2-turb} is the same as \texttt{S2} but including different degree of turbulence in the GMF ($A_{\rm{turb}} = 0,\,0.3,\,0.9$) (see Sect.~\ref{sec:qualityrGMFturbana}).

The maps are shown in Fig.~\ref{fig:HR_models} for $N_{\rm{side}} = 2048$. 
We give the input values of the regular-model parameters in Table~\ref{tab:paramVal}.
An illustration of the input GMF is given in Fig.~\ref{fig:inputWMAP_xyz}.

To construct these simulated maps, we adopted a sampling of the Galactic space defined by $N_{\rm{side}} = 1024$ and a radial step of 0.2 kpc.\footnote{Using the \texttt{gpempy} software this realization takes up to 300 gigabyte of the RAM before the line-of-sight integration performed.}
We then upgraded the maps at $N_{\rm{side}} = 2048$ to directly calibrate the Galactic thermal dust emission of our simulated maps to the measured one at 353 GHz by \textit{Planck} through a linear fit.
To add realistic \textit{Planck} noise, we consider the one derived from the difference between the Stokes parameter maps obtained from the data corresponding to the first and the second half mission of the \textit{Planck} individual pointings. \\

In summary, the simulated \textit{Planck} maps, $M,$ of the polarized thermal dust emission at 353-GHz are given by:
\begin{equation}
M_X =  {\rm{A_{cal}}} S_X^{\rm{dust}} + N_X^{\rm{Planck}} ,
\end{equation}
where $X$ refers either to ${I_{\rm{d}},\,Q_{\rm{d}},\,U_{\rm{d}}}$, $N$ is the noise map, $S$ is computed using \texttt{gpempy} to estimate Eqs.~\ref{eq:DUSTEMISSION} and~\ref{eq:Idust}, and
$\rm{A_{cal}}$ is the global calibration factor that is computed by a linear fit to the \textit{Planck} data independently for the intensity and the polarization data. The same $\rm{A_{cal}}$ is applied to both $Q_{\rm{d}}$ and $U_{\rm{d}}$ to preserve the polarization position angle.
In the remainder of this paper, we drop the subscript $_{\rm{d}}$ as we are only dealing with the thermal dust emission, apart from Sect.~\ref{sec:synchrotronproof}.

\section{Fitting procedure in intensity and polarization for the Galactic thermal dust emission}
\label{sec:GMFrecons}
In this section, we describe the fitting procedure that we use to infer large-scale GMF features using the dust Galactic thermal dust emission. The fit to the (simulated) data is performed in two steps. First, we fit for the intensity in order to recover the dust density distribution.
Second, we use the best-fit model of the dust density distribution and fit for the polarization maps ($q,\,u$) to recover the GMF.
We rely on MCMC technique to recover both models of the dust density distribution and of the GMF.

\subsection{Choice of fitted-map resolution}
\label{subsec:resolbias}
Although we have optimized our codes, producing simulated maps of the diffuse Galactic emission at the instrument resolution is extremely time consuming.
To effectively adjust models to data and efficiently explore the different parameter spaces, a large number of simulations is needed.
Inevitably we thus have to work at a lower resolution than the native one of the data.
In Appendix~\ref{sec:resolutionBias} we explore in detail the consequences of this requirement both at the map level and in term of the model-parameter posterior distribution reconstruction. We demonstrate that the necessity to work at lower resolution induces a bias in the parameter space. However, this bias is currently, and for many practical cases, not the dominant source of uncertainty in the kind of modeling tackled in this study. Nonetheless, it is best to ensure that this source of uncertainty is sub-dominant by working at the highest resolution made possible by the computing facilities.

Inside the  fitting procedure, we chose to implement the model maps at $N_{\rm{side}} = 64$ and with a radial bin of 0.2 kpc.
For the models we consider in this paper, the computation time of a full set of Stokes parameters maps is always on the order of few tenths of a second. The requirement to run a MCMC-based exploration of the parameter spaces are thus fulfilled.

\subsection{Likelihood definition}
\label{sec:likelihood}
The best-fit parameters are obtained by maximizing a Gaussian likelihood function $\mathcal{L}$, defined as
\begin{equation}
\ln \mathcal{L}  = -\frac{1}{2} \ \left( D - \alpha M \right) \  C^{-1}_{D} \ \left( D - \alpha M \right)^{T}
\label{eq:chi2}
,\end{equation}
where $D$ and $M$ represent either the intensity, $I$, or the concatenated
reduced polarization parameters $(q,\,u)$ for all pixels in the data (or simulated data) and the
model maps, respectively;
$C_{D}$ is the covariance matrix associated to the uncertainties in the
data and $\alpha$ is an overall normalization factor. It is estimated for
each model, and independently for I and (q,u), at each MCMC step via a simple linear fit:
$\alpha = (\sum_i (D_i \, M_i/\sigma^2_i))/(\sum_i (M_i^2/\sigma^2_i))$.

\subsection{Estimation of the covariance matrix}
\label{sec:uncertainties}
The choice of the exact form for the noise covariance matrix is not trivial when we think about the properties of the \textit{Planck} data at 353~GHz.
In the case of the intensity we are in a highly signal-dominated case over most of the sky and there is significant intrinsic dispersion in the signal with respect to the low degree of complexity allowed by the models currently used in the literature. With regard to polarization, the situation is slightly more complex as we are mainly in a signal-dominated case close to the Galactic equator, while at high Galactic latitudes we are mainly in a noise-dominated regime. To account for these peculiarities, we developed a hybrid approach considering both statistical instrumental uncertainties and intrinsic signal dispersion.

\smallskip

In the case of the instrumental 
uncertainties in the Stokes parameters, $\{I,Q,U \}$, we consider the block diagonal per-pixel covariance matrix maps released by the \textit{Planck} collaboration at $N_{\rm{side}} = 2048$. We neglect the off-diagonal terms\footnote{To cross-check this working simplification, we evaluated the full noise covariance matrix for the reduced Stokes parameters $q$ and $u$, relying on MC simulations. We find that they distribute around zero and that for 90 percent of the pixels we have $\frac{|C_{qu}|}{ C_{qq}} \leq 16.4\%$ and $\frac{|C_{qu}|}{ C_{uu}} \leq 16.3\%$.}
and propagate the uncertainties at low resolution according to
\begin{equation}
\sigma^2_{D_{64};\rm{stat},i} = \frac{1}{N_{[2048,64]}^2} \sum_{j \in i} \sigma^2_{D_{2048};\rm{stat},j} \;,
\end{equation}
where $\sigma^2_{D_{N_{\rm{side}}};\rm{stat},i}$ is the noise variance for a pixel $i$ in the data maps at the resolution given by $N_{\rm{side}}$; and, $N_{[2048,64]}$ is the number of pixels in a $N_{\rm{side}} = 2048$ map corresponding to a given pixel in a $N_{\rm{side}} = 64$ map.
Furthermore, a first estimate of the signal intrinsic dispersion can be obtained using the normalized dispersion of the data at high resolution ($N_{\rm{side}} = 2048$) corresponding to pixel, $i$, of the low resolution ($N_{\rm{side}} = 64$) map, as:
\begin{equation}
\sigma^2_{D_{64};\rm{disp},i} = \frac{1}{N_{[2048,64]}^2} \sum_{j \in i} \left( D_{2048;j}  -  D_{64;i} \right)^2
,\end{equation}
where $D=\{I,Q,U\}$\footnote{Notice that, we do not consider the mixing of $Q$ and $U$ due to parallel transport when averaging over pixels.}.
Let us note that in the synchrotron literature (e.g. \citealt{Jan2012a,Jan2012b}), the signal dispersion is often taken as simply the dispersion of the data in low-resolution pixels.
The two estimates of the variance due to signal dispersion differ by a factor of $N_{[2048,64]} = 1024$. We adopt the normalized dispersion, also referred to as the standard error on the mean, in order to obtained a reduced $\chi^2$ on the order of one when the data are fitted with the appropriate models (see Table~\ref{tab:chi2values}) and when the uncertainties are dominated by instrumental noise.
The drawback of this choice, however, is that the reduced $\chi^2$ values rapidly increases as soon as the models do not fully correspond to the fitted data. \\

Finally, we can define "pseudo" uncertainties in the low resolution $\{I,Q,U\}$ maps, which
would account for the intrinsic signal dispersion and the noise, as:
\begin{equation}
\sigma_{D_{64}} = \max \left(\sigma_{\rm{stat}},\, \sigma_{\rm{disp}} \right) \;.
\label{eq:error_max}
\end{equation}

\noindent These pseudo uncertainties are then propagated to the reduced Stokes parameters, $q$ and $u$ as:
\begin{equation}
\sigma^2_{{x}} = \frac{1}{I^2} \sigma^2_{{X}} + \frac{X^2}{I^4} \sigma^2_{{I}}
\label{eq:sig_pol}
,\end{equation}
where $x = X/I$ and $X=\{Q,U\}$. And so, we write the covariance matrix for $\{ q,\, u \}$ as $C_{xx'} = \rm{diag} \{\sigma^2_{x} \}$.

\begin{figure}
\centering
\begin{tabular}{c}
\includegraphics[trim={.0cm 0cm .0cm 0.0cm},clip,width=.99\linewidth]{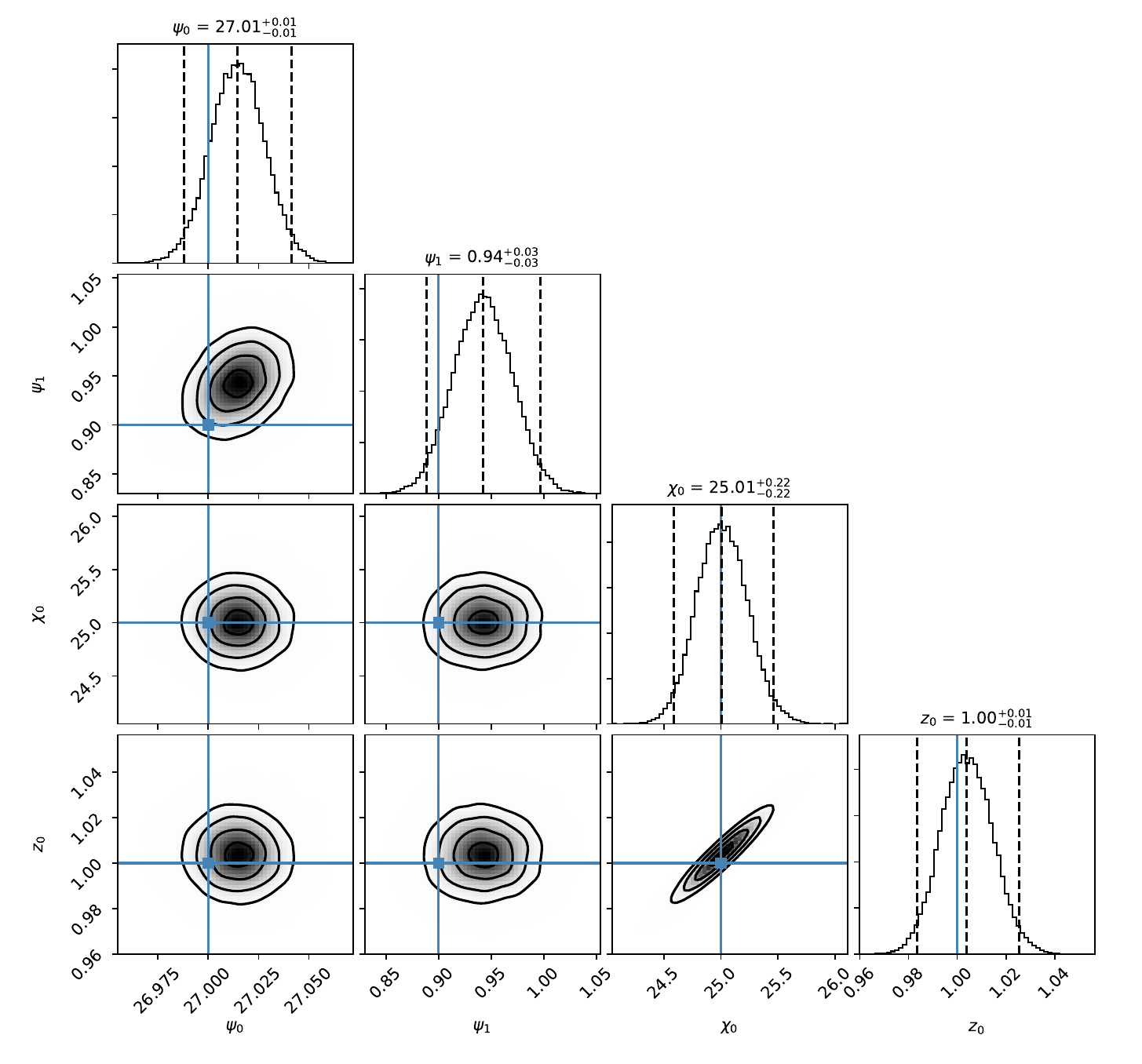} \\
\end{tabular}
\caption{
Corner plot showing the marginalized 1D and 2D
posterior probability distributions for the parameters of the regular LSA GMF model in the case where the simulated data and the fit are performed at $N_{\rm{side}}=64$, see Sect.~\ref{sec:mcmcalgoval}.
The vertical and horizontal light-blue lines mark the input parameter values. Angles are given in degrees and the scale height ($z_0$) in kpc.
\label{fig:qu64GMF_corner}}
\end{figure}

\subsection{Implementation of the MCMC experiment}
In order to explore the parameter space, determine the posterior distributions of the parameters, and find their best-fit values,
we used the \texttt{emcee} MCMC Python software written by \cite{For2013}, who implemented the Affine-Invariant sampler proposed by \cite{Goo2010}.
We run the MCMC code for several Markov chains until the convergence criteria proposed by \cite{Gel1992} is fulfilled for all the model parameters with a threshold value of 1.03.
We tested for the convergence every 100 MCMC steps.
Due to the complex nature of the problem at hand, local minima can be encountered by some of the chains. The reason is that the $\chi^2$ hyper-surface may exhibit multiple minima that might be sharp in the case of thermal dust emission (mainly for dust intensity).
To remedy this problem, we stopped the MCMC after several thousand steps and relaunched the experiment from the location of the minimum $\chi^2$ and waited for the convergence criteria to be fulfilled to reach well-sampled posterior distributions.

We used 250 Markov walkers for the intensity fits and 100 for the polarization fits. The Markov chains are initialized according to uniform distributions.
For the exploration of the parameter space, we considered non-informative priors, adopting top-hat distributions. The explored ranges of values of the free parameters are given in Table~\ref{tab:paramVal} for the different fitted models. The results of different adjustments are presented in the next sections for the simulated and true data sets. 
In the more evolved cases, it was necessary to run the MCMC for several thousands of steps. This implied the generation of millions of magnetized dusty Milky Ways.

We optimized the codes such that a MCMC realization of a map is always a fraction of a second at $N_{\rm{side}} = 64$. We are limited by the efficiency of the basic Python functions to compute the functional forms of the models, such as the hyperbolic cosine. The computing time required for such a converged MCMC fit to be reached is at the day scale running on twelve CPU cores.

\begin{figure*}
\begin{tabular}{ccc}
\includegraphics[width=.3\linewidth]{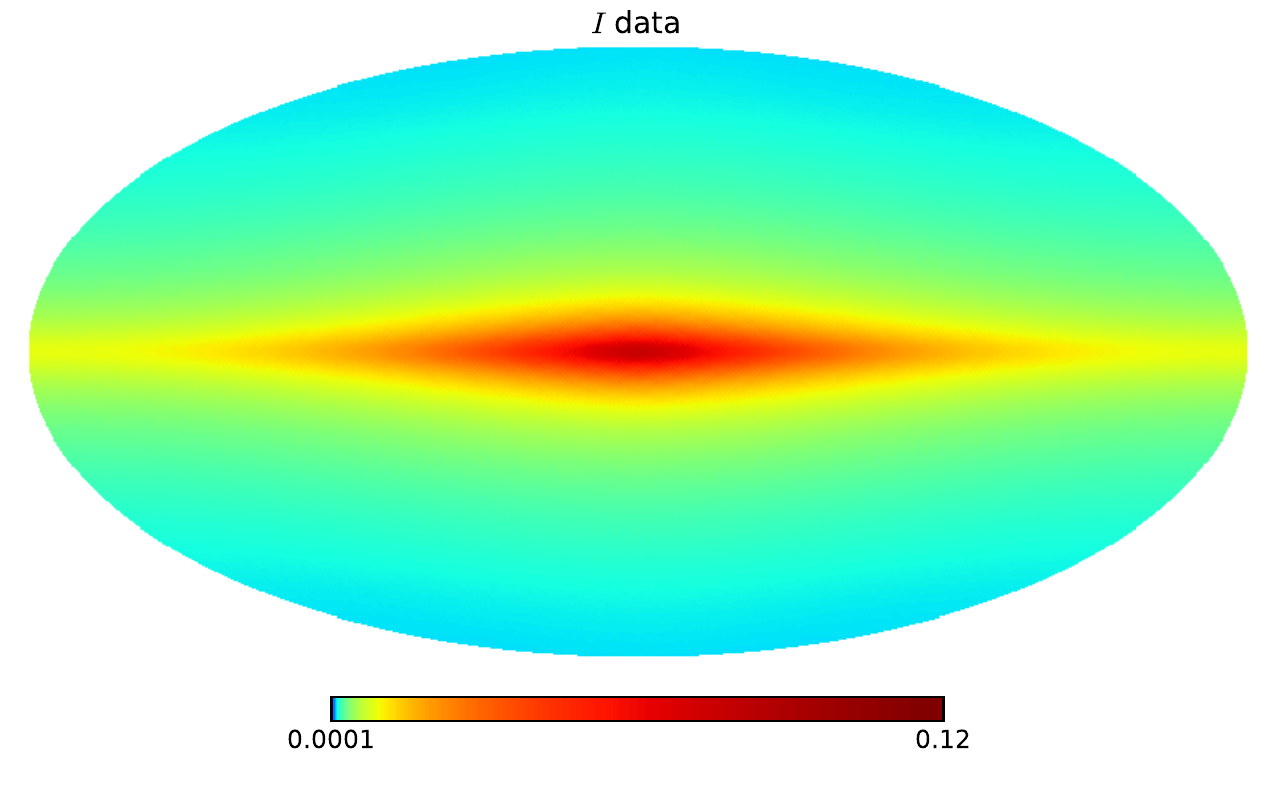}
        &       \includegraphics[width=.3\linewidth]{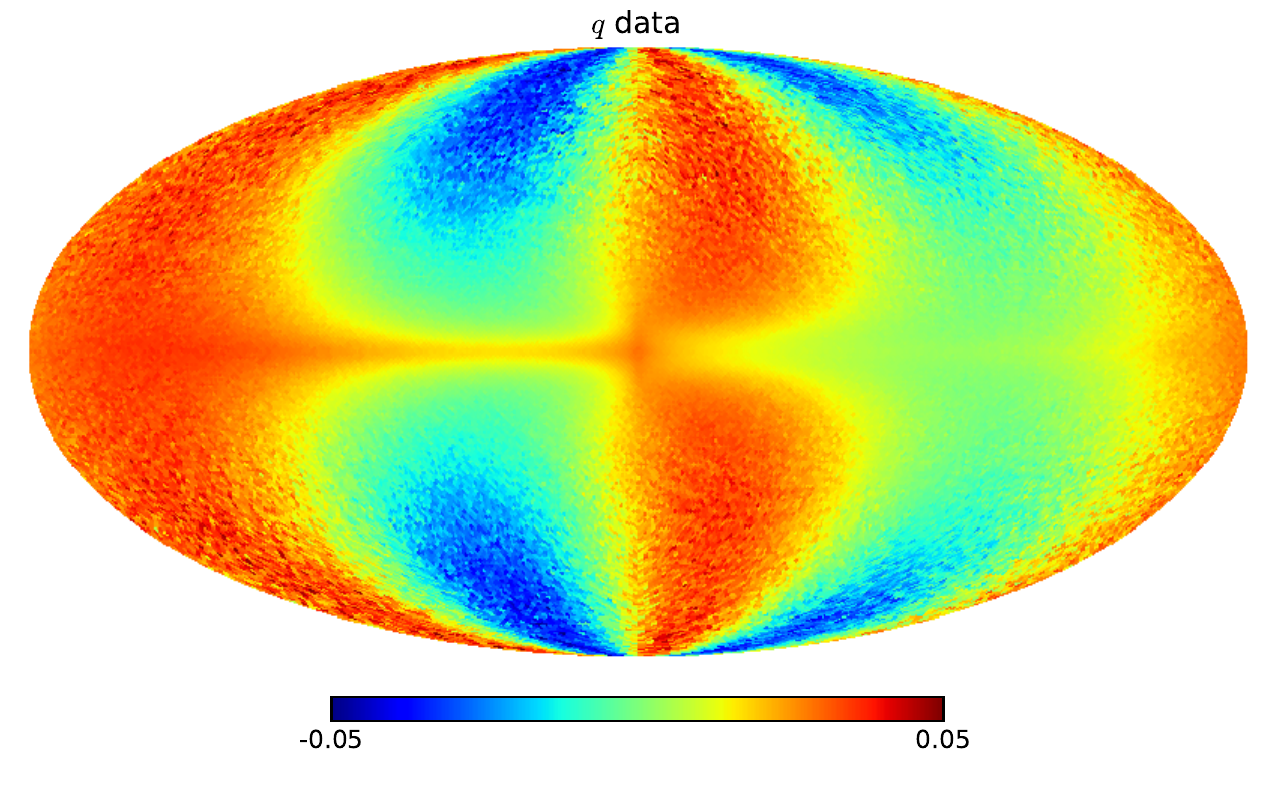} 
                &       \includegraphics[width=.3\linewidth]{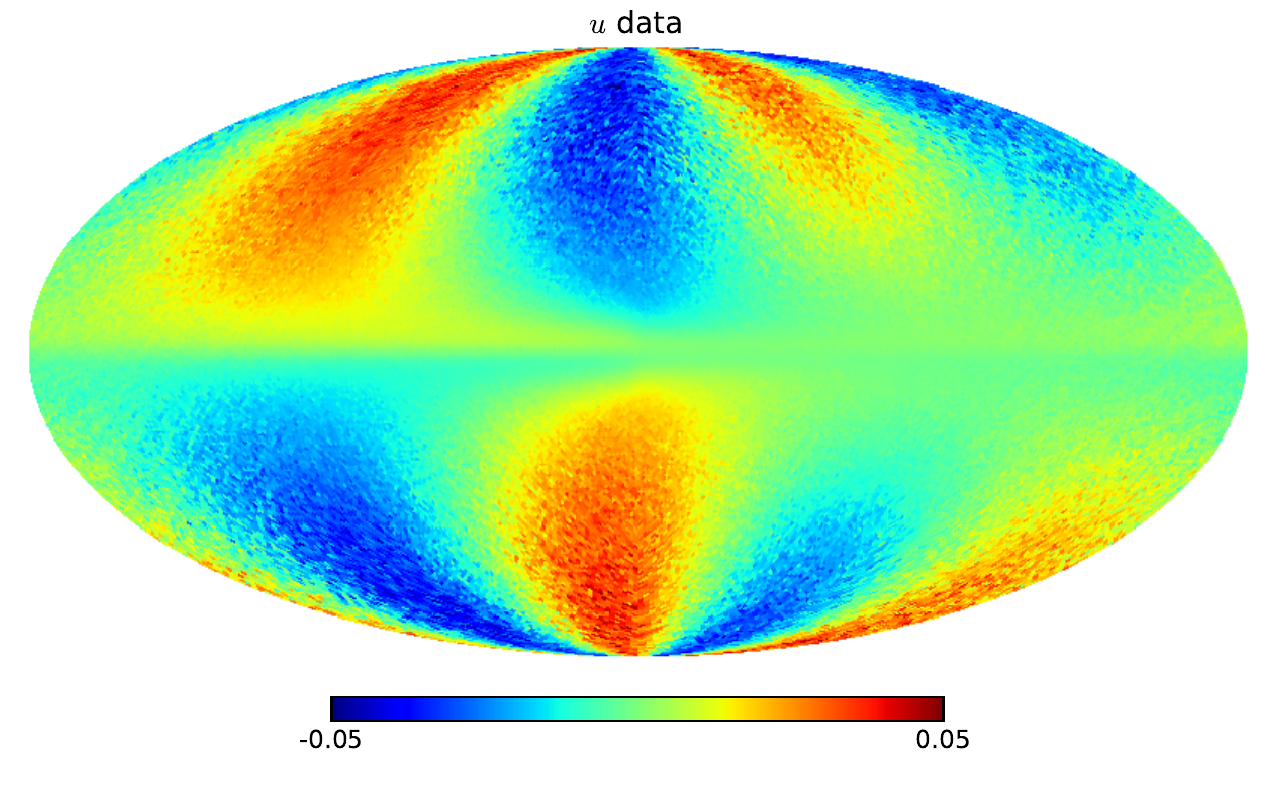} \\
\hline
\includegraphics[width=.3\linewidth]{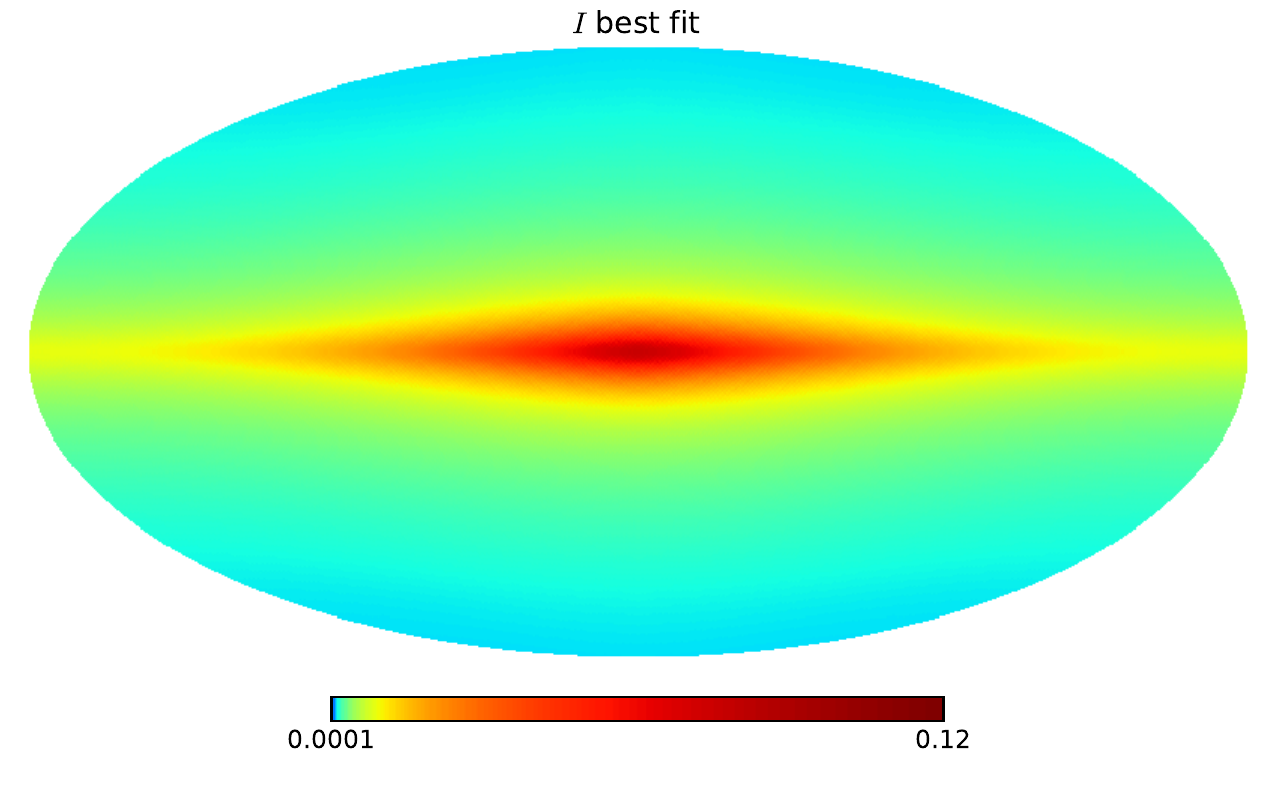}
        &       \includegraphics[width=.3\linewidth]{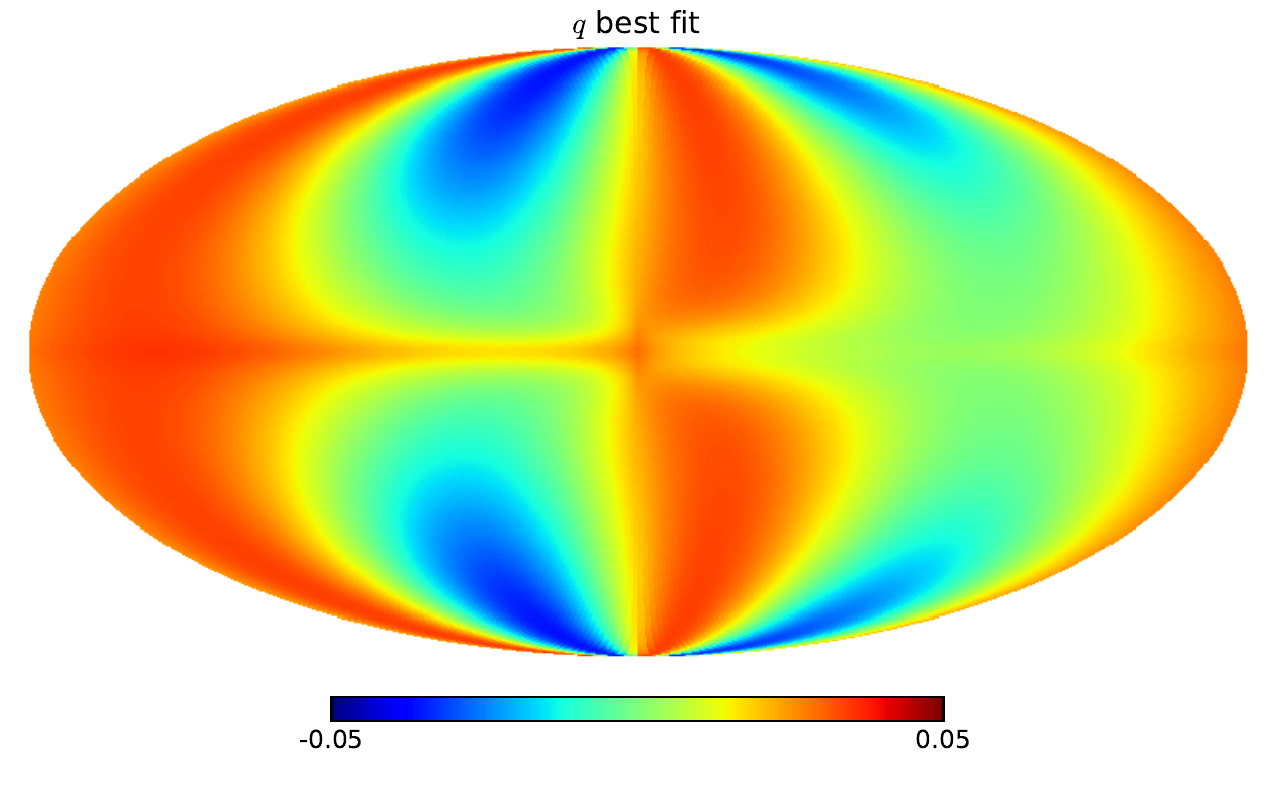} 
                &       \includegraphics[width=.3\linewidth]{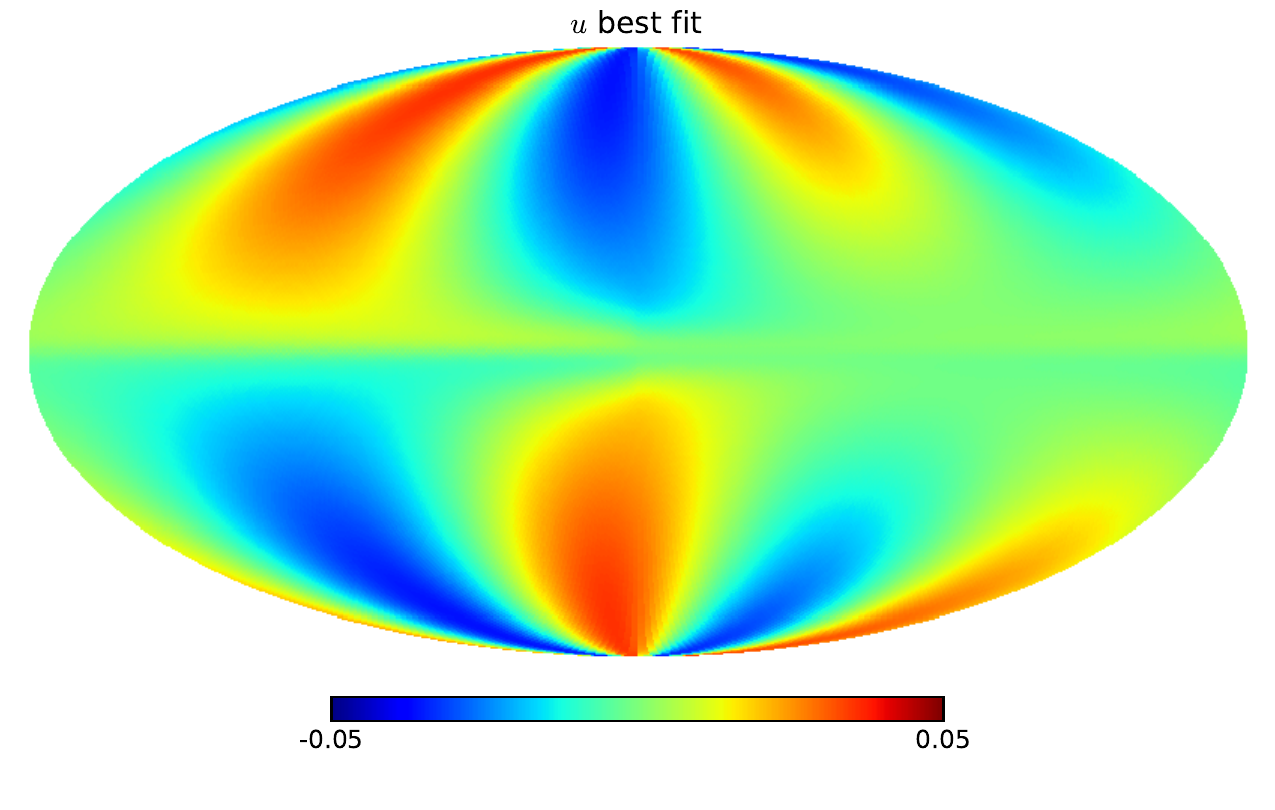} \\
\hline
\includegraphics[width=.3\linewidth]{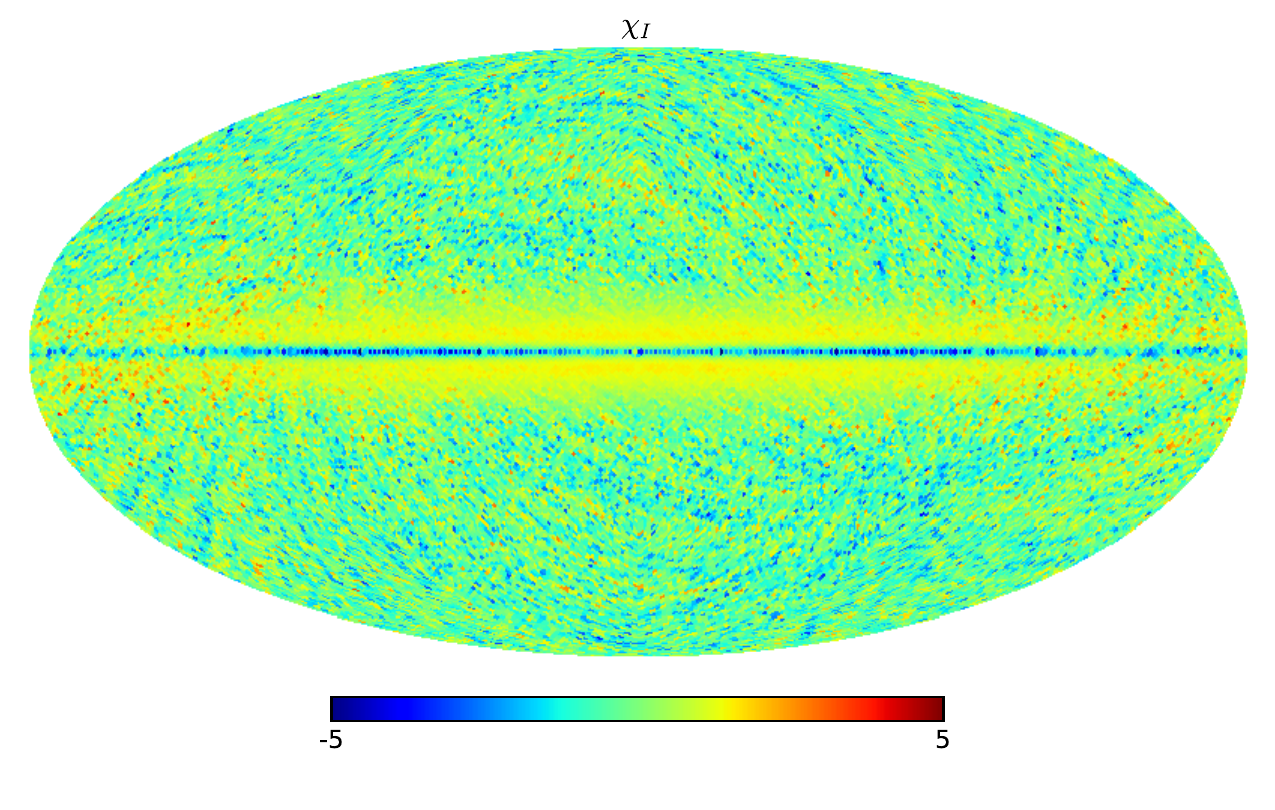}
        &       \includegraphics[width=.3\linewidth]{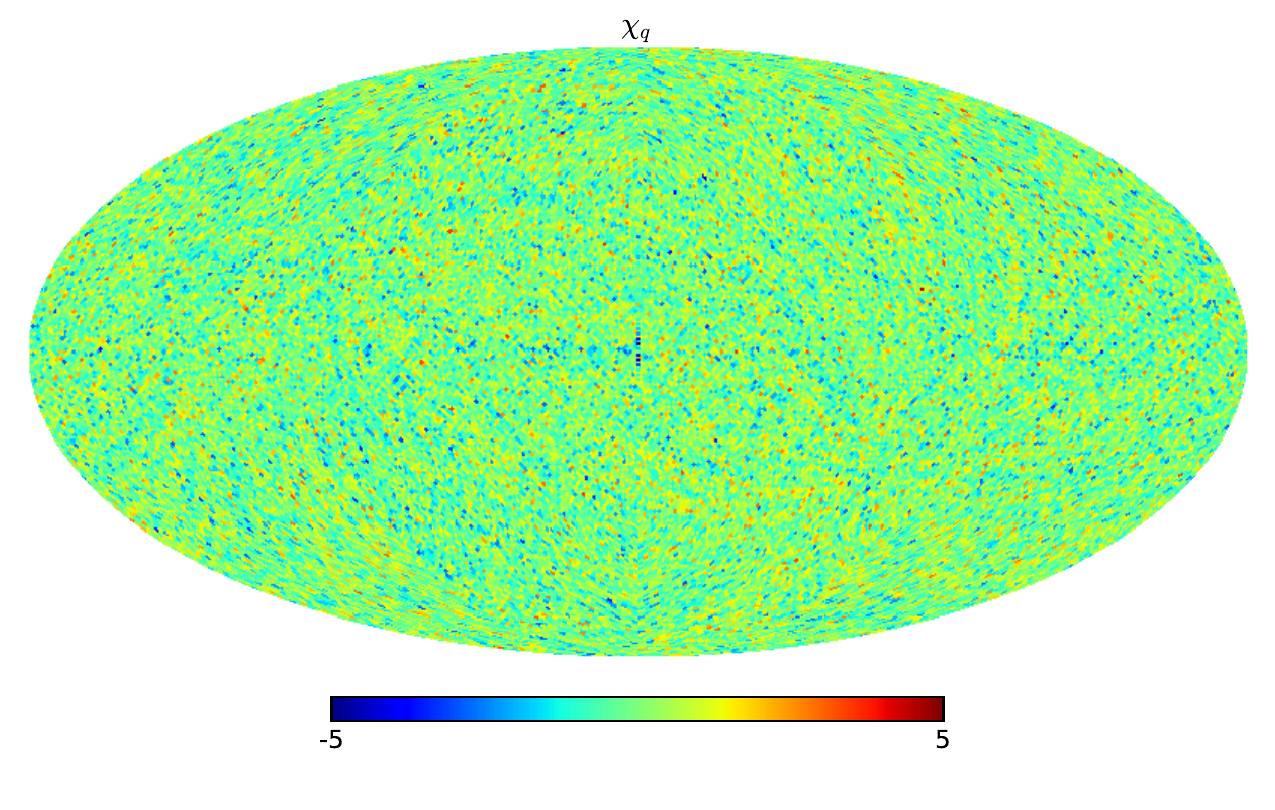} 
                &       \includegraphics[width=.3\linewidth]{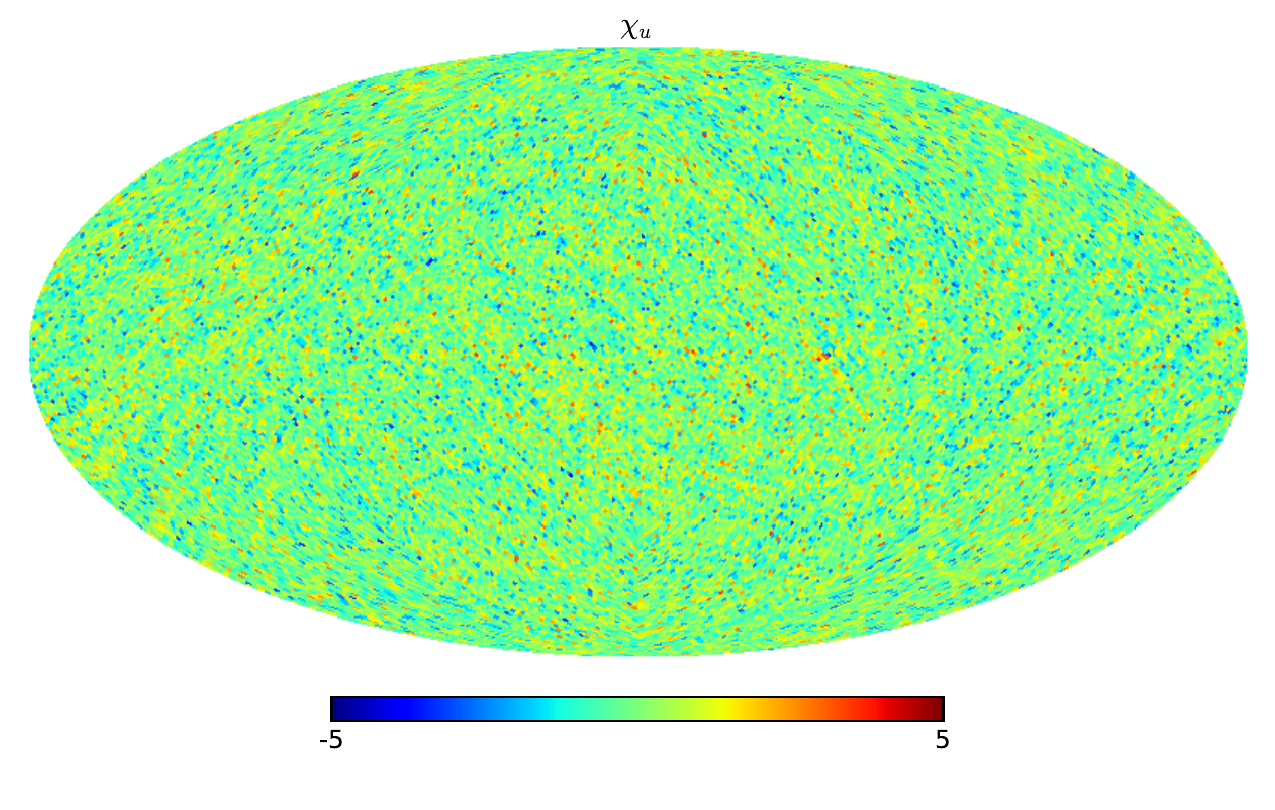} \\
\end{tabular}
\caption{shown from left to right.\ 
(\textit{Top}) Data from the \texttt{S1} simulation downgraded to $N_{\rm{side}} = 64$,
(\textit{middle}) best-fit maps obtained while assuming the dust distribution follows an ED model and the GMF the LSA model (Case A in the text),
and (\textit{bottom}) significance of the residuals. The color scale of the last row ranges from -5 to 5.}
\label{fig:model1_fits}
\end{figure*}

\begin{figure*}
\begin{tabular}{ccc}
\includegraphics[width=.3\linewidth]{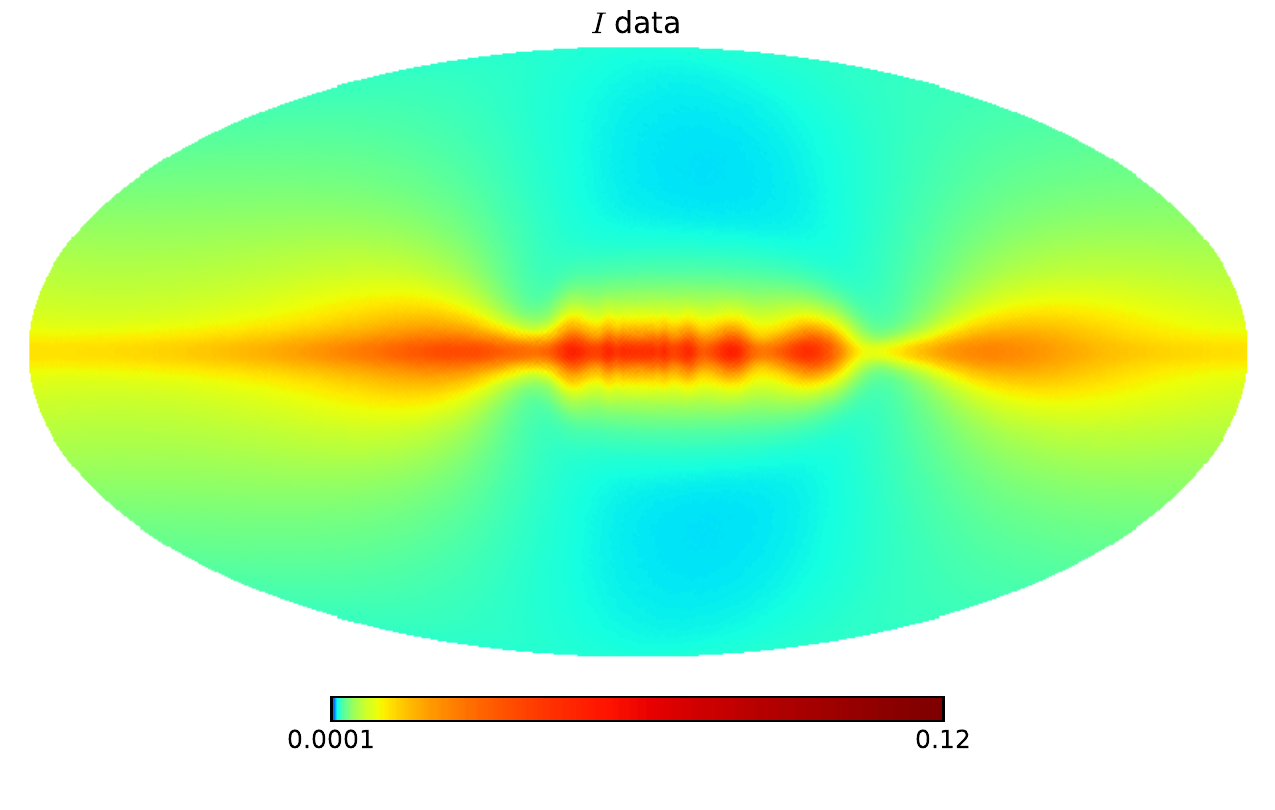}
        &       \includegraphics[width=.3\linewidth]{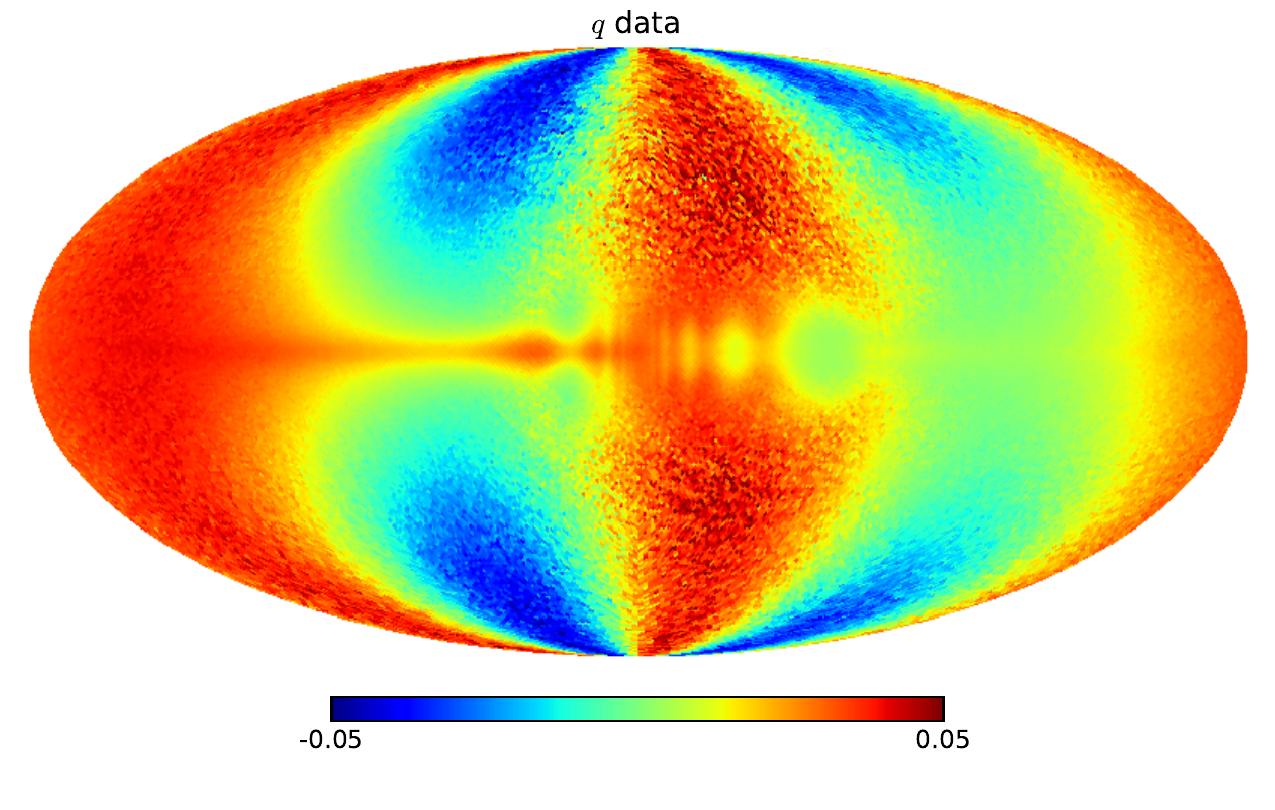} 
                &       \includegraphics[width=.3\linewidth]{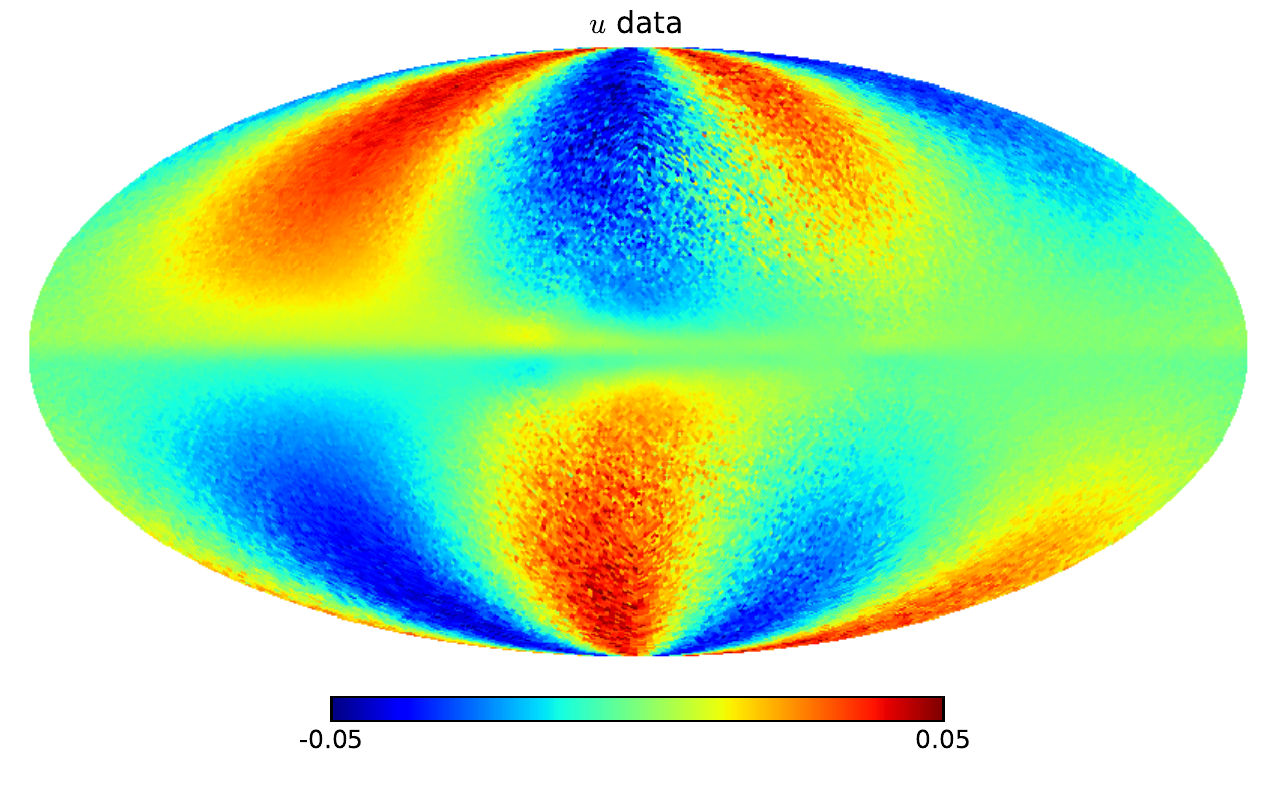} \\
\hline
\includegraphics[width=.3\linewidth]{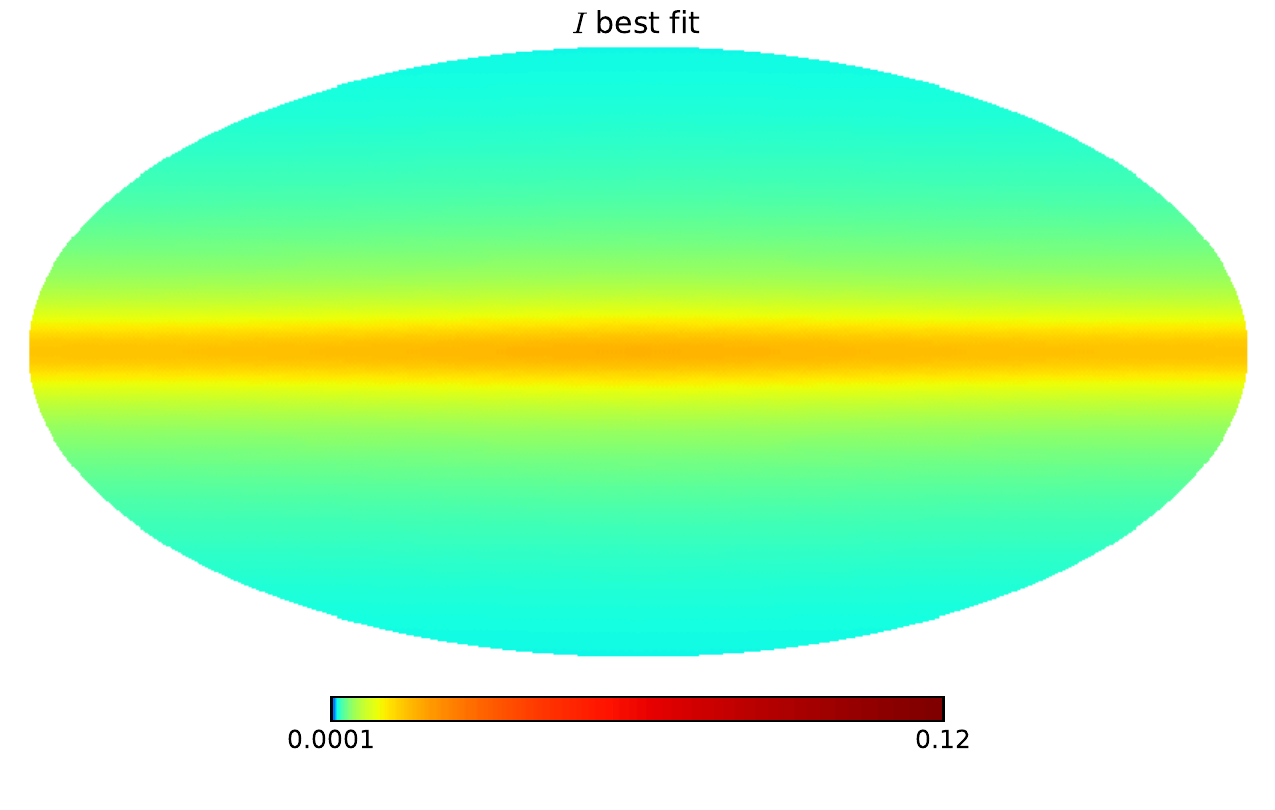}
        &       \includegraphics[width=.3\linewidth]{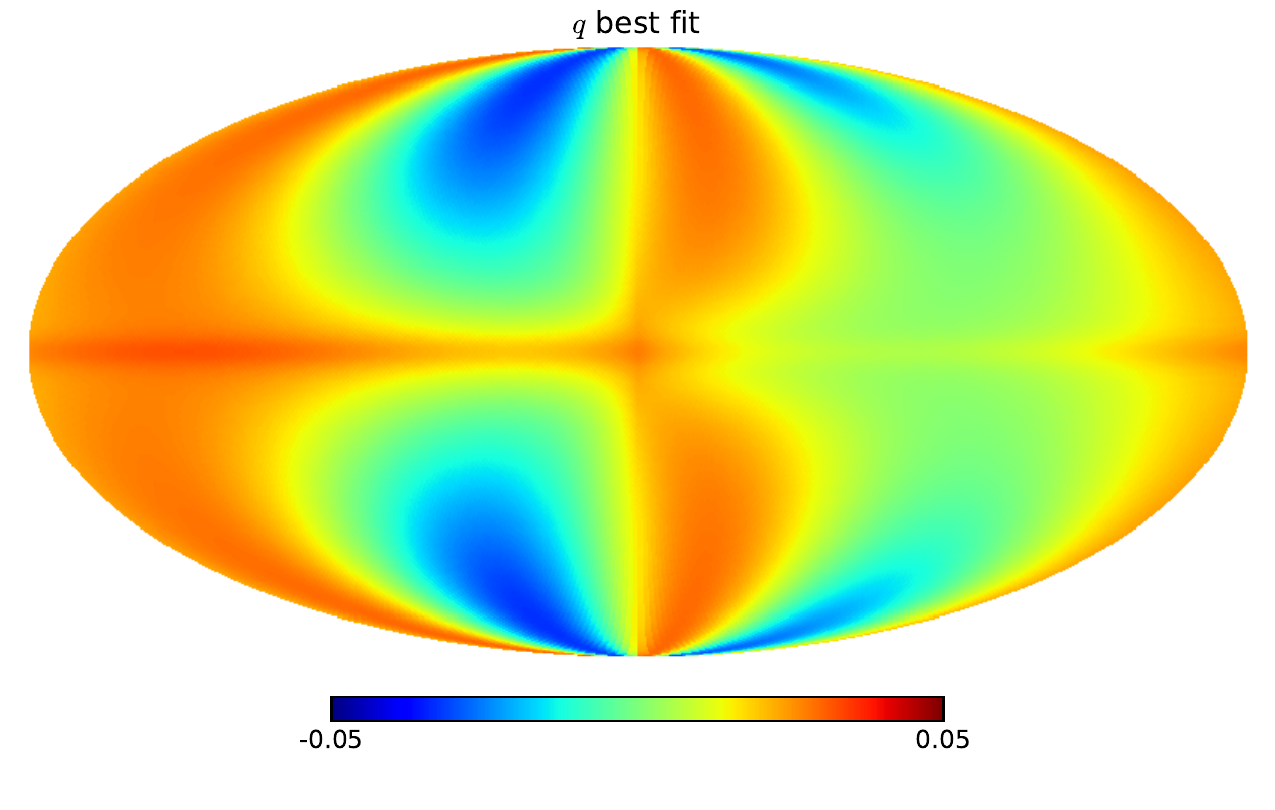} 
                &       \includegraphics[width=.3\linewidth]{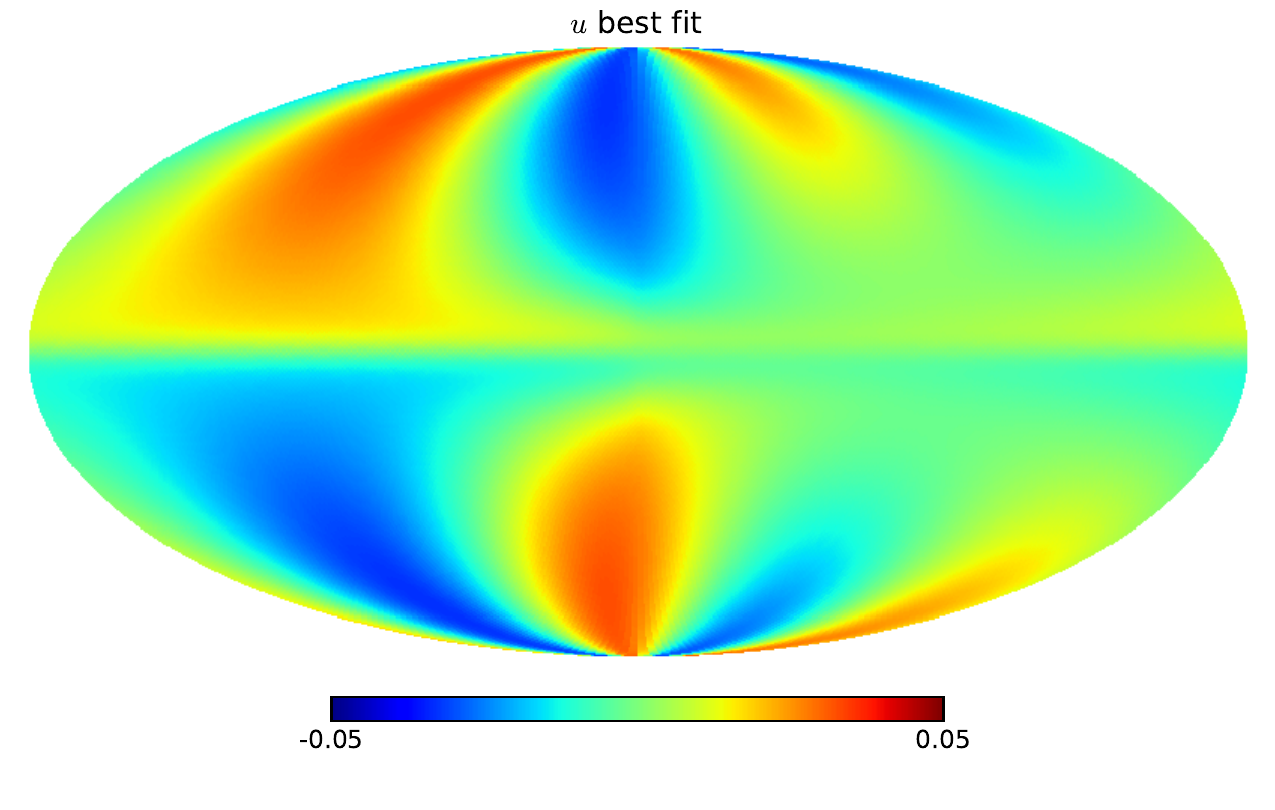} \\
\includegraphics[width=.3\linewidth]{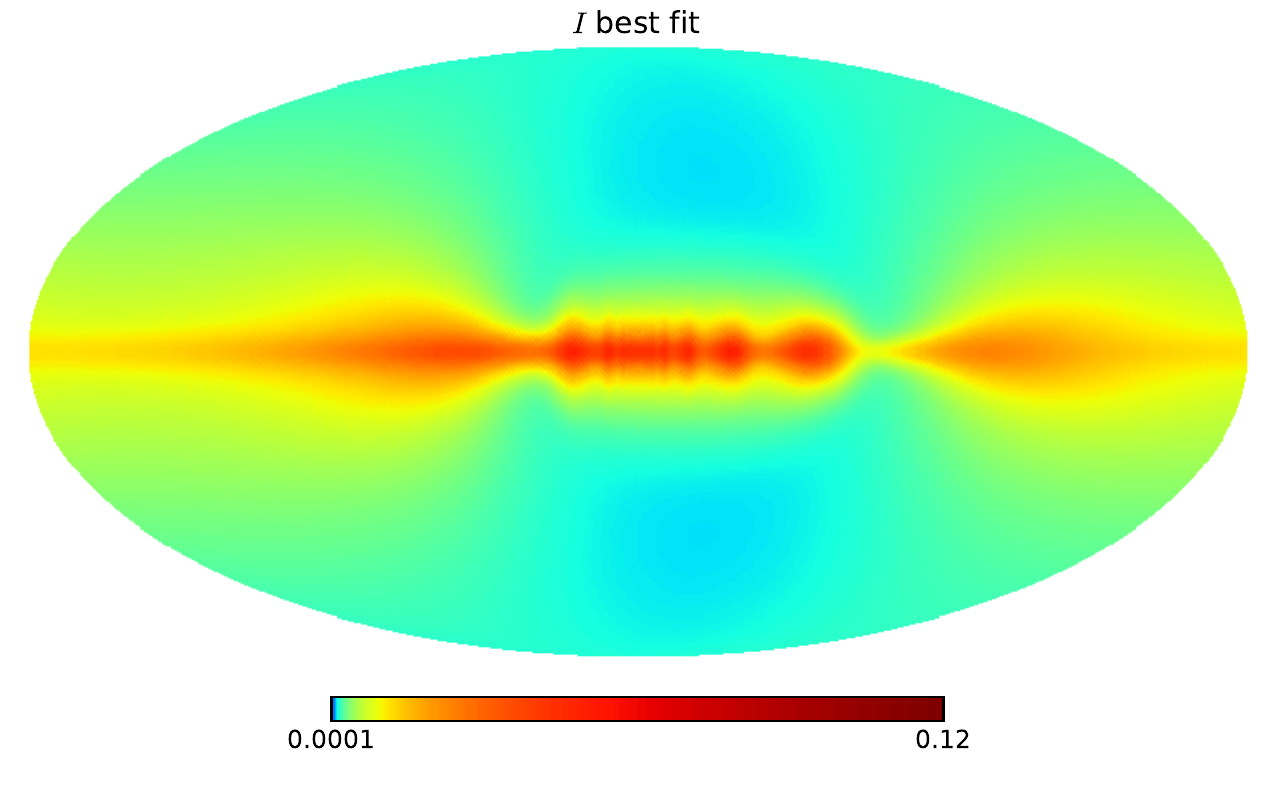}
        &       \includegraphics[width=.3\linewidth]{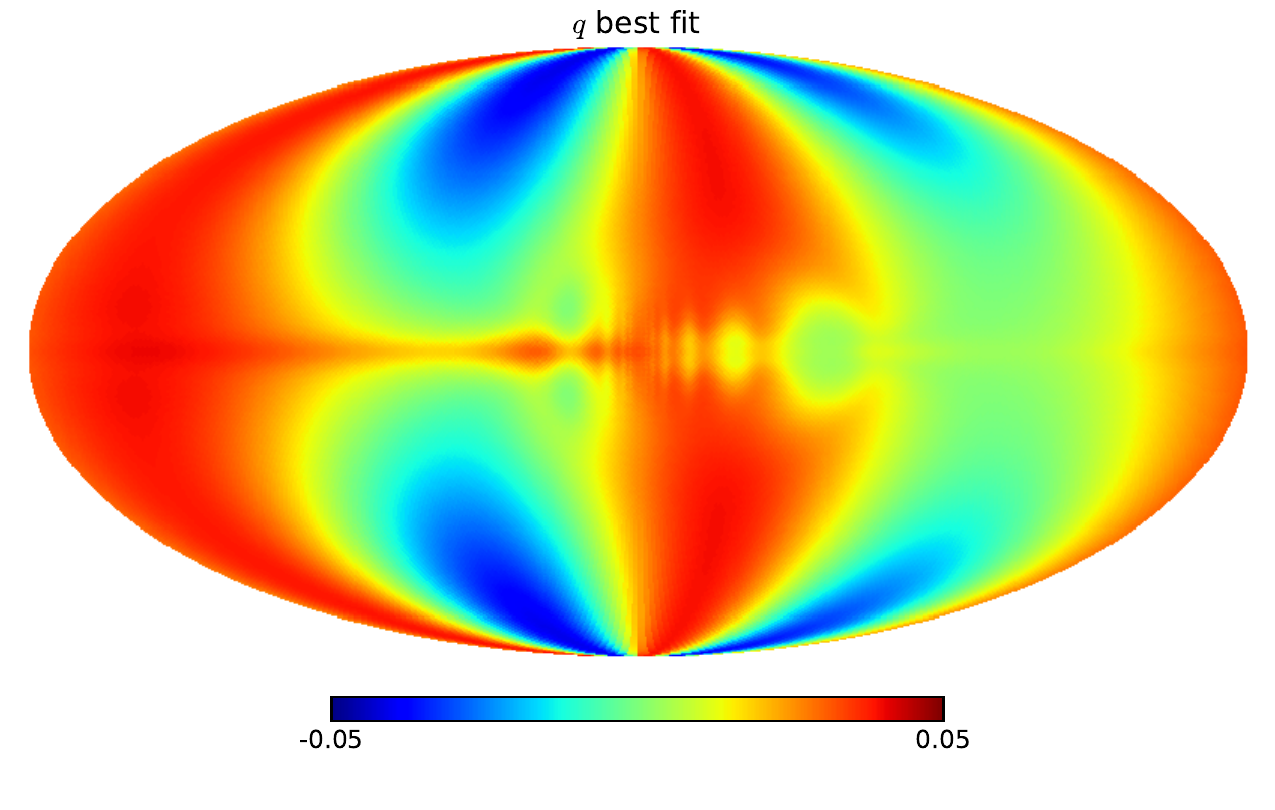} 
                &       \includegraphics[width=.3\linewidth]{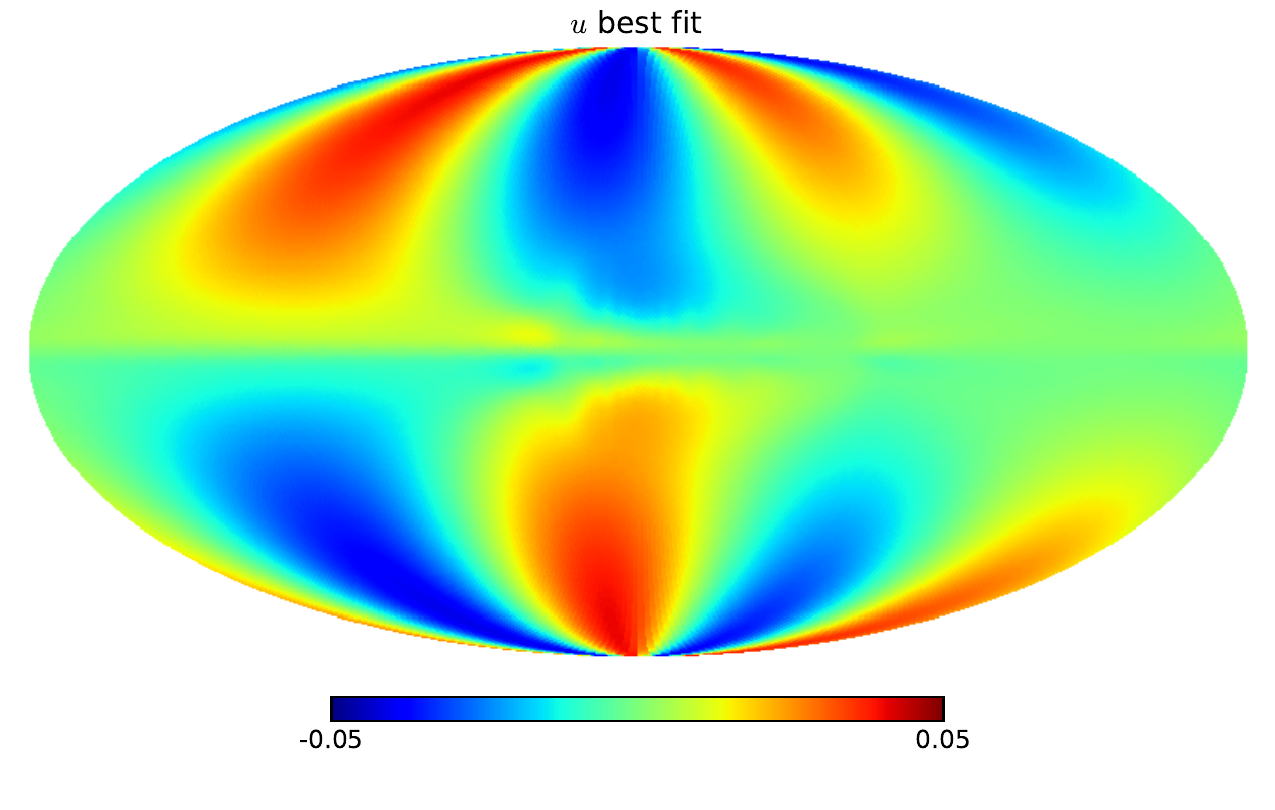} \\
\hline
\includegraphics[width=.3\linewidth]{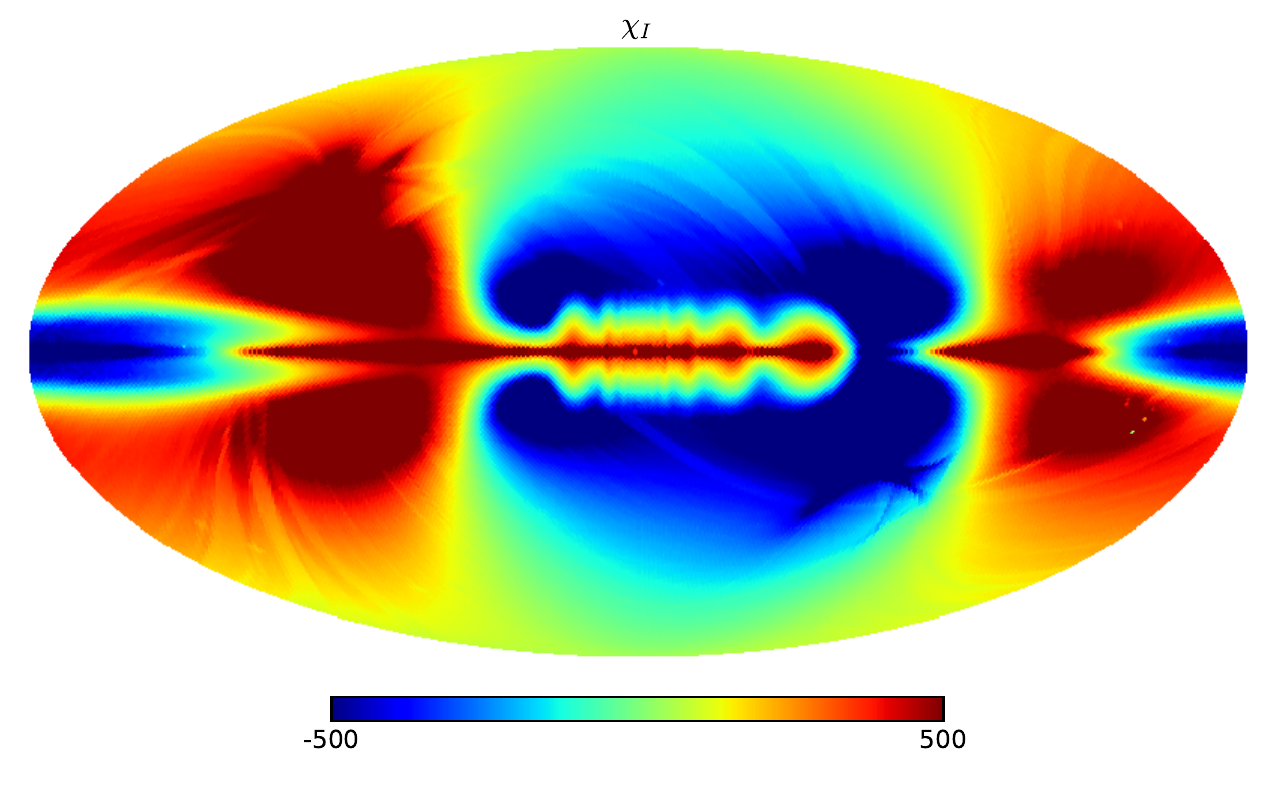}
        &       \includegraphics[width=.3\linewidth]{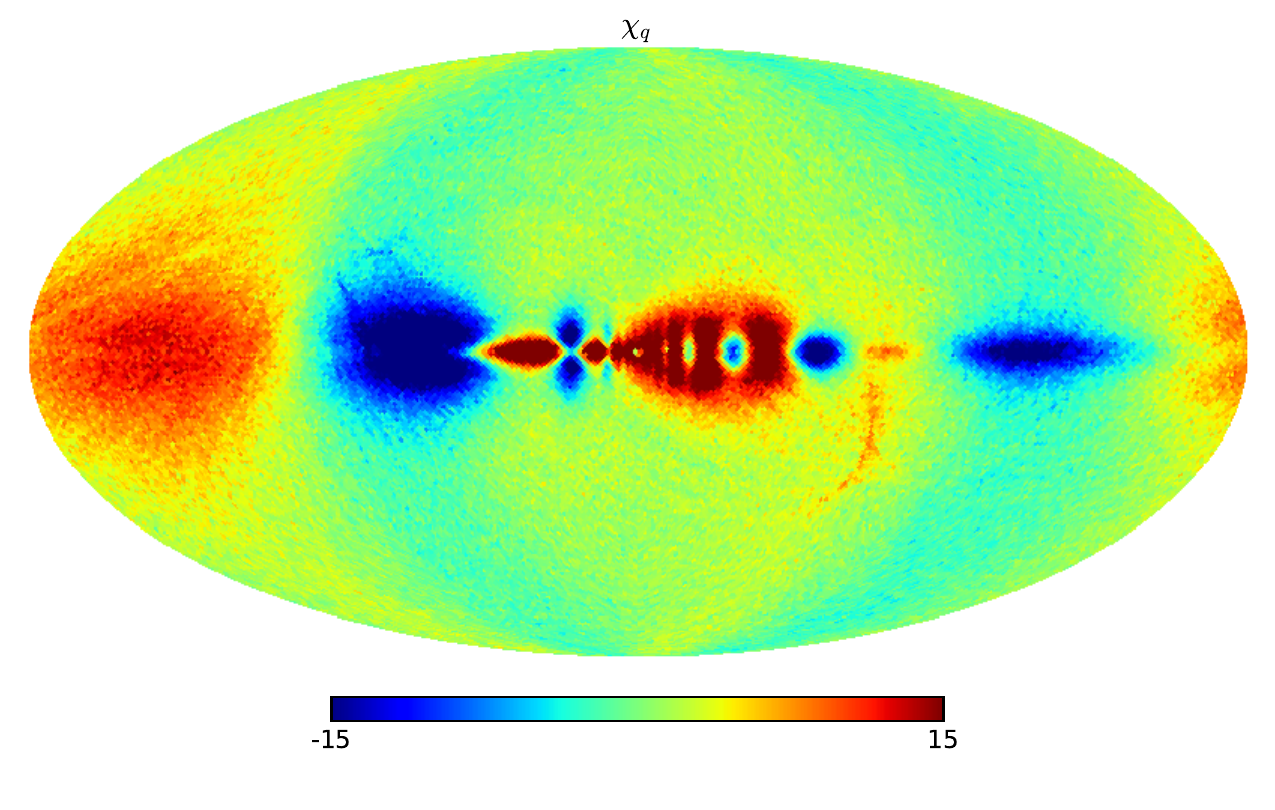} 
                &       \includegraphics[width=.3\linewidth]{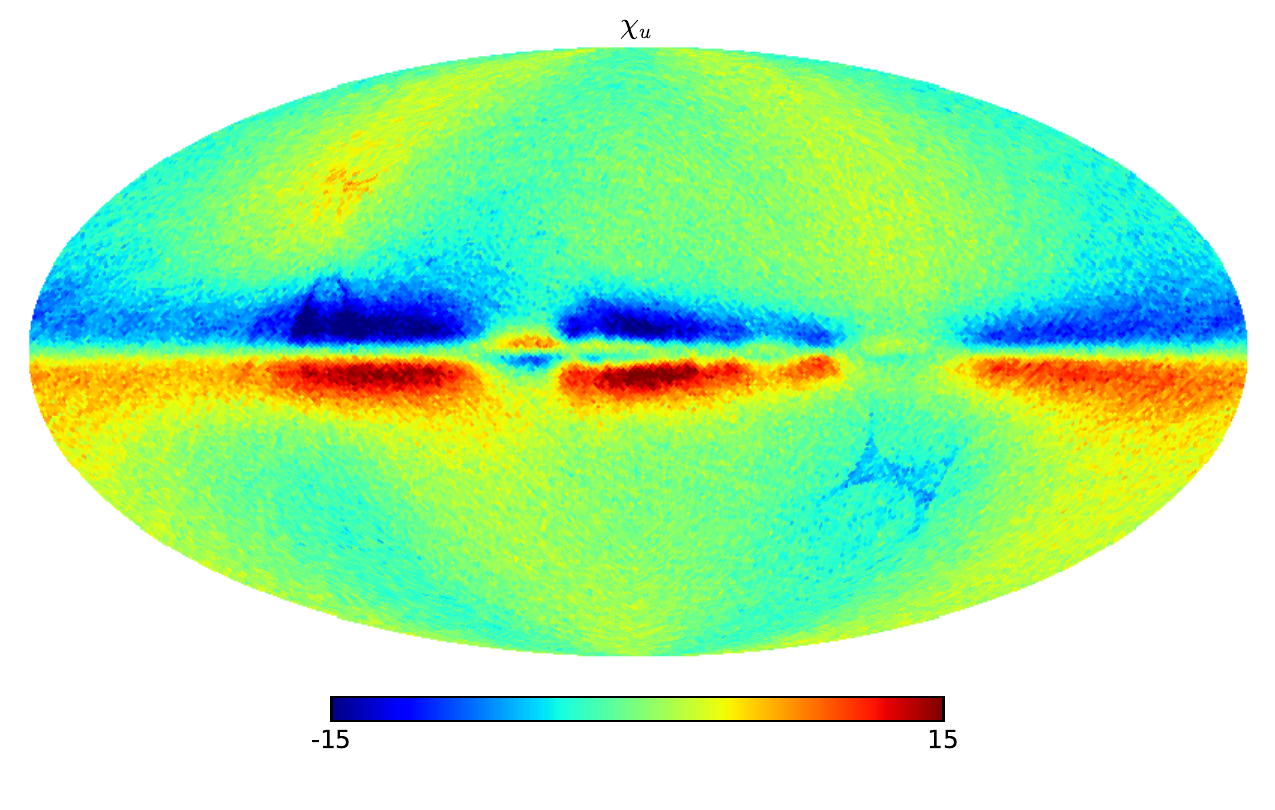} \\
\includegraphics[width=.3\linewidth]{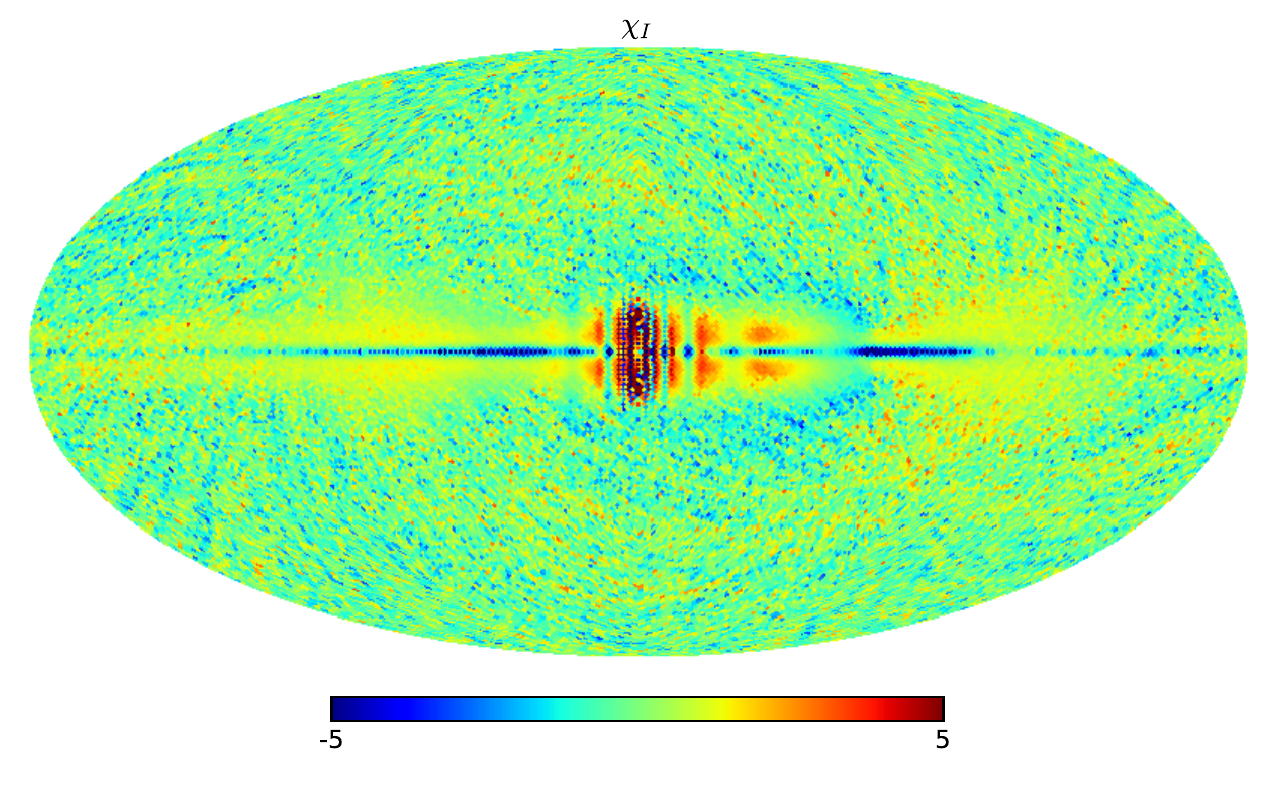}
        &       \includegraphics[width=.3\linewidth]{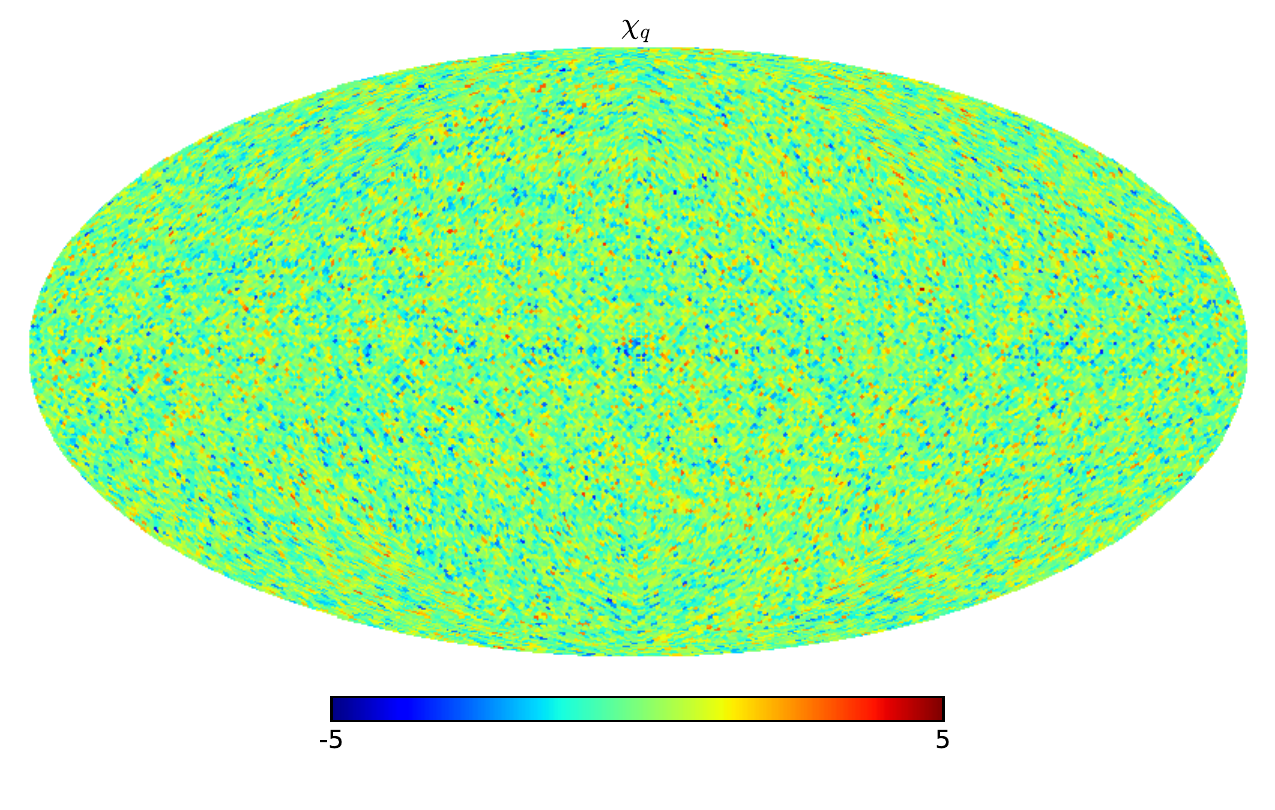} 
                &       \includegraphics[width=.3\linewidth]{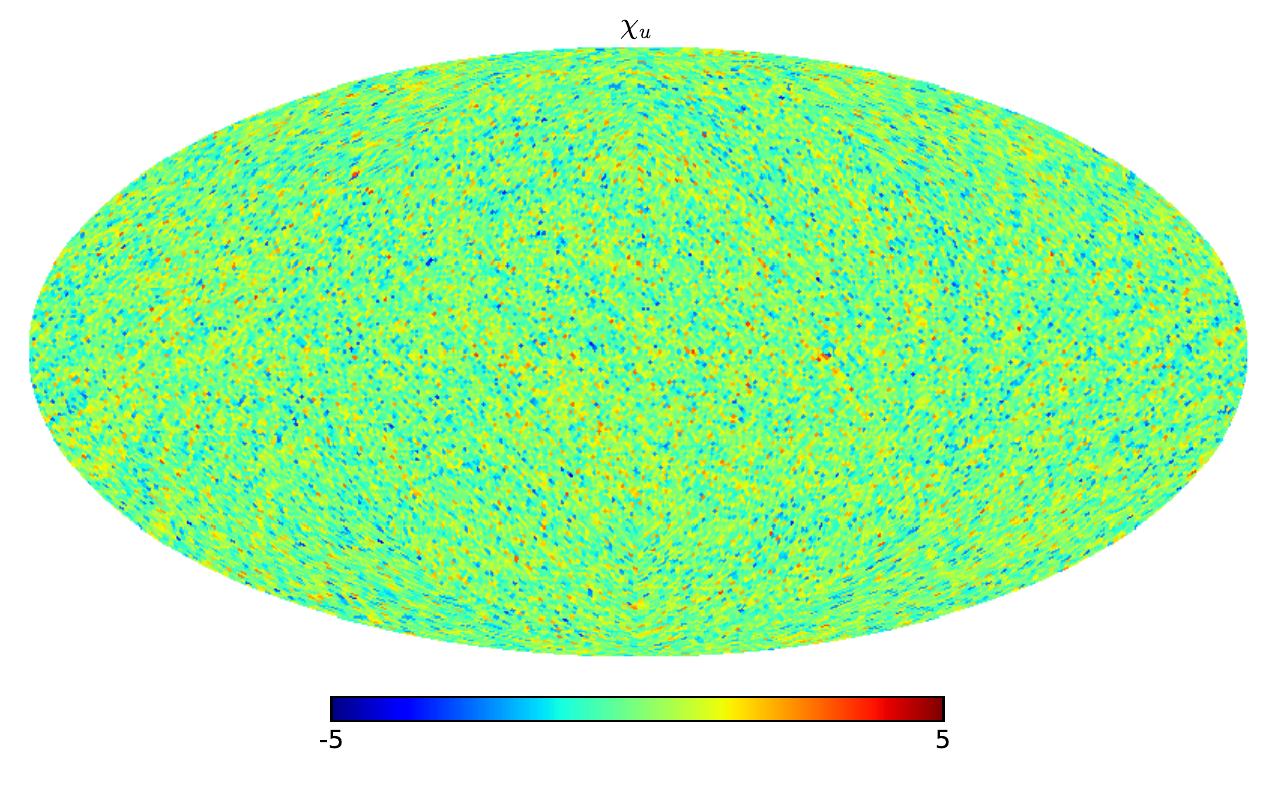} \\
\end{tabular}
\caption{Maps corresponding to the modified Stokes $I$,
$q,$ and $u$ shown from left to right.
(\textit{Top}) Data from the \texttt{S2} simulation downgraded to $N_{\rm{side}} = 64$
(\textit{second row}) best-fit maps obtained while assuming the dust distribution follows an ED model (case B)
(\textit{third row}) best-fit maps obtained while assuming the dust distribution follows an ARM4 model (case C).
The two last rows show the significance of the residuals for the case B (4th row) and the case C (5th row).
For the case B, the color scales range from -500 and 500 and from -15 to 15 for the intensity and the polarization, respectively.
For the case C, color scales range from -5 to 5, as in Fig.~\ref{fig:model1_fits}.
}
\label{fig:model2_fits}
\end{figure*}

\subsection{Validation of the MCMC algorithm}
\label{sec:mcmcalgoval}
We validated our MCMC implementation on simple simulations of the thermal dust emission. We used the same 3D dust density distribution and regular GMF models that those discussed in Sect.~\ref{sec:simu} (specifically \texttt{S1} and \texttt{S2}).
However, for this test, the $I$, $Q,$ and $U$ maps are simulated and fitted at the same HEALPix $N_{\rm{side}} = 64$ resolution. The results of the fitting procedure are summarized in Fig.~\ref{fig:qu64GMF_corner} which shows the marginalized 1D and 2D posterior probability distributions for the parameters of the regular LSA GMF model. It turns out that our MCMC reconstruction algorithm performs correctly with the input model parameters being well within the 95\% C.L. contours. Similar to better results are obtained while fitting for the dust density distribution.

\section{Reconstruction accuracy for the dust density and regular GMF component}
\label{sec:Iqu_fit}
In this section, we use simulations \texttt{S1} and \texttt{S2} to demonstrate that we can retrieve the input GMF model from the ``observed'' thermal dust emission using the approach described above. In other words, we show that the first step of our global attempt to infer the large-scale GMF features from the Galactic polarized diffuse emission is achievable.
Thus, a first fit is performed on the intensity map in order to obtain the best-fit model of the dust density distribution ($n_{\rm{d}}$).
Then, we use the best-fit $n_{\rm{d}}$ model as an input to constrain the GMF model by fitting the ($q,\,u$) polarization maps.

We consider the following cases.
First, {\bf Case A}, we fit the \texttt{S1} simulations using the ED model for the dust density and the LSA model for the GMF. 
Second, {\bf Case B}, we fit the \texttt{S2} simulations using the ED model for the dust density and the LSA model for the GMF.
And finally, {\bf Case C}, we fit the \texttt{S2} simulations using the ARM4 model for the dust density and the LSA model for the GMF.

Case A allows us to validate the methodology in a relatively simple model and is thus appropriated to diagnose possible issues.
Case B is of interest because it helps us to evaluate the impact of the poor knowledge of the dust density distribution on the GMF reconstruction. This situation is likely to occur when tackling real data given the convoluted nature of the real data set.
Case C helps us to evaluate the possible effect of the loss of information due to line-of-sight integration (\textit{i}) on the modeling of the intensity map and
(\textit{ii}) to evaluate the effect of the propagation of this source of uncertainty on the reconstruction of the 3D GMF.

The best-fit parameter values that we obtained are reported in Table~\ref{tab:paramVal}, both for the intensity and the polarization fits.
The obtained values of the reduced $\chi^2$ for each case are reported in Table~\ref{tab:chi2values}.

\begin{table}
\centering
\caption{Reduced $\chi^2$ values for the best-fit in intensity and in polarization
for the three study cases.}
\label{tab:chi2values}
\begin{tabular}{l c c c}
\hline
\hline
\\[-.5ex]
fitted data     &       case A  &       case B  & case C \\
\hline
\\[-1.ex]
$I$     & $1.02$        & $1.9\,10^5$           & $1.33$        \\

$(q,\,u)$       & $1.935$       & $9.38$        & $1.937$       \\
\\[-1ex]
\hline
\end{tabular}
\end{table}

\subsection{Reconstruction of the dust density}
The first steps in our fitting procedure is to provide best-fit parameter
values for the dust density distribution model from a fit on the intensity
map using the MCMC procedure for intensity described in
Sect.~\ref{sec:GMFrecons}.

\smallskip

The main results for the intensity fit for case A are shown in the first
column of Fig.~\ref{fig:model1_fits}.
We find that we obtained a good fit to the data  with and overall reduced 
$\chi^2 = 1.02 \ $
and residuals consistent with noise at high Galactic latitudes.
Nevertheless, we also see that the significance of the residuals increases (first towards positive values and suddenly towards negative values) while getting closer to the Galactic equator. This behavior actually takes its origin in the resolution bias discussed in Appendix~\ref{sec:resolutionBias}.
It is indeed close to the Galactic plane that the functional forms of the dust density are allowed to vary quickly. Therefore, it is from those locations (then projected on the sky) that the loss in space-sampling resolution has larger effects.

\smallskip

For cases B and C, the intensity fit results are shown in Fig.~\ref{fig:model2_fits}.
The best-fit model and residuals are shown in rows 2 (3)  and 4 (5) for case B (C), respectively.
For case B, the fitting to the intensity data is very poor with a reduced $\chi^2$ of $1.9\,10^5$.
This is as expected because of the very small error bars, the large number of data points and principally because the density distribution model cannot reproduce the complexity of the intensity signal injected in the \texttt{S2} simulation.
The uncertainties used to computed the reduced $\chi^2$ do not account for mis-modeling.
The significance of the residuals are large both in and
outside the Galactic equatorial region. 

For case C, we obtain a good fit to the data with reduced $\chi^2 = 1.33$. The residuals are consistent with noise at high Galactic latitudes. In the Galactic plane similar comments as for case A hold.
The significance of the residuals even shows some structures and it is slightly larger than for case A due to the somewhat larger complexity of the underlying model. 

\smallskip

Finally, from the results for case A and C we conclude that, despite the line-of-sight integration, and assuming we have the right model at hand, we are able to retrieve the parameters of a complex dust density distribution with reasonable confidence. In contrast, as shown in case B, if the applied dust density model does not reflect reality, large residuals can be found. Although we cannot directly extrapolate the results of case B, it allows us to anticipate on the kind of issues we would face with real data sets, as discussed in Sect.~\ref{sec:fit2Planck}.

\subsection{Reconstruction of the GMF}
\label{sec:reconstructionGMF}

Having found the best-fit models for the intensity map, we use the
best-fit values of the parameters to populate the Galaxy with dust density
distribution and then to constrain the underlying GMF model using MCMC
fit on the polarization maps.

In practice, we use the best-fit parameter values obtained above to evaluate
the dust density at all locations of the sampled space. Then, using a MCMC sampling,
we vary the four parameters of the GMF model (given in Eqs.~\ref{eq:LSAmodel}) and
integrate, along the line-of-sight, the elemental emitting volumes following
Eqs.~\ref{eq:DUSTEMISSION} to produce the $Q$ and $U$ maps. Then we normalize
these simulated maps by the best-fit intensity map to obtain the $q$ and $u$ maps.
Finally, we proceed to a comparison of the simulated data and the computed model
by maximizing the likelihood function described in Sect.~\ref{sec:likelihood}.
The best-fit maps are presented in the second and third columns of Fig.~\ref{fig:model1_fits}, second row, for case A, and on Fig.~\ref{fig:model2_fits}, second and third rows, for cases B and C, respectively. The maps of the significance of the residuals are also shown in the last rows of the figures.

\smallskip

From the comparison of the input and best-fit $(q,\,u)$ maps of case A and case C, it appears that we are able to provide good fit to the data sets. This is confirmed by the maps of the significance of the residuals, and by the reduced $\chi^2$ values of Table~\ref{tab:chi2values}. The best-fit values of the GMF parameters are reported in Table~\ref{tab:paramVal} along with their marginalized uncertainties at the 1$\sigma$ level. It is seen that the best-fit parameter values and the input ones are very close in terms of relative differences even if the latter do not stand within the  1$\sigma$ confidence interval.
Again, the observed shift is most likely due to the resolution bias which turns out to be significant in this case since it is the only important source of uncertainty (see Appendix~\ref{sec:resolutionBias}).
Unlike what we obtained for the intensity fit, we do not observe any particular trend or feature in the residual maps near the Galactic equator.
At the map level, the fitting issue from the resolution bias is likely dimmed by the fact that we consider the reduced Stokes parameters.

\smallskip

For case B, we note that the residuals are significantly larger than for the other two cases. We observe a lack in the signal amplitude of the best-fit polarization maps with respect to the input ones.
Nevertheless, we see that we are able to capture the correct sky location of the extrema of the reduced Stokes parameters.
An inspection of Table~\ref{tab:paramVal} shows us that the best-fit values of the GMF parameter are close to the input ones.
While the agreement is globally worse here than in case A and C, we see that the agreement is better for the parameters of the spiral shape of the GMF than those of the X-shape (see Appendix~\ref{sec:GMFmodel}).

\smallskip

\begin{figure}
\centering
\includegraphics[trim={0cm 0cm 4.5cm 0cm},clip,width=.72\linewidth]{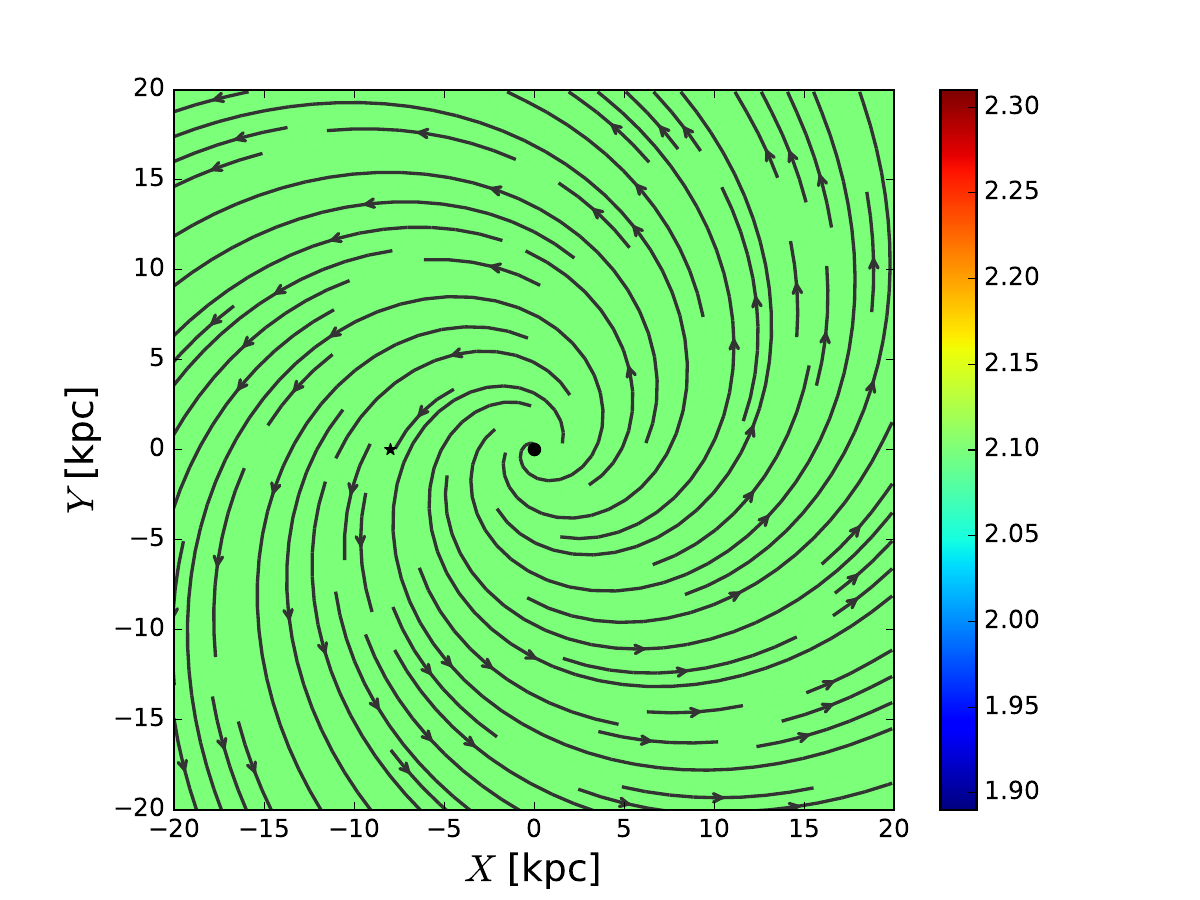} \\[-3.9ex]
        \hspace{.1cm}
        \includegraphics[trim={0cm 4cm 4.5cm 5.3cm},clip,width=.696\linewidth]{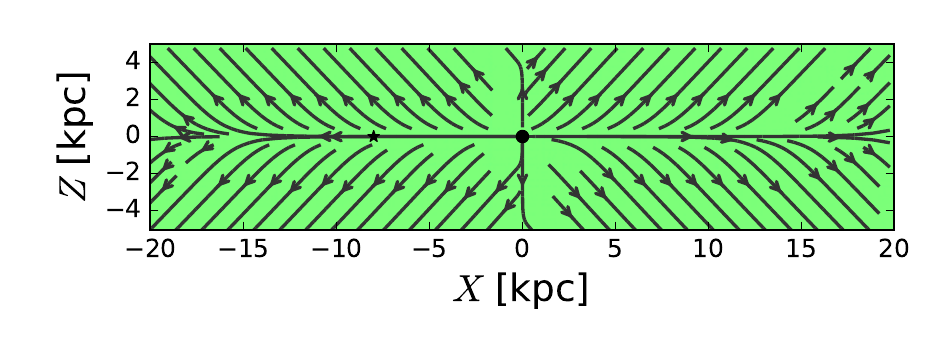} \\[-2.ex]
\includegraphics[trim={0cm 0cm 4.5cm 0cm},clip,width=.72\linewidth]{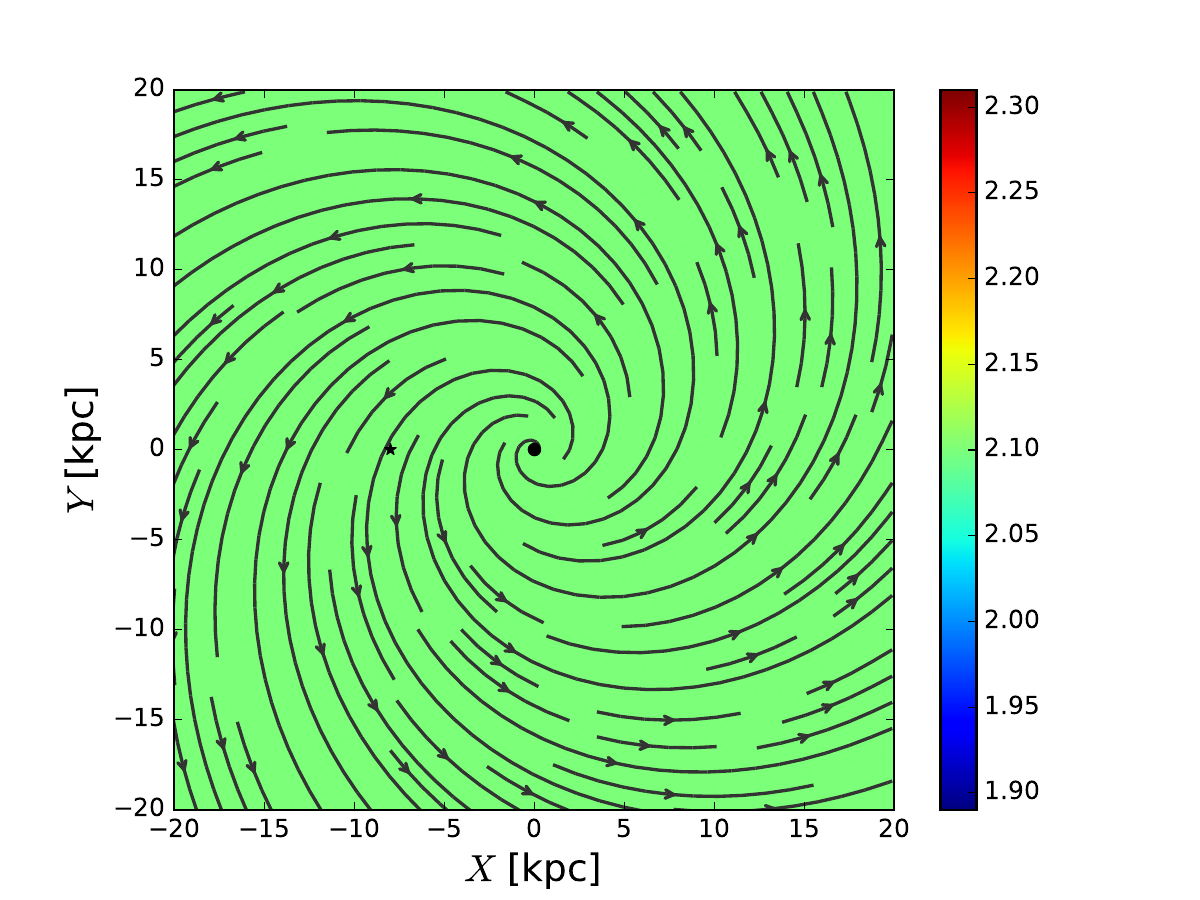} \\[-3.9ex]
        \hspace{.1cm}
        \includegraphics[trim={0cm 4cm 4.5cm 5.3cm},clip,width=.696\linewidth]{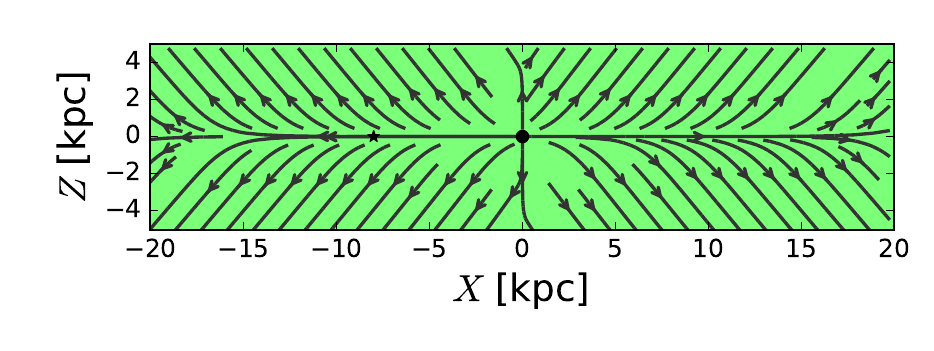} \\ [-2.ex]
\includegraphics[trim={0cm 0cm 4.5cm 0cm},clip,width=.72\linewidth]{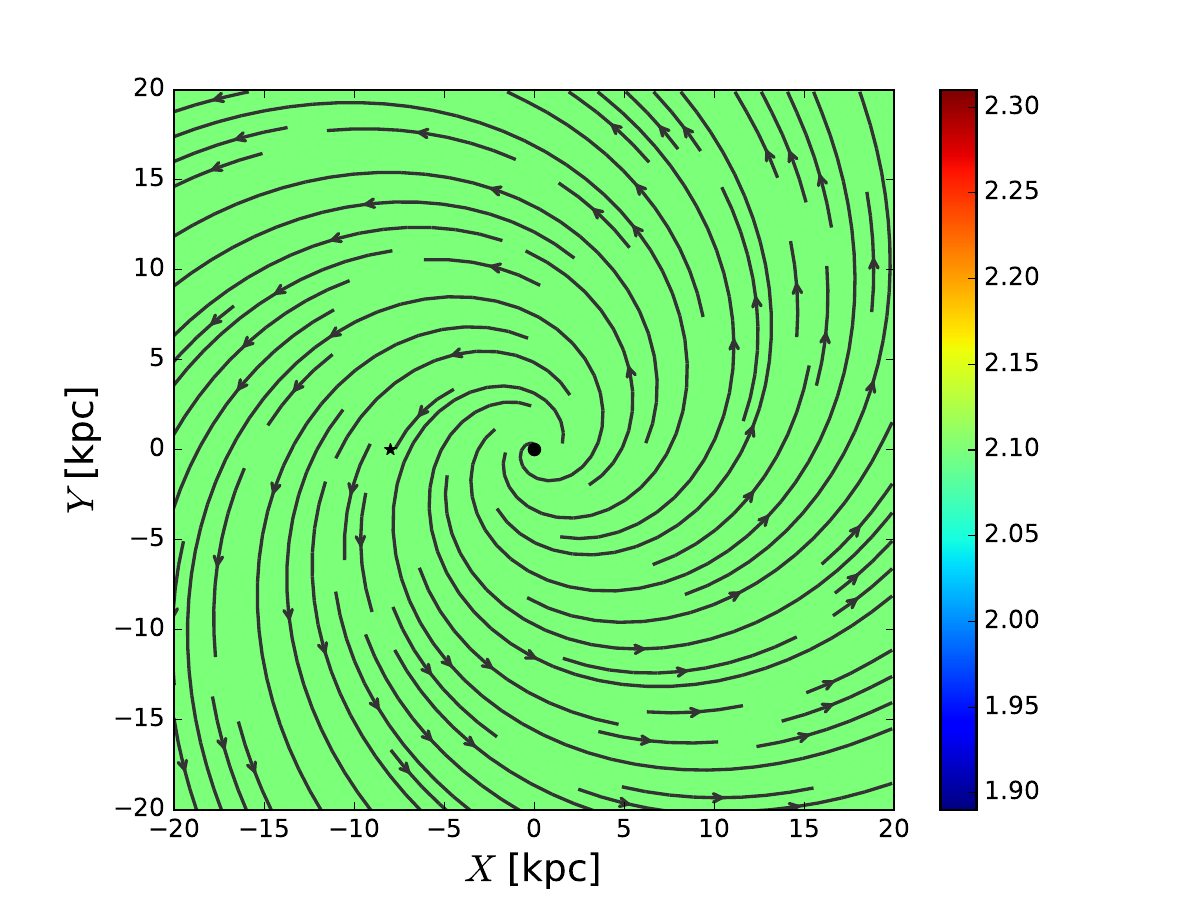} \\
        \hspace{.1cm}
        \includegraphics[trim={0cm 4cm 4.5cm 5.3cm},clip,width=.696\linewidth]{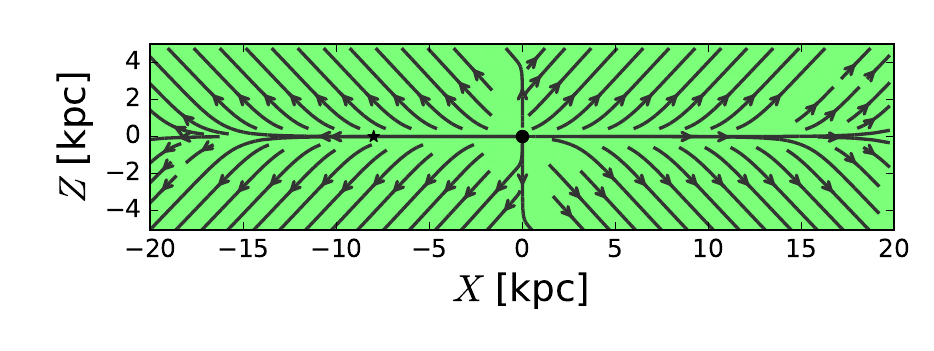} \\[-2.ex]
\caption{ Reconstructed GMF structures in the
XY plane (top) and XZ plane (bottom)
in case A, case B, and case C, respectively,  shown from top to bottom. See text for details. Similar plot for the input GMF model is shown in Fig.~\ref{fig:inputWMAP_xyz}.
\label{fig:GMFmodel2_fits}}
\end{figure}

The reconstructed geometrical structures of the GMF corresponding to the cases A, B, and C best-fit parameters
are shown in Fig.~\ref{fig:GMFmodel2_fits}, and can be compared to the input one shown in Fig.~\ref{fig:inputWMAP_xyz}. These figures  qualitatively illustrate the fact that we are indeed able to reconstruct (i.e., constrain) the geometry of the GMF from the observation of the polarized thermal dust emission in the three cases discussed above.

\begin{figure*}
\centering
\begin{tabular}{cc}
\includegraphics[trim={0cm 0cm 0cm 1cm},clip,height=.22\textheight]{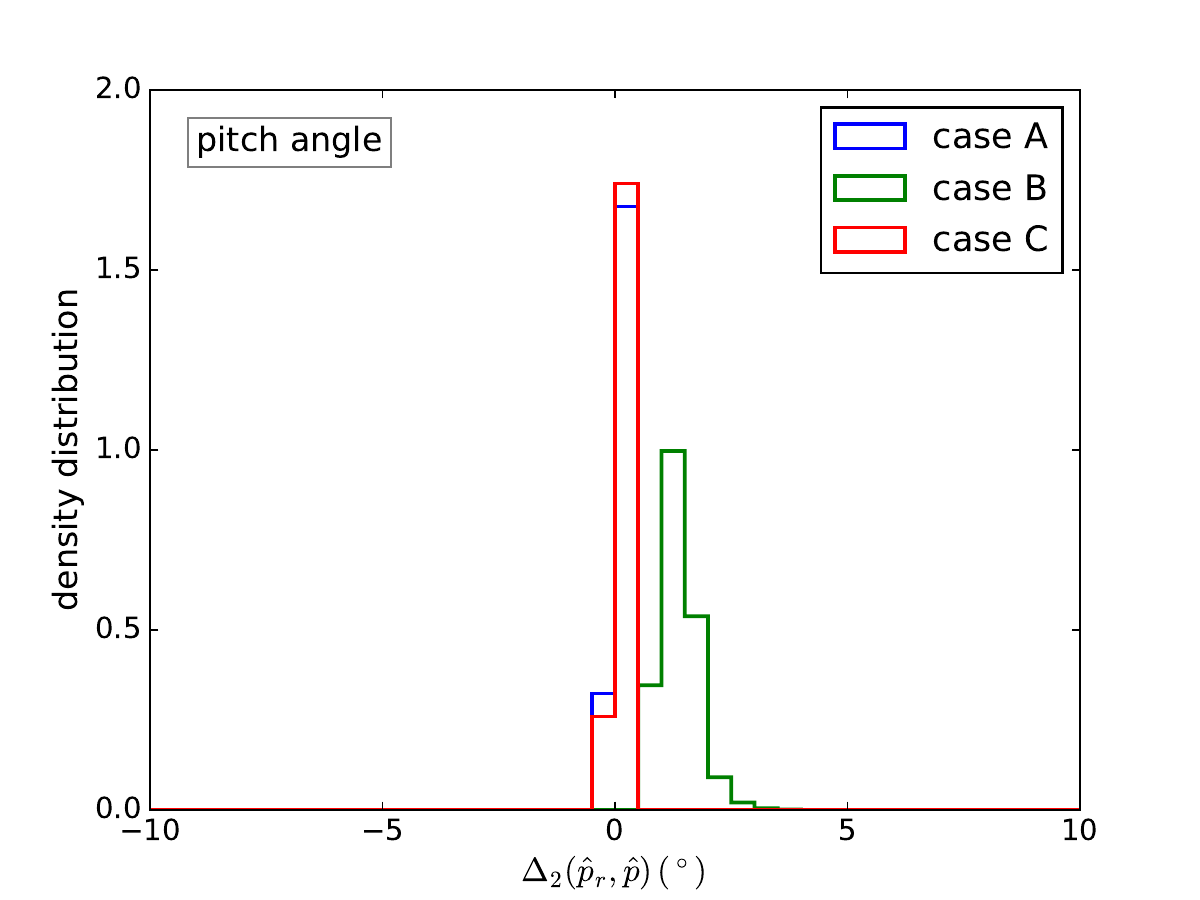} &
        \includegraphics[trim={0cm 0cm 0cm 1cm},clip,height=.22\textheight]{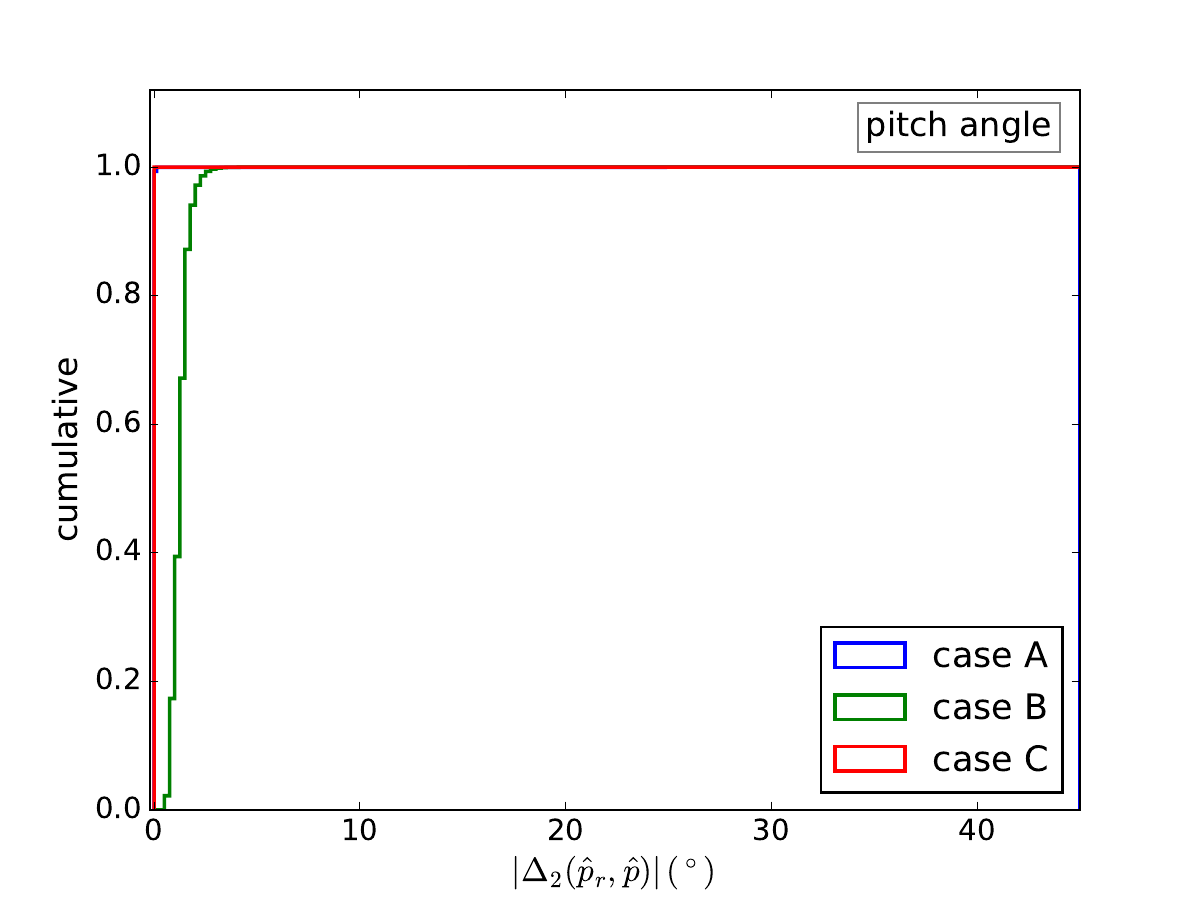} \\
\includegraphics[trim={0cm 0cm 0cm 1cm},clip,height=.22\textheight]{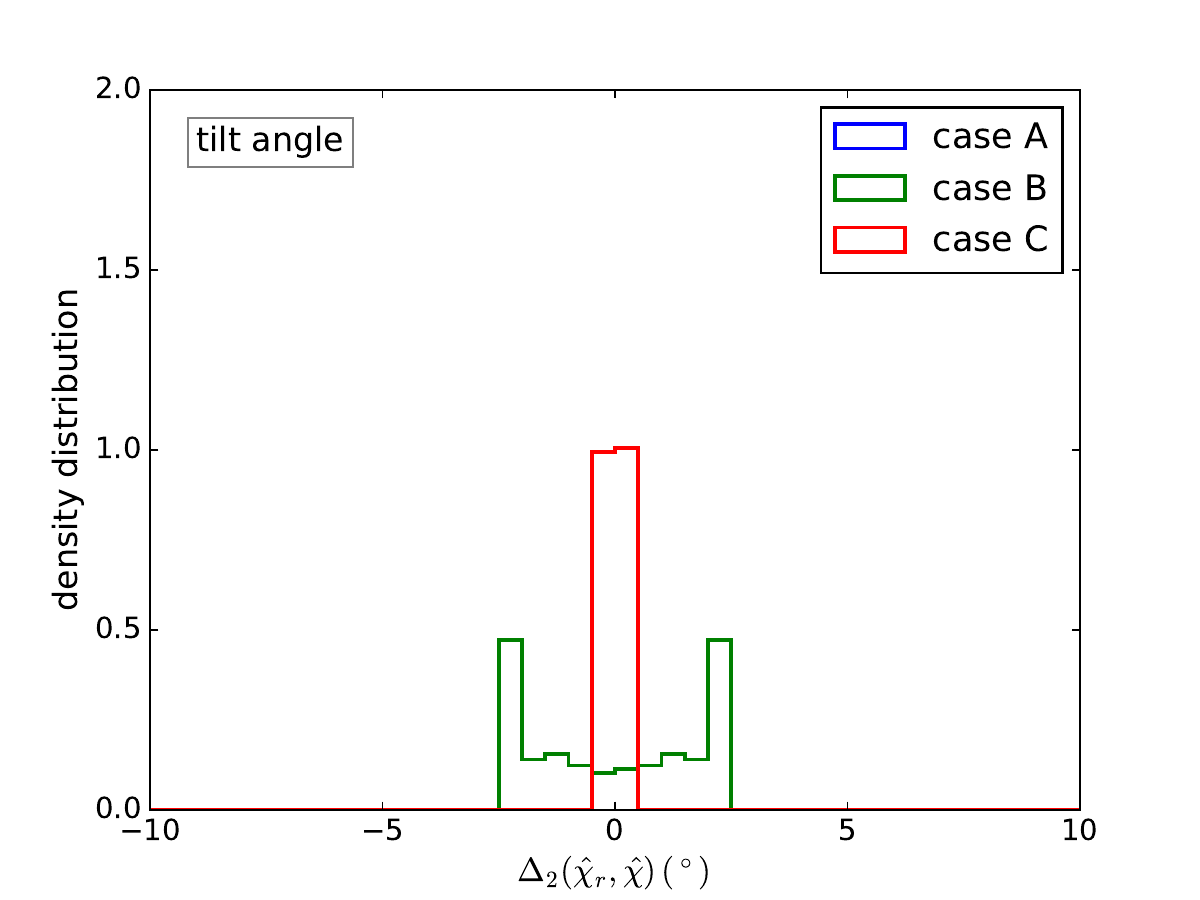} &
        \includegraphics[trim={0cm 0cm 0cm 1cm},clip,height=.22\textheight]{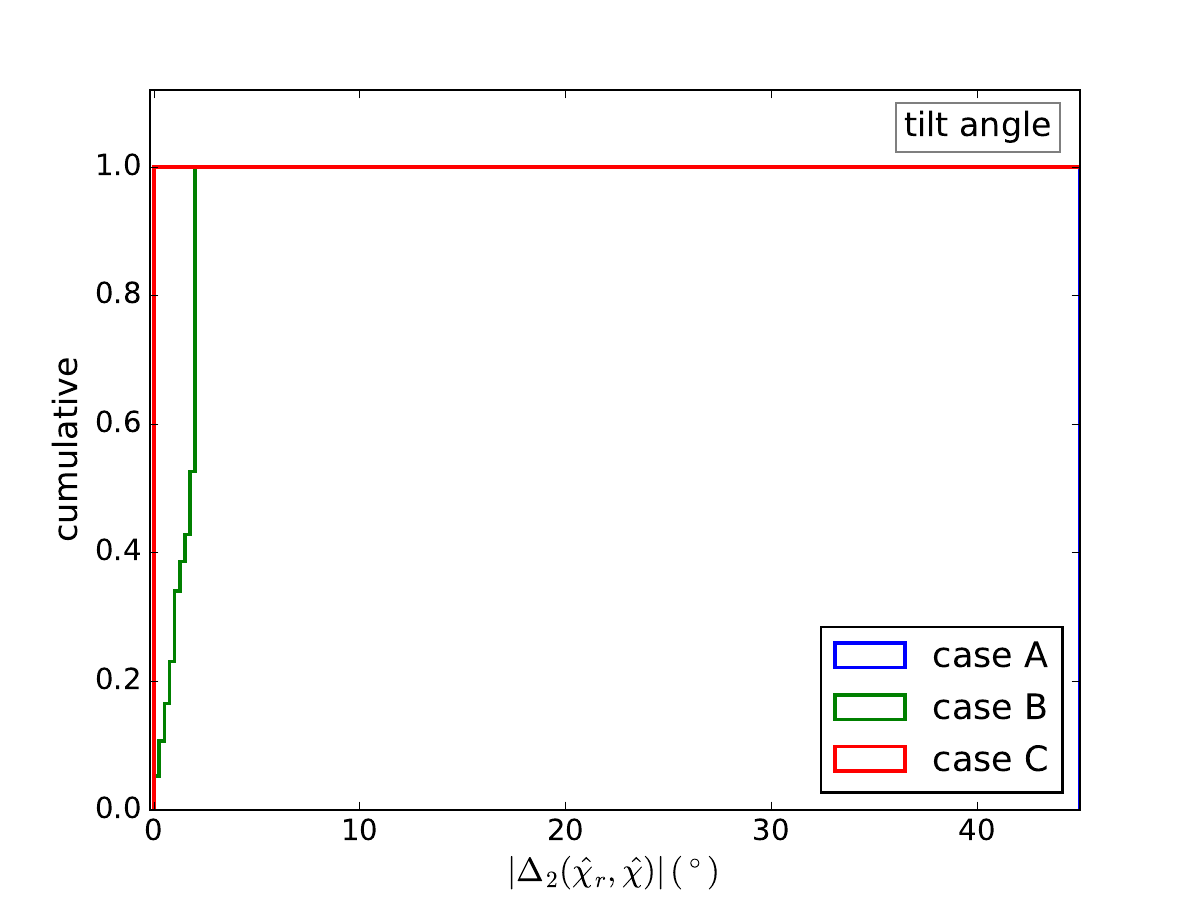} \\
\includegraphics[trim={0cm 0cm 0cm 1cm},clip,height=.22\textheight]{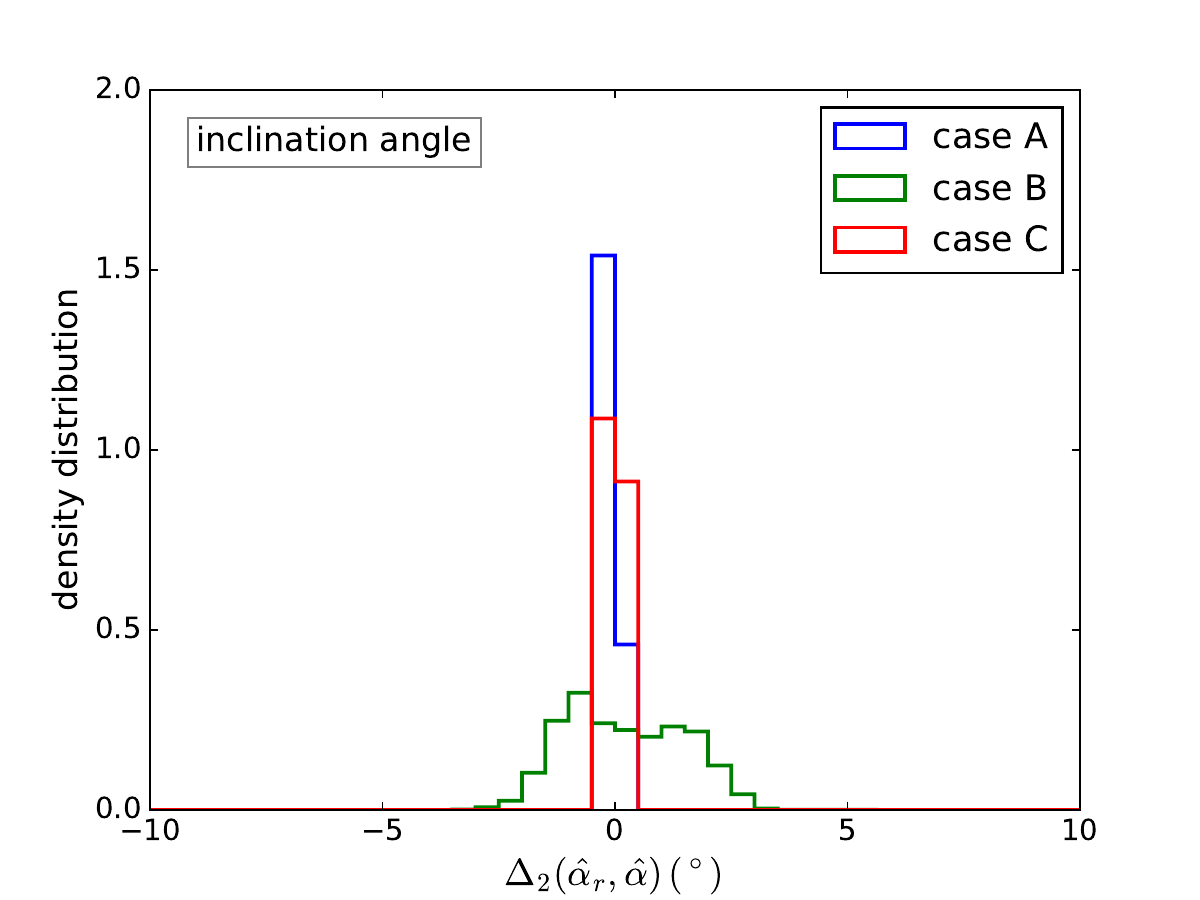} &
        \includegraphics[trim={0cm 0cm 0cm 1cm},clip,height=.22\textheight]{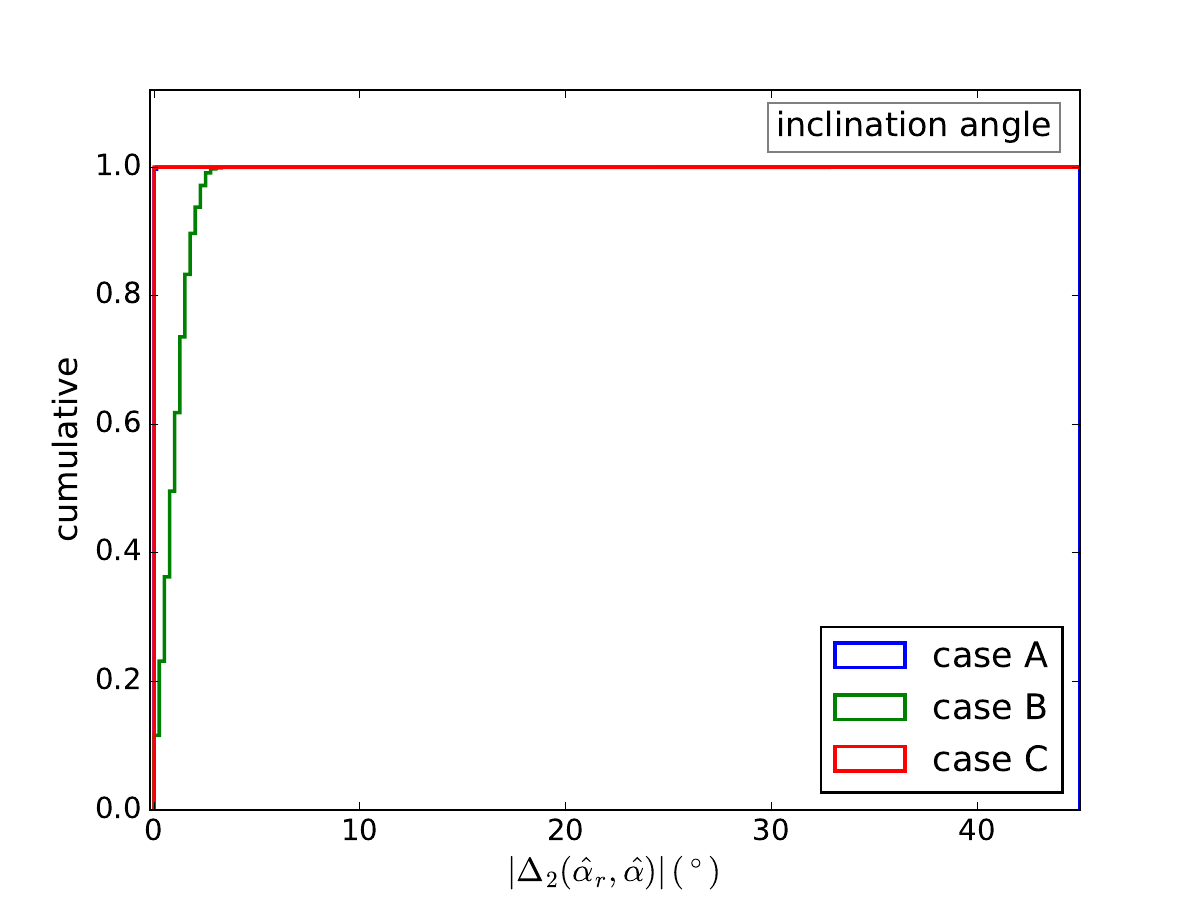} \\
\includegraphics[trim={0cm 0cm 0cm 1cm},clip,height=.22\textheight]{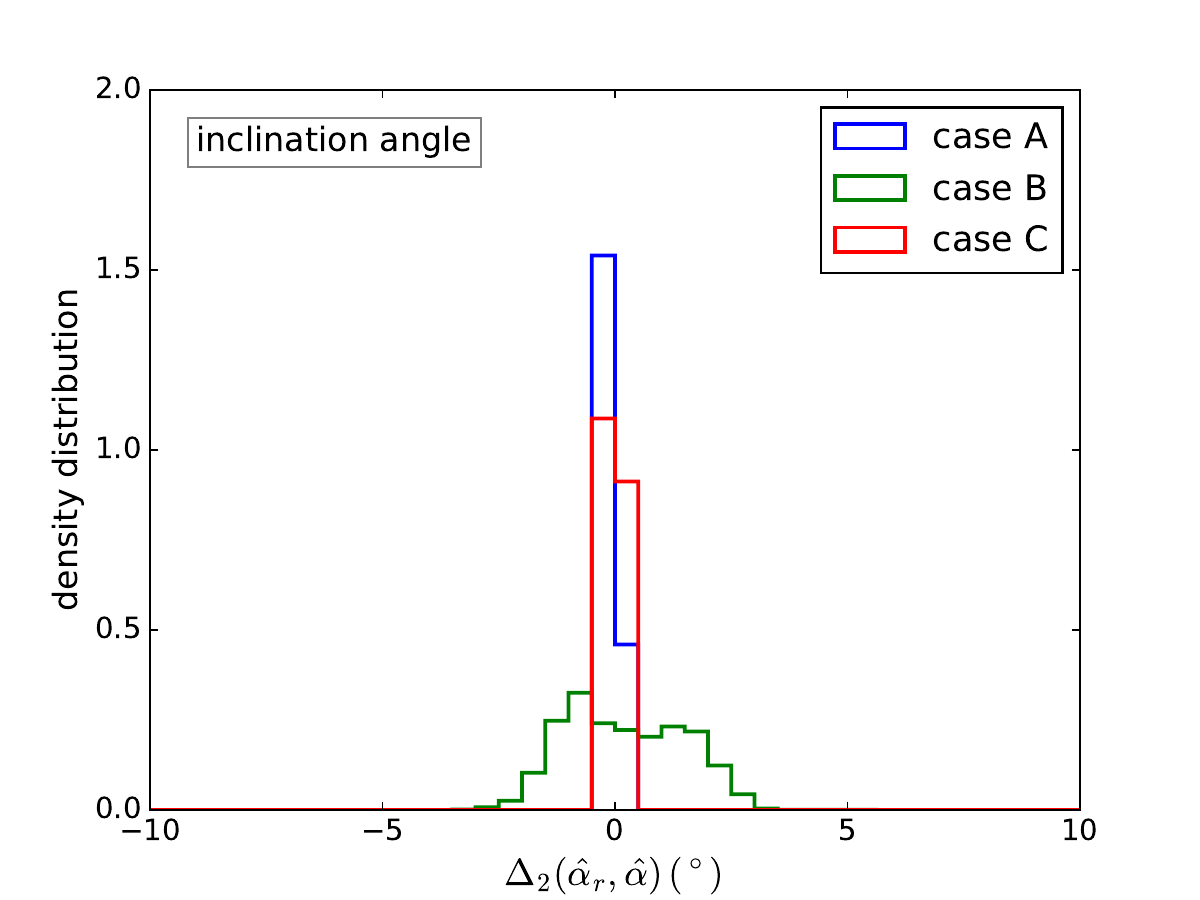} &
        \includegraphics[trim={0cm 0cm 0cm 1cm},clip,height=.22\textheight]{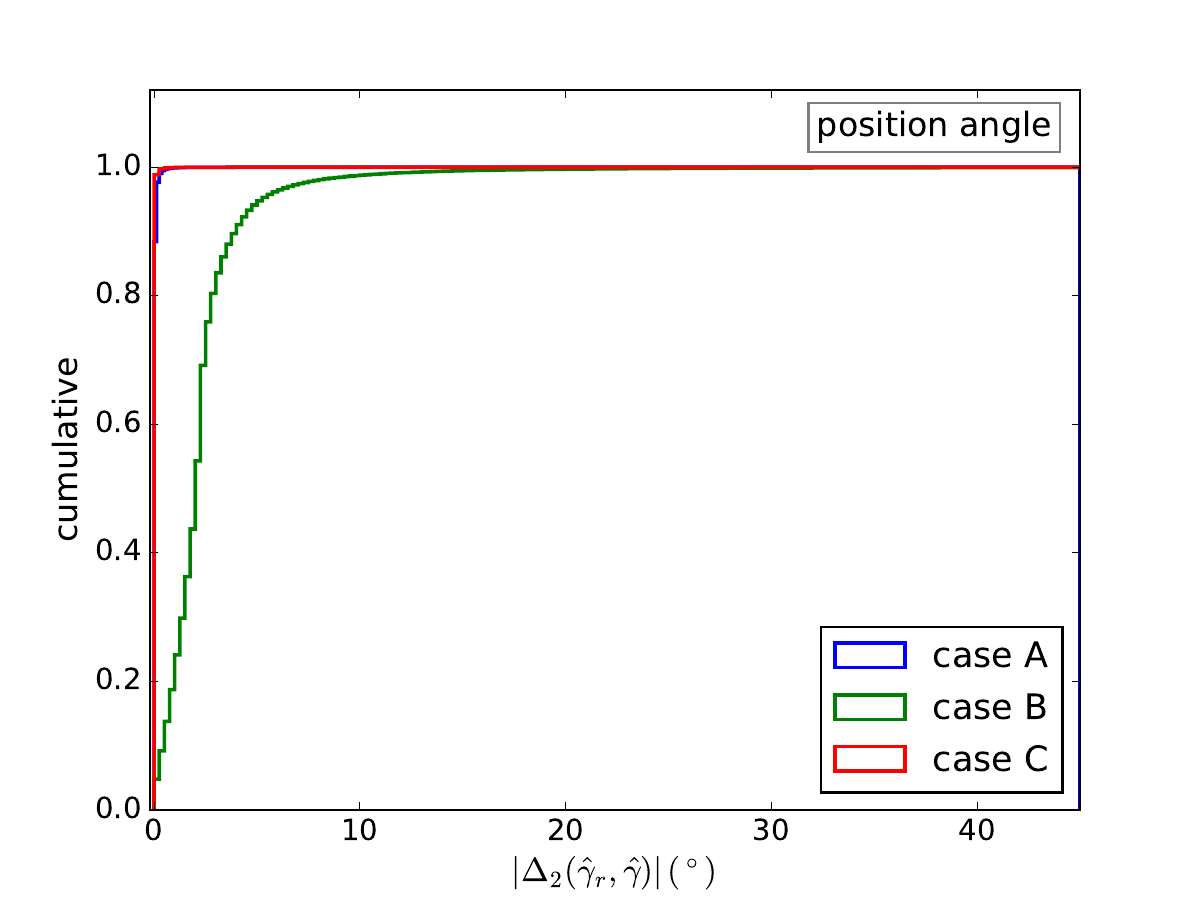} \\
\end{tabular}
\caption{Histograms of the angle differences (in degree) between the best reconstructions of the GMF and the input one computed at every location of the sampled Galactic space (left). From top to bottom are shown the differences of the pitch angles, the tilt angles, the inclination angles, and the position angles (see text). Case C overlaps case A in tilt and position angle differences.
(right) Same as for left column but we present the cumulative distributions of the absolute values of the angle differences.
\label{fig:GMFmodel2_fits_angles}}
\end{figure*}

\begin{figure}
\centering
\includegraphics[width=1\linewidth]{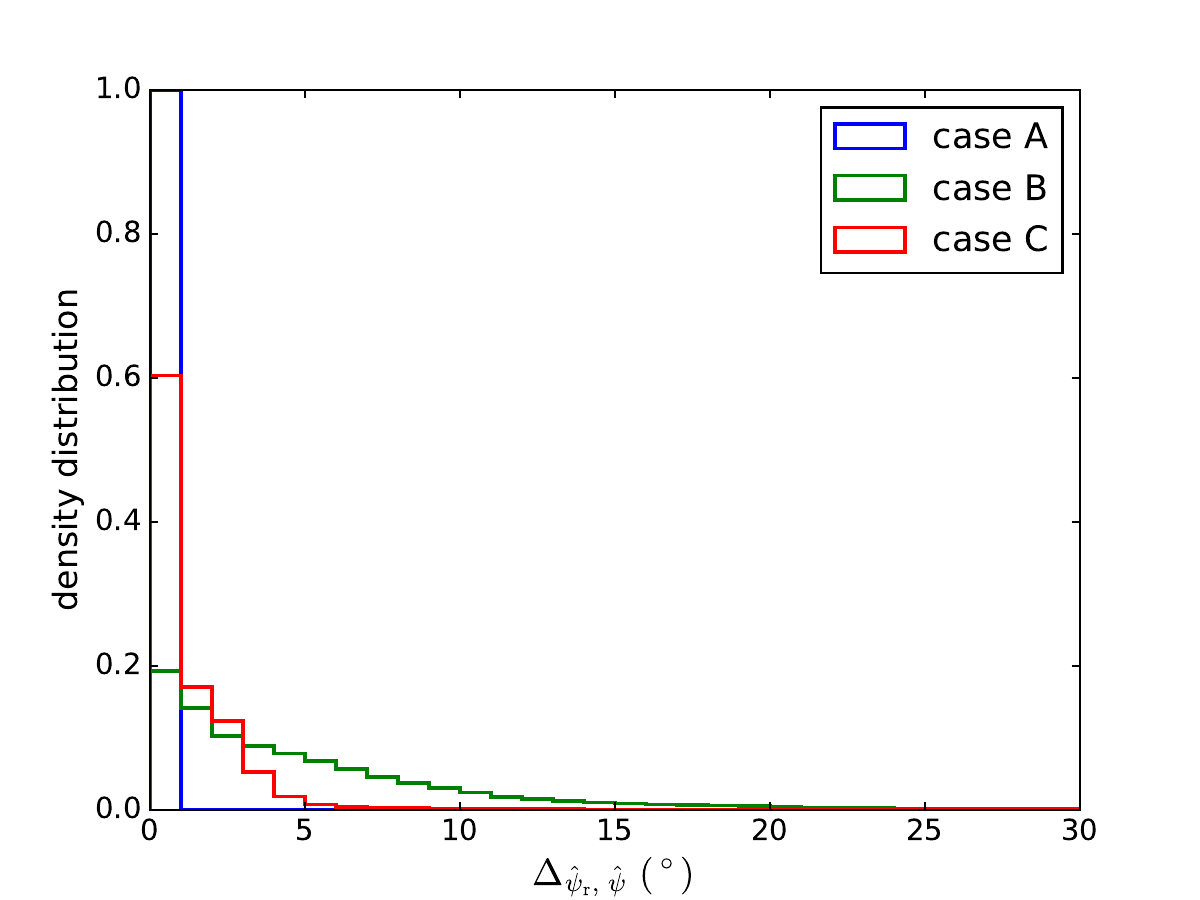} \\
\caption{Histogram of the differences of the polarization position angles 
between the data and the best fits. The polarization position angles are
deduced from the $q$ and $u$ maps both from the downgraded
simulations (\texttt{S1} and \texttt{S2} but without the \textit{Planck}
noise) and from the best-fit maps.
\label{fig:GMFmodel2_fitsPPA}}
\end{figure}

\subsection{Quality of the reconstructed GMF}
\label{sec:qualreconstructionGMF}
In this subsection, we  quantitatively compare the GMF geometrical
structures that we reconstructed by fitting the polarization maps for
case A, B, and C (see Fig.~\ref{fig:GMFmodel2_fits}) to the input
model shown in Fig.~\ref{fig:inputWMAP_xyz}.
At each location of the Galactic space considered when computing
the polarization observables (see Sect.~\ref{sec:ModAndImp}), we
consider two sets of angles that determine the orientation of the GMF
lines. First, in the cylindrical coordinate system centered on the Galactic
center (with the $z$ axis perpendicular to the Galactic equator), we
define the angles $p$ and $\chi$, and second in the spherical coordinate
system centered on the Sun, the angles $\alpha$ and $\gamma$.
In the first scheme, $p$, the pitch angle, is the angle that makes the
GMF line with $\mathbf{e}_\phi$, and $\chi$, the tilt angle, is the angle
that makes the field line with planes parallel to the Galactic plane ($z$
constant); $\chi$ is the complementary to the angle made by the GMF
line with $\mathbf{e}_z$ and characterizes the out-of-plane GMF
component.
In the second scheme, $\alpha$ is the inclination angle that makes the
field line relative to the line of sight and $\gamma$ the position angle in
the plane orthogonal to it.
Let $\hat{\xi}_{\rm{r}}$ be any of the angles defined above and measured
for the input GMF model and $\hat{\xi}$ the same angle but for
the reconstructed GMF models.
We define the signed difference angle $\Delta_2(\hat{\xi}, \hat{\xi}_{\rm{r}})$
as the difference of the two angles that range between -90 and 90
degrees (-90 and 90 corresponding to the same configuration).

\medskip

To quantify the accuracy of the reconstructed GMF structures, we compute at each location the $\Delta_2(\hat{\xi}, \hat{\xi}_{\rm{r}})$ for the angles between the field lines of a best-fit GMF model and the field lines of the input GMF model. We generate histograms of the angle differences.
An hypothetical perfect reconstruction would lead to a single bin centered in zero. The histograms for the three reconstruction cases are presented in Fig.~\ref{fig:GMFmodel2_fits_angles} for the four angles discussed above. Cases A, B, and C are shown on the same plots for comparison.

For case A and case C, the difference between the pitch and the tilt angles measured at all locations of the sampled Galaxy never exceeds one degree. The two reconstructions are very good.
For case B, the departure is slightly larger. The distributions for the whole sampled space peak at about two degree and are centered on zero, and can go up to four degrees.
We consider the reconstruction of the GMF geometry in case B as being fair given the large difference between the modeled dust density distribution
and the recovered one. Such an achievement is possible because we fit the GMF parametric model to the data using reduced Stokes parameters.

\smallskip

We can also compare the best-fit models to the input one via the differences of the polarization position angles in the maps. For this, we use the acute angle, defined between 0 and 90 degree as
\begin{equation}
\Delta_{\hat{\psi}_{\rm{r}},\, \hat{\psi}} = 90^\circ - | 90^\circ - |\hat{\psi} - \hat{\psi}_{\rm{r}} | |,
\label{eq:DPPA}
\end{equation}
where the consecutive absolute values take into account the axial nature of the polarization vectors, that is, that only their orientation are relevant. In Fig.~\ref{fig:GMFmodel2_fitsPPA} we show the difference of polarization position angles as defined before. We observe that the reconstruction is excellent for case A and very good for case C, with a distribution peaked at zero and about 95 \% of the pixels below 0.1 and 4 degrees, respectively. For case B, the results are worse with a distribution also peaked at zero but with a larger scatter (95  \% of the pixels below 20 degrees).

\begin{figure*}
\begin{tabular}{ccc}
\includegraphics[trim={0cm .5cm 0cm .7cm},clip,width=.31\linewidth]{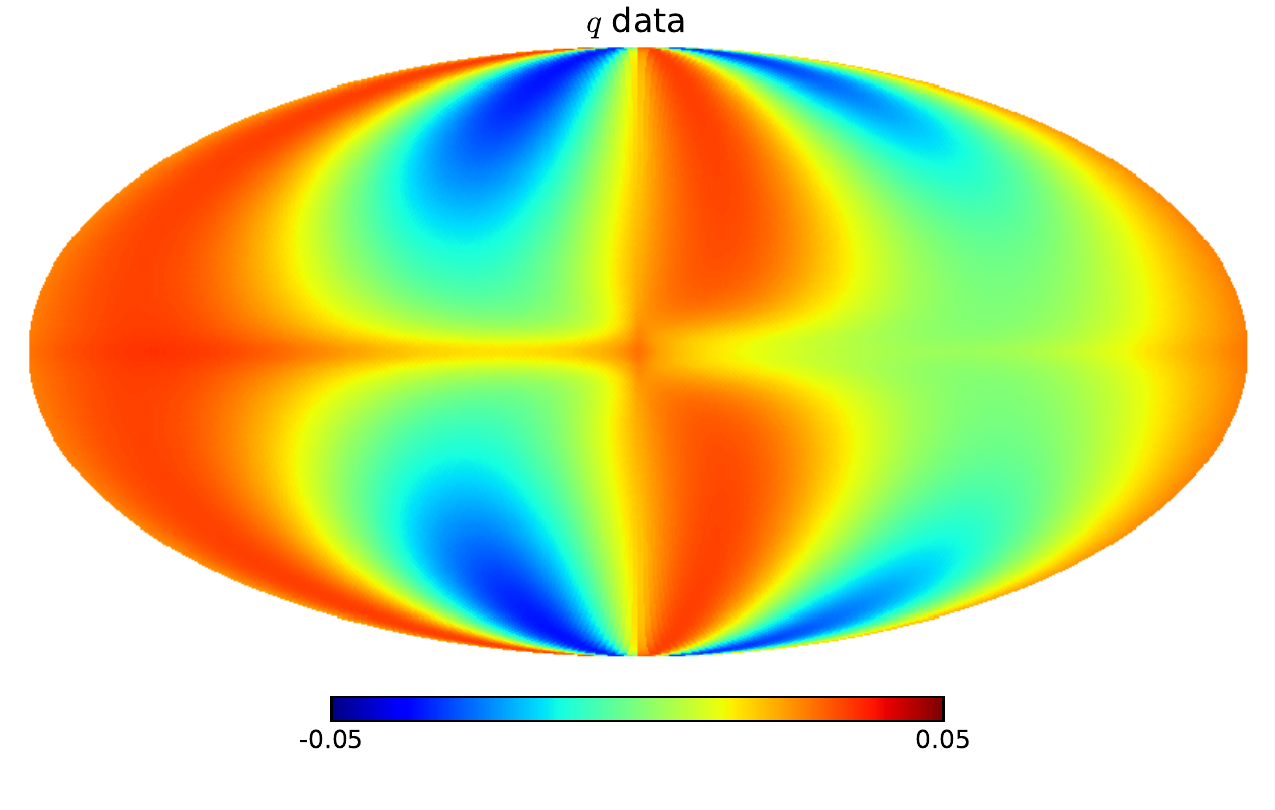}
        & \includegraphics[trim={0cm .5cm 0cm .8cm},clip,width=.31\linewidth]{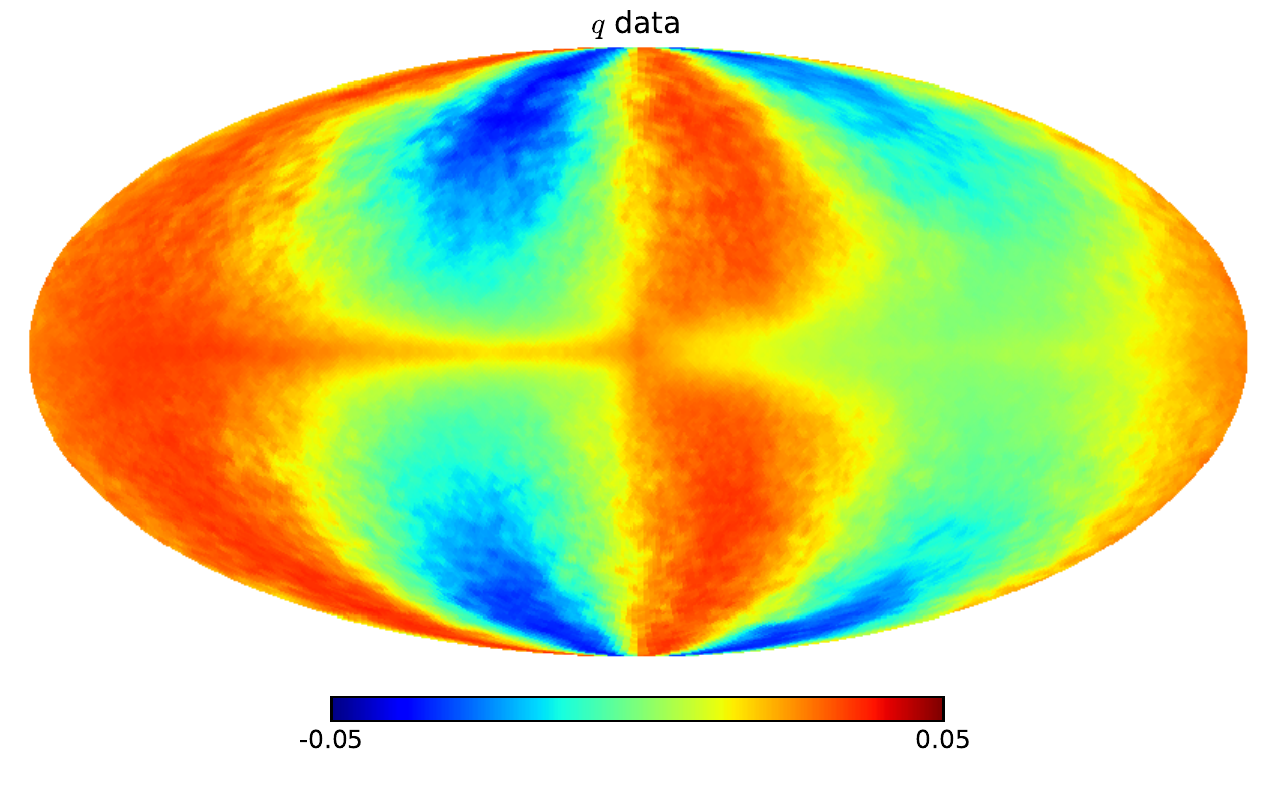}
        & \includegraphics[trim={0cm .5cm 0cm .8cm},clip,width=.31\linewidth]{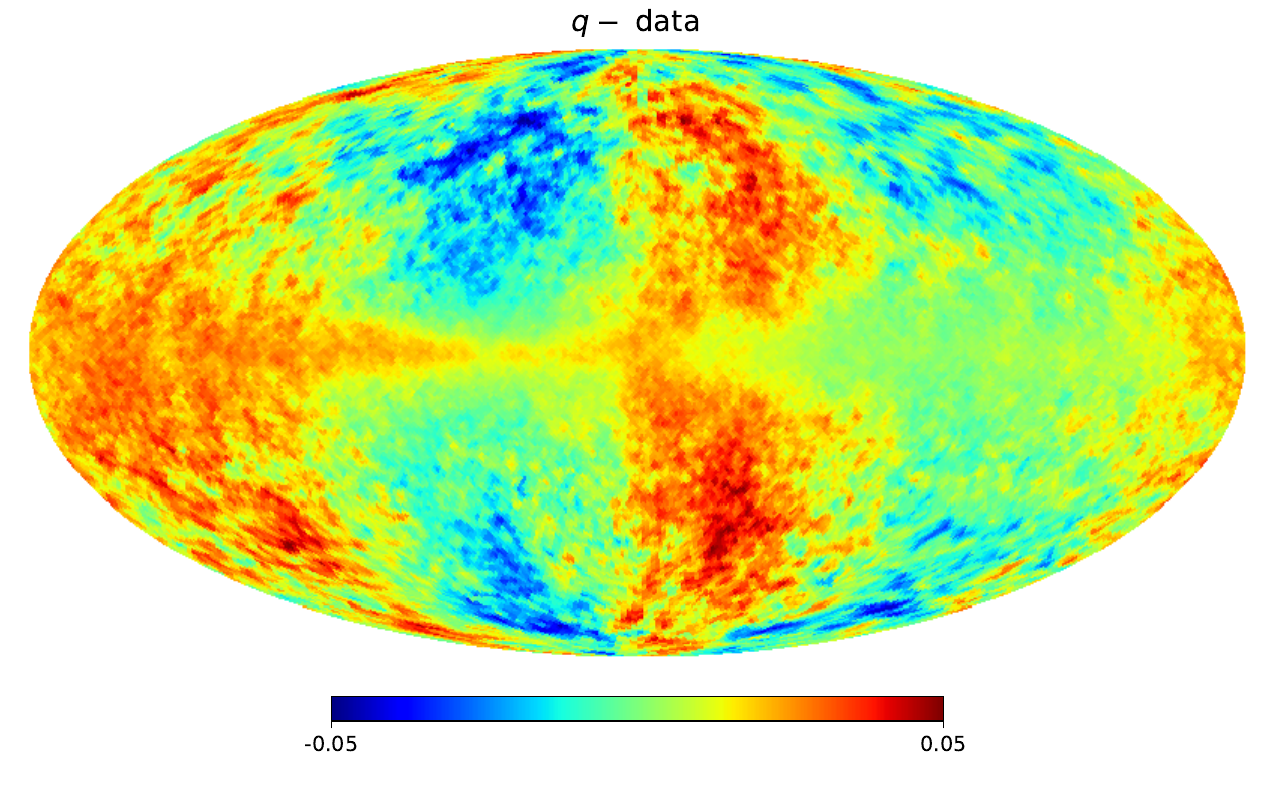} \\

\includegraphics[trim={0cm .5cm 0cm .7cm},clip,width=.31\linewidth]{_figs/1EDWMAP-q.pdf}
        &       \includegraphics[trim={0cm .5cm 0cm .8cm},clip,width=.31\linewidth]{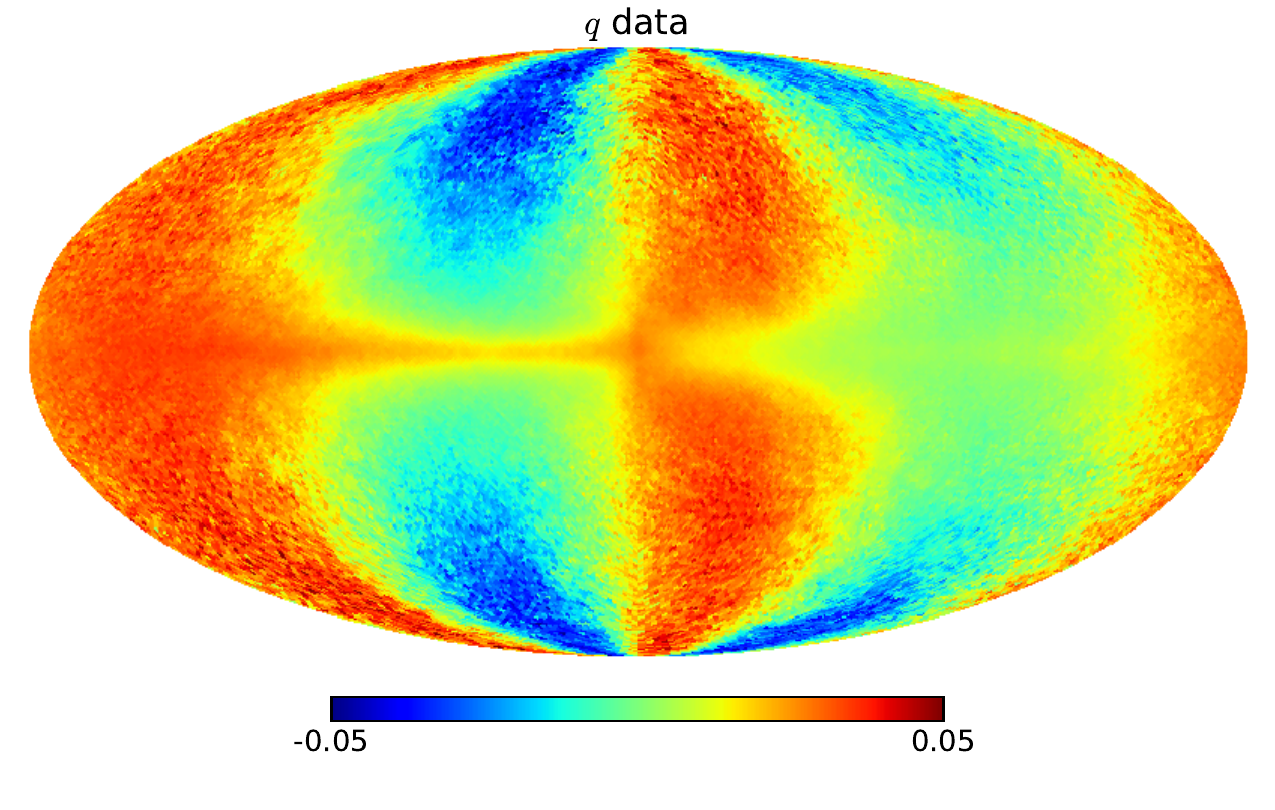}
        &       \includegraphics[trim={0cm .5cm 0cm .8cm},clip,width=.31\linewidth]{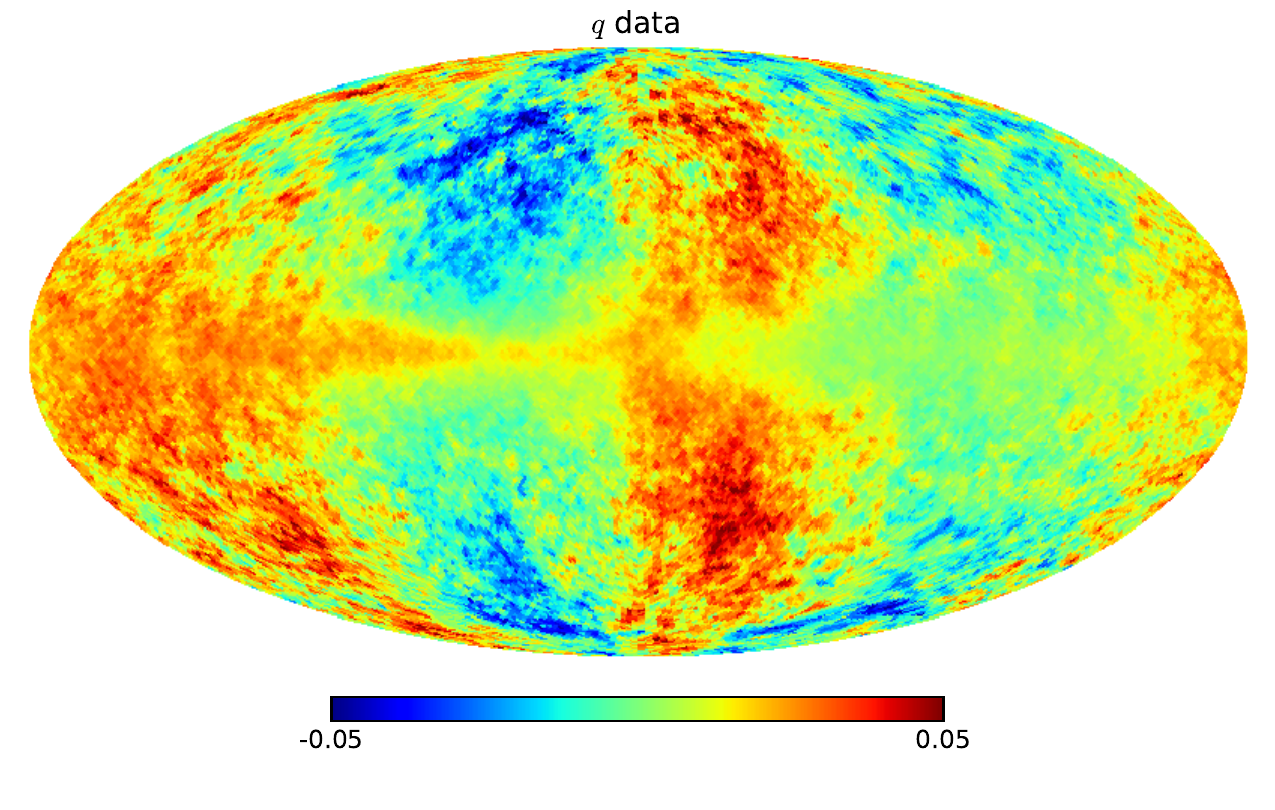}
\end{tabular}
\caption{For the \texttt{S1} simulation, we show the map of the
reduced Stokes $q$ downgraded at $N_{\rm{side}} = 64$ for the
cases with $A_{\rm{turb}} = 0.0$, $A_{\rm{turb}} = 0.3$ and
$A_{\rm{turb}} = 0.9$, from left to right. Top panels results from
regular and turbulent field; \textit{Planck} noise is added in the
bottom panels.
}
\label{fig:model1_noise-vs-turb}
\end{figure*}

\section{Impact of turbulence in the reconstruction of the regular GMF}
\label{sec:turbinthegame}

In the context of a first study, we consider in this paper only the reconstruction of the regular GMF and do not explicitly account for a stochastic component in the likelihood function. Nevertheless, in this section, we study the consequences of ignoring the turbulent component of the GMF in our analysis.

\subsection{Modeling of the turbulent component}
We consider an overall turbulent contribution so that the total 3D GMF can be written as \citep{Cha1953,Hil2009}:
\begin{equation}
\mathbf{B}^{\rm{tot}}(\mathbf{r}) =  \mathbf{B}^{\rm{reg}}(\mathbf{r}) + A_{\rm{turb}} \mathbf{B}^{\rm{turb}}(\mathbf{r}),
\label{eq:Btot_BregBturb}
\end{equation}
where the regular ($\mathbf{B}^{\rm{reg}}$) and turbulent ($\mathbf{B}^{\rm{turb}}$) GMF components are normalized so that $A_{\rm{turb}}$ represents the relative amplitude between them (e.g. $A_{\rm{turb}}=1$ means that contributions to the GMF of the regular and turbulent components are equal).
In this paper $\mathbf{B}^{\rm{reg}}$ is assumed to be one of the regular models depicted in Appendix~\ref{sec:GMFmodel} and $\mathbf{B}^{\rm{turb}}$ is assumed to be a Gaussian realization of a random isotropic 3D vector field that follows a two-slopes power-law spectrum as described in detailed in Appendix~\ref{sec:turbModel}.

Using polarization data only, \cite{Fau2011} obtained
an upper limit for the turbulent relative amplitude, $A_{\rm{turb}} \lesssim 0.25$, by combining synchrotron and dust polarization data. In more recent studies (e.g., \citealt{PlanckXIX2015}; \citealt{PlanckXLIV2016} and references therein) which include also data from pulsars, it is commonly admitted that the turbulent component might be of the same order as the regular one, $A_{\rm{turb}} \lesssim 1.0$.
Accounting for these results we limit our investigations
to the comparison of results for three different cases: $A_{\rm{turb}}=0$ (regular component only as in Sect.~\ref{sec:reconstructionGMF}), 0.3 (comparable to the upper limit of $A_{\rm{turb}}$ from \cite{Fau2011}), and 0.9 (comparable regular and turbulent components as indicated by the more recent studies). In all the cases, the turbulent component will not be considered in the fits described below. \\

In practice, assuming a model for the regular part of the GMF and a realization of our turbulent model, the regular and turbulent components of the GMF are combined at every sampled positions in the Galaxy according to Eq.~\ref{eq:Btot_BregBturb}. Then, the thermal dust emission $I$, $Q$ and $U$ Stokes parameters are computed by integrating Eq.~\ref{eq:DUSTEMISSION} considering a thermal dust density distribution model and the total GMF.
For illustration, we show in Fig.~\ref{fig:model1_noise-vs-turb} the simulated $q$ maps (downgraded at $N_{\rm{side}} = 64$) of the thermal dust emission as observed by {\it Planck} at 353~GHz when considering the turbulent component of the GMF, and, for the \texttt{S1} simulation regular GMF component. We represent the signal-only (top) and signal plus noise maps (bottom) for $A_{\rm{turb}} = 0.0$ (top), $0.3$ (middle), and $0.9$ (bottom). 

\begin{figure*}
\centering
\begin{tabular}{cccc}
        & $A_{\rm{turb}} = 0.00$        & $A_{\rm{turb}} = 0.30$        & $A_{\rm{turb}} = 0.90$ \\
        \rotatebox{90}{ \hspace{5em} Case I} &
\includegraphics[trim={0cm 0cm 0cm 0cm},clip,width=.3\linewidth]{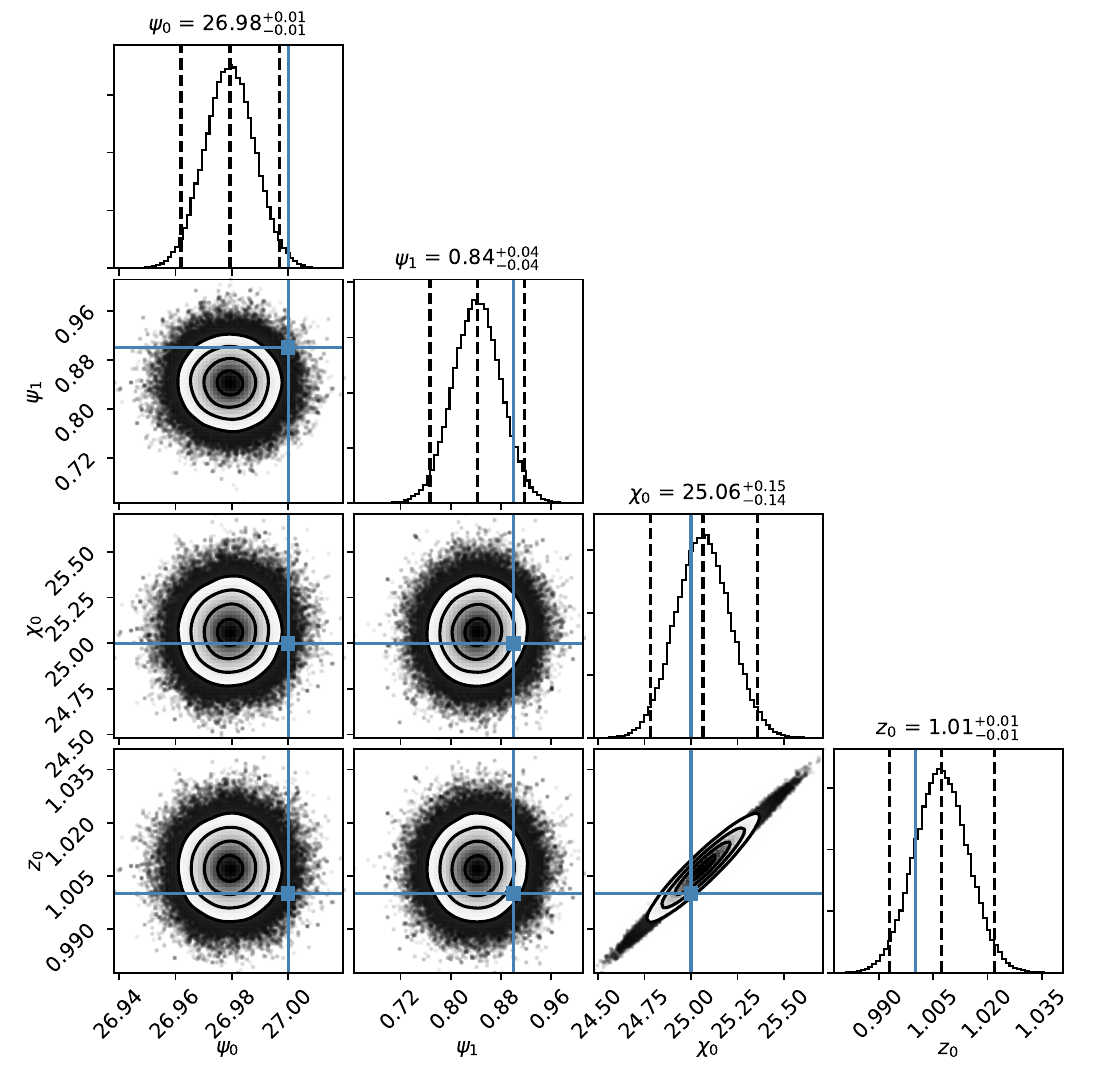} &
        \includegraphics[trim={0cm 0cm 0cm 0cm},clip,width=.3\linewidth]{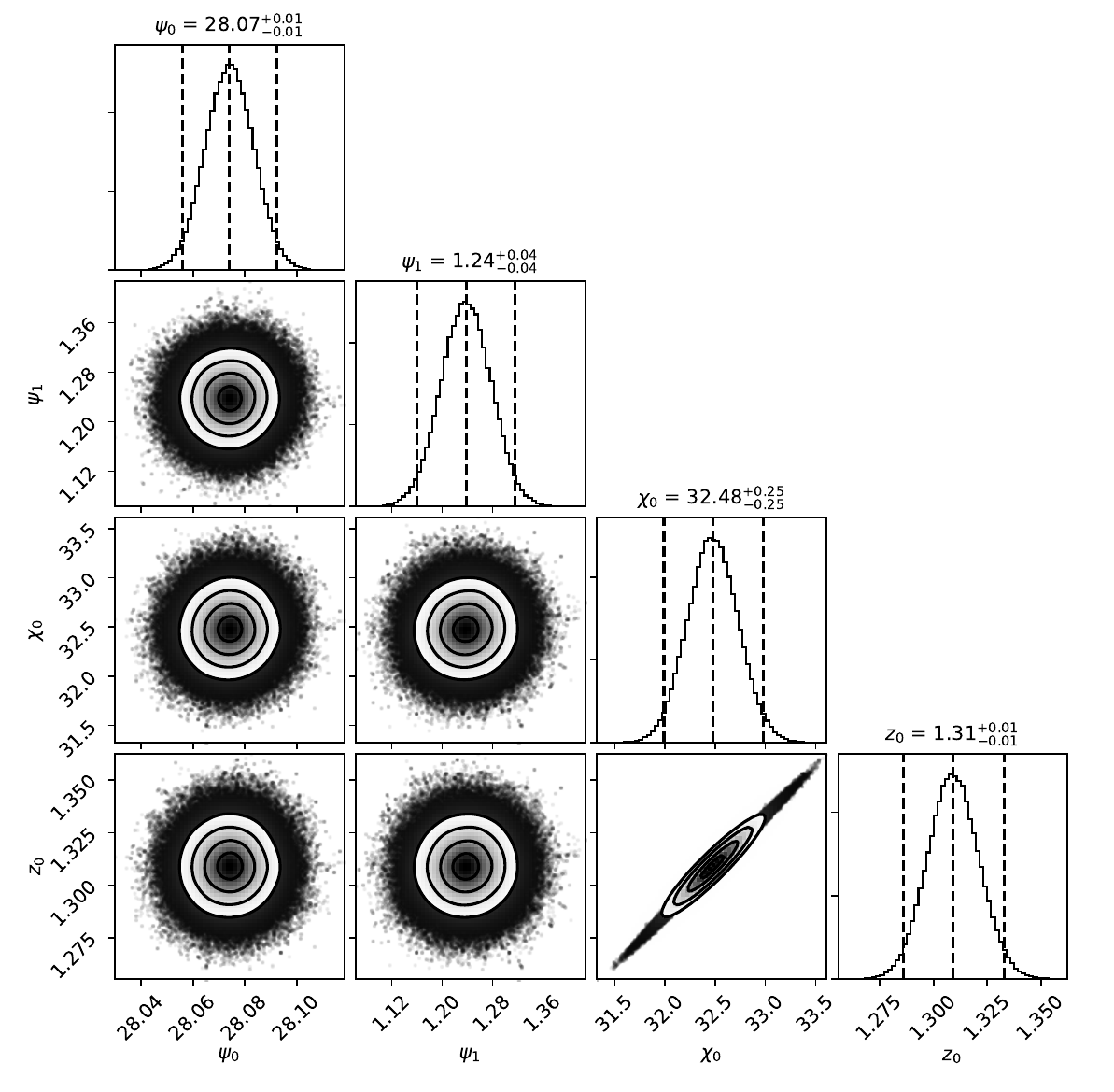} &
                \includegraphics[trim={0cm 0cm 0cm 0cm},clip,width=.3\linewidth]{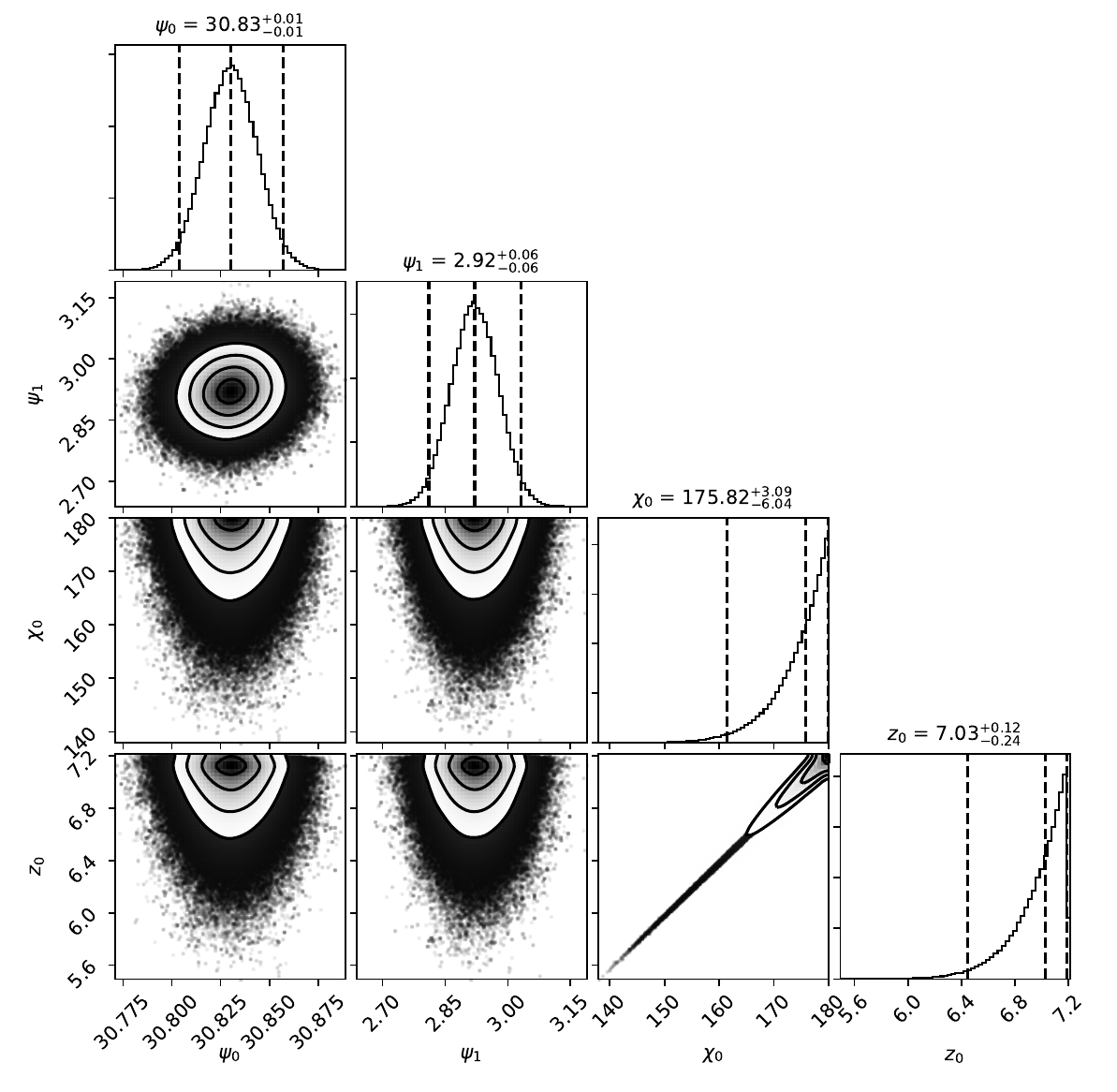}
\\
        \rotatebox{90}{ \hspace{5em} Case II}  &
\includegraphics[trim={0cm 0cm 0cm 0cm},clip,width=.3\linewidth]{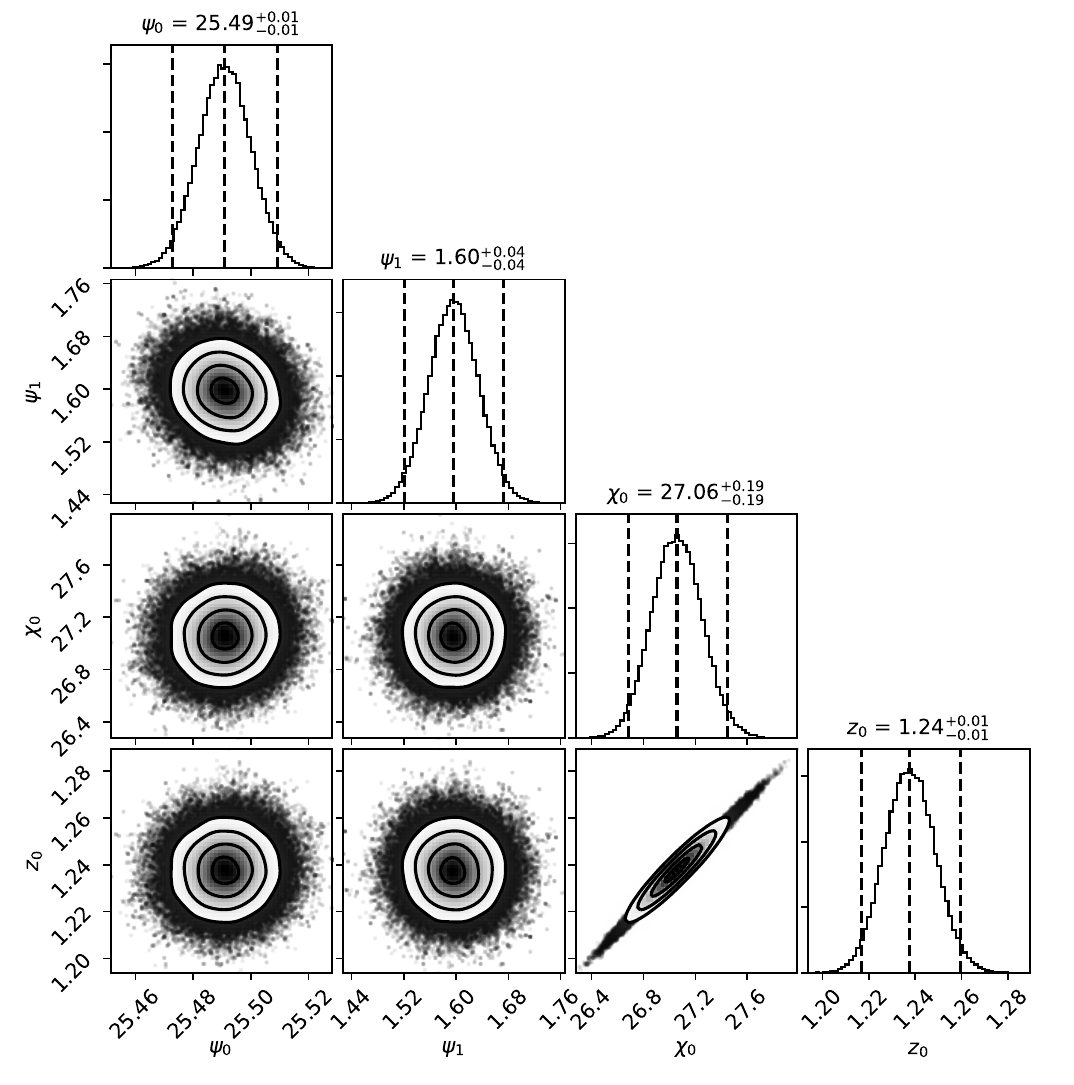} &
        \includegraphics[trim={0cm 0cm 0cm 0cm},clip,width=.3\linewidth]{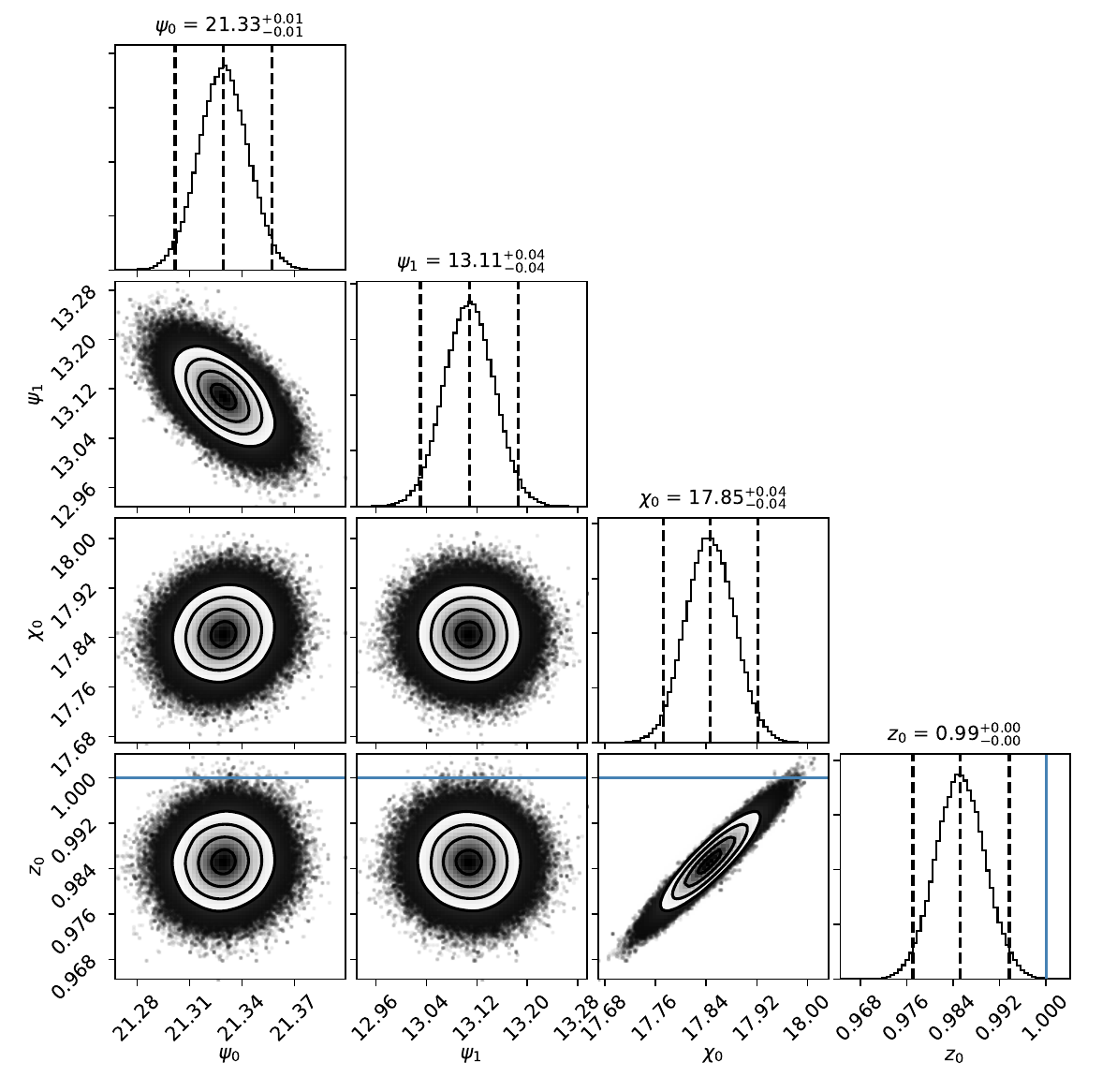} &
                \includegraphics[trim={0cm 0cm 0cm 0cm},clip,width=.3\linewidth]{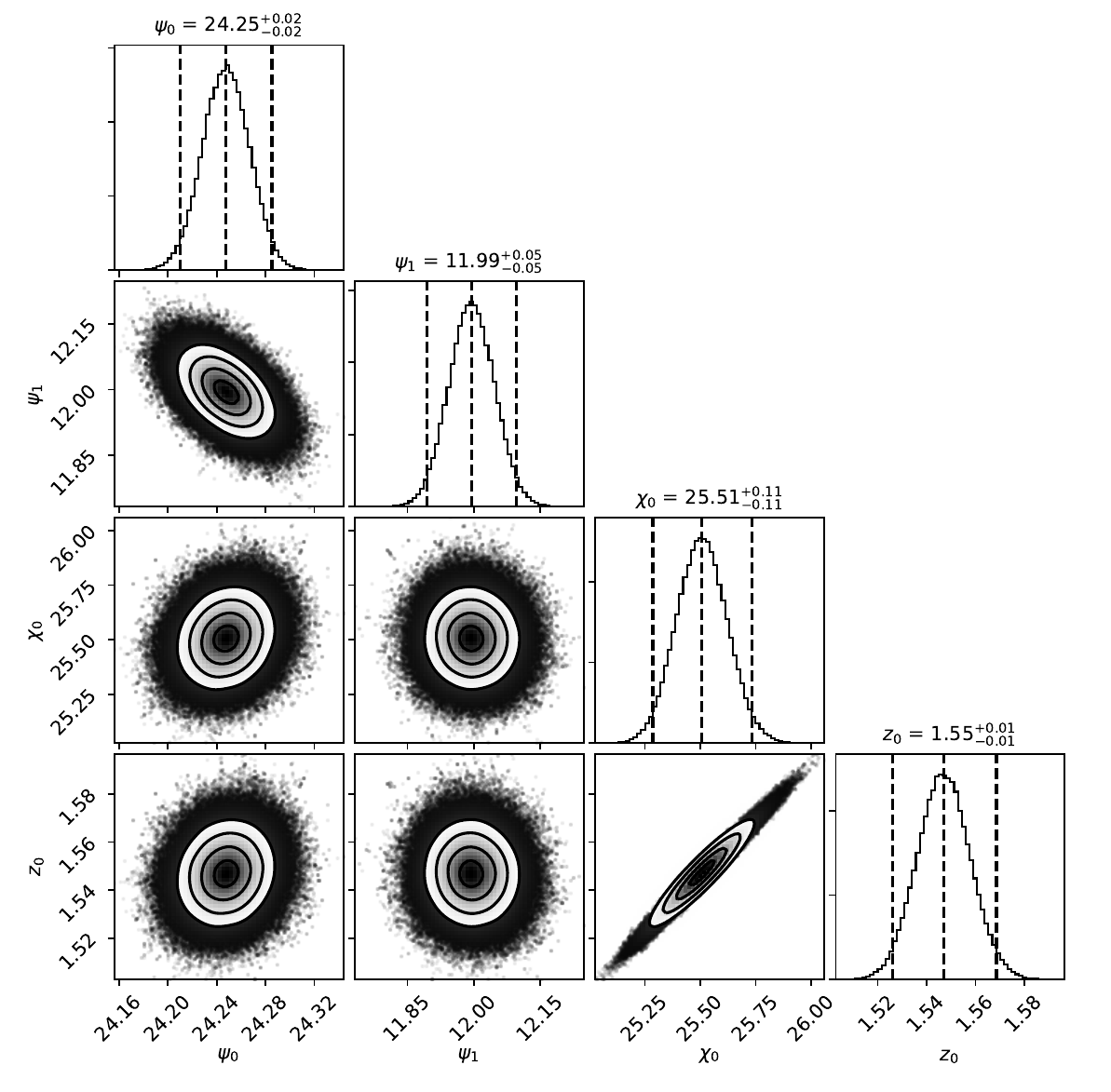}
\\
        \rotatebox{90}{ \hspace{5em} Case III}  &
\includegraphics[trim={0cm 0cm 0cm 0cm},clip,width=.3\linewidth]{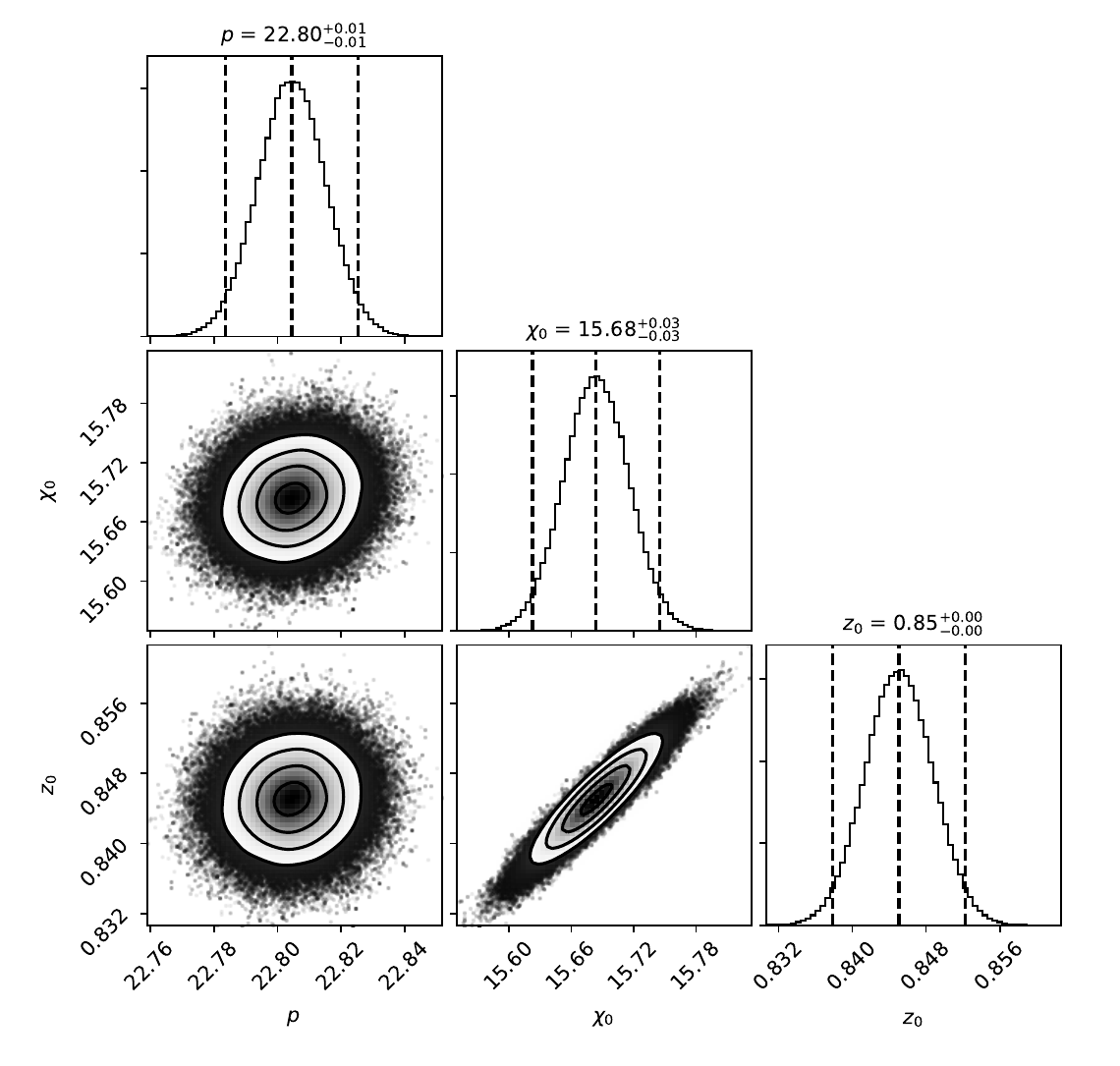} &
        \includegraphics[trim={0cm 0cm 0cm 0cm},clip,width=.3\linewidth]{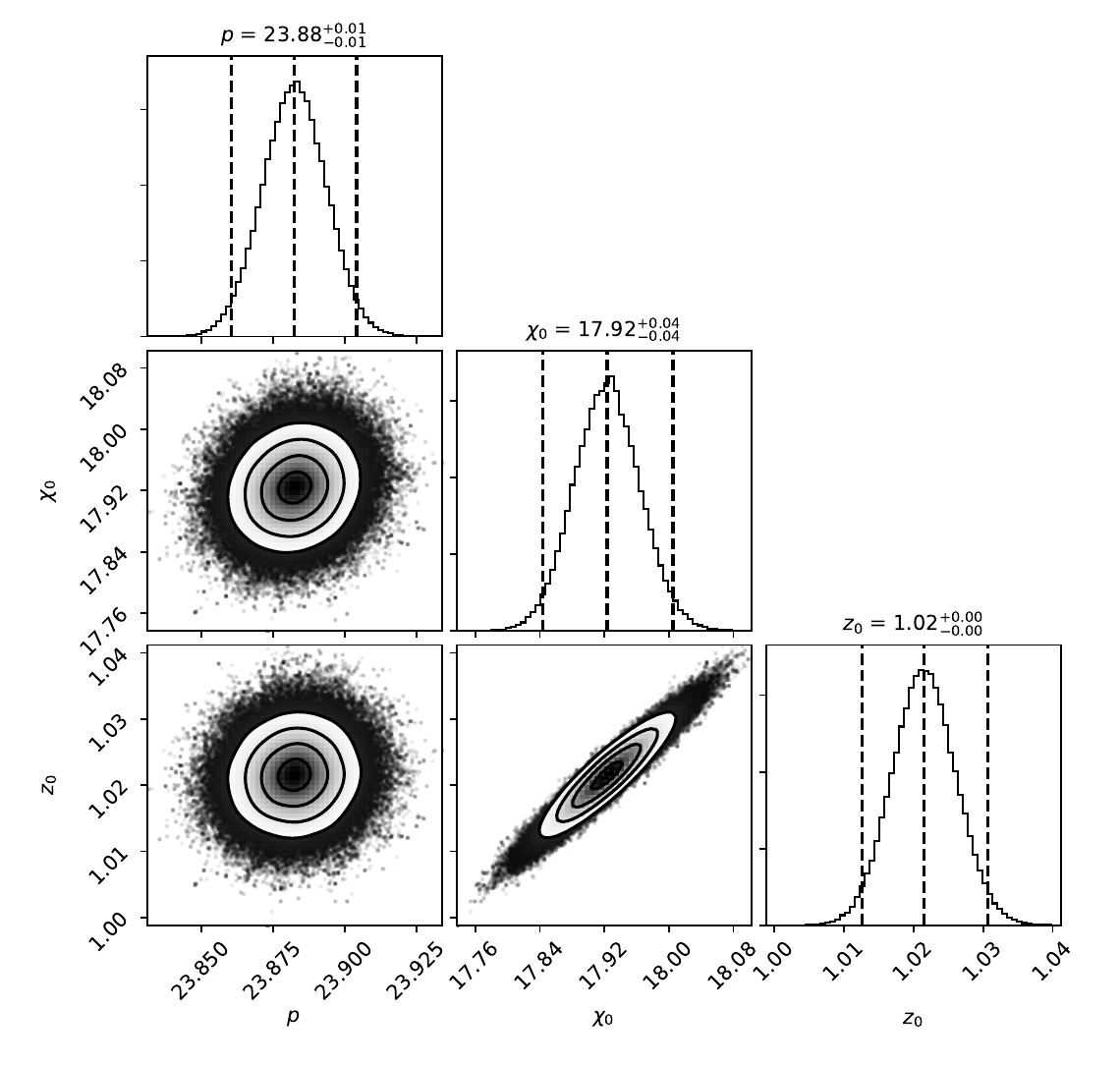} &
                \includegraphics[trim={0cm 0cm 0cm 0cm},clip,width=.3\linewidth]{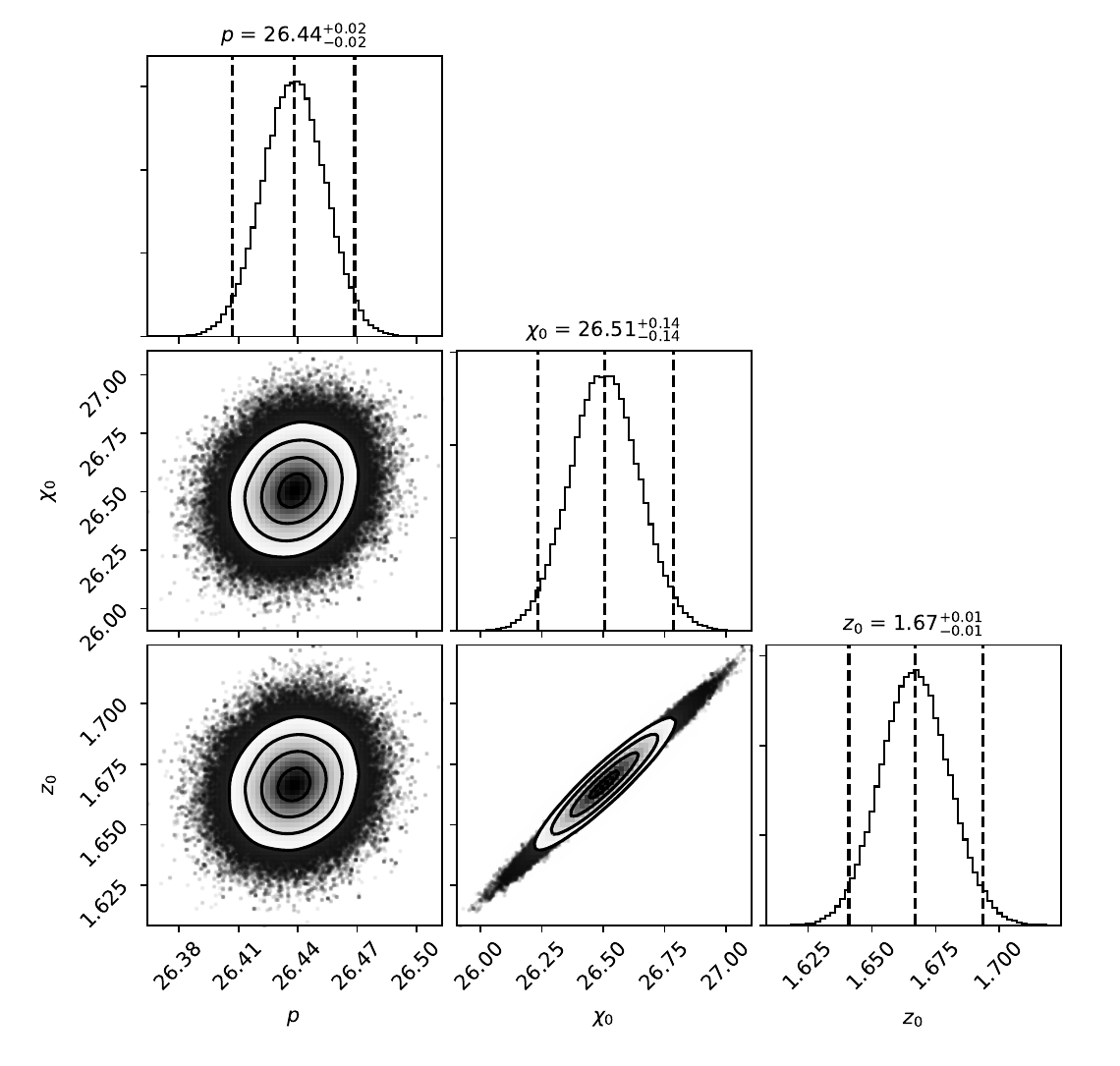}
\\
\end{tabular}
\caption{Corner plots of the 1D and 2D marginalized posterior distributions of the regular GMF parameter for the three reconstruction cases discussed in Sect.~\ref{sec:qualityrGMFturbana} applied to the \texttt{S2-turb} simulations. Angles are given in degree and the scale heights in kpc.
From left to right: Columns correspond to $A_{\rm{turb}} = 0.0$, $A_{\rm{turb}} = 0.3$ and $A_{\rm{turb}} = 0.9$, respectively.
From top to bottom for cases I to III defined as:
case I: $n_{\rm{d}} \equiv ARM4$ and $\rm{gmf} \equiv$ LSA
(both models underlying the data);
case II: $n_{\rm{d}} \equiv ED$ and $\rm{gmf} \equiv$ LSA; and
case III: $n_{\rm{d}} \equiv ED$ and $\rm{gmf} \equiv$ ASS.
The vertical and horizontal light-blue lines mark input parameter values.
}
\label{fig:GMFmodelTall_corner}
\end{figure*}

\begin{figure*}
\centering
\begin{tabular}{cccc}
        & $A_{\rm{turb}} = 0.0$         & $A_{\rm{turb}} = 0.3$         & $A_{\rm{turb}} = 0.9$ \\
        \rotatebox{90}{ \hspace{5em} pitch} &
\includegraphics[trim={.5cm 0cm .3cm 1.2cm},clip,width=.3\linewidth]{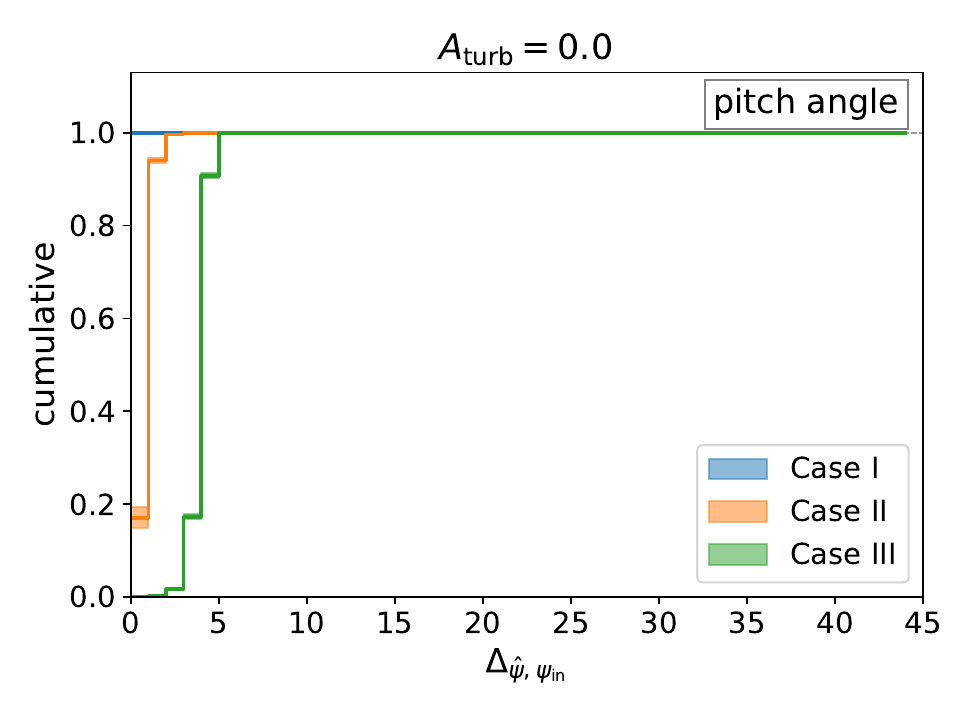} &
        \includegraphics[trim={.5cm 0cm .3cm 1.2cm},clip,width=.3\linewidth]
        {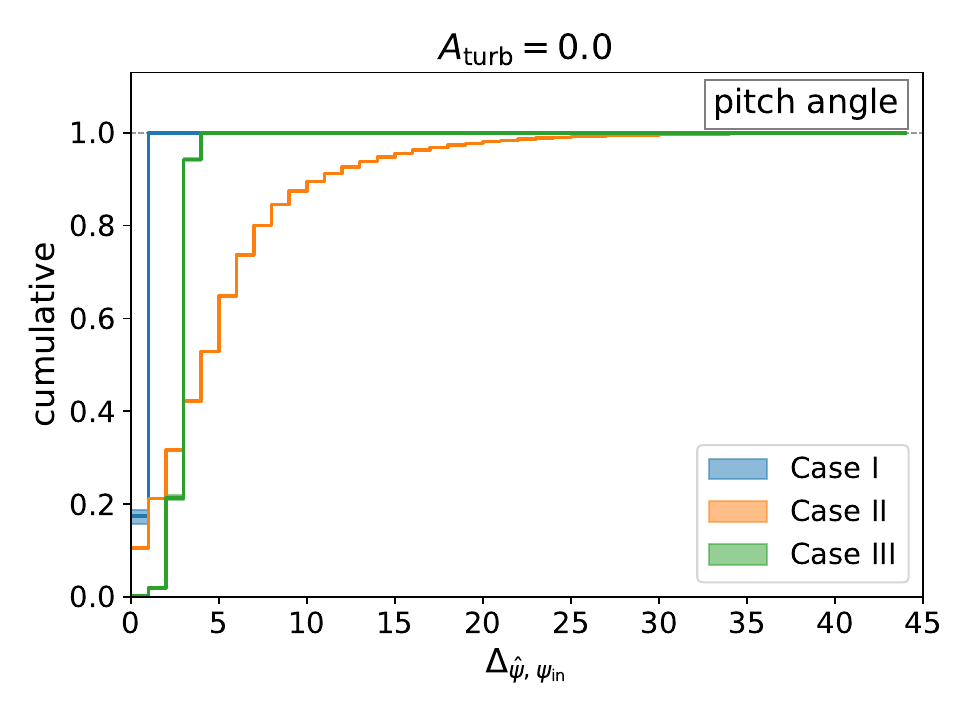} &
                \includegraphics[trim={.5cm 0cm .3cm 1.2cm},clip,width=.3\linewidth]{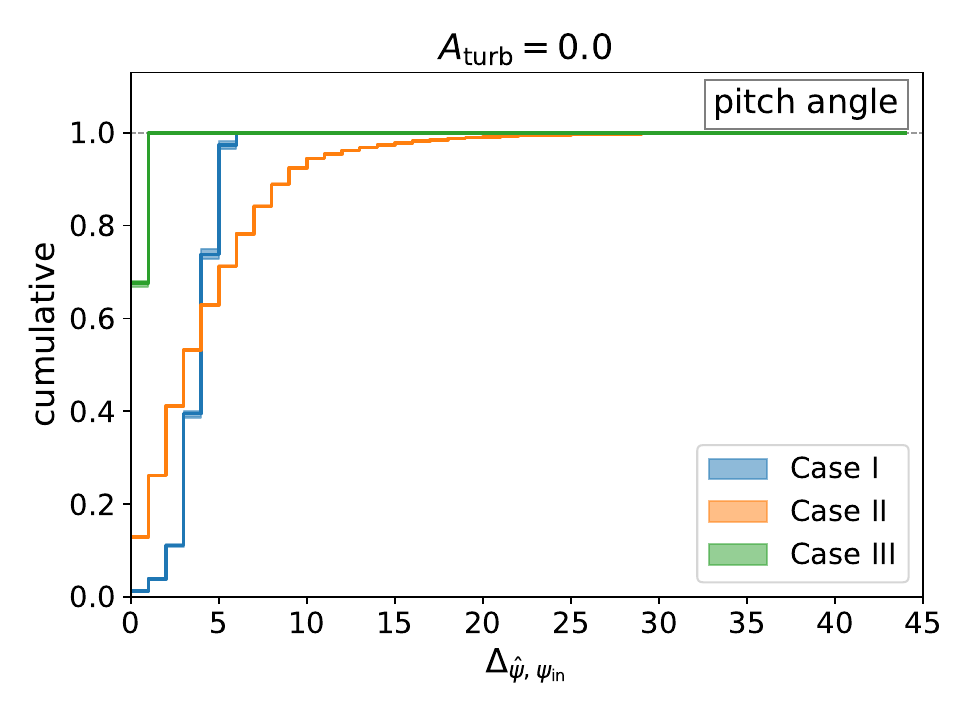}
\\
        \rotatebox{90}{ \hspace{5em} tilt }  &
\includegraphics[trim={.5cm 0cm .3cm 1.2cm},clip,width=.3\linewidth]{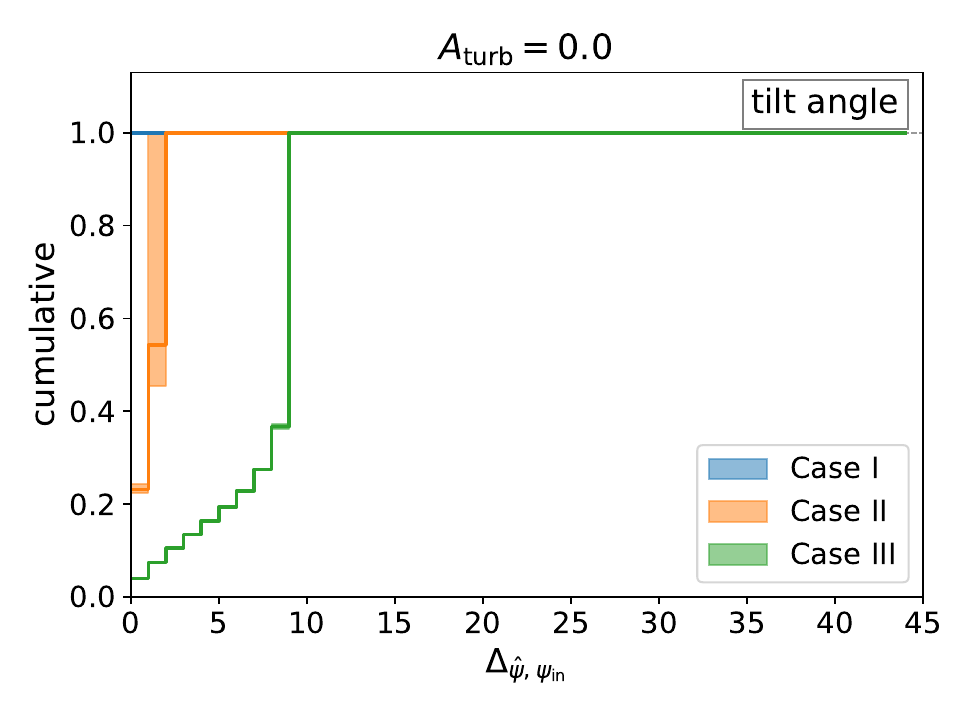} &
        \includegraphics[trim={.5cm 0cm .3cm 1.2cm},clip,width=.3\linewidth]
        {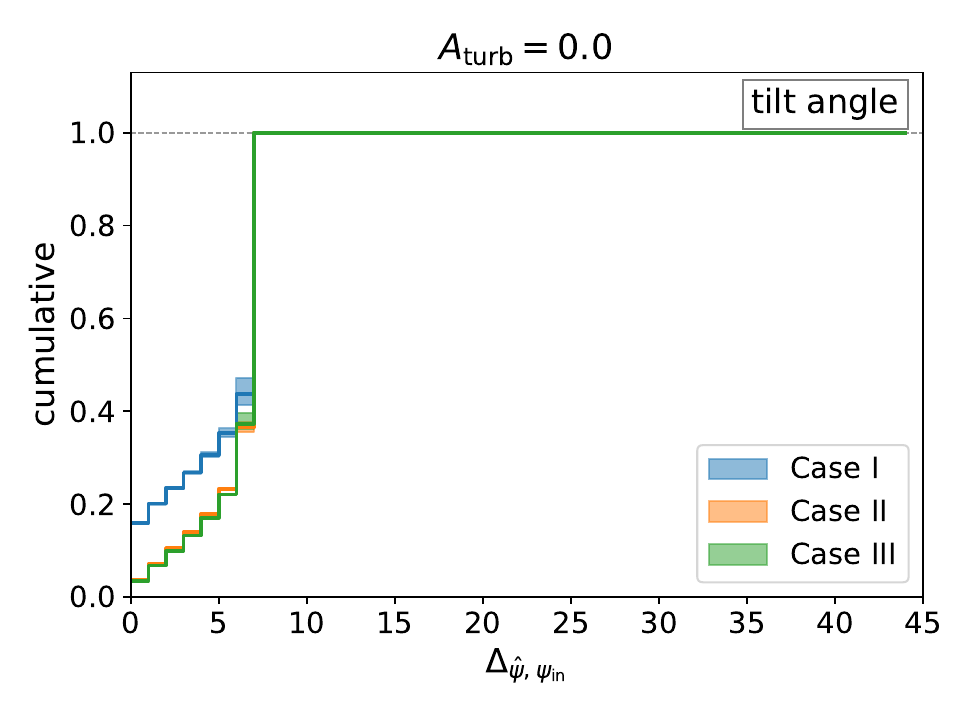} &
                \includegraphics[trim={.5cm 0cm .3cm 1.2cm},clip,width=.3\linewidth]{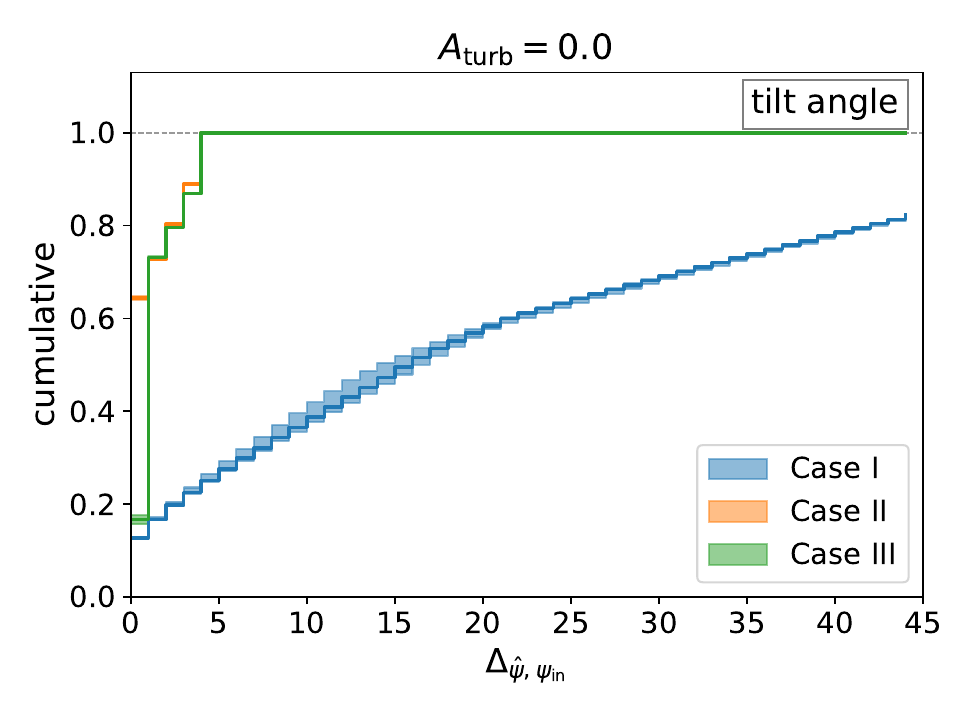}
\\
        \rotatebox{90}{ \hspace{5em} inclination }  &
\includegraphics[trim={.5cm 0cm .3cm 1.2cm},clip,width=.3\linewidth]{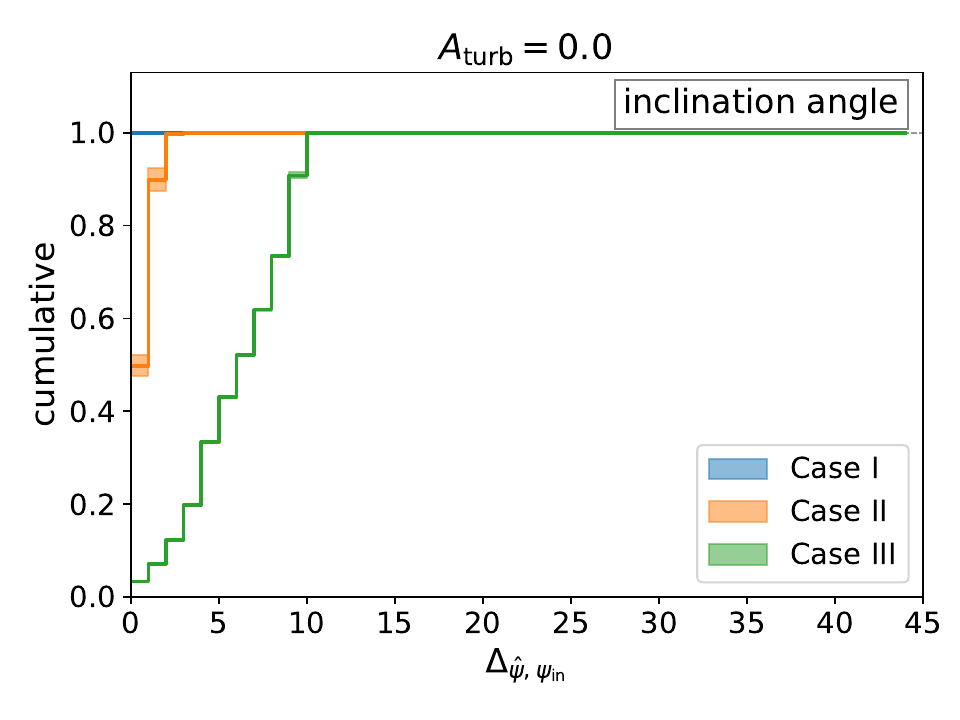} &
        \includegraphics[trim={.5cm 0cm .3cm 1.2cm},clip,width=.3\linewidth]
        {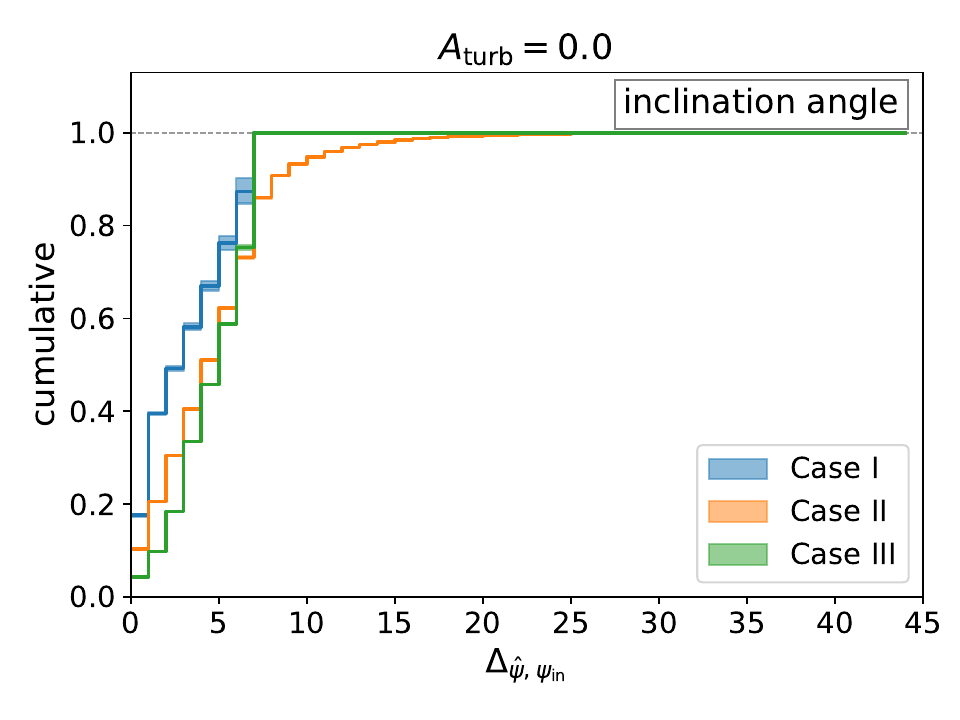} &
                \includegraphics[trim={.5cm 0cm .3cm 1.2cm},clip,width=.3\linewidth]{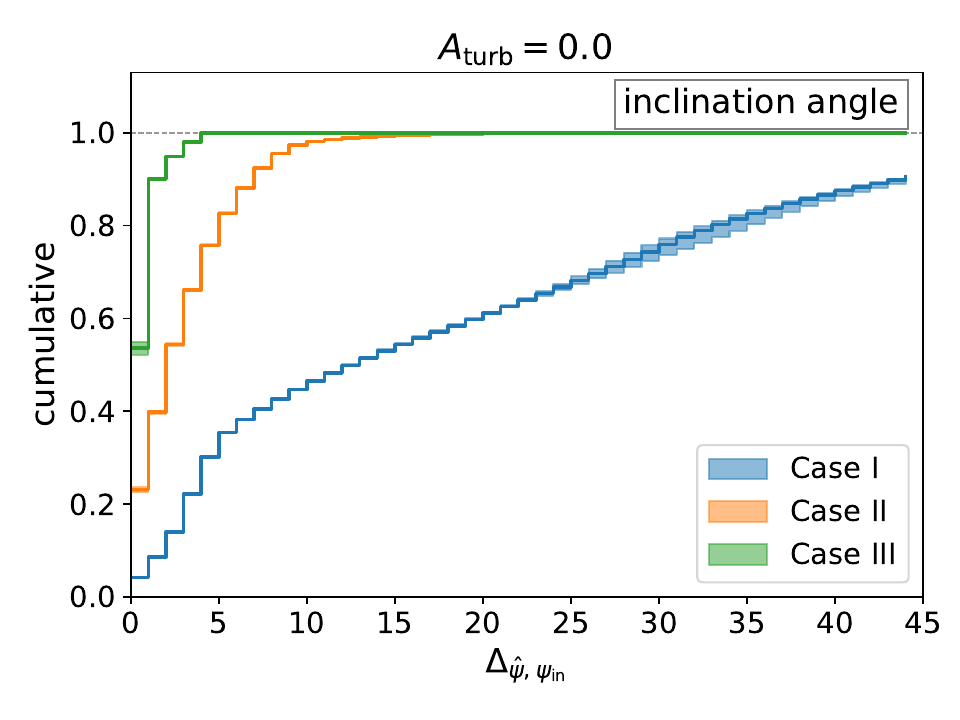}
\\
        \rotatebox{90}{ \hspace{5em} position }  &
\includegraphics[trim={.5cm 0cm .3cm 1.2cm},clip,width=.3\linewidth]{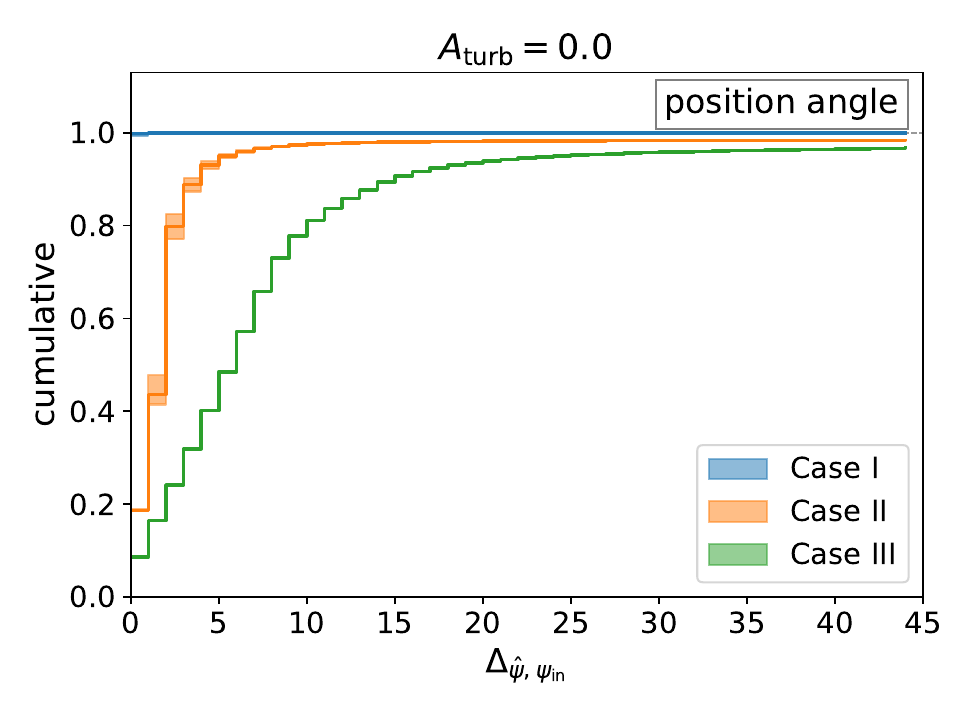} &
        \includegraphics[trim={.5cm 0cm .3cm 1.2cm},clip,width=.3\linewidth]
        {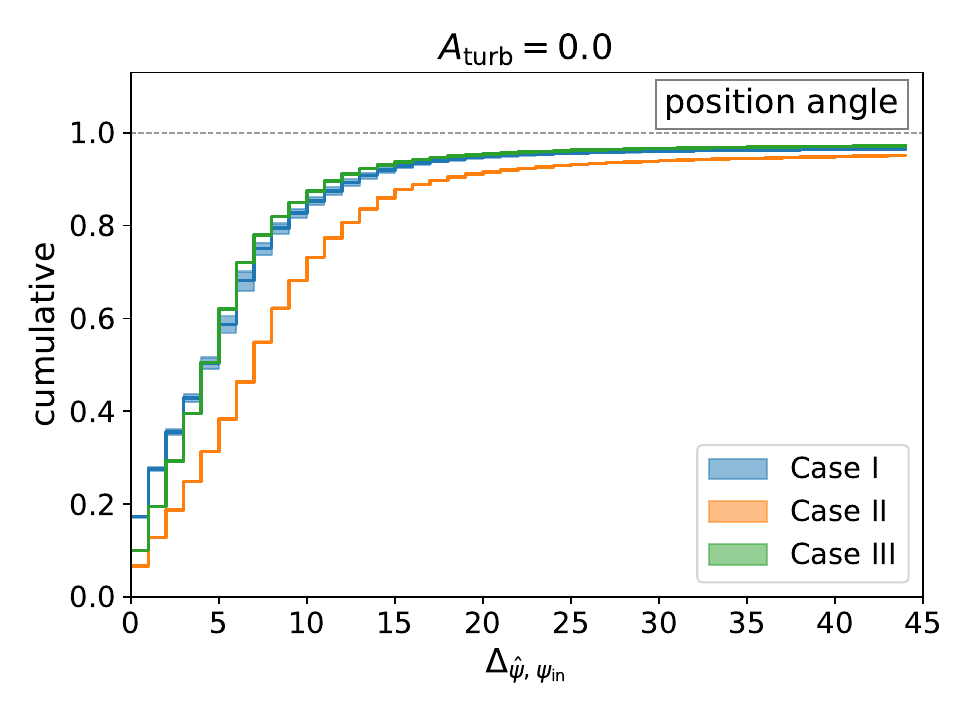} &
                \includegraphics[trim={.5cm 0cm .3cm 1.2cm},clip,width=.3\linewidth]{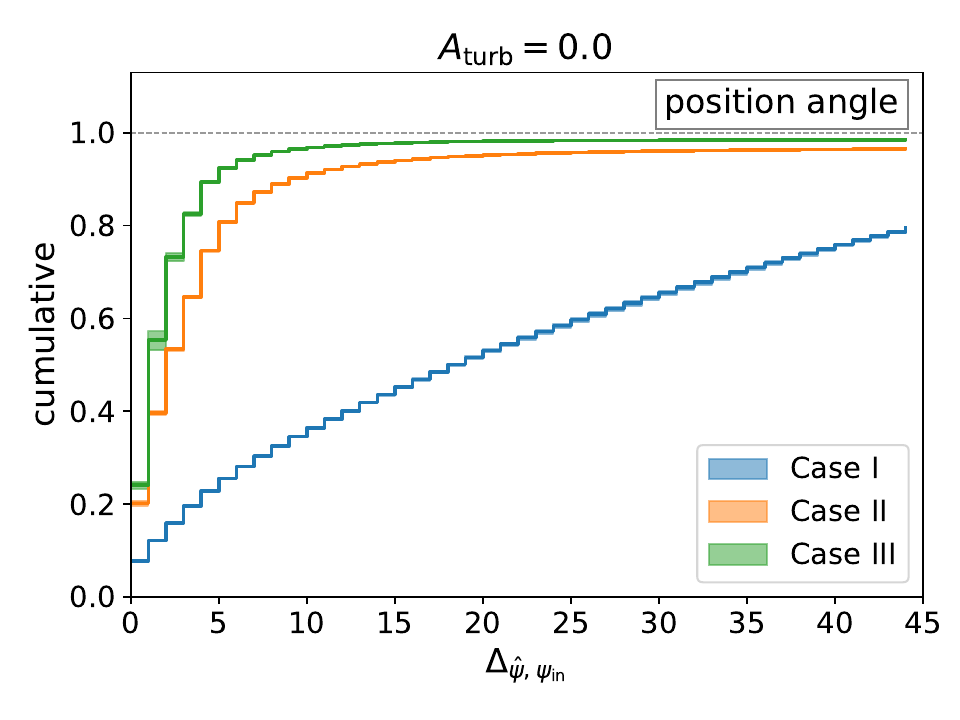}
\\
\end{tabular}
\caption{
Cumulative distributions the difference angles between the regular
GMF model in simulations \texttt{S2-turb} and the best-fit reconstructed
models using the MCMC likelihood analysis. From top to bottom for the
pitch angles, the tilt angles, the inclination angles, and the position angles.
The three reconstruction cases corresponds to different models assumed
to model the data:
case I: $n_{\rm{d}} \equiv ARM4$ and $\rm{gmf} \equiv$ LSA (both models
underlying the data);
case II: $n_{\rm{d}} \equiv ED$ and $\rm{gmf} \equiv$ LSA; and
case III: $n_{\rm{d}} \equiv ED$ and $\rm{gmf} \equiv$ ASS.
From left to right, the columns correspond to $A_{\rm{turb}} = 0.00$,
$A_{\rm{turb}} = 0.30,$ and $A_{\rm{turb}} = 0.90$. The shaded area
corresponds to the 68 \% CL extracted from the posterior of the
probability distribution as sampled from the MCMC analysis.
\label{fig:GMFmodelTall_fits_angles_cum}}
\end{figure*}

\subsection{Quality of the reconstruction of the regular GMF in the presence of turbulence}
\label{sec:qualityrGMFturbana}
We rely on the \texttt{S2-turb} (see Sect.~\ref{sec:simu}) sets of simulated maps to estimate the uncertainties induced in the reconstruction of the large-scale regular component of the GMF by ignoring the turbulent component.
We apply our MCMC procedure, which does not account for the turbulent component, on these three sets of polarization maps for three different cases: 
\begin{enumerate}[]
\item   Case I:  $n_{\rm{d}} \equiv ARM4$ and $\rm{GMF} \equiv$ LSA,
\item   Case II:  $n_{\rm{d}} \equiv ED$ and $\rm{GMF} \equiv$ LSA,
\item   Case III:  $n_{\rm{d}} \equiv ED$ and $\rm{GMF} \equiv$ ASS,
\end{enumerate}
where ASS is a more constrained version of the LSA model for which the opening angle of the spiral is fixed to a constant (see Appendix~\ref{sec:GMFmodel}). \\

In Fig.~\ref{fig:GMFmodelTall_corner}, we show the 1D and
2D marginalized posterior distribution obtained while fitting the \texttt{S2-turb} simulations following cases I, II, and III.
From left to right, we show the results for the three turbulent contributions $A_{\rm{turb}} = 0.0,\,0.3,\,0.9$.

For case I (first row) without turbulence ($A_{\rm{turb}} = 0.0$, thus corresponding the case C of Sect.~\ref{sec:Iqu_fit}), the input parameters for the regular GMF model are recovered well inside the 95\% CL of the posterior distribution.
However, this is not the case when turbulence is included and we observe a bias in the recovered regular GMF model parameters with respect to the input ones.
Significant biases are also observed for case II and case III both for the cases with and without turbulence.
Interestingly, though, we see that differences in the $\psi_{0}$ angle values are only of few degrees for all cases.
Inspection of Fig.~\ref{fig:GMFmodelTall_corner} also reveals that the biases in recovered parameter values induced by the non-consideration of the turbulence is roughly of the same order of the one induced by the mismodeling of the density distribution or of the mismodeling of the regular part of the GMF. In this paper, we concentrate in the latter effect and do not attempt to fit for the turbulence component in the likelihood.

In Fig.~\ref{fig:GMFmodelTall_fits_angles_cum}, we present the
cumulative histograms of the angle differences (in degrees) measured at every location of the Galactic sampled space between the input regular GMF model in the \texttt{S2-turb} simulations and the reconstructed ones obtained from the Cases I, II, and III (top to bottom).
From left to right we show results for the three turbulent contributions $A_{\rm{turb}} = 0.0,\,0.3,\,0.9$. Uncertainties at the 68 \% C.L. are overplotted as shaded regions and are computed by sampling the posterior MCMC probability distributions.
We observe that the direction of the regular GMF component can be reconstructed to a good precision, better than 10 degrees, in 70-80\% of the space even in the presence of a moderate amount of turbulence and for the three cases considered.
However, when the fractional amount of turbulence is high, $A_{\rm{turb}} = 0.9$, the reconstruction degrades significantly. 
Nevertheless, for cases II and III we observe a surprisingly good
reconstruction when comparing to case I.
This may point to the conclusion that using an incorrect model for the dust density allows to better reconstruct the 3D geometry of the large-scale regular part of the GMF. Such a result requires further investigation and might be an unexpected advantage of reconstructing the regular GMF using the reduced Stokes parameters.
However, the fact that the reconstruction in case III is better than in case II is expected. Indeed, the LSA model is a generalization of the ASS model and the LSA model realization in the simulations is not very different from an ASS model. Therefore, the regular GMF model that is adjusted in case III leaves less degrees of freedom for the field lines but constrains the field to be close to the simulated one.
We verify this interpretation by fitting another regular GMF model that is intrinsically different from the LSA. In that case we obtained a poor reconstruction of the 3D geometry of the GMF. Nevertheless, we observe that reliable constraints can be obtained for the pitch angle even for the case of comparable regular and turbulent components, which is favored by current studies.

\smallskip

As expected the effect of not considering the turbulent part in the modeling induces a bias in the parameter space. This bias appears to be very significant due to the large number of data points, the relatively small error values and, more importantly,
that uncertainties from mismodeling or incomplete modeling are not included in our log-likelihood definition and so, not included in our posterior distributions.
This stresses the need for introducing the turbulent component in the modeling.
Furthermore, inspection of both the marginalized posterior distributions and the cumulative histograms reveals that the bias in the model-parameter space that is induced by the non-consideration of the turbulence is roughly of the same order of the one induced by the mismodeling of the density distribution or of the mismodeling of the regular part.
This shows that improvements in the modelings are mandatory both on the regular and turbulent contributions to the GMF. \\

\begin{figure*}[t]
\centering
\begin{tabular}{cccc}
\includegraphics[width=.22\linewidth]{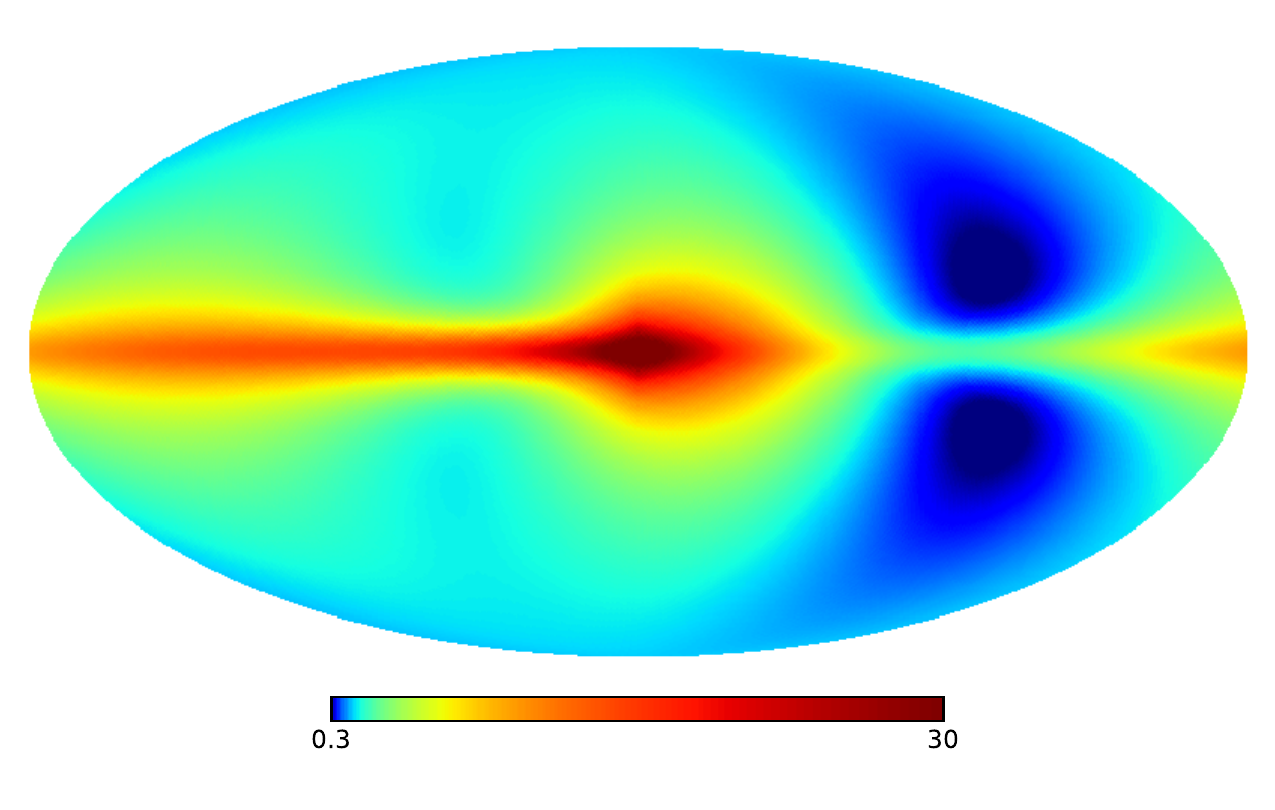} &
        \includegraphics[width=.22\linewidth]{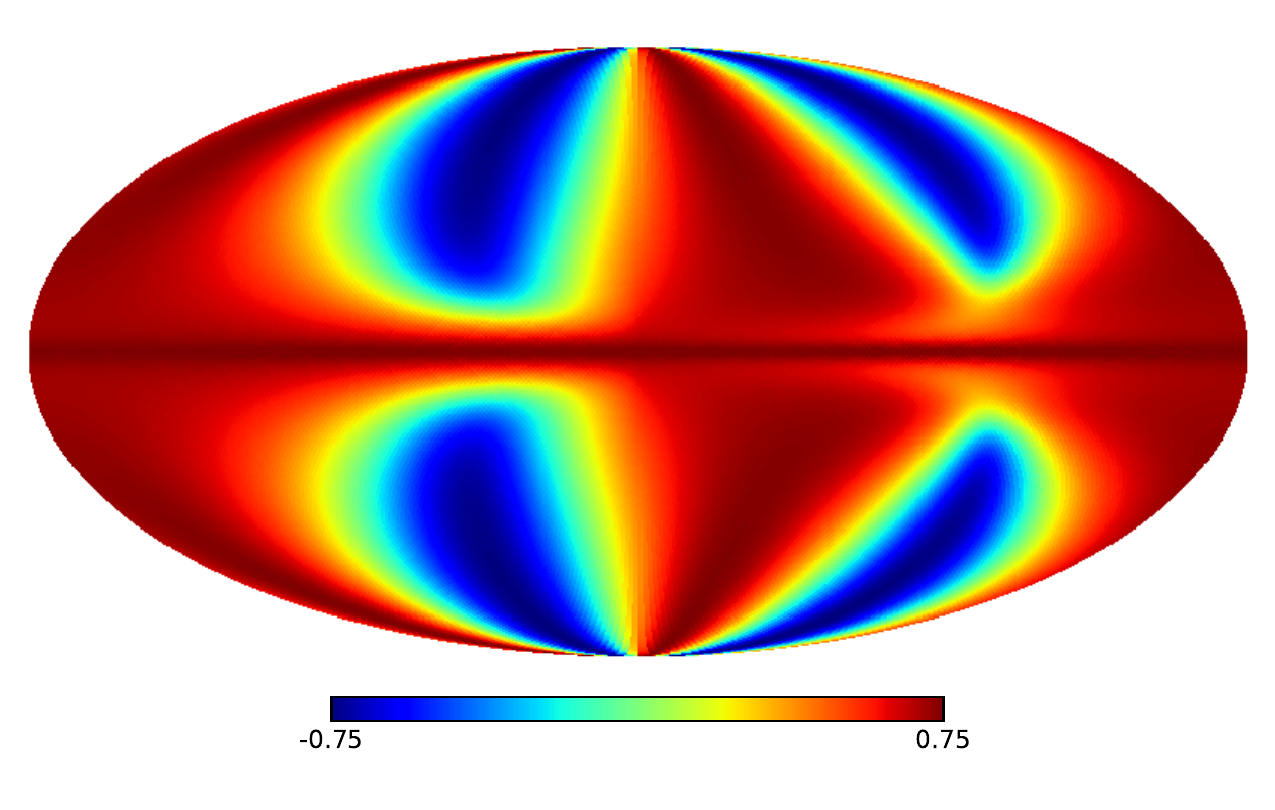} &
                \includegraphics[width=.22\linewidth]{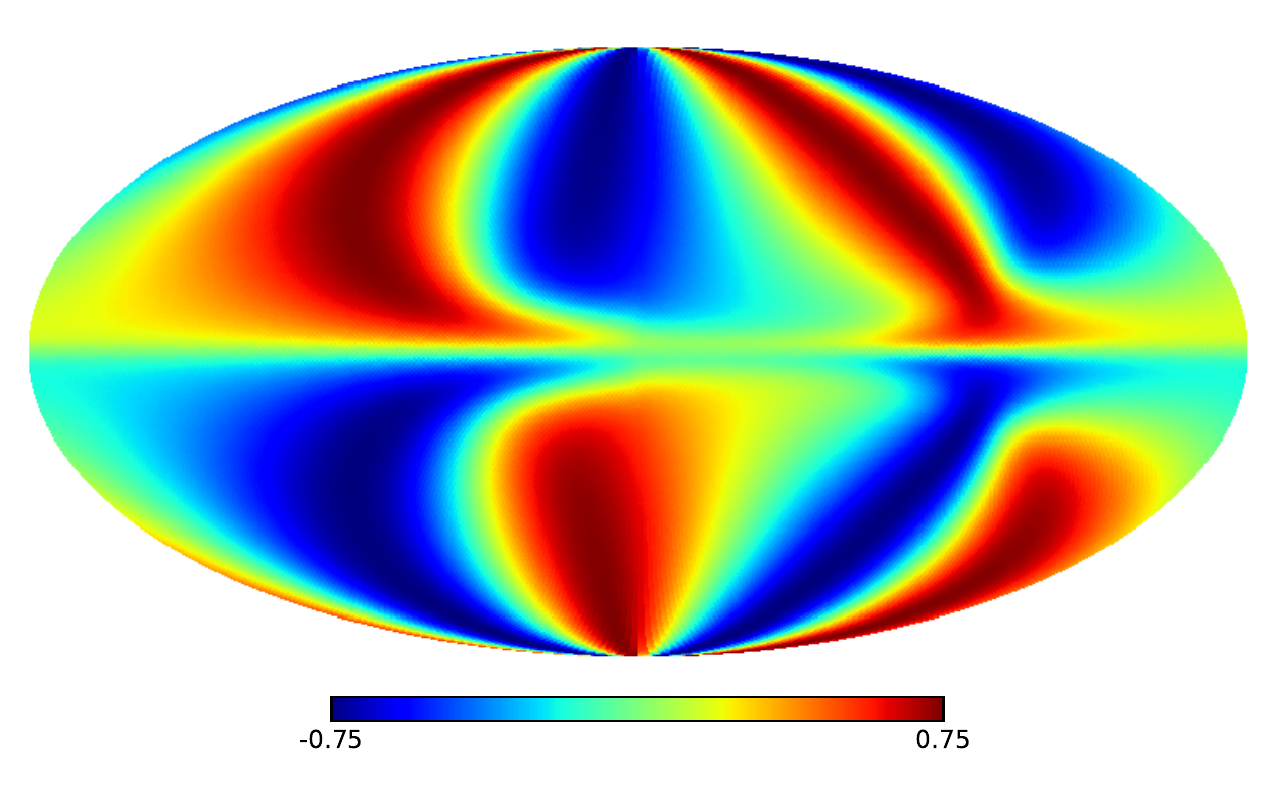} &
                        \includegraphics[width=.22\linewidth]{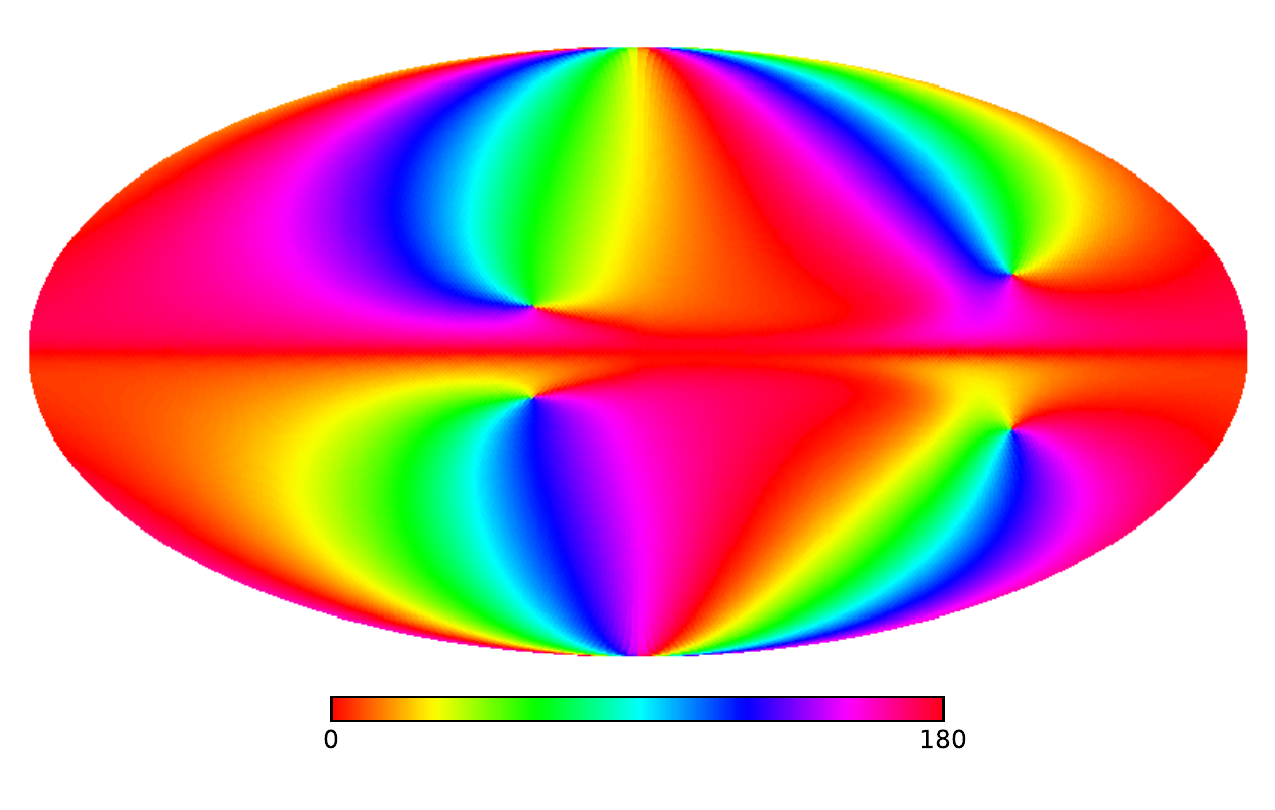}\\

\includegraphics[width=.22\linewidth]{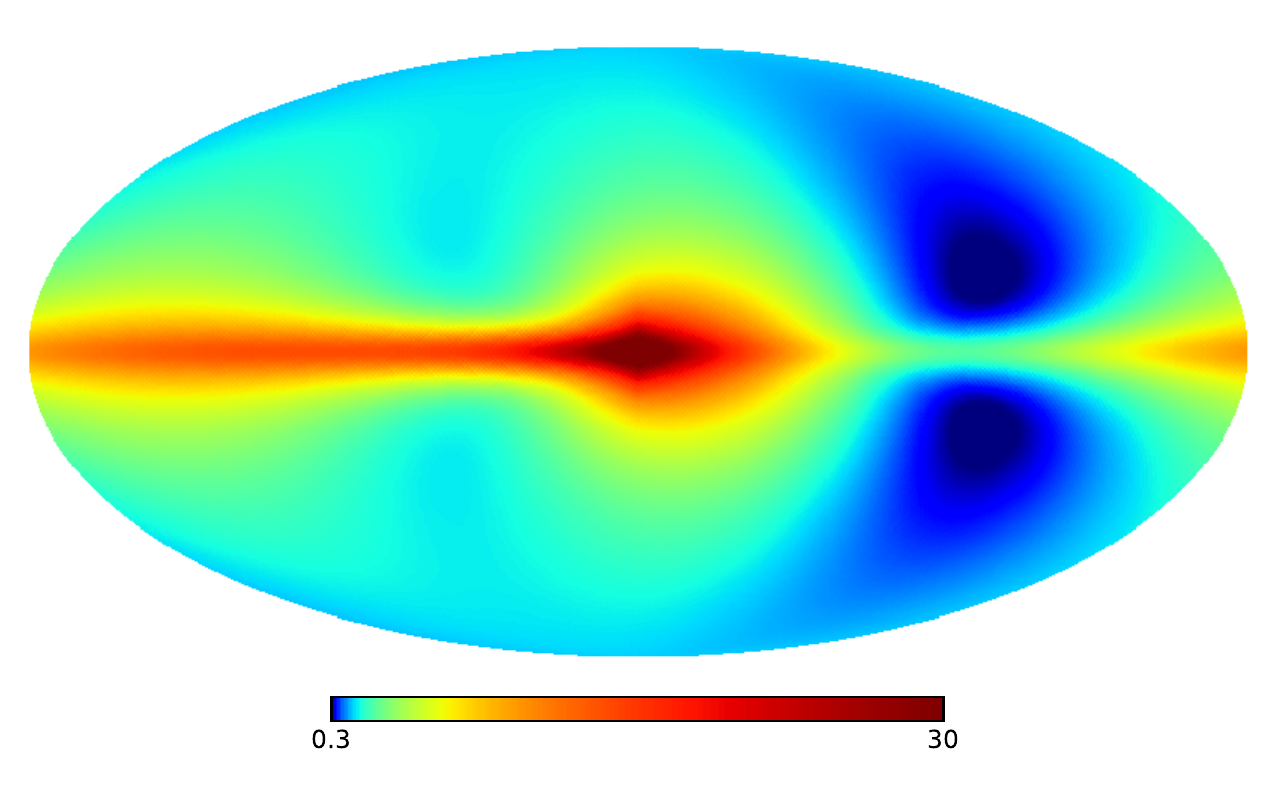} &
        \includegraphics[width=.22\linewidth]{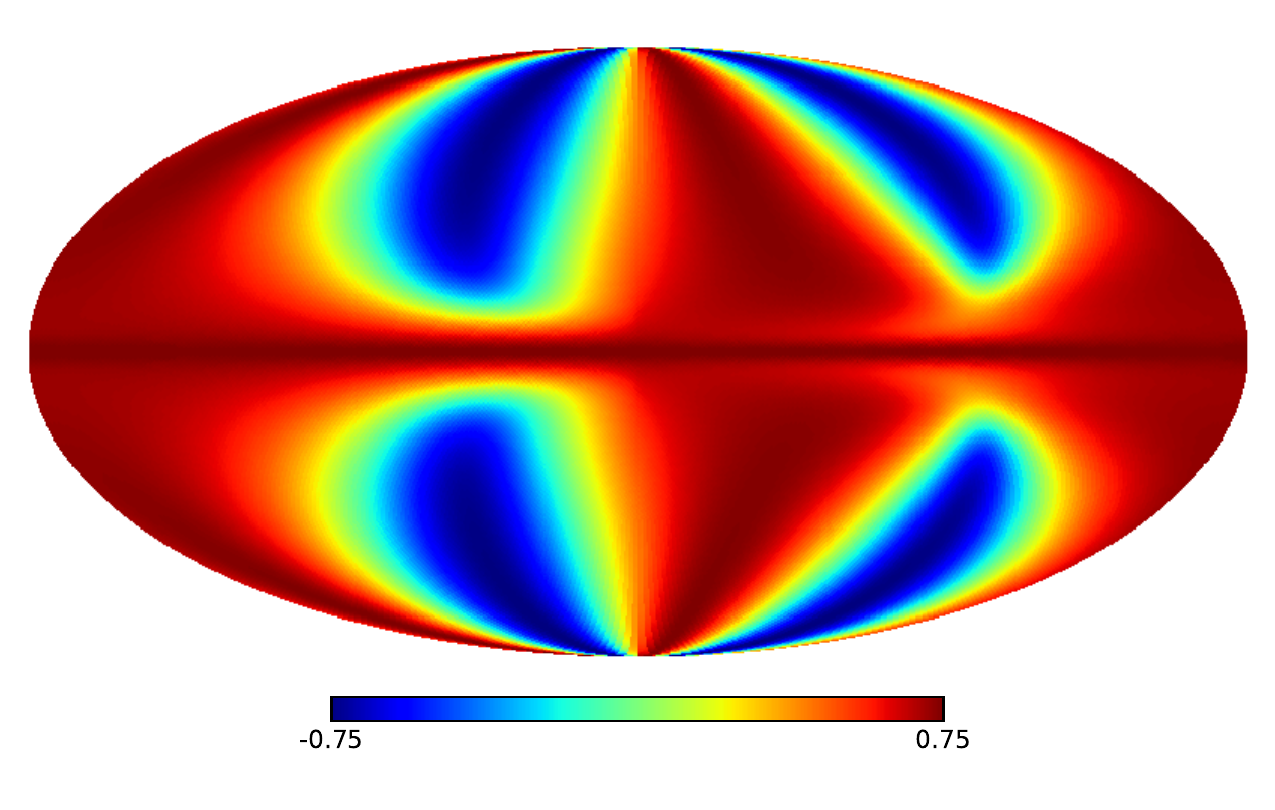} &
                \includegraphics[width=.22\linewidth]{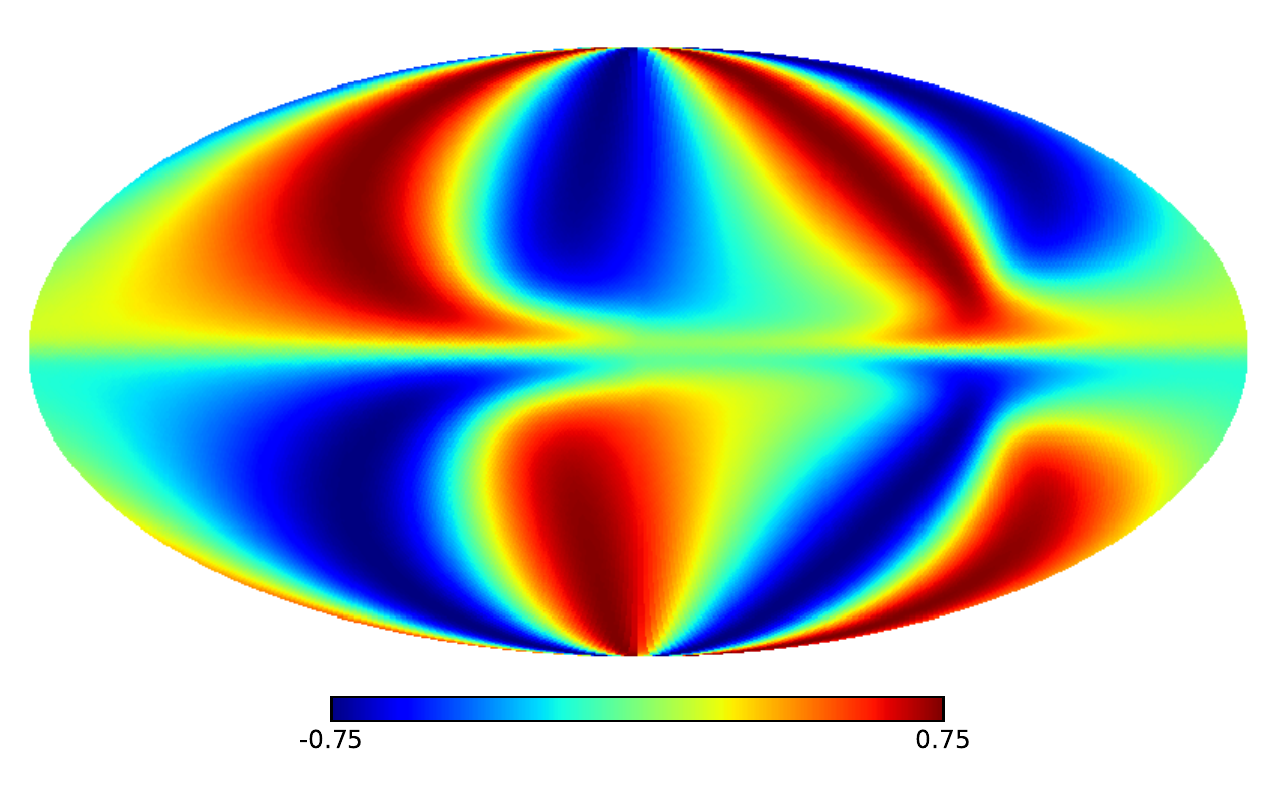} &
                        \includegraphics[width=.22\linewidth]{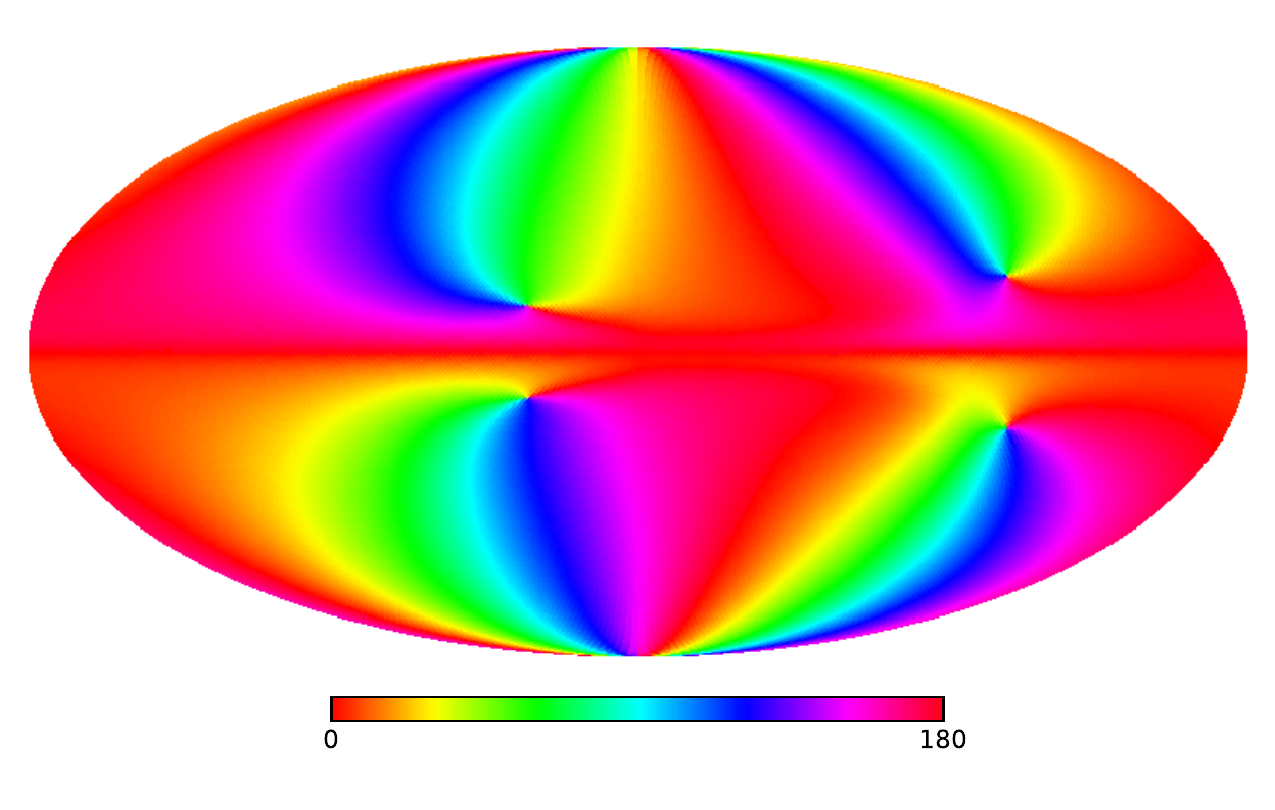}\\

\includegraphics[width=.22\linewidth]{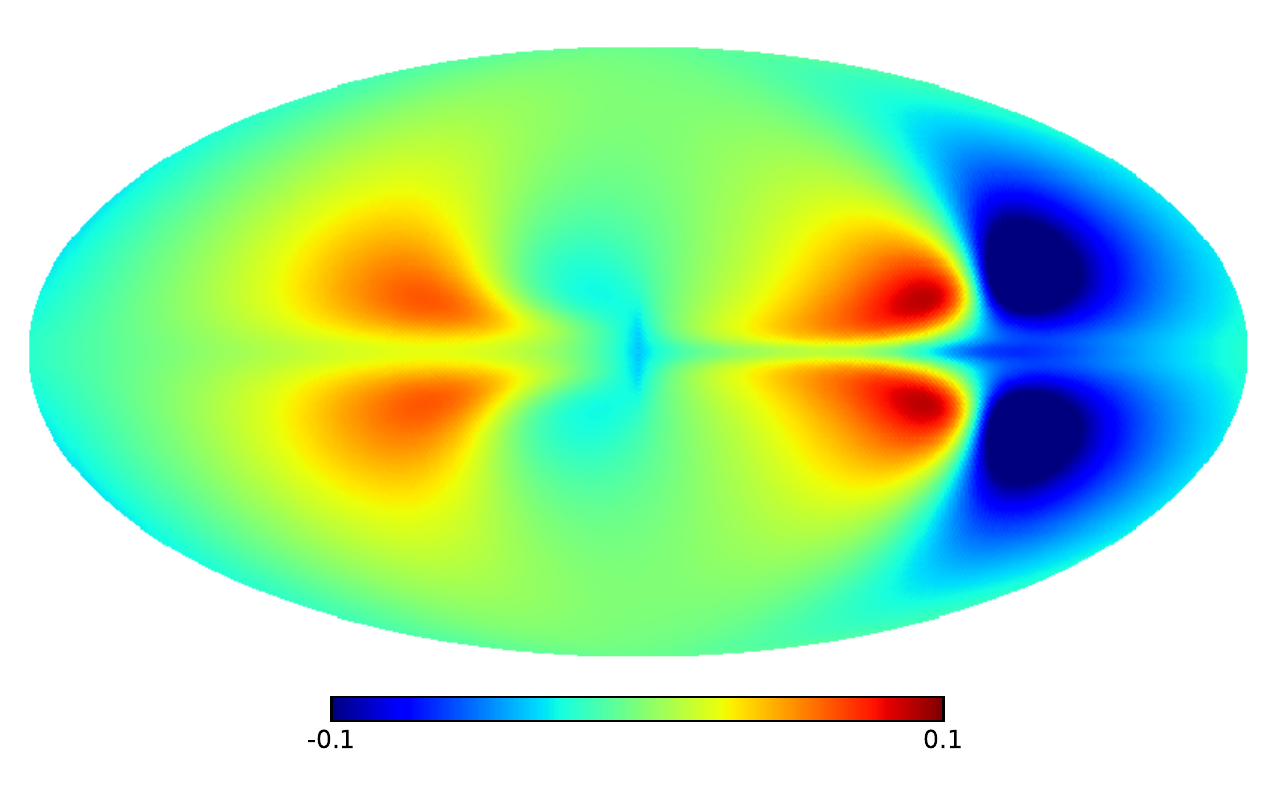} &
& 
& 
        \includegraphics[width=.22\linewidth]{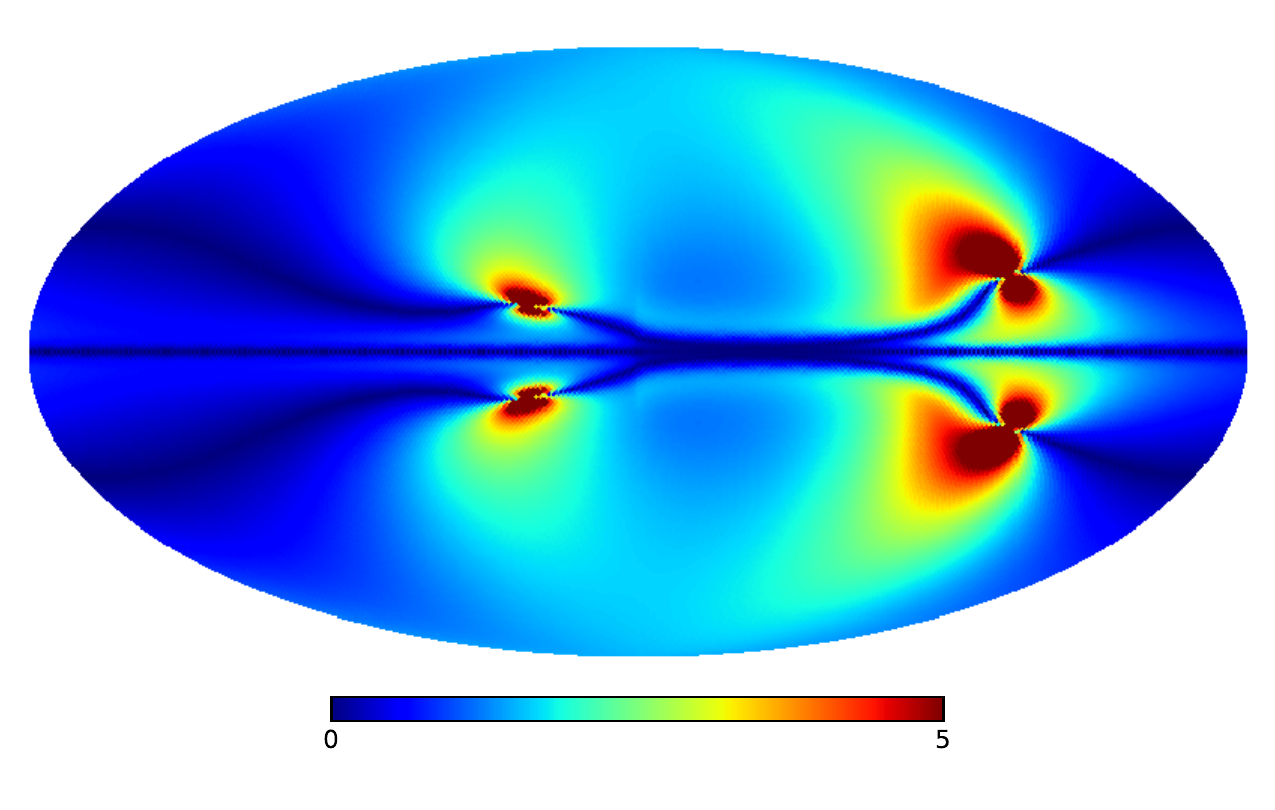} \\
\end{tabular}
\caption{Maps of the synchrotron emission. Columns correspond to the maps of $I$, $q$, $u$, and polarization position angles (in the IAU convention). The first row correspond to synchrotron emission when the original GMF model is adopted. For the second row, the adopted GMF corresponds to the best-fit obtained from the analysis of dust maps, case B in Sect.~\ref{sec:Iqu_fit}. In the third row, we present the relative difference of the intensity maps ($(I_{\rm{in}} - \hat{I})/ (I_{\rm{in}} + \hat{I})$). For the polarization angles, we display the acute angles between the polarization vectors.
\label{fig:SYNCmaps}}
\end{figure*}

\begin{figure}
\includegraphics[width=1\columnwidth]{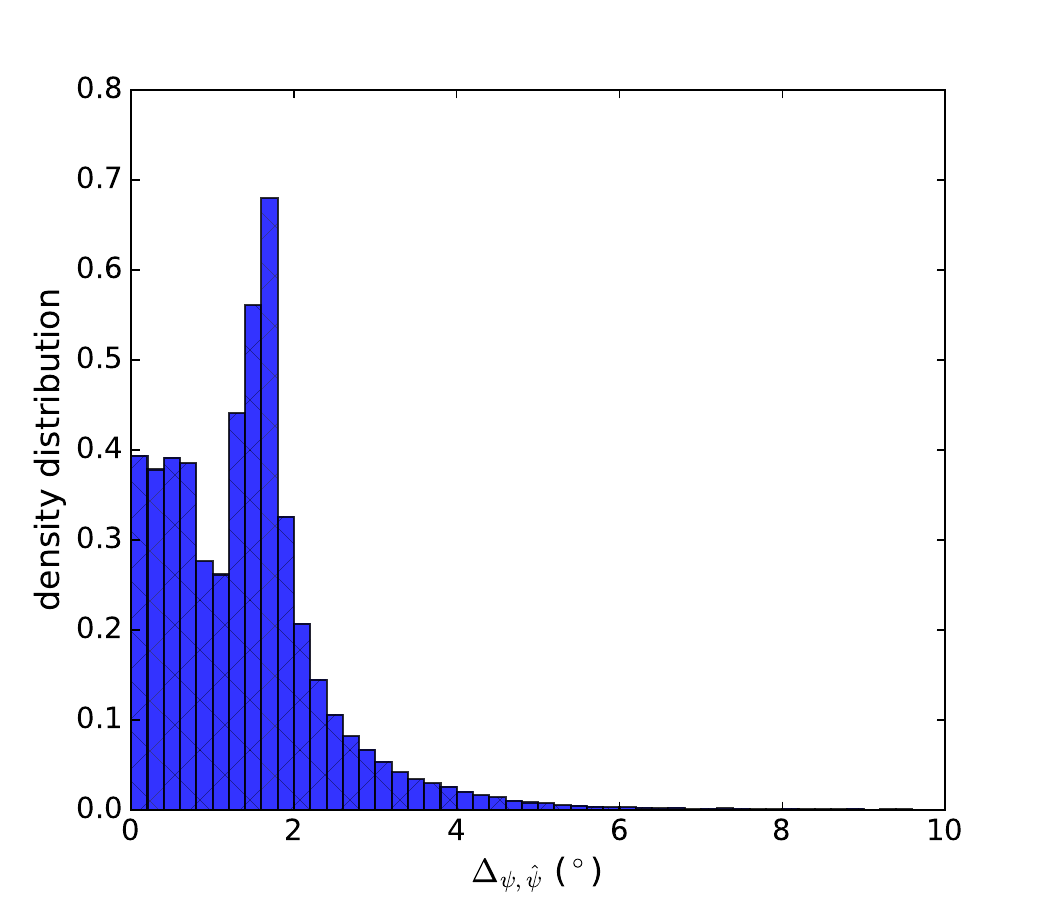} \\
\caption{Acute angle histogram comparing the polarization vector orientations from the synchrotron emission assuming the original GMF model and the worst one from dust analysis (case B). Angles are in degree.\label{fig:SYNCmaps-histDPPA}}
\end{figure}

\section{Proof of concept for future synchrotron analysis}
\label{sec:synchrotronproof}

As discussed in Sect.~\ref{sec:syncmodeling} for the case of the synchrotron emission, the matter density distribution and the GMF are deeply intertwined.
Therefore, they have to be constrained simultaneously along with the synchrotron emission spectral index.
Here, relying on simulations, we investigate in a toy model if we can safely use the constraints obtained on the geometry of the GMF from the Galactic thermal dust emission analysis to help the fit of the Galactic synchrotron emission maps. 

\smallskip

We assume a relatively simple density distribution for the Galactic relativistic electrons.
As the density distribution of the relativistic electron is expected to be smoother than the dust density distribution, for instance, due to the diffusion mechanism, we adopt the same exponential disk model as the one used to produce the \texttt{S1} thermal dust simulations. We adopted the most used values for the two scale lengths of the model $(\rho_0,\,z_0) = (8.0,\,1.0)$ kpc in the case of synchrotron emission.
We then produce, directly at $N_{\rm{side}} = 64$, the intensity and polarization maps of the synchrotron emission at a reference frequency according to Eqs.~\ref{eq:SYNCEMISSION}.
We consider two cases in terms of the GMF model parameters. In the first one, we use the input parameter of the GMF given in Table~\ref{tab:paramVal}, and in the second, the best-fit parameters from the fit of the thermal dust emission discussed in Sect.~\ref{sec:reconstructionGMF}.
We use the GMF reconstruction from case B, where the incorrect dust density distribution model was used, which it is more likely to correspond to what we would obtain in the case of real data sets and  produces the worst reconstruction of the GMF geometrical structure of Sect.~\ref{sec:reconstructionGMF}.

\smallskip

Using \texttt{gpempy}, we computed the intensity and polarization maps corresponding to the Galactic synchrotron emission in the two cases presented above.
We show them in Fig.~\ref{fig:SYNCmaps}. In that figure we also show the position angles of the polarization vectors deduced from the Stokes $Q$ and $U$ parameters. The maps obtained from the two GMF look very similar.
In pixel space, the relative difference of the intensity exceeds the ten percent threshold for only five percent of the sky. The relative difference of the polarization degree (not shown) exceeds the ten percent threshold for less than one percent of the sky.
The agreement in the polarization position angle is also remarkable as also inferred from Fig.~\ref{fig:SYNCmaps-histDPPA} where we show the histogram of the acute angles (Eq.~\ref{eq:DPPA}) between the polarization vectors corresponding to the two realizations of the synchrotron sky. It shows that the position angles agree at the 3 degree level almost for 95\% of the lines of sight.

\smallskip

From these results, we conclude that the GMF model reconstructed from the Galactic thermal dust emission analysis, even when using an oversimplified model for the dust density distribution, can be used at lower frequencies as an input, that is, a prior, to model the Galactic synchrotron emission and obtain constraints on the population of relativistic electrons.
This constitutes a proof of concept of the overall methodology proposed in Sect.~\ref{sec:Methodo}. Full validation and application of this methodology will be presented elsewhere.

\begin{table}[t]
\centering
\caption{Free parameters of the dust density distribution models are given for the three explored modelings (ED, ARM4, and ARM4$\oplus$ED). Column three gives the ranges that we explored. Column four gives the best-fit parameter values.}
\label{table:param-ndust}
{\tiny{
\begin{tabular}{clll}
\hline
\hline
models & parameters & ranges & best-fit                 \\
\hline
\\[-1.ex]
ED      & $\rho_0$ (kpc)        & $]0,\, 100]$  &       $21.199 \pm 0.007$      \\
                & $z_0$  (kpc)          & $]0,\, 10]$           &       $0.7514 \pm 0.0001$     \\
\\[-1.5ex]
ARM4
                & $\sigma_\rho$ (kpc)   & $1,\, 45]$    &       $8.400 \pm 0.003$  \\
                & $\rho_c$ (kpc)                        & $]0,\, 20]$   &         $5.974 \pm 0.003$       \\
                & $\sigma_z$ (kpc)              & $]0,\, 8]$            &       $0.7331 \pm 10^{-4}$    \\
                & $p$ ($^\circ$)                        & $]0,\, 45]$   & $27.67 \pm 0.003$       \\
                & $\phi_{\rm{00}}$ ($^\circ$)   & $[-90,\, 360]$        & $213.33 \pm 0.03$       \\
                & $\phi_0$ ($^\circ$)   & $]0,\, 45]$                                   & $ 26.030 \pm 0.002$     \\
                & $A_{i = 2, 3, 4}$     & $[10^{-4},\, 10^4]$   &       $\{1.102, 2.284, 3.747 \}$\\
                &                                                               &                                                       &  $\pm \, \{0.001, 0.001, 0.003\} $\\
\\[-1.5ex]
ARM4$\oplus$ED
                & $\rho_0$ (kpc)                & $]0,\, 100]$  &       $100.0 \pm 0.4\,10^{-3}$       \\
                & $z_0$  (kpc)                  & $]0,\, 10]$   &       $10.0 \pm 0.4\,10^{-5}$               \\
                & $\sigma_\rho$ (kpc)   & $1,\, 45]$    &       $5.93 \pm 0.003$  \\
                & $\rho_c$ (kpc)                        & $]0,\, 20]$   &         $6.0 \pm 0.003$         \\
                & $\sigma_z$ (kpc)              & $]0,\, 8]$            &       $0.4691 \pm 0.1\,10^{-3}$               \\
                & $p$ ($^\circ$)                        & $]0,\, 45]$   & $25.595 \pm 0.005$      \\
                & $\phi_{\rm{00}}$ ($^\circ$)   & $[-90,\, 360]$        & $-45.45 \pm 0.06$\\
                & $\phi_0$ ($^\circ$)   & $]0,\, 45]$   & $23.863 \pm 0.002$    \\
                & $A_{i = 1,2, 3, 4}$   & $[10^{-4},\, 10^4]$           &         $\{ 28.25, 94.15, 212.05, 57.2\}$\\
                &                                                               &                                                                       &  $\pm \, \{0.01, 0.07, 0.13, 0.04 \} $\\
\hline
\end{tabular}
}}
\end{table}

\begin{figure}[t]
\centering
\begin{tabular}{cc}
\includegraphics[width=.47\linewidth]{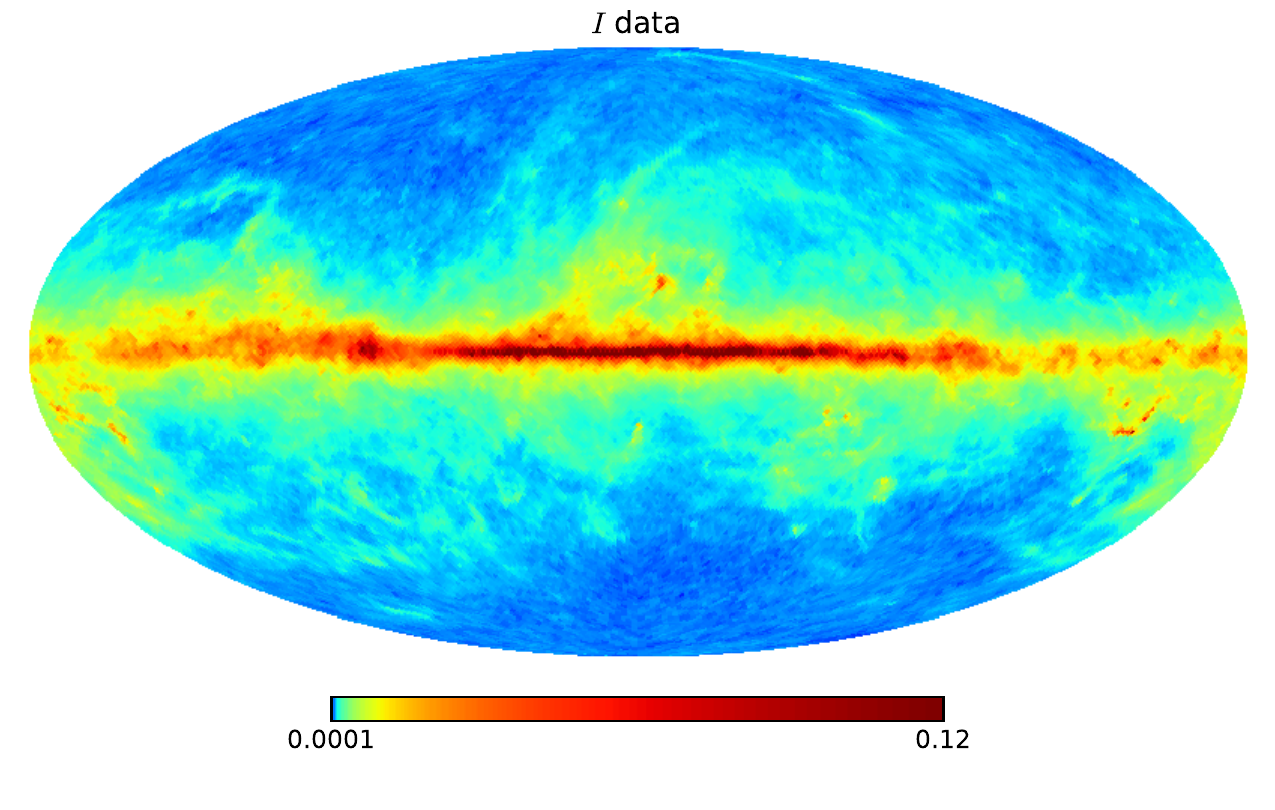} &
        \includegraphics[width=.47\linewidth]{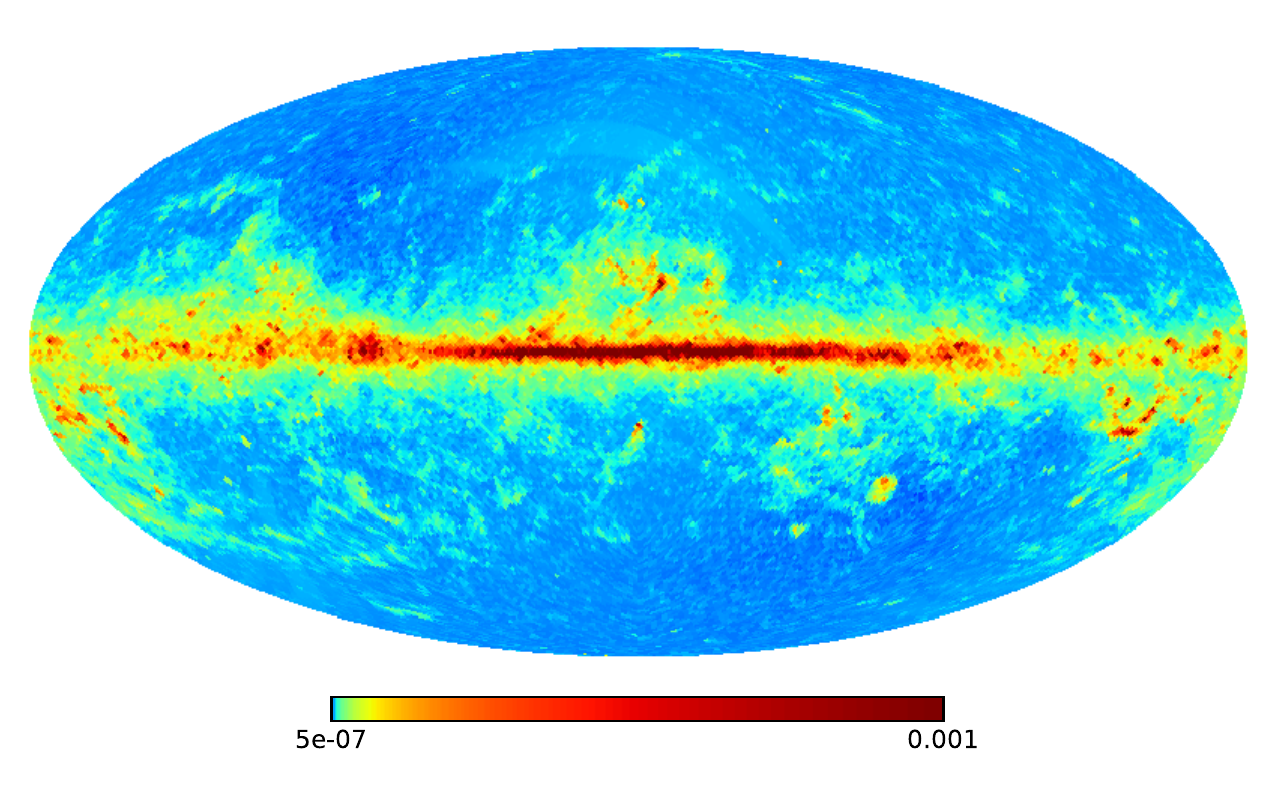} \\

\includegraphics[width=.47\linewidth]{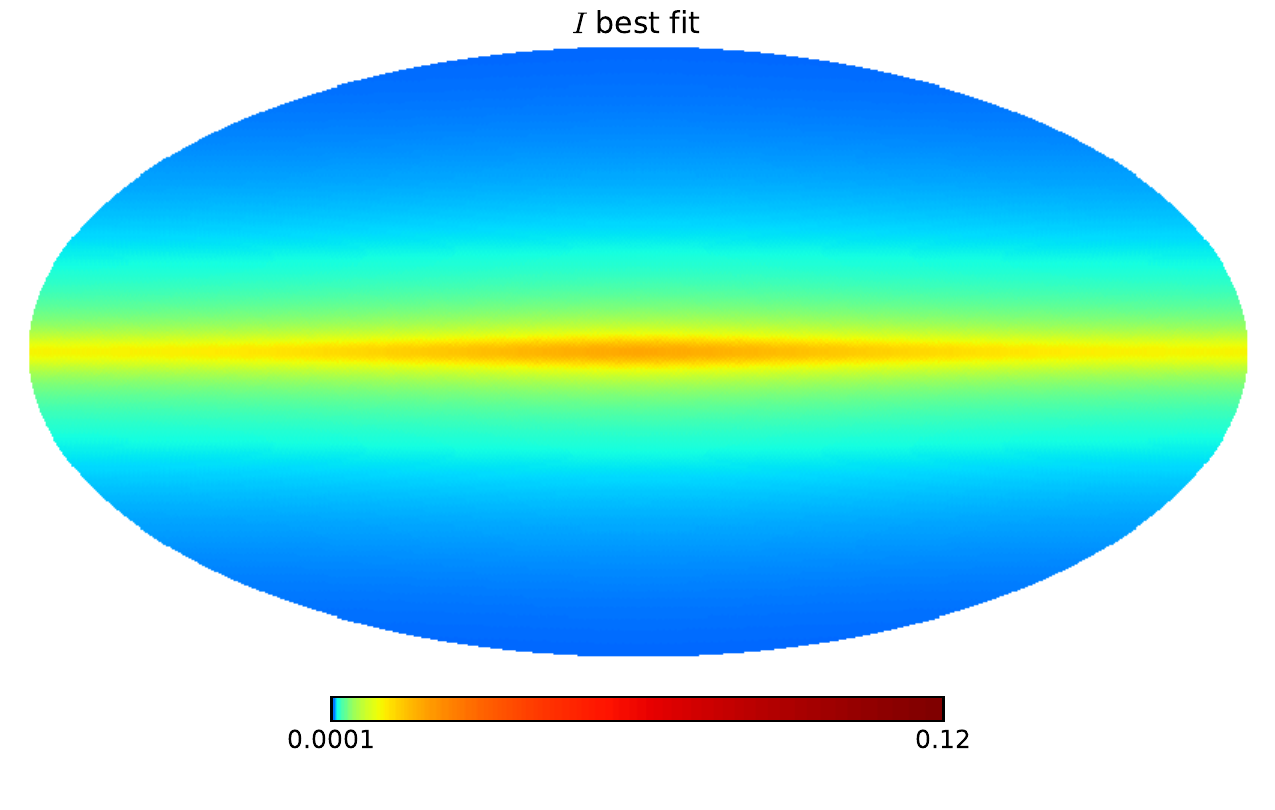} &
        \includegraphics[width=.47\linewidth]{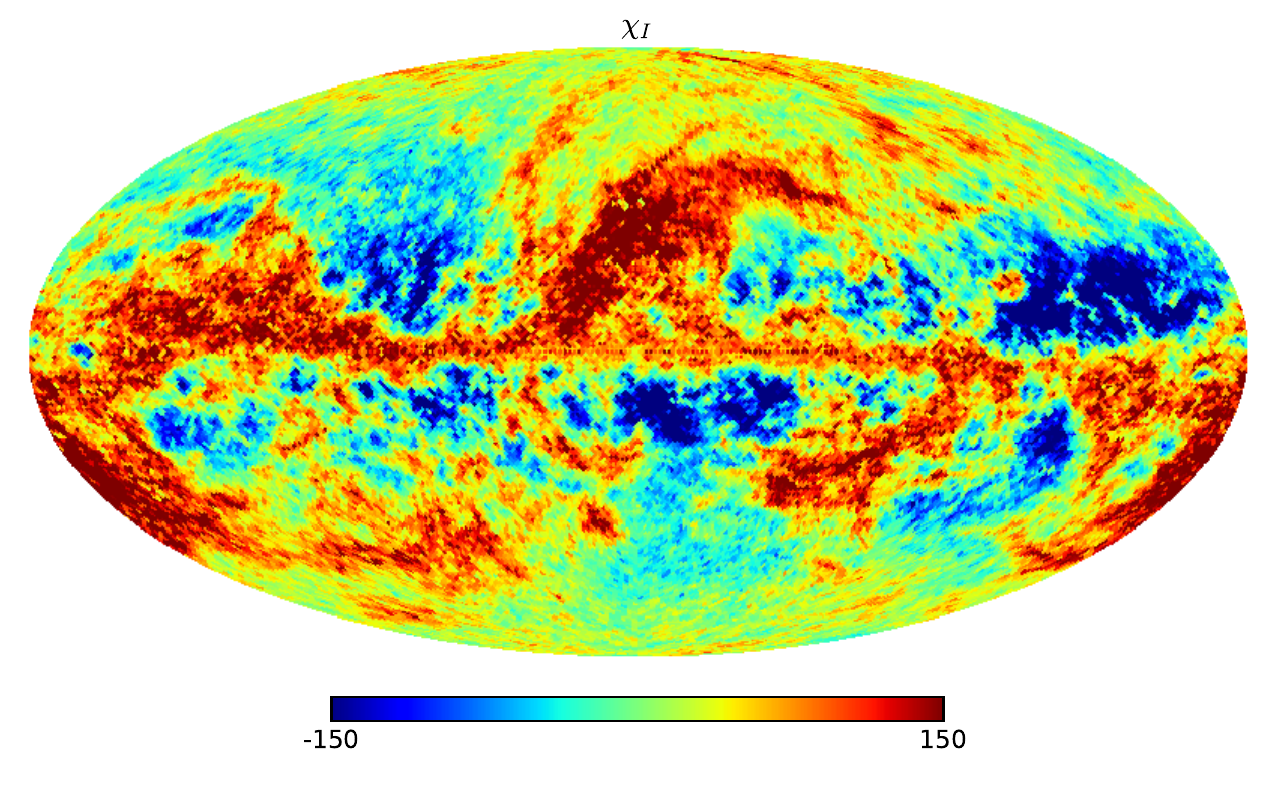} \\

\includegraphics[width=.47\linewidth]{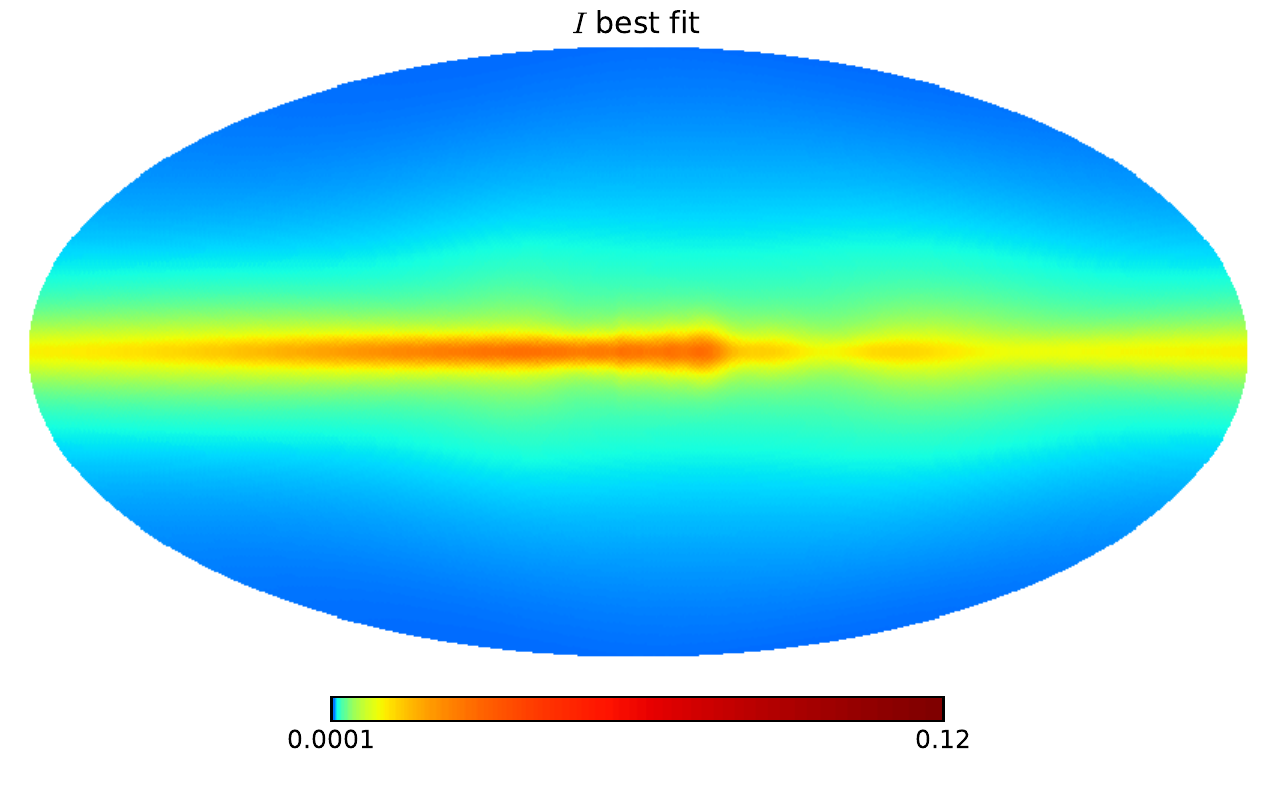} &
        \includegraphics[width=.47\linewidth]{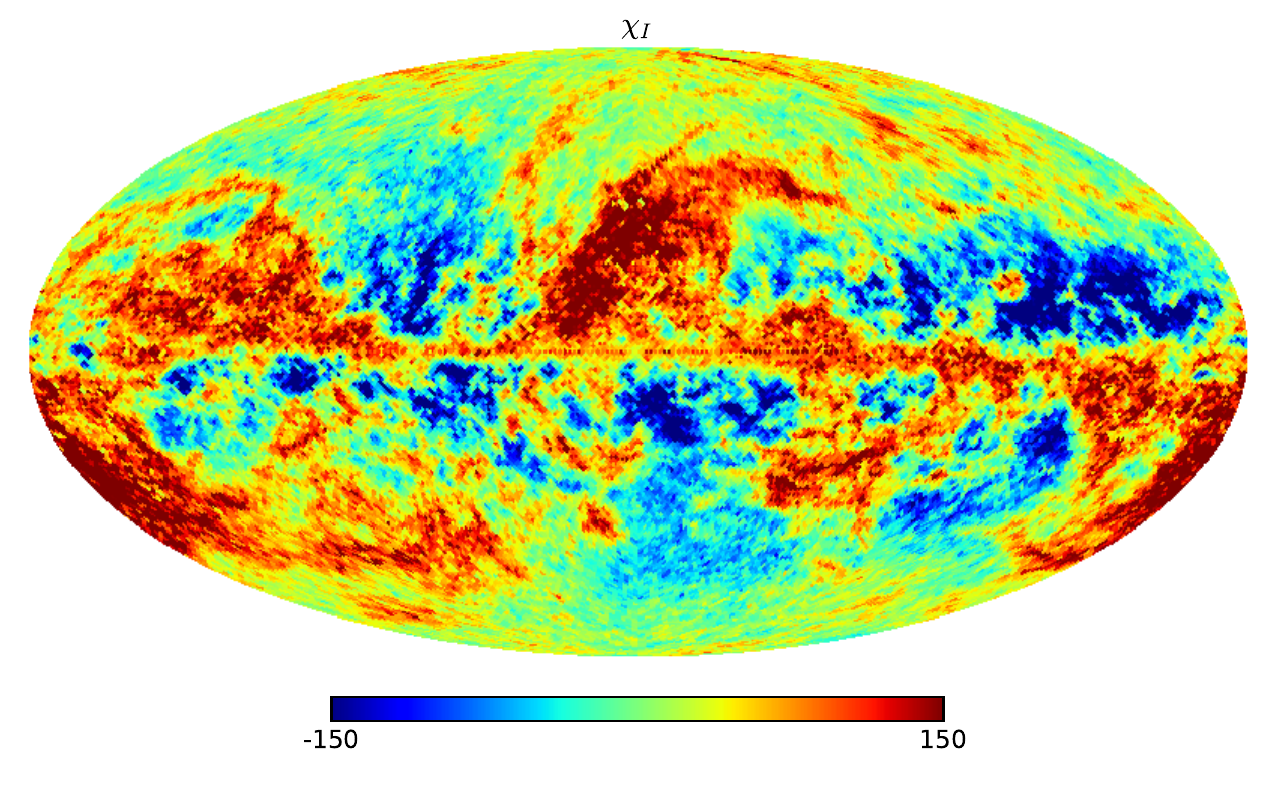} \\

\includegraphics[width=.47\linewidth]{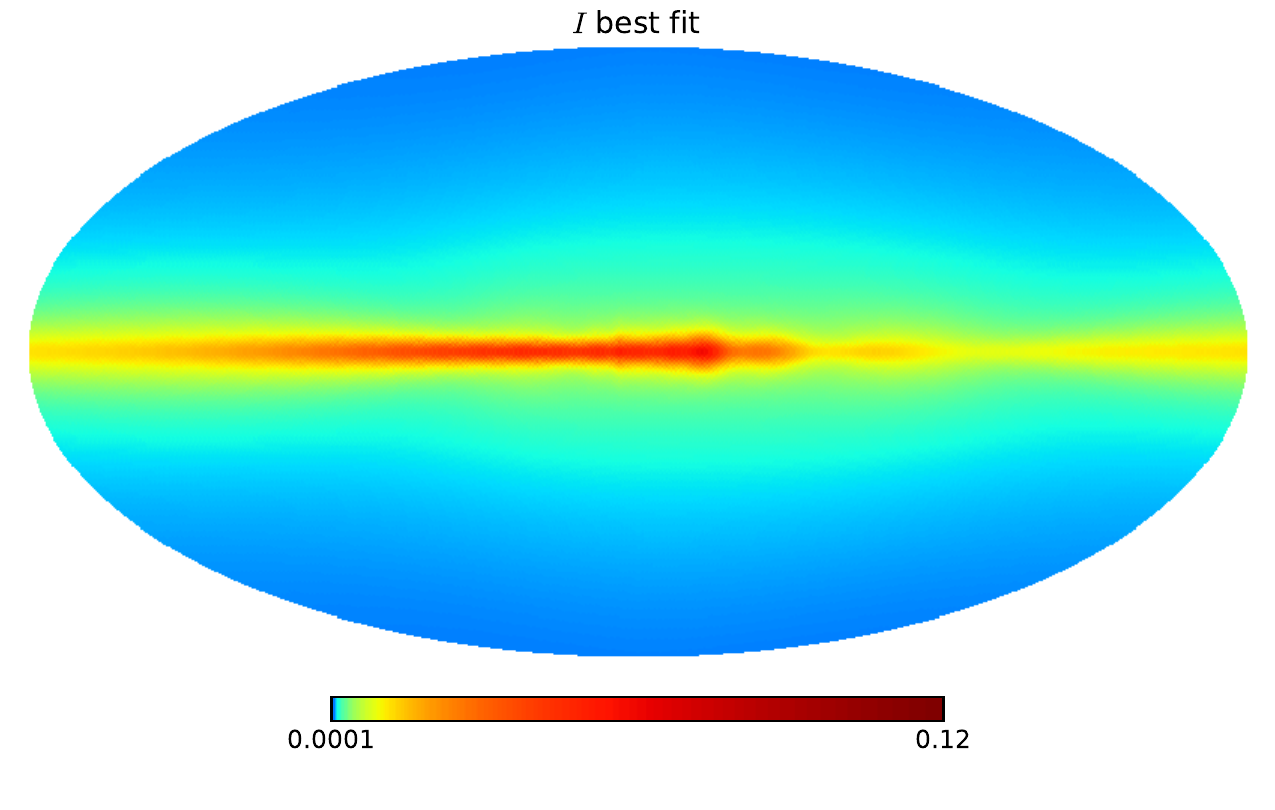} &
        \includegraphics[width=.47\linewidth]{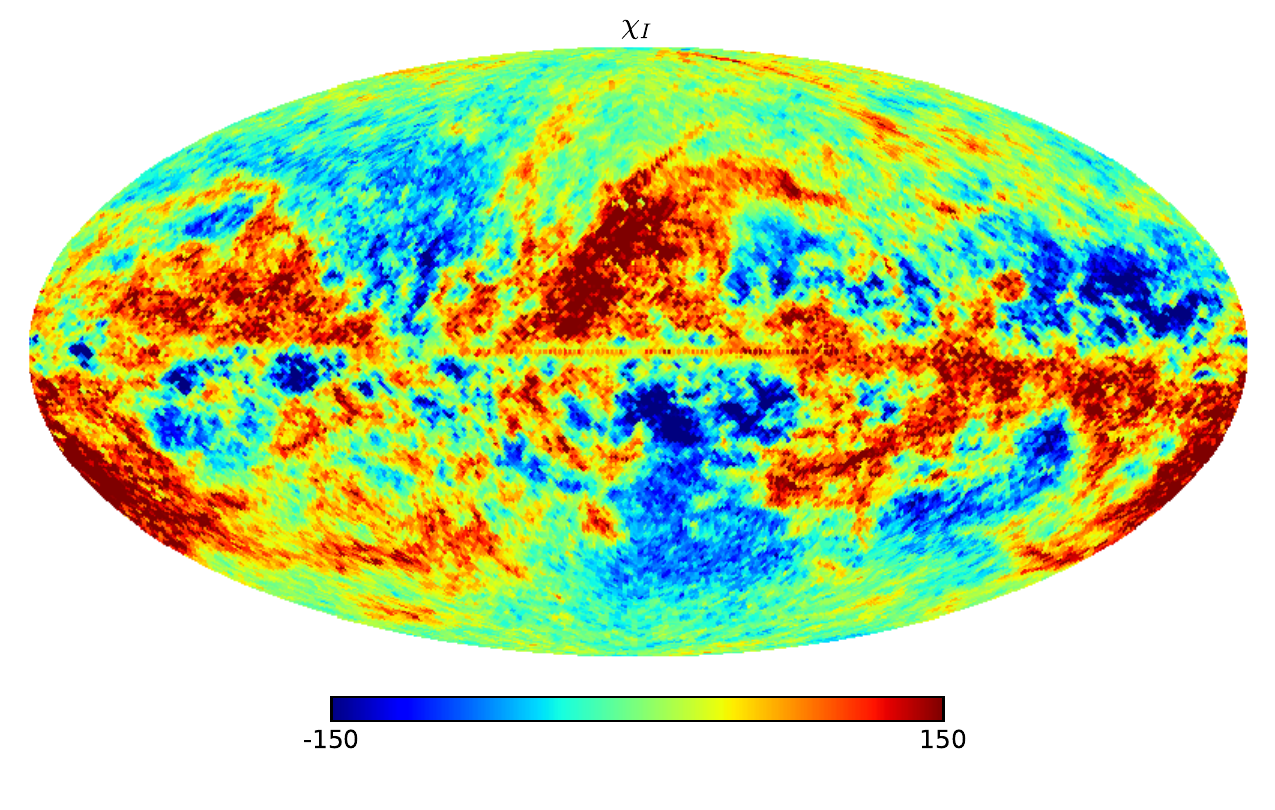} \\
\end{tabular}
\caption{Intensity maps. First row: 353-GHz map from \textit{Planck} downgraded at $N_{\rm{side}} = 64$ and the corresponding map of uncertainties that we use to compute the $\chi^2$. Rows two to four correspond to dust density distribution models labeled ED, ARM4, and ARM4$\oplus$ED, respectively. The obtained best-fits are shown in the first column and the statistical significance of the residual, per-pixel, are shown the second column.}
\label{fig:I_fit}
\end{figure}

\section{Reconstruction of the GMF structure from the Planck 353~GHz data}
\label{sec:fit2Planck}

In this section, we apply the MCMC-fitting procedure described in Sect.~\ref{sec:GMFrecons} to the \textit{Planck} polarization data at 353-GHz (see  Sect.~\ref{sec:data}) to infer the large-scale structure of the GMF.
We first rely on three dust density distribution models to fit for the intensity map. Then, we use those best-fit models to constrain four different models of the regular component of the GMF through a fit to the reduced Stokes parameter maps. Both the dust density and GMF distribution models are detailed in Appendix~\ref{sec:AppendixA} and encode different degree of complexity whereas sharing geometrical features.
In this first attempt, the contribution of the turbulent component of the GMF is not accounted for, as discussed above, and local structures of the ISM are not included.

The fits, as discussed below, are performed at the resolution of $N_{\rm{side}} = 64$. We thus downgrade the $I$, $Q$ and $U$ 353~GHz Planck maps from $N_{\rm{side}} = 2048$ to $N_{\rm{side}} = 64$ and compute the reduced-Stokes parameters $q$ and $u$ and their uncertainties (see Eq.~\ref{eq:sig_pol}). These data are shown in the first row of Fig.~\ref{fig:qu_fit}.
Maps at $N_{\rm{side}} = 32$ are also used in the case of the intensity.

\subsection{Dust density reconstruction}
\label{sec:ndust}

\subsubsection{Fitting procedure}
For the fit of the intensity map, we first fit the data at low resolution, $N_{\rm{side}} = 32$ to explore the full parameter space more quickly and determine regions of the parameter space favored by the data. At this resolution, the Markov chains are initialized according to a uniform distribution for each of the parameters. Then, when a converged solution is reached at that low resolution, we start a new exploration of the parameter space by comparing simulations to data at $N_{\rm{side}} = 64$.
In this case, the Markov chains are initialized according to Gaussian centered on the best-fit parameter values obtained at $N_{\rm{side}} = 32$ and with a width of a few percent (typically 10\%).

\smallskip

The initially explored ranges of values of the free parameters are given in Table~\ref{table:param-ndust} for the different fitted models. We consider the two dust density distribution models used
for the simulation data (see Sect.~\ref{sec:Iqu_fit}), ED and ARM4, plus and an extra  one formed from the combination of these two: ARM4$\oplus$ED (see Appendix~\ref{sec:AppendixA}).

\subsubsection{Best-fit models}
The best-fit parameters are reported in Table~\ref{table:param-ndust}.
In Fig.~\ref{fig:I_fit}, we compare the data and the best-fit intensity maps obtained for the three dust density distribution models fitted to $I_{\rm{353}}$. We also show $\chi$-maps, which give the statistical significance of the residuals.
We observe that the best-fit models cannot account for all the complexity and the richness contained in the data. This is not surprising since our models do not include patterns such as clumps, filaments, or bones that can be directly spotted by eye and that require dedicated extra modeling (see e.g. \citealt{Zuc2017}). Accounting for such features would significantly increase the number of parameters and will not be easily tractable within a MCMC approach.

\begin{figure}[t]
\centering
\includegraphics[width=\columnwidth]{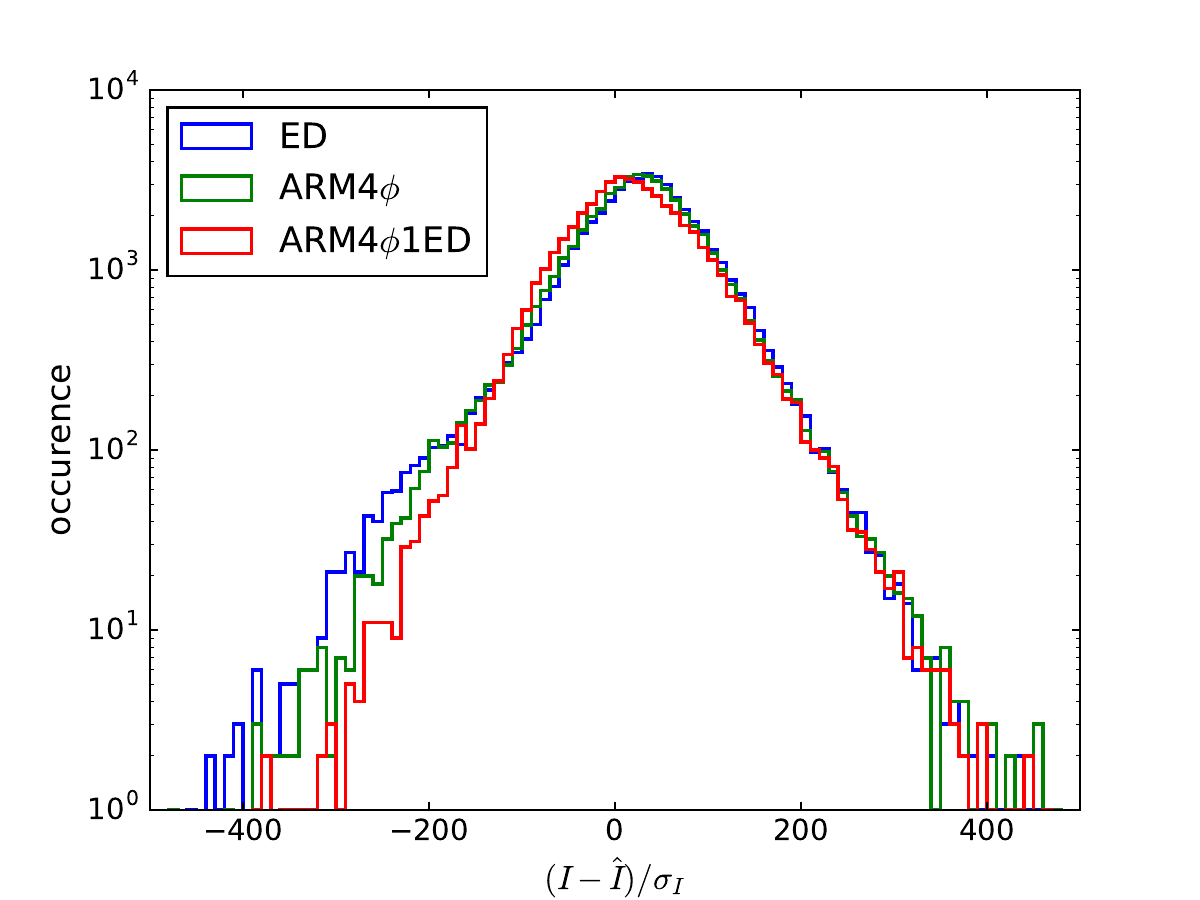}
\caption{Histograms of the signed significance of the residuals of the
best-fits obtained by adjusting the three dust density distribution models
(ED, ARM4, and ARM4$\oplus$ED) to the thermal dust intensity map.}
\label{fig:Ifits_chi-hist}
\end{figure}

In Fig.~\ref{fig:Ifits_chi-hist}, we show the histograms of the $\chi_i$ corresponding to the three best-fits to the intensity map. The distributions are nearly Gaussian with very large spread. This is again due  to the fact that the uncertainties in the data (even including the intrinsic dispersion) cannot overcome the residuals to such simplistic parametric models  as discussed before.
As a result, any small-scale feature turns out to be significant and the values of the reduced $\chi^2$ are $\gg1$ (see Table~\ref{table:chi2}). 
This was to be expected from our simulation-based study where we show that such high values of reduced $\chi^2$ are reached as soon as we do not have the right parametric model at hand (see Sect.~\ref{sec:Iqu_fit}).

\smallskip

\smallskip

To verify the robustness of our best-fit models of the dust density distribution ($n_{\rm{d}}$) with respect to contamination from other emission components (e.g., point sources, CIB), we also fit for the dust optical depth ($\tau_{353}$) map presented in \cite{PlanckXI2014}. That observable, deduced from the multi-frequency analysis, is known to trace the integrated dust density distribution at least as well as the intensity map does. The best-fit maps agree globally with those obtained from $I_{\rm{353}}$ fits. This, in turn, also justifies the assumption around Eq.~\ref{eq:Idust}.

\smallskip

In conclusion, none of the best-fit models is satisfying on its own.
Nevertheless, as discussed in Sect.~\ref{sec:GMFrecons}, it is still possible to constrain the GMF geometry even in the case of any mismodeling of the Galactic dust grain density.
Therefore, in the following, we use the recovered dust density distribution models to constrain the geometry of the large-scale, regular, GMF models.
In order to propagate the uncertainties due to the assumed $n_{\rm{d}}$ model, we find relevant to fit the GMF models with all the three $n_{\rm{d}}$ models and to take the dispersion on the parameters of the reconstructed GMF as a conservative estimate of the uncertainties on those parameters.

\subsection{3D regular GMF reconstruction}
\label{sec:gmf}

\begin{table*}[t]
\centering
\caption{Free parameters of the GMF models are given for the three
explored modelings (ASS, LSA, BSS, and QSS). Column two and three give
the model parameters and the ranges that are explored. Columns 4-6 give the best-fit parameter values while assuming the dust density distribution
to be given by the best-fit of the three $n_{\rm{d}}$ models: ED, ARM4,
and ARM4$\oplus$ED.}
\label{table:param-GMF}
\begin{tabular}{lll  ccc}
\hline
\hline
model   & parameter     & explored range & \multicolumn{3}{c}{best-fit values (full-sky)} \\
\multicolumn{3}{c}{}    &       $n_{\rm{d}} \equiv$ ED &        $n_{\rm{d}} \equiv$ ARM4 &  $n_{\rm{d}} \equiv$ ARM4$\oplus$ED      \\
\hline
\\[-1.ex]
ASS             & $p$ ($^\circ$)                        &$]0,\, 65]$                    &       $26.67 \pm 0.02$       &       $26.53 \pm 0.02$        &       $25.74 \pm 0.02$         \\
                        & $\chi_0$ ($^\circ$)   & $]-180,\, 180]$       & $-1.27 \pm 0.02$        &       $-2.49 \pm 0.02$        &       $-1.85 \pm 0.02$   \\
\\[-1.5ex]
WMAP    & $\psi_0$ ($^\circ$)   &$]0,\, 65]$                    & $32.61 \pm 0.02$       &  $34.84 \pm 0.02$     & $35.40 \pm 0.02$      \\
                        & $\psi_1$ ($^\circ$)   & $]-180,\, 360]$       & $-34.28 \pm 0.04$       & $-44.47 \pm 0.05$     & $-56.18 \pm 0.06$     \\
                        & $\chi_0$ ($^\circ$)   & $]-180,\, 180]$       & $-4.68 \pm 0.02$                & $-5.82 \pm 0.02$              & $-6.94 \pm 0.02$               \\
\\[-1.5ex]
BSS     & $p$ ($^\circ$)                        &$]0,\, 65]$                    & $24.586 \pm 0.002$              &  $21.955 \pm 0.001$   & $17.514 \pm 0.002$            \\
                & $\chi_0$ ($^\circ$)   & $]-180,\, 180]$       & $-3.073 \pm 0.008$              &       $-7.43 \pm 0.04$                & $-5.01 \pm 0.01$                       \\
                & $\rho_0$ (kpc)                & $]4,\, 10]$                   & $4.2544 \pm 0.0003$     &  $4.7355 \pm 0.0002$  & $5.1553 \pm 0.0003$   \\
                & $z_0$  (kpc)                  & $]0,\, 3]$                    & $0.318 \pm 0.001$               &  $0.789 \pm 0.005$            & $0.268 \pm 0.001$              \\
\\[-1.5ex]
QSS     & $p$ ($^\circ$)                        & $]0,\, 65]$                   & $25.706 \pm 0.002$              &  $29.817 \pm 0.004$   &  $25.843 \pm 0.002$           \\
                & $\chi_0$ ($^\circ$)   & $]-180,\, 180]$       & $3.217 \pm 0.006$              &  $3.5687 \pm 0.008$   &       $4.29 \pm 0.01$          \\
                & $\rho_0$ (kpc)                & $]4,\, 10]$                   & $5.8503 \pm 0.0002$     &  $5.6058 \pm 0.0004$  &  $5.8389 \pm 0.0003$          \\
                & $z_0$  (kpc)                  & $]0,\, 3]$                            & $0.0981 \pm 0.0004$     &  $0.1770 \pm 0.0007 $ &  $0.1712 \pm 0.0007$          \\
\\[-1.5ex]
\hline
\end{tabular}
\end{table*}

\begin{figure*}[h]
\centering
\begin{tabular}{cccc}
\includegraphics[width=.23\linewidth]{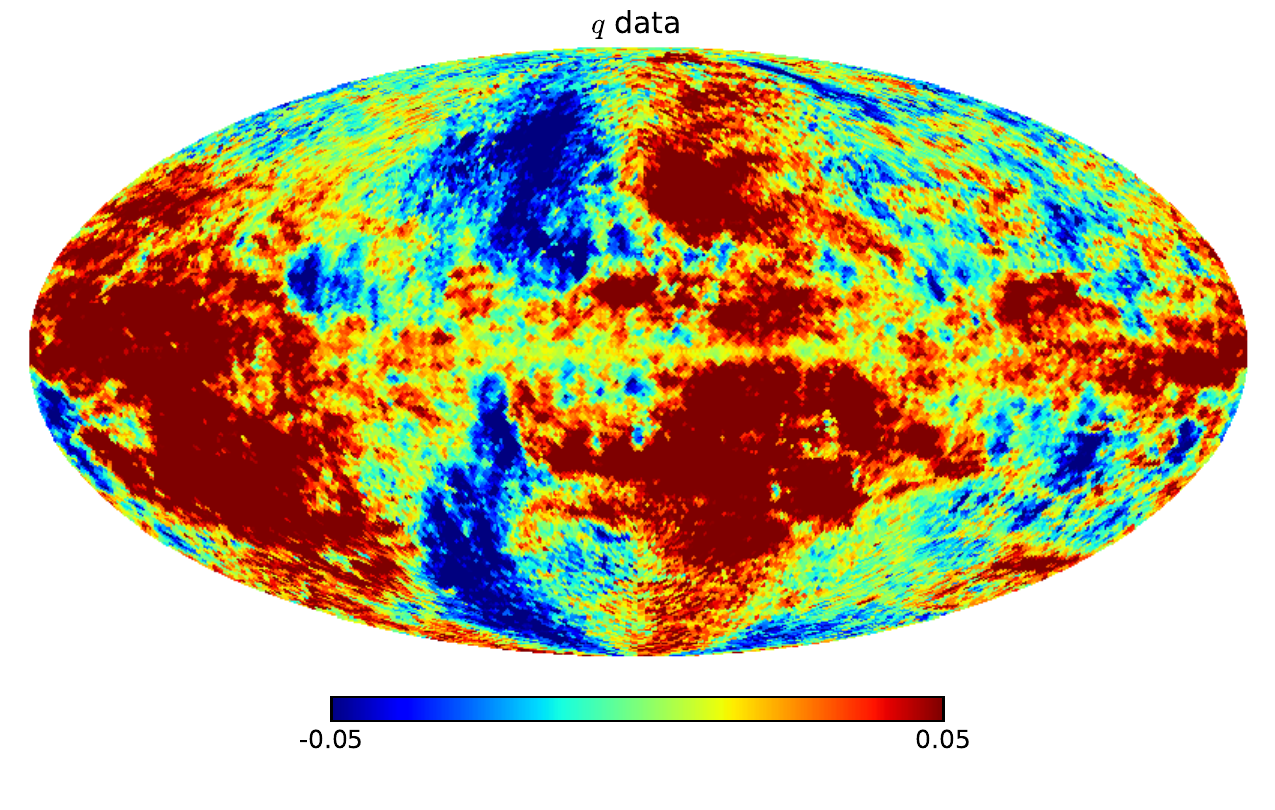} &
        \includegraphics[width=.23\linewidth]{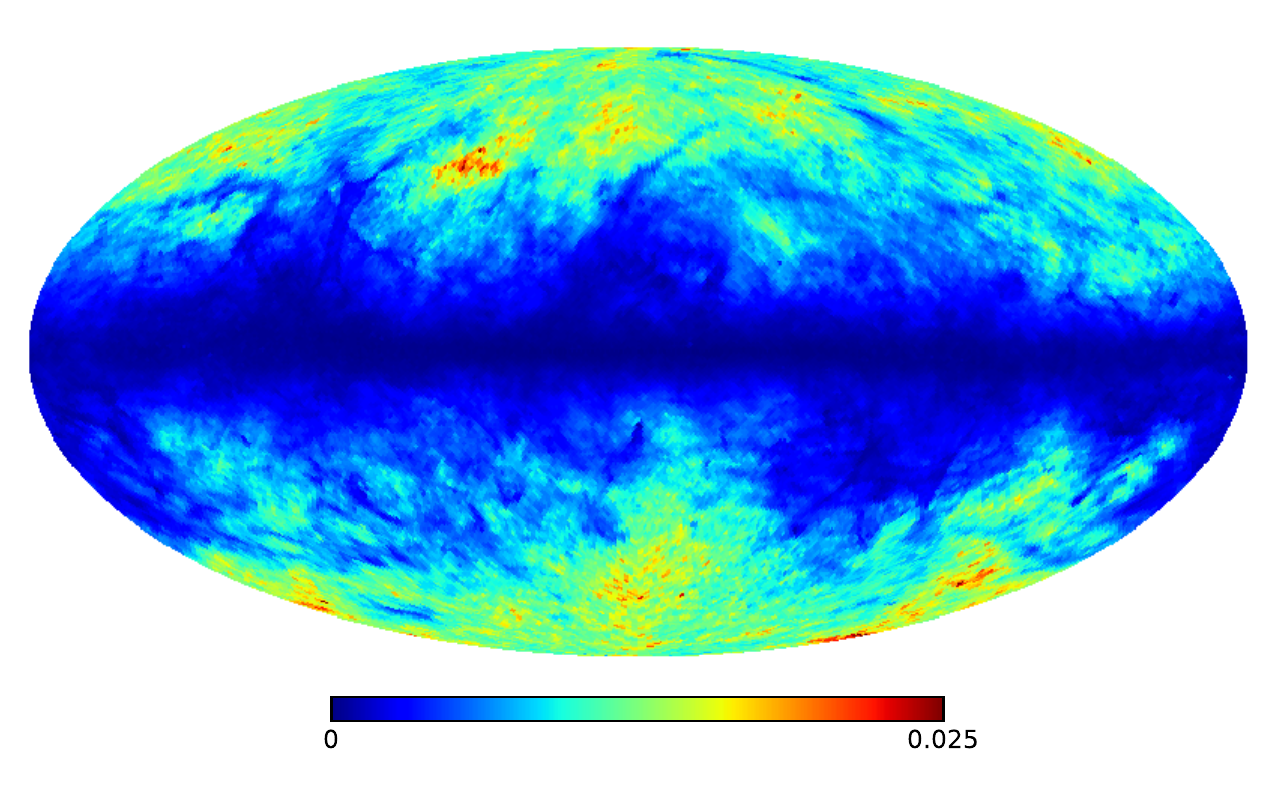} &
                \includegraphics[width=.23\linewidth]{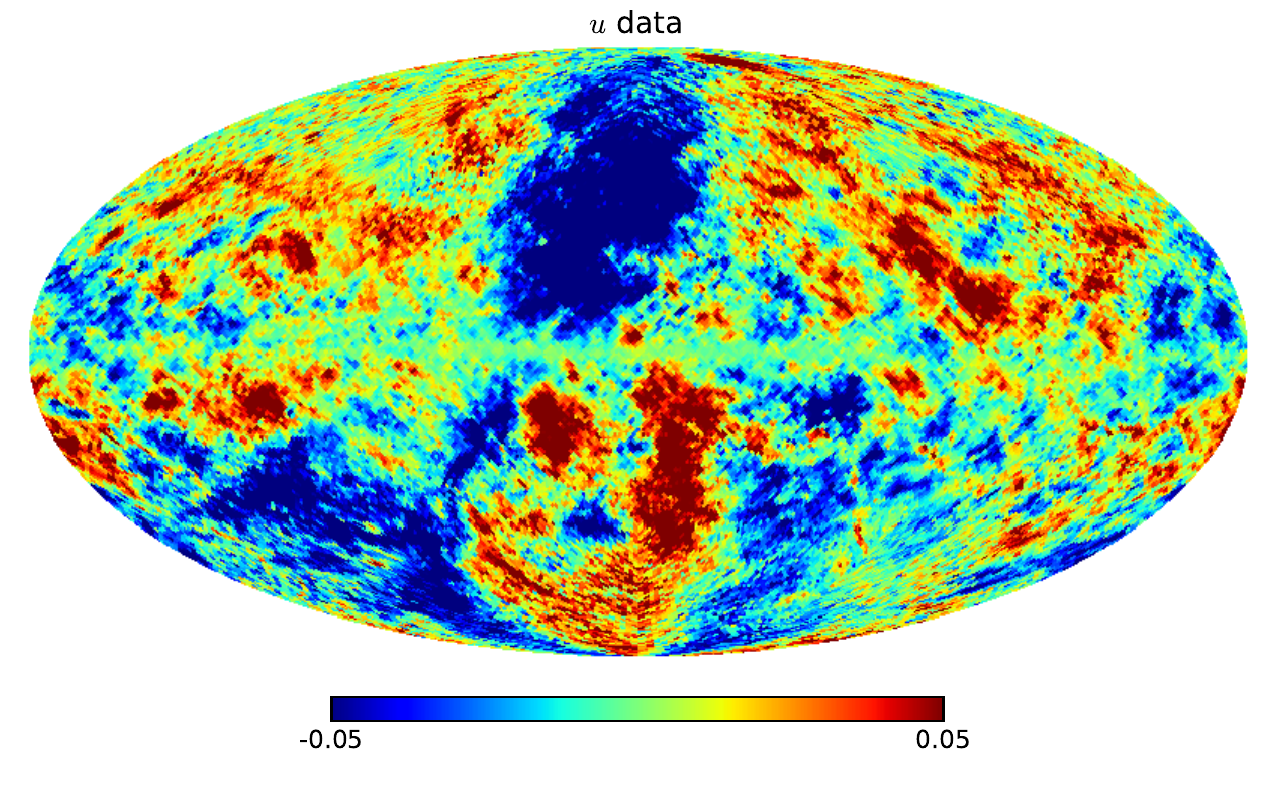} &
                        \includegraphics[width=.23\linewidth]{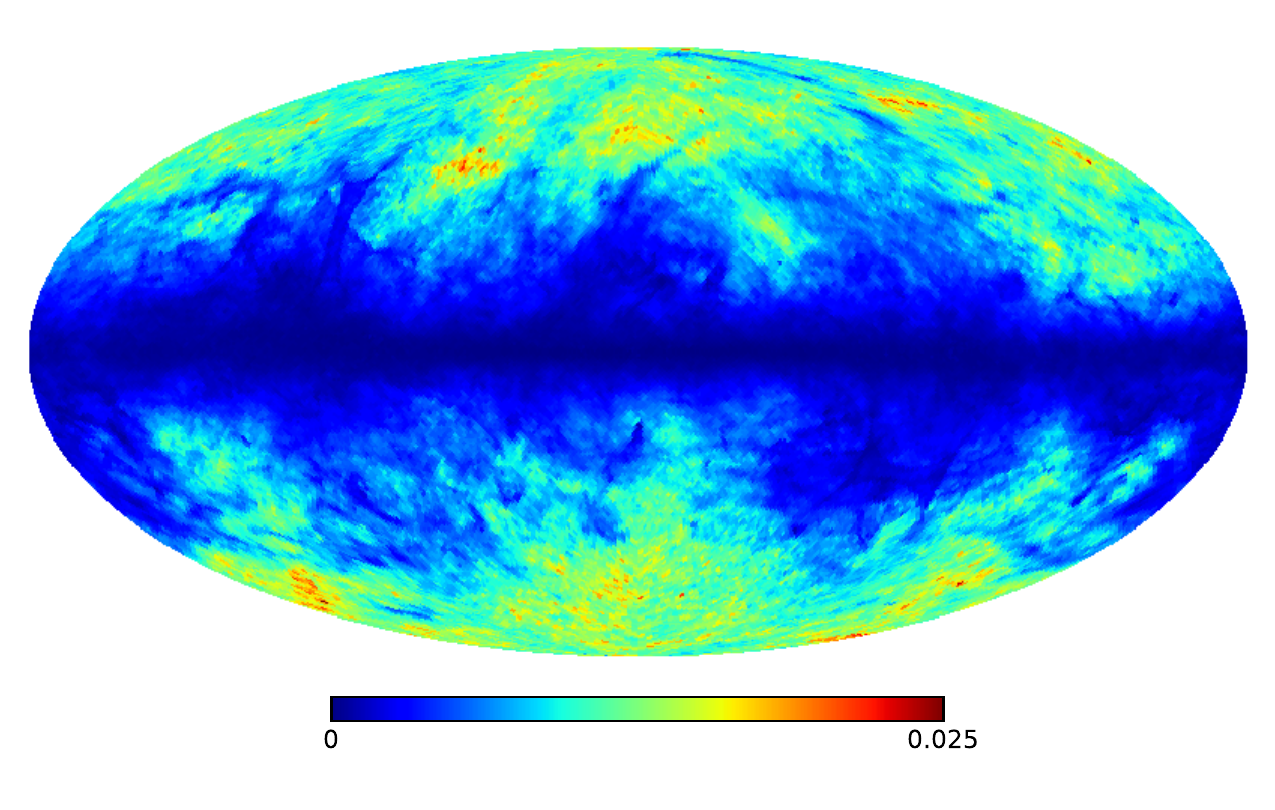} \\

\includegraphics[width=.23\linewidth]{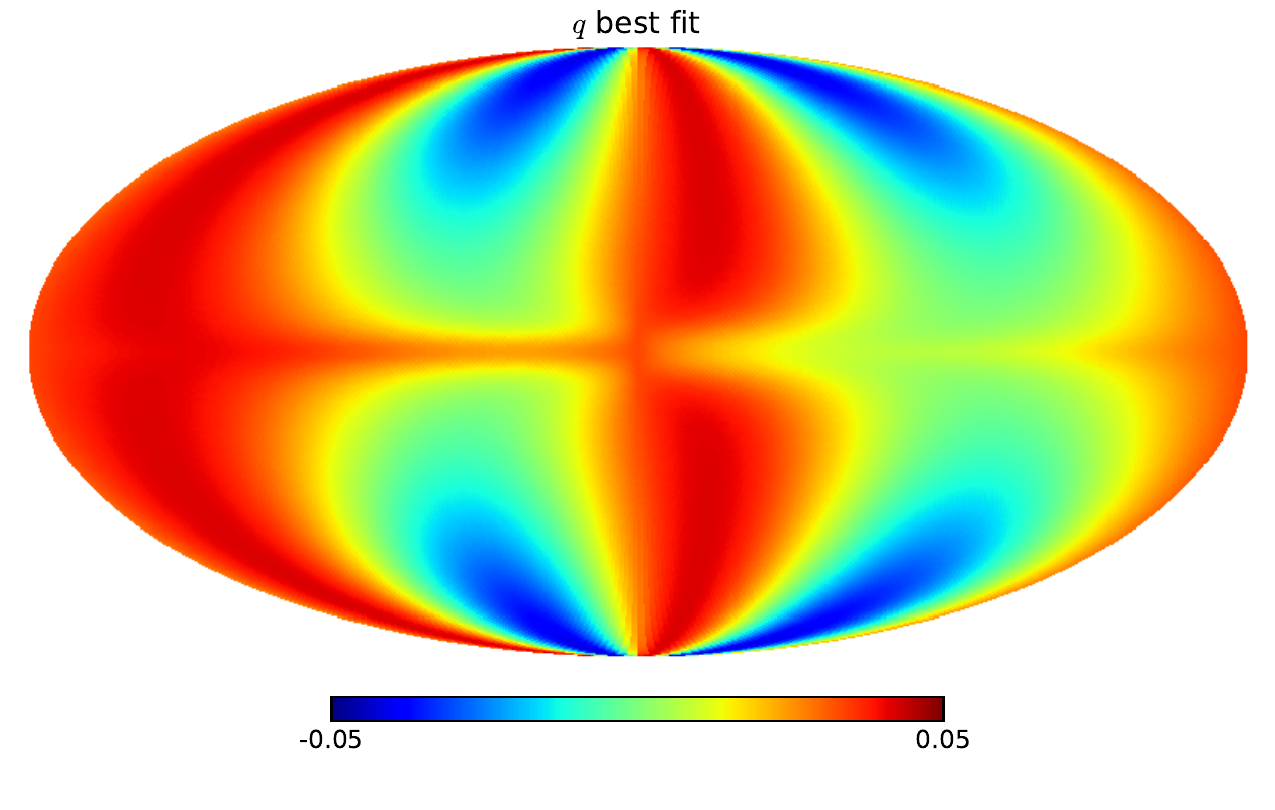} &
        \includegraphics[width=.23\linewidth]{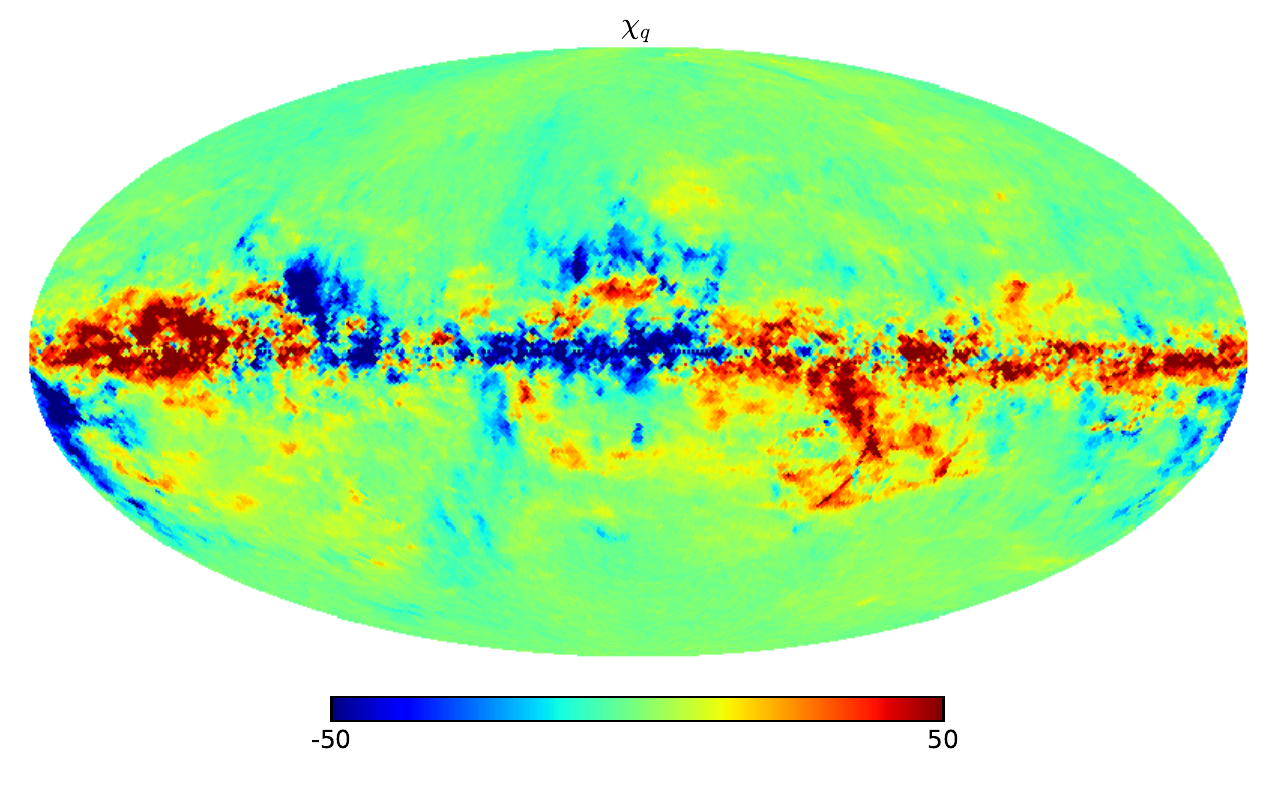} &
                \includegraphics[width=.23\linewidth]{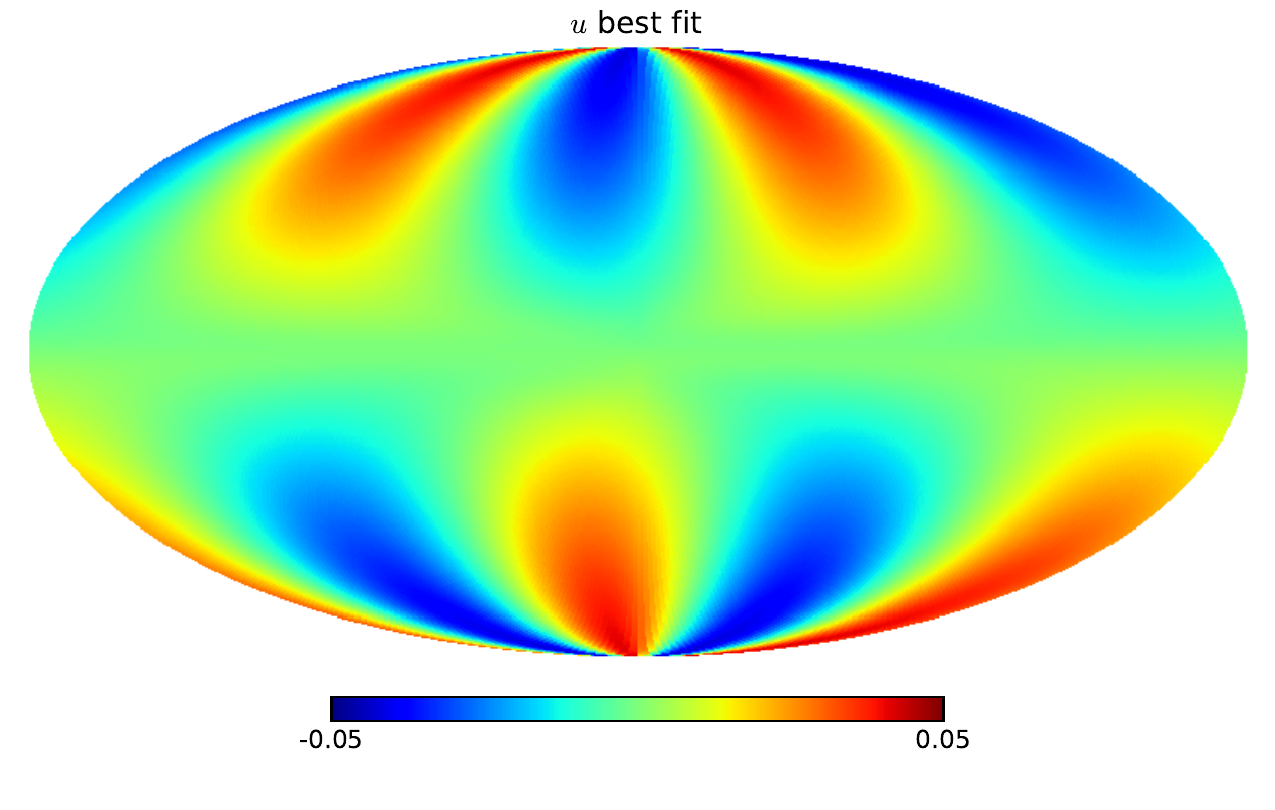} &
                        \includegraphics[width=.23\linewidth]{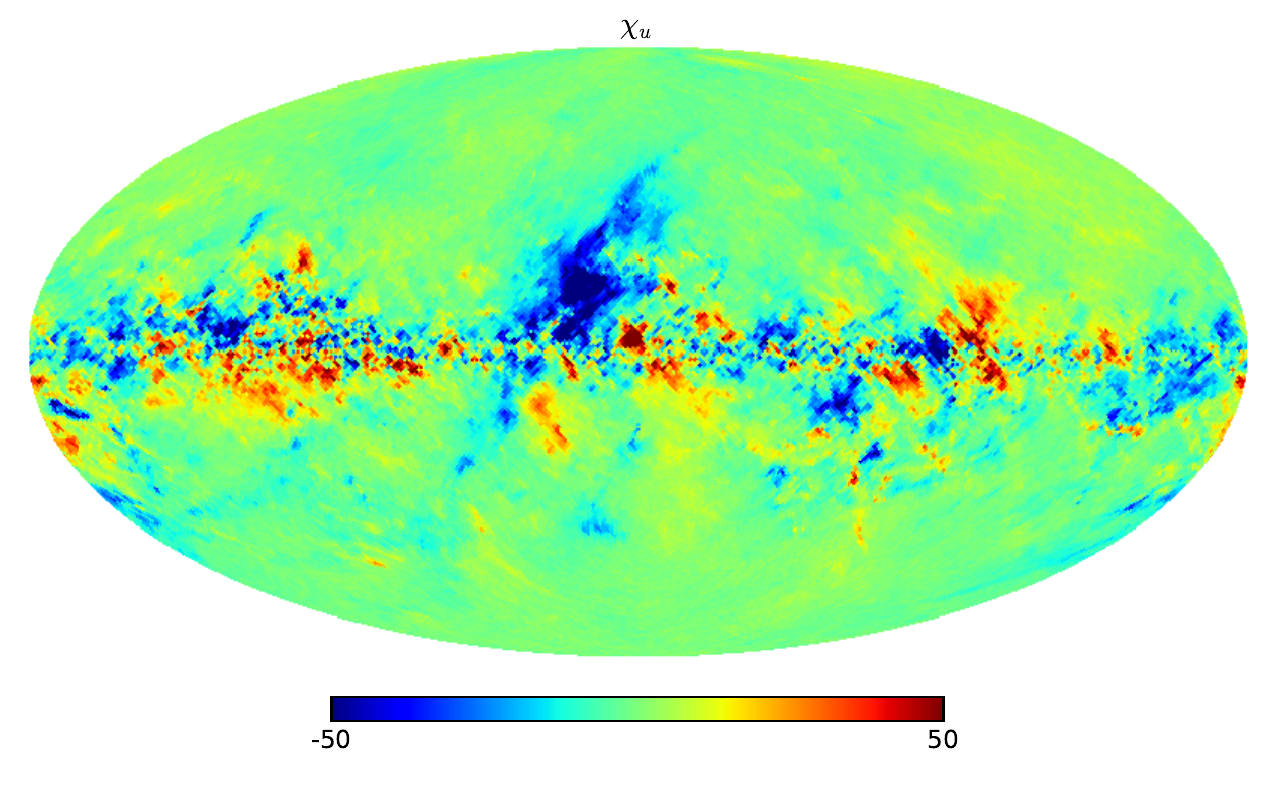} \\

\includegraphics[width=.23\linewidth]{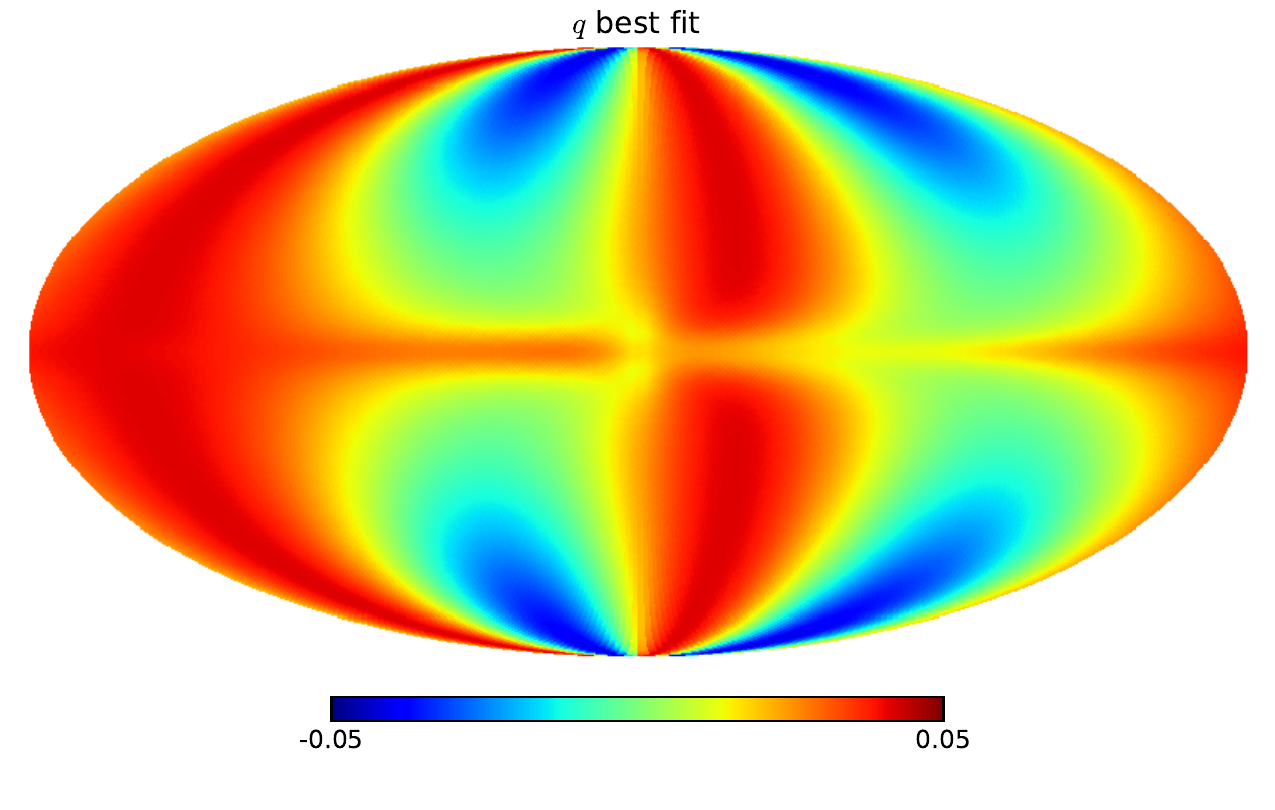} &
        \includegraphics[width=.23\linewidth]{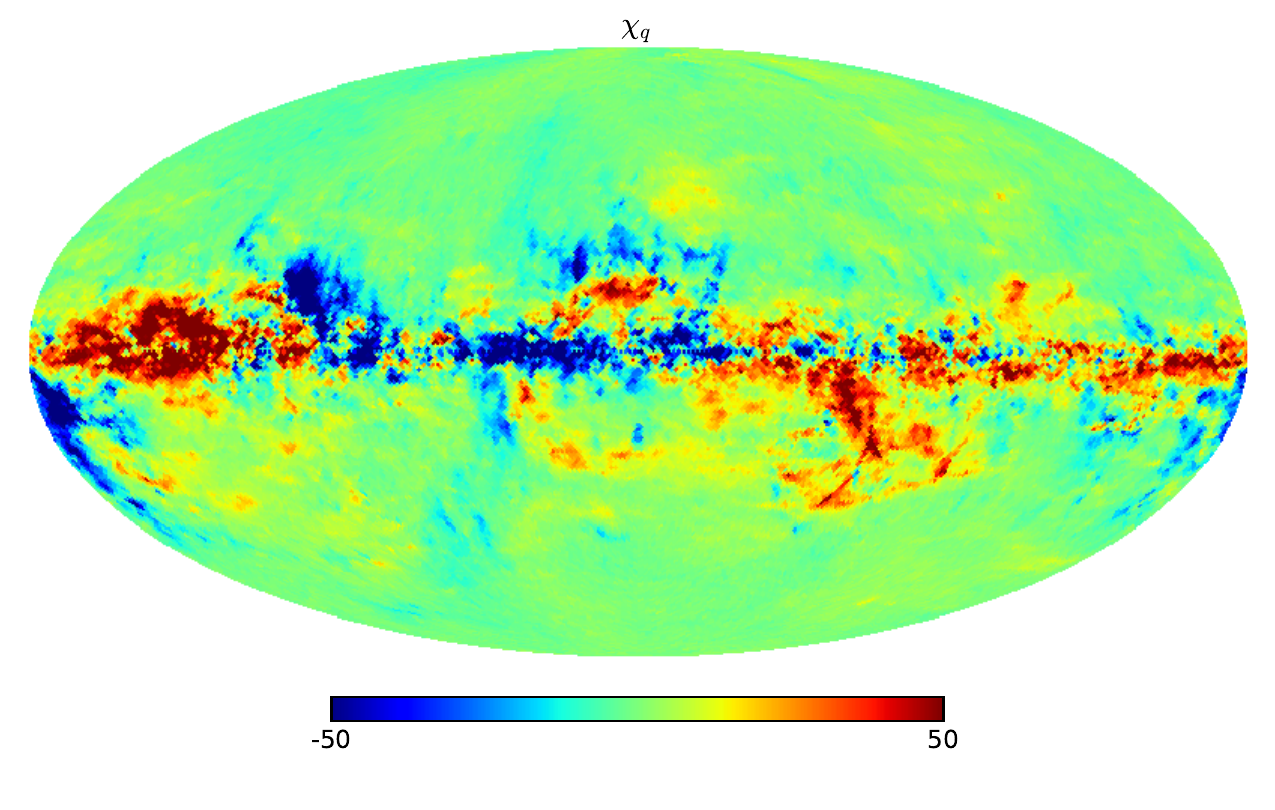} &
                \includegraphics[width=.23\linewidth]{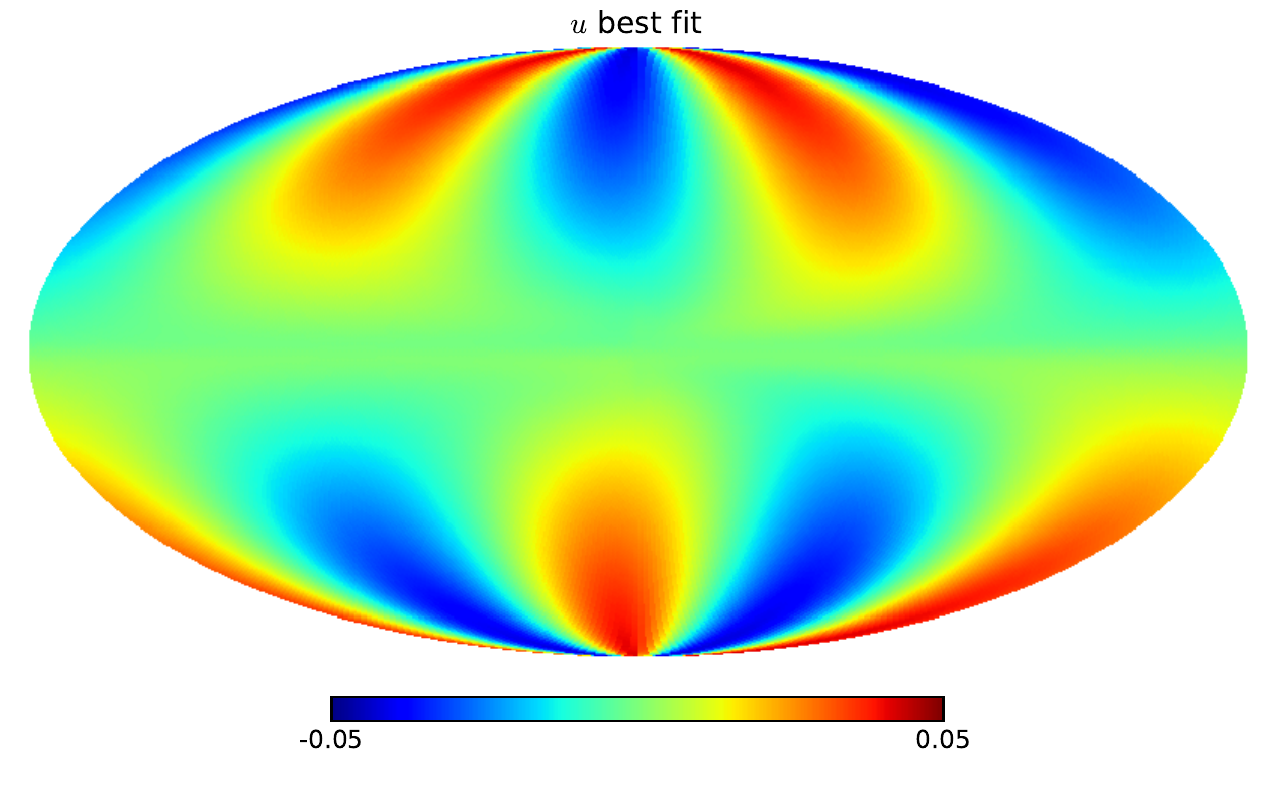} &
                        \includegraphics[width=.23\linewidth]{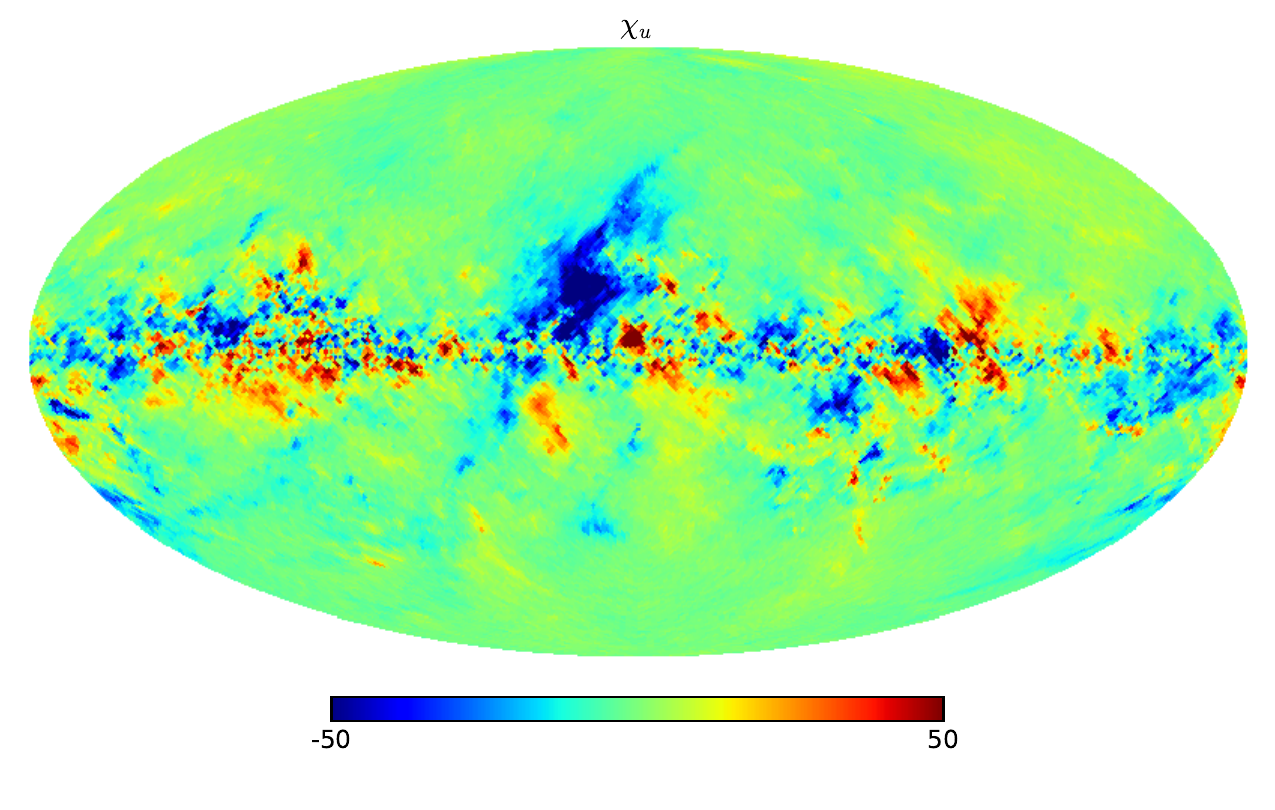} \\

\includegraphics[width=.23\linewidth]{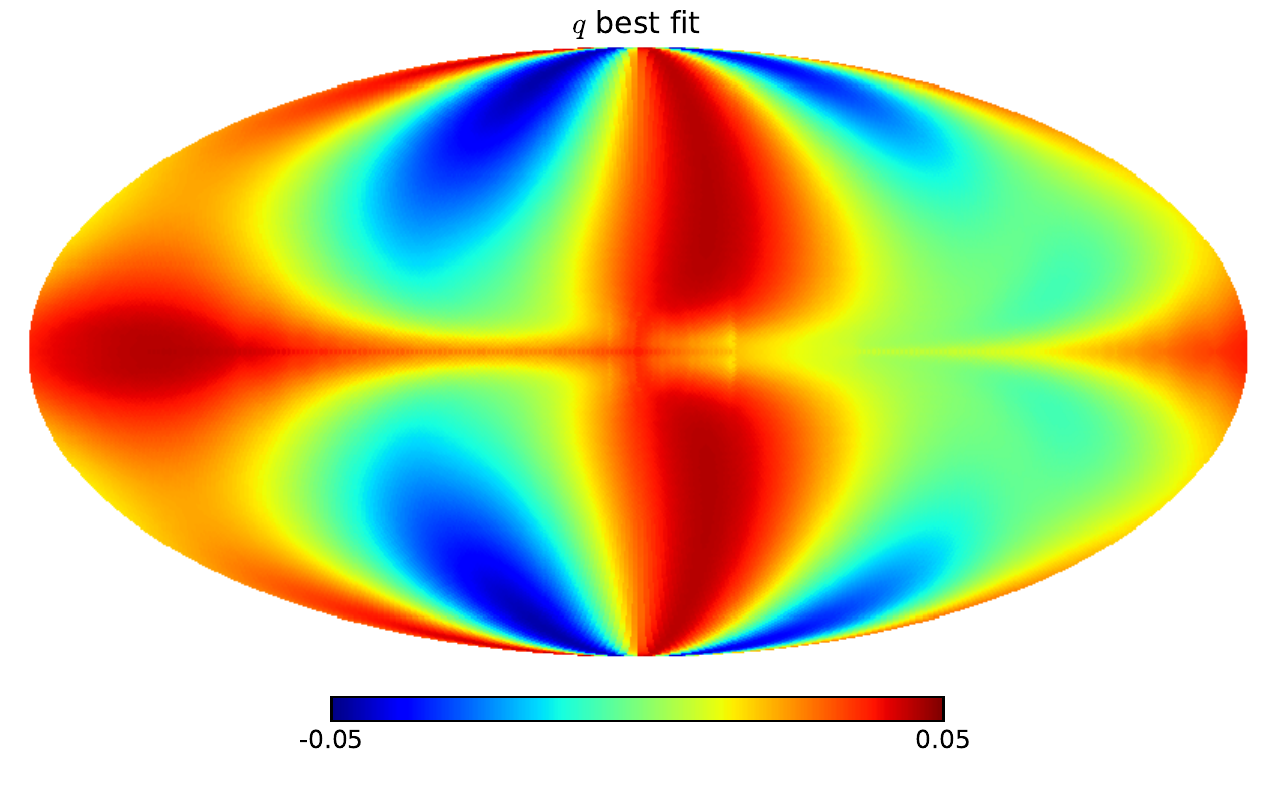} &
        \includegraphics[width=.23\linewidth]{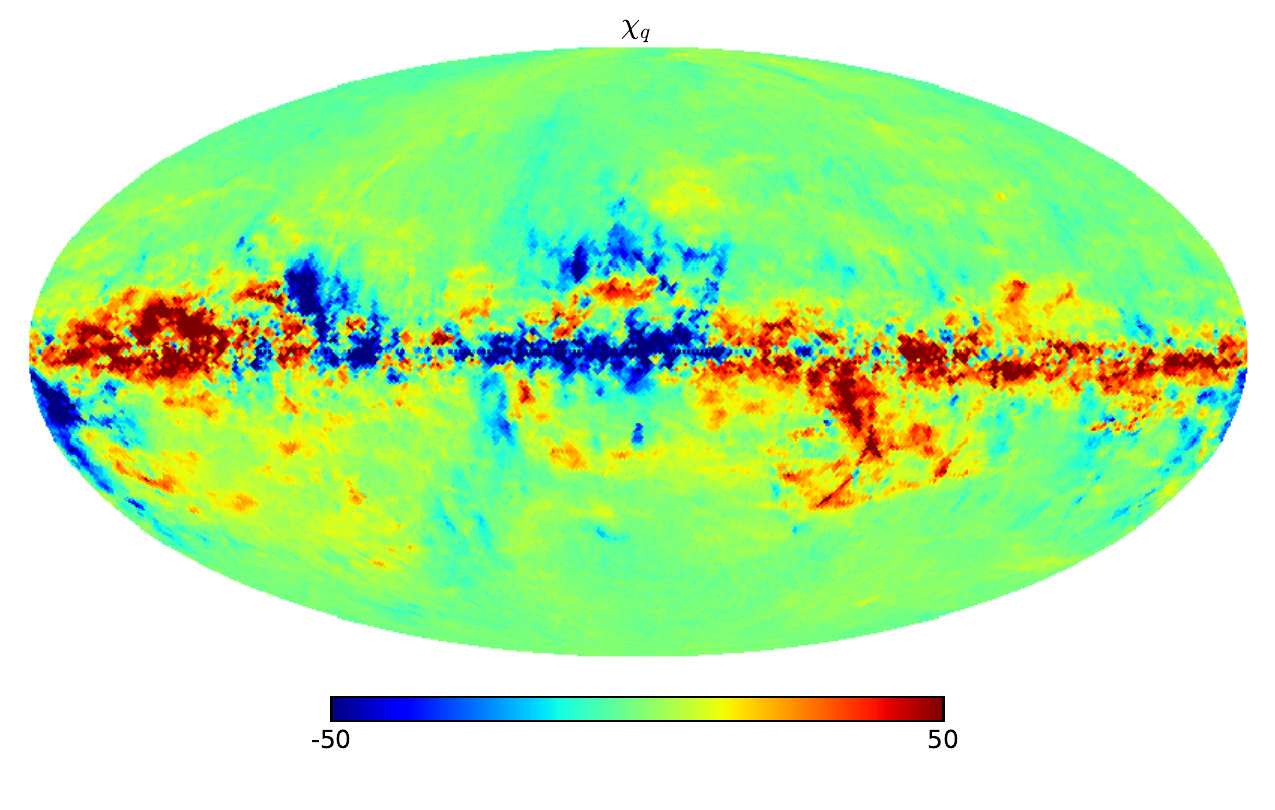} &
                \includegraphics[width=.23\linewidth]{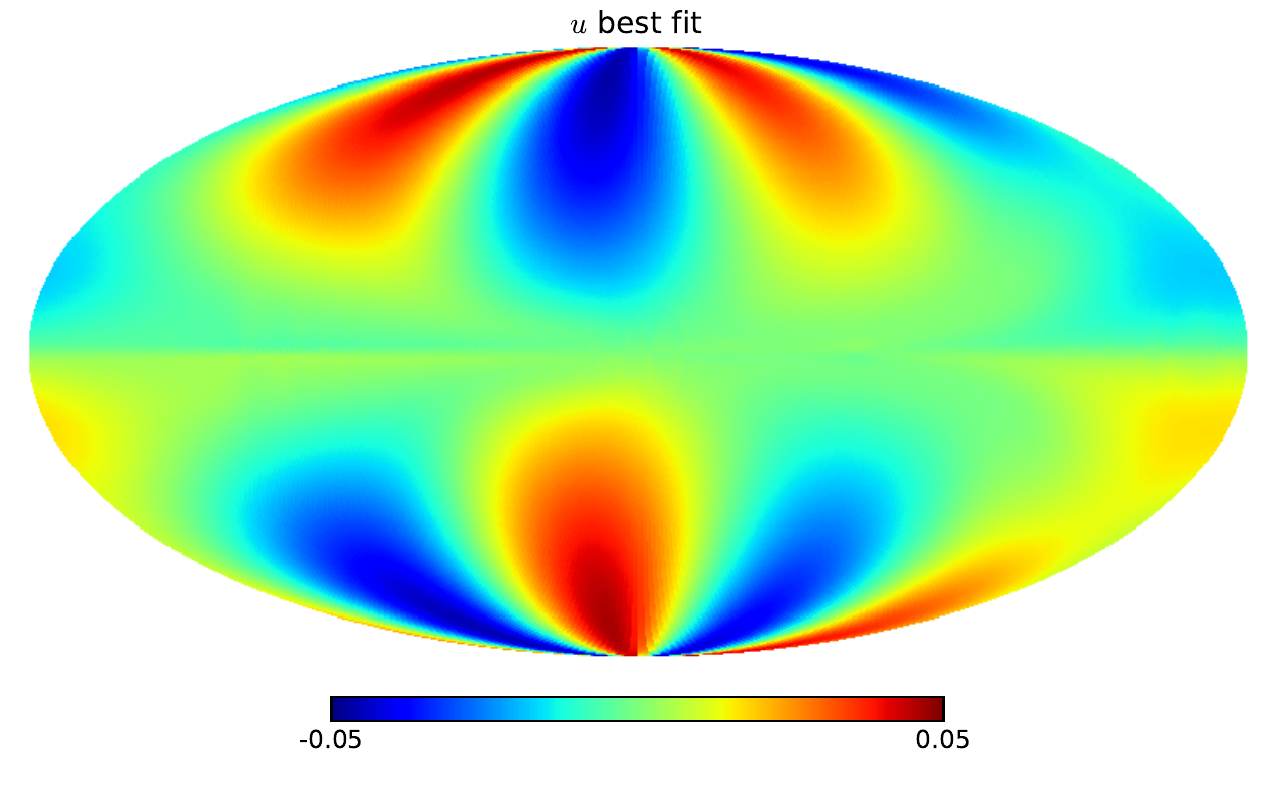} &
                        \includegraphics[width=.23\linewidth]{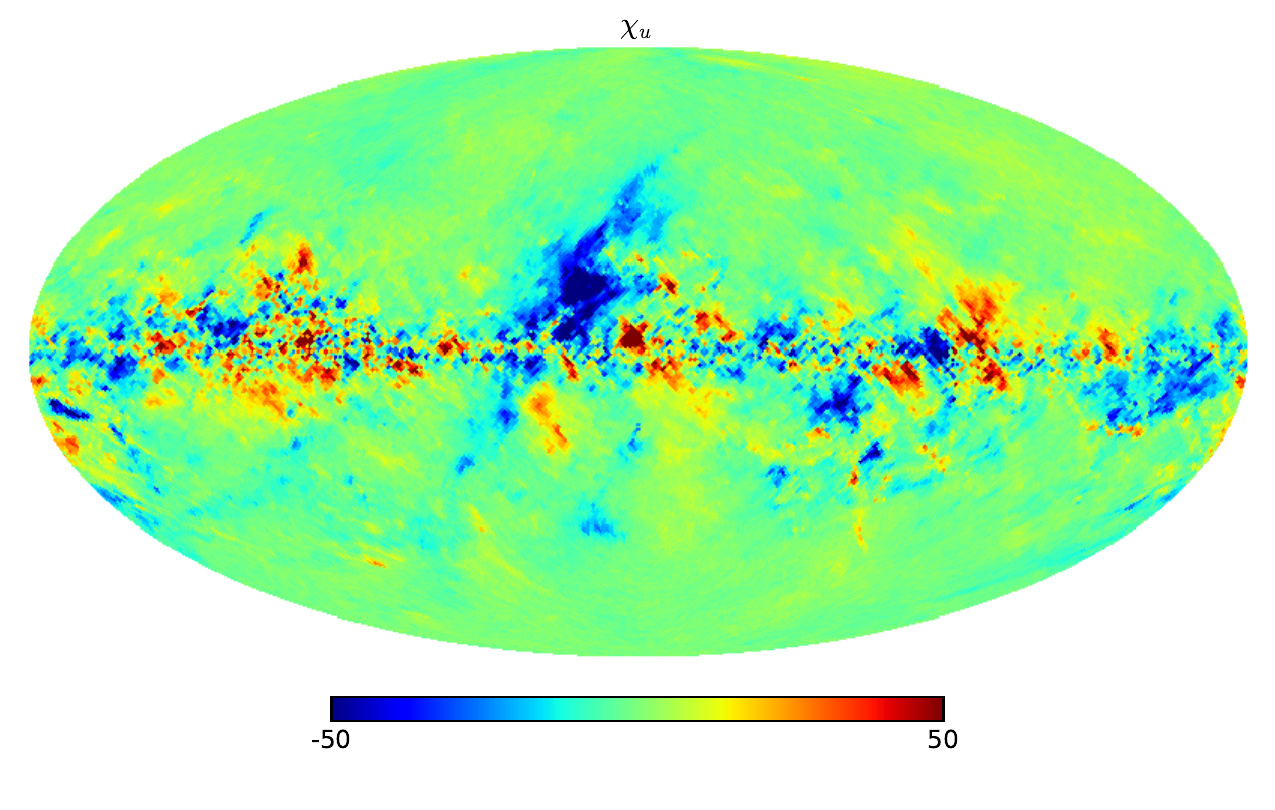} \\

\includegraphics[width=.23\linewidth]{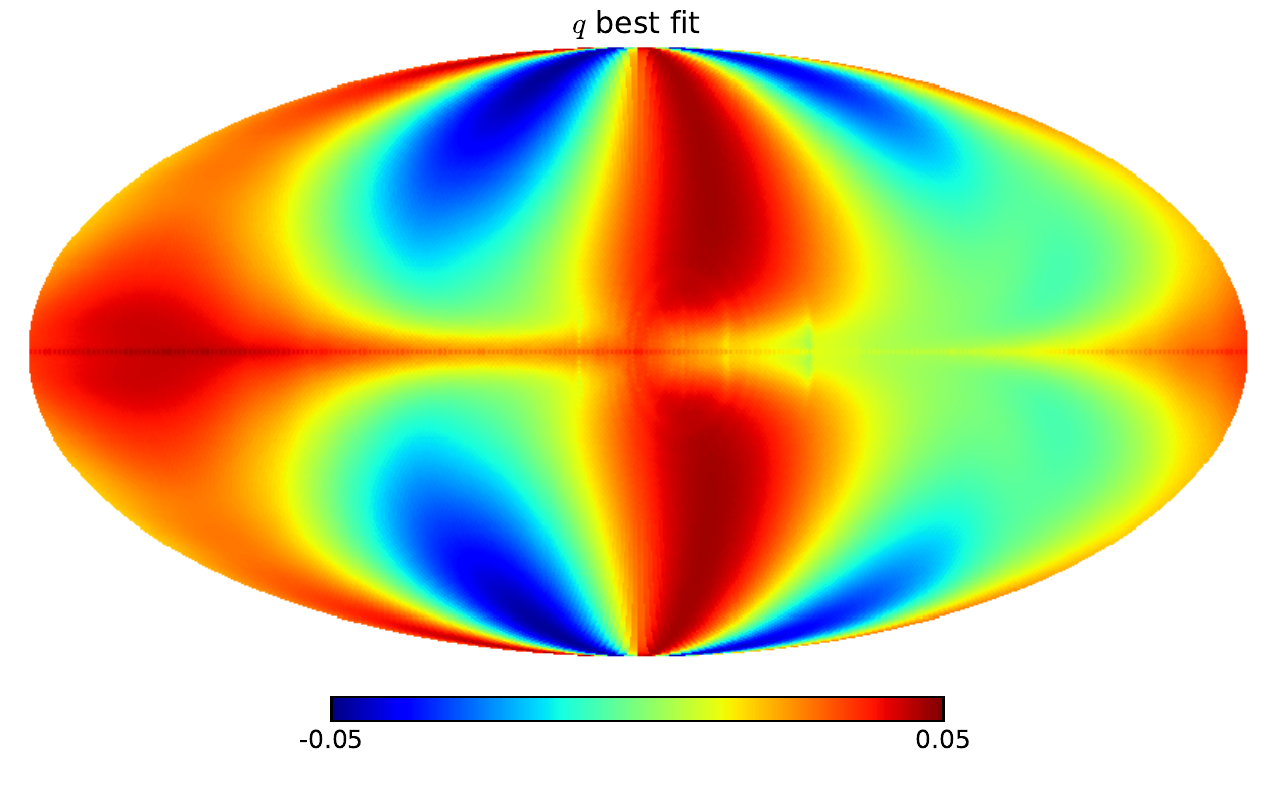} &
        \includegraphics[width=.23\linewidth]{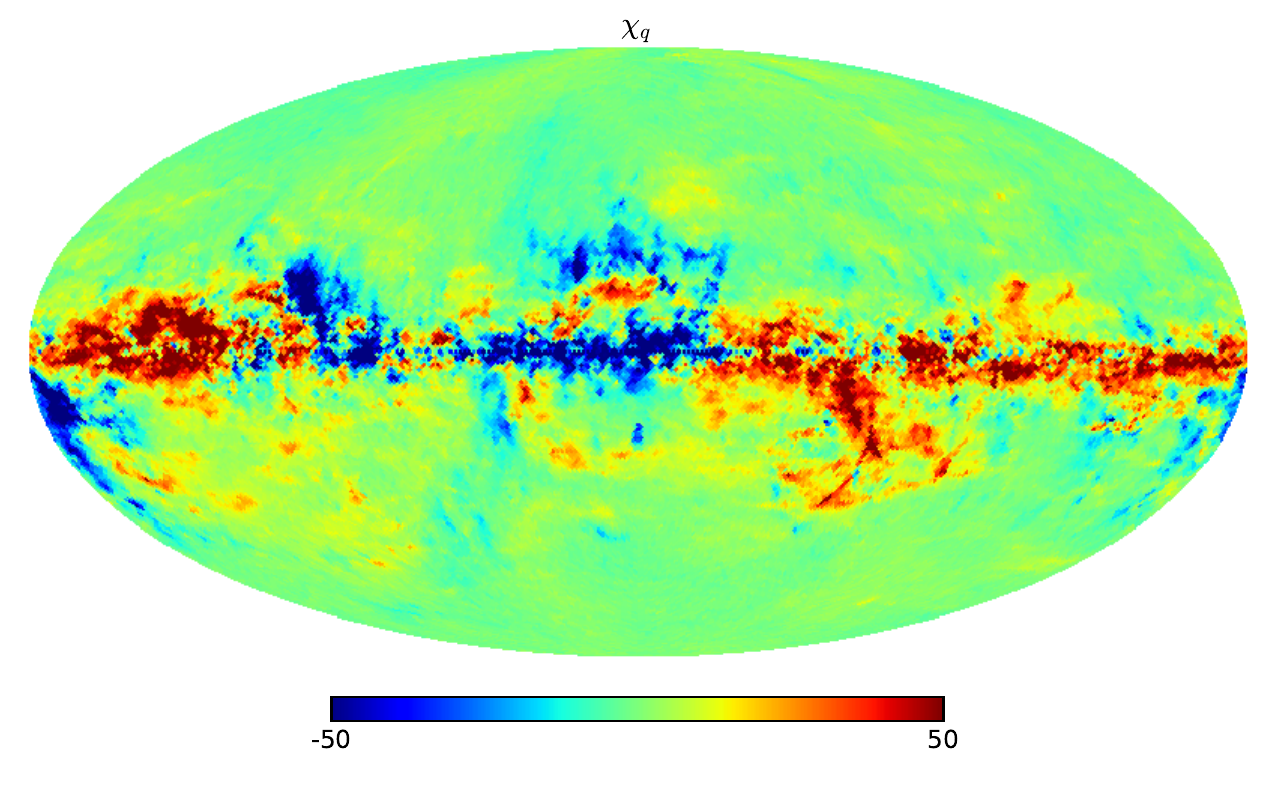} &
                \includegraphics[width=.23\linewidth]{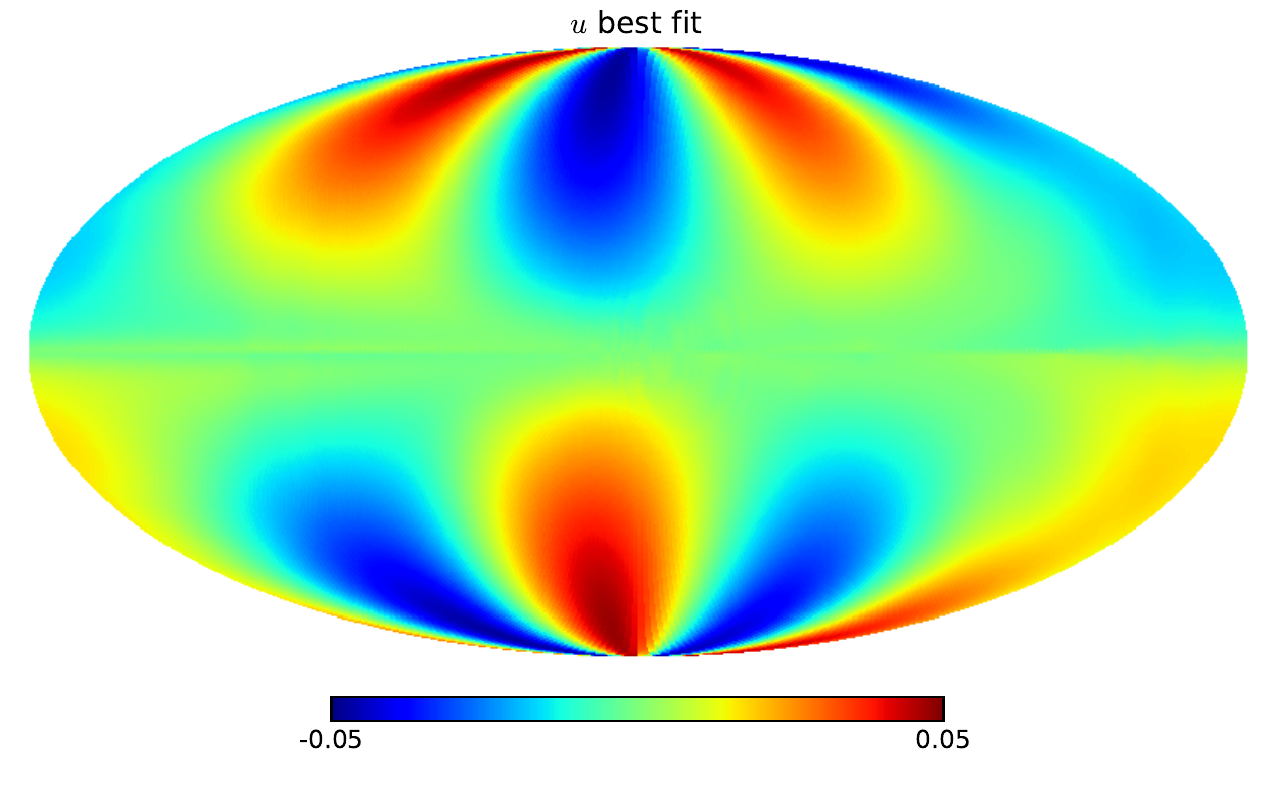} &
                        \includegraphics[width=.23\linewidth]{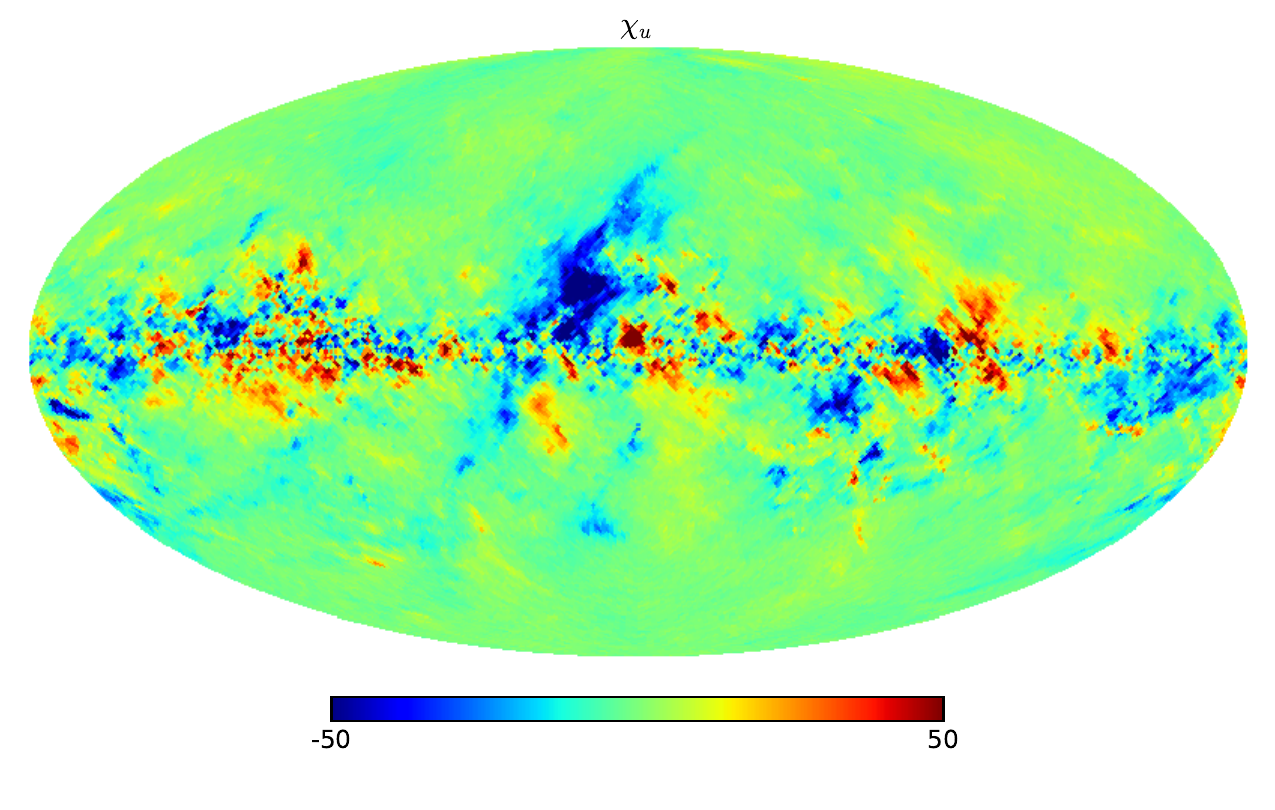} \\

\end{tabular}
\caption{Polarization maps. First row show the 353-GHz maps of the reduced Stokes parameters from \textit{Planck} downgraded at $N_{\rm{side}} = 64$ and the corresponding map of uncertainties that we use to compute the $\chi^2$ ($q$, $\sigma_{\rm{q}}$, $u$, $\sigma_{\rm{u}}$). Rows two to five correspond to GMF models labeled ASS, LSA, BSS and QSS using the best-fit of the ED $n_{\rm{d}}$ model. The obtained best-fits are shown in the first and third columns and the statistical significance of their residuals, per-pixel, are shown in the second and fourth columns.}
\label{fig:qu_fit}
\end{figure*}
\begin{figure*}[t]
\centering
\begin{tabular}{ccc}
\includegraphics[trim={0.5cm 0cm 2.cm 0cm},clip,width=.3\textwidth]{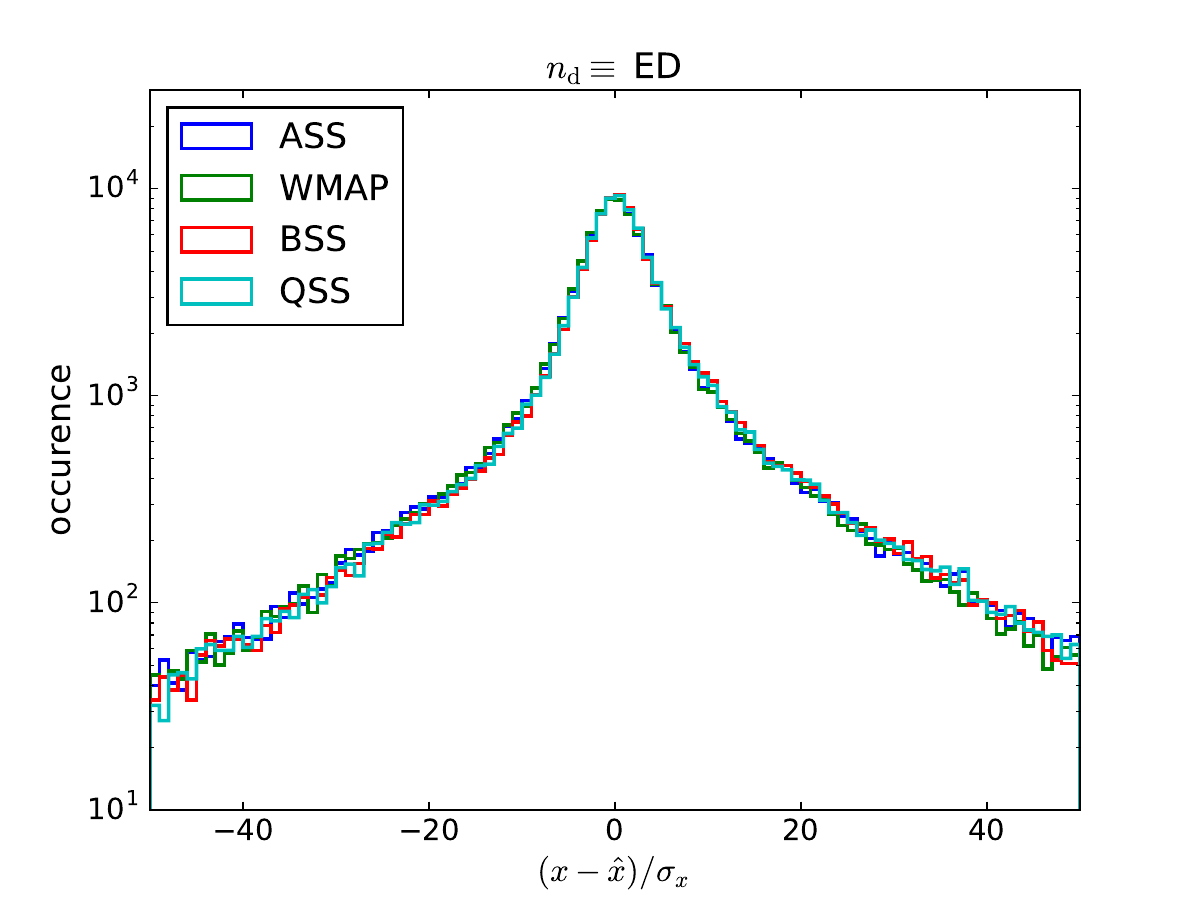} &
\includegraphics[trim={0.5cm 0cm 2.cm 0cm},clip,width=.3\textwidth]{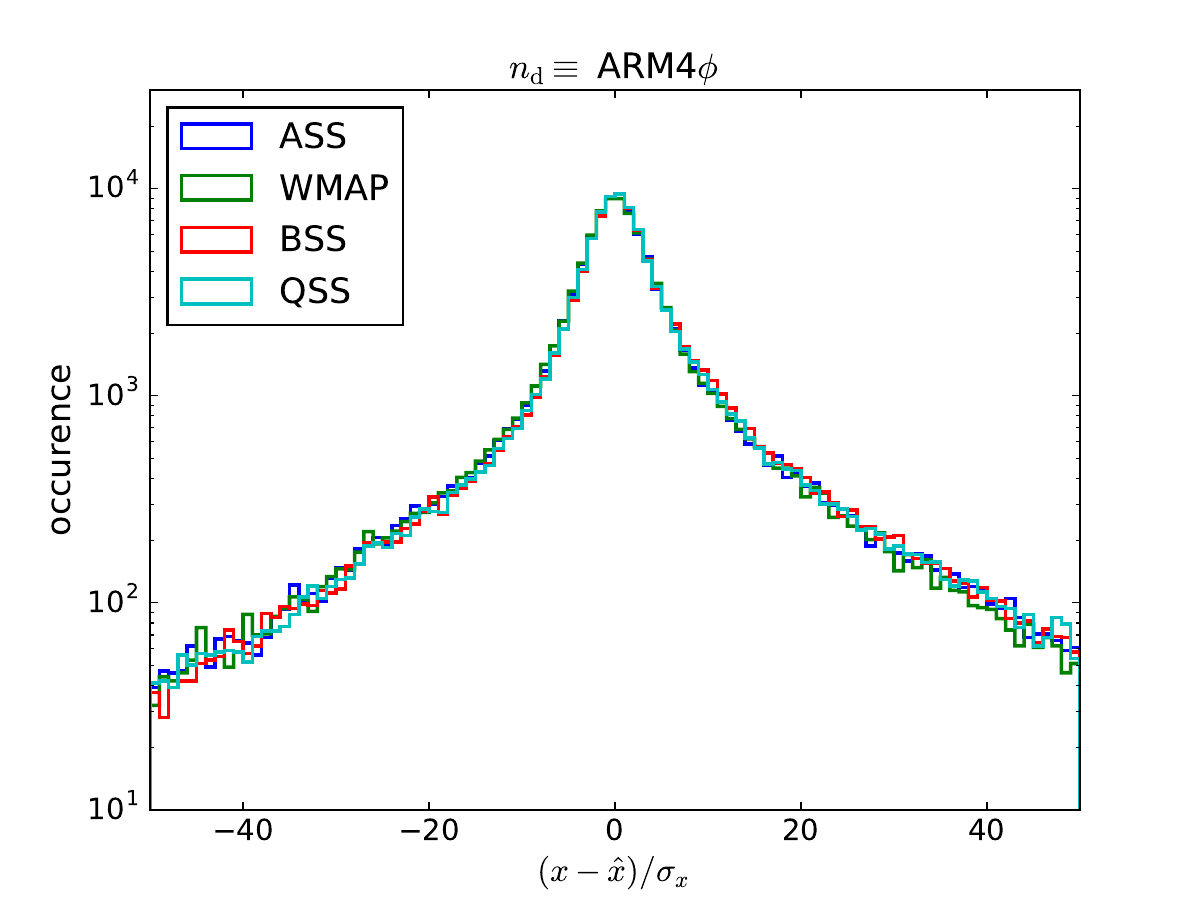} &
\includegraphics[trim={0.5cm 0cm 2.cm 0cm},clip,width=.3\textwidth]{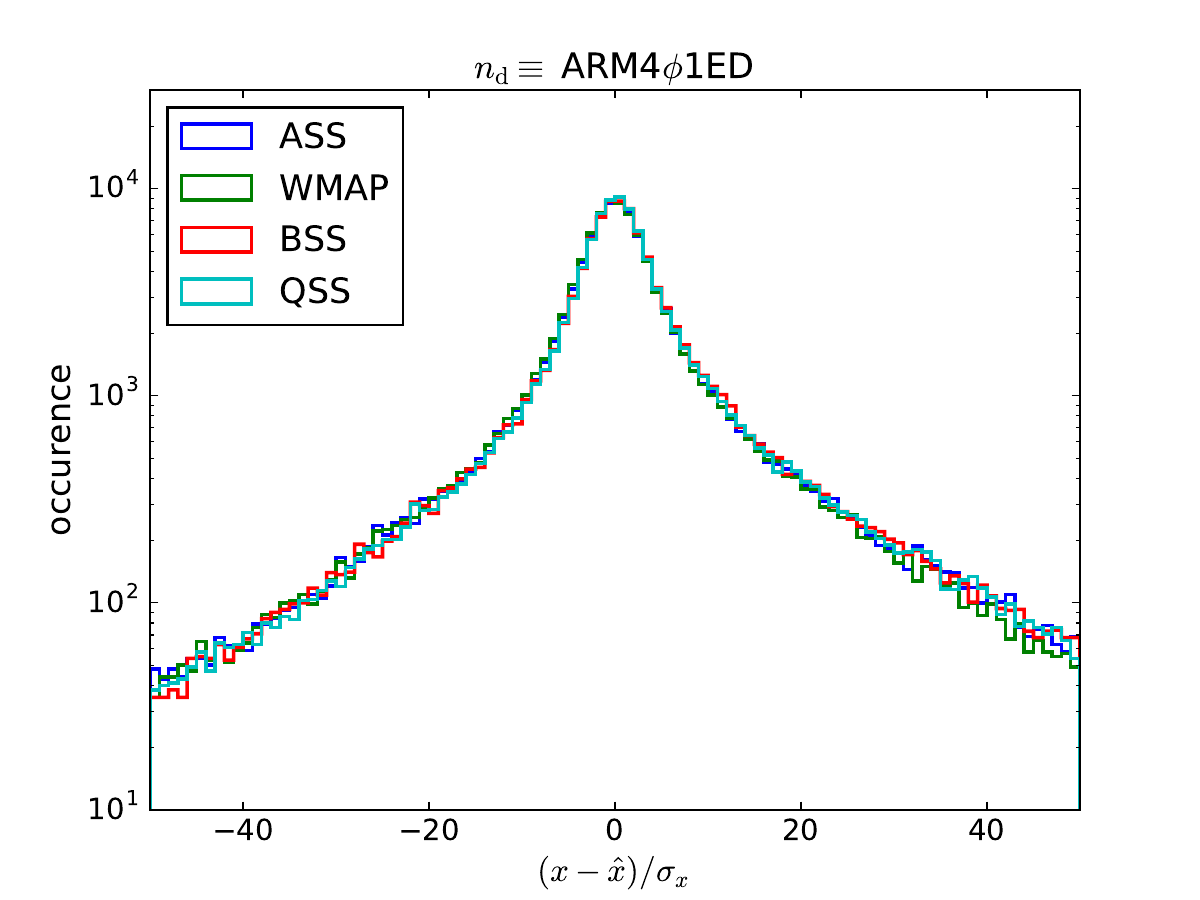}
\end{tabular}
\caption{Histograms of the significance of the residuals. $x$ stands for the concatenation
of the $q$ and $u$ parameters.
Each GMF model is represented by a different color and each panel
corresponds to a different underlying $n_{\rm{d}}$ model: ED, ARM4,
and ARM4$\oplus$ED from left to right.}
\label{fig:qufits_chi-hist}
\end{figure*}

\subsubsection{Fitting procedure}
Similar to what we did to fit the intensity map, we adopted flat priors on the free-parameters of the different models of the GMF, which are presented in Table~\ref{table:param-GMF}. Details on the parametrization of these GMF models are given in Appendix~\ref{sec:GMFmodel}. We initialized the several hundred walkers according to uniform distributions and let the system evolve until convergence is reached. The parameter space for each class of model is also given in Table~\ref{table:param-GMF}.
Unlike the case of the intensity fit, we work only at the $N_{\rm{side}} = 64$ because our implementation is fast enough and  there are only a few free parameters of the GMF models that it can make vary.

\subsubsection{Best-fit models}
In Fig.~\ref{fig:qu_fit}, we show the maps corresponding to the best-fit of each GMF and the maps corresponding to the significance of the residuals in the case where the $n_{\rm{d}}$ model is the simplest (ED).
In that figure, the different degree of complexity of the pattern appearing in the maps between the different best-fit models is simply due to the GMF model.
It is quite remarkable how the simple modeling adopted here are capable of roughly reproducing the largest patterns observed in the data. However, an inspection of the residuals maps calls for further developments in the polarization map modeling, especially with regard to tackling low and intermediate Galactic latitudes.
The agreement between the models and the data are indeed visually found to be better at high Galactic latitudes. This is further discussed in the next section.

\smallskip

The parameter values of all the best-fits of the different GMF models considered are reported in Table~\ref{table:param-GMF} for each of the $n_{\rm{d}}$ models that we investigated in Sect.~\ref{sec:ndust}.
\begin{table}[t]
\centering
\caption{Reduced $\chi^2$ values of the best-fit models (\textit{i}) of the dust density distribution on the intensity map and (\textit{ii}) of the GMF models on the reduced-jointed-Stokes maps ($q,\,u$) for each of the best-fit of the $n_{\rm{d}}$ model.}
\label{table:chi2}
\begin{tabular}{ll ccc}
\hline
\hline
data    & GMF model     & \multicolumn{3}{c}{$n_{\rm{d}}$ model}                        \\
                &               &       ED      &       ARM4    &       ARM4$\oplus$ED  \\
\hline
\\[-1.ex]
$I$     &               & 6720.0   &   6206.6 &   5608.3        \\
\\[-.5ex]
$(q,\,u)$       &       ASS     & 186.9 & 195.1 & 199.6 \\
                        &       LSA     & 179.1 & 185.5 & 188.3 \\
                        &       BSS     & 181.0 & 188.4 & 193.8 \\
                        &   QSS & 182.6 & 191.6 & 195.8 \\
\\[-.5ex]
\hline
\end{tabular}
\end{table}
In Table~\ref{table:chi2}, we also present the values of the reduced $\chi^2$ that quantify the goodness of our fits to the GMF models.
In Fig.~\ref{fig:qufits_chi-hist}, we show the histograms of the $\chi_i$ corresponding to the best adjustments of the four investigated GMF models and for the three $n_{\rm{d}}$ models fitted to the $I_{\rm{353}}$ map.
These distributions are nearly Gaussian. Again, their widths are large as the uncertainties in the data can not entirely cope with the residual to such simplistic models.
We note, however, that the values of the reduced $\chi^2$ of the best-fit models are one order of magnitude better than those obtained for the fits of the intensity map.

\medskip

Furthermore, the obtained best-fits of the GMF are generally not compatible within the uncertainties when moving from one dust density distribution to another. The main reason for this is
that the data are complex and signal dominated in some regions and that our models are far from being able to account for all the complexity and the richness contained in the maps. Indeed, despite the conservative determination of the errors that we adopt (see Eq.~\ref{eq:error_max}),
the posterior distributions of the model parameters are too narrow, as they do not reflect the uncertainties that come from mismodeling.
The use of a MCMC technique, however, is fully justified by the fact that the multidimensional $\chi^2$ surfaces, overall, have convoluted geometries and present a large number of local minima that would compromise the performance of other minimization technique.
For now, we consider that the scatter of the parameter values obtained with the different $n_{\rm{d}}$ models is more representative of the uncertainties on those best-fit parameters. This again stresses the need for more complex and realistic models of the dust density distribution and of the GMF in order to optimally exploit the currently available data.

\smallskip

To test the robustness of our best-fit models, we also constrain the GMF parameters using the $n_{\rm{d}}$ best-fit models obtained from the fits on the dust optical-depth map.
The fitted ($q,\,u$) maps are obtained simply by replacement of the intensity map by the $\tau_{353}$ map, with the corresponding uncertainties. The best-fit polarization maps and the geometrical structure of the GMF are found to be in good agreement with those obtained when the modeling of $n_{\rm{d}}$ is from the $I_{\rm{353}}$ fits.
This test makes us confident that the GMF constraints obtained through the fit of the reduced Stokes parameter maps are robust against possible residuals from point sources or other components of the diffuse Galactic emission.

In order to test further the stability of our GMF reconstruction against systematic effects, such as the leakage of the intensity towards the $Q$ and $U$ polarization channels, we apply  the \textit{Global Generalized Fit} correction to the data, as suggested by \cite{PlanckVIII2015}. We proceed to the fit of the GMF model for the least evolved and the most evolved modelings (ED~+~ASS and ARM4$\oplus$ED~+~QSS) with and without applying the quoted corrections.
The best-fit parameter values that we obtain are in agreement at the ten percent level with the previous values. These two tests make us confident that our GMF reconstruction with the \textit{Planck} data is not strongly biased by instrumental and map-making systematics and shows that those systematics in \textit{Planck} polarization maps have a significant contribution to the modeling uncertainty budget whereas not accounted for in the covariance matrix. This conclusion was already reached in \cite{Alv2018} and \cite{Pel2020}.

\subsubsection{Constraints on the GMF geometry}
\begin{figure}[t]
\centering
\includegraphics[width=.95\linewidth]{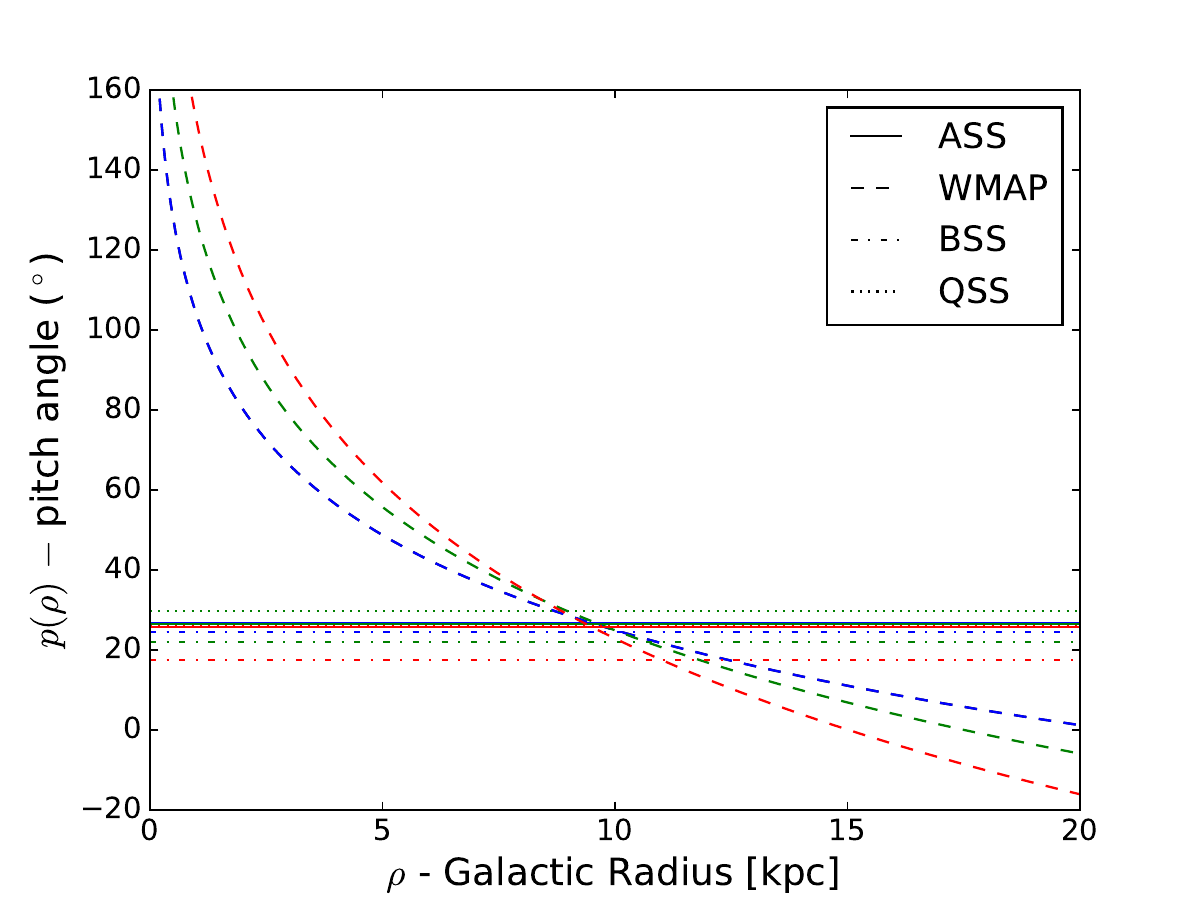} \\
\includegraphics[width=.95\linewidth]{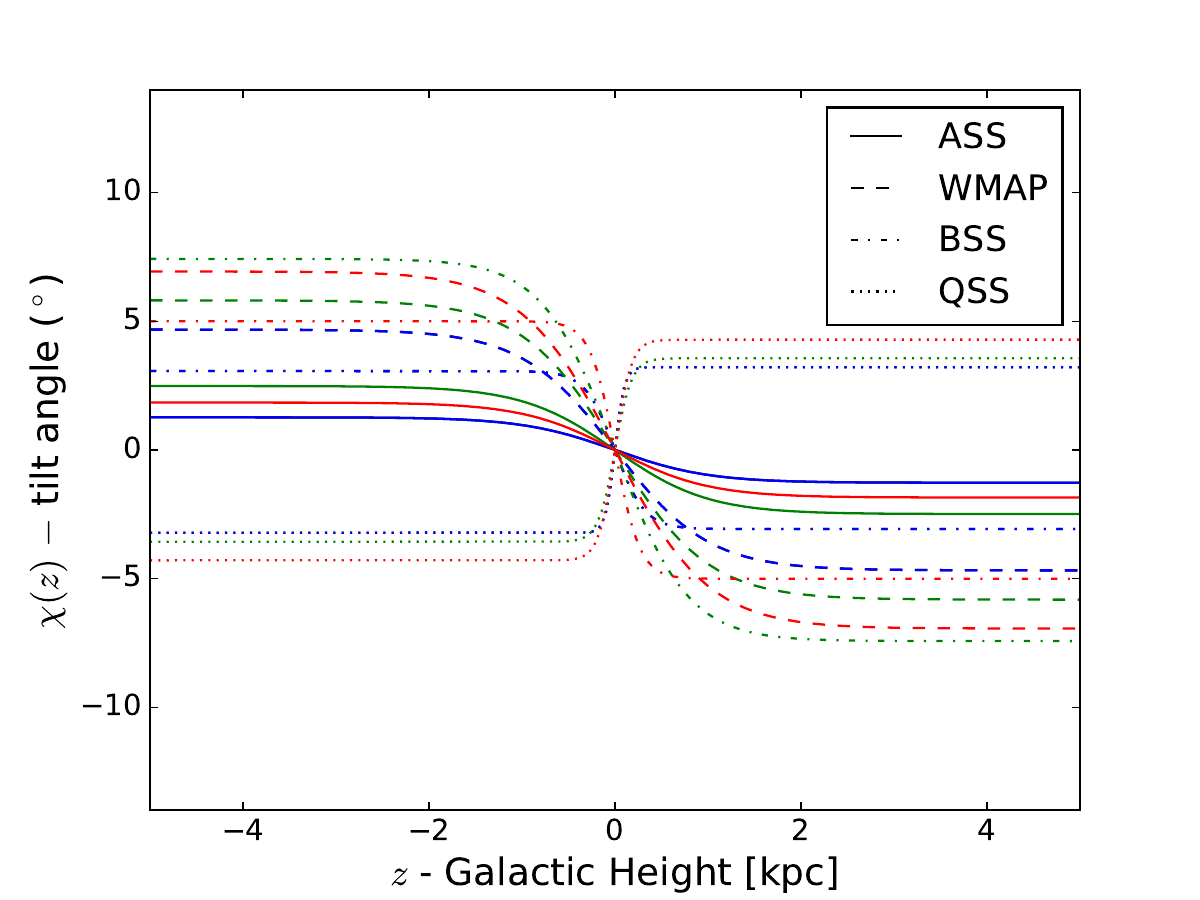}
\caption{Functional forms of the pitch (top) and tilt (bottom) angles as a function of the Galactic cylindrical coordinate; $\rho$ and $z$, respectively. Each line corresponds to a best-fit model. Solid, dashed, dash-dotted, and dotted lines correspond to the ASS, LSA, BSS, and QSS GMF models. Blue, green, and red correspond to the best-fits obtained while assuming the $n_{\rm{d}}$ model to be ED, ARM4 or ARM4$\oplus$ED, respectively. The pitch angle is constant for the ASS, BSS, and QSS models.}
\label{fig:GMF_pitchandtilt}
\end{figure}

The radial and height evolution of the pitch and the tilt angles that correspond to the best-fit parameters for all the  models considered are shown on Fig.~\ref{fig:GMF_pitchandtilt}.
\begin{figure*}[h]
\centering
\begin{tabular}{ll}
\includegraphics[trim={0cm 0cm 4.5cm 0cm},clip,width=.36\linewidth]{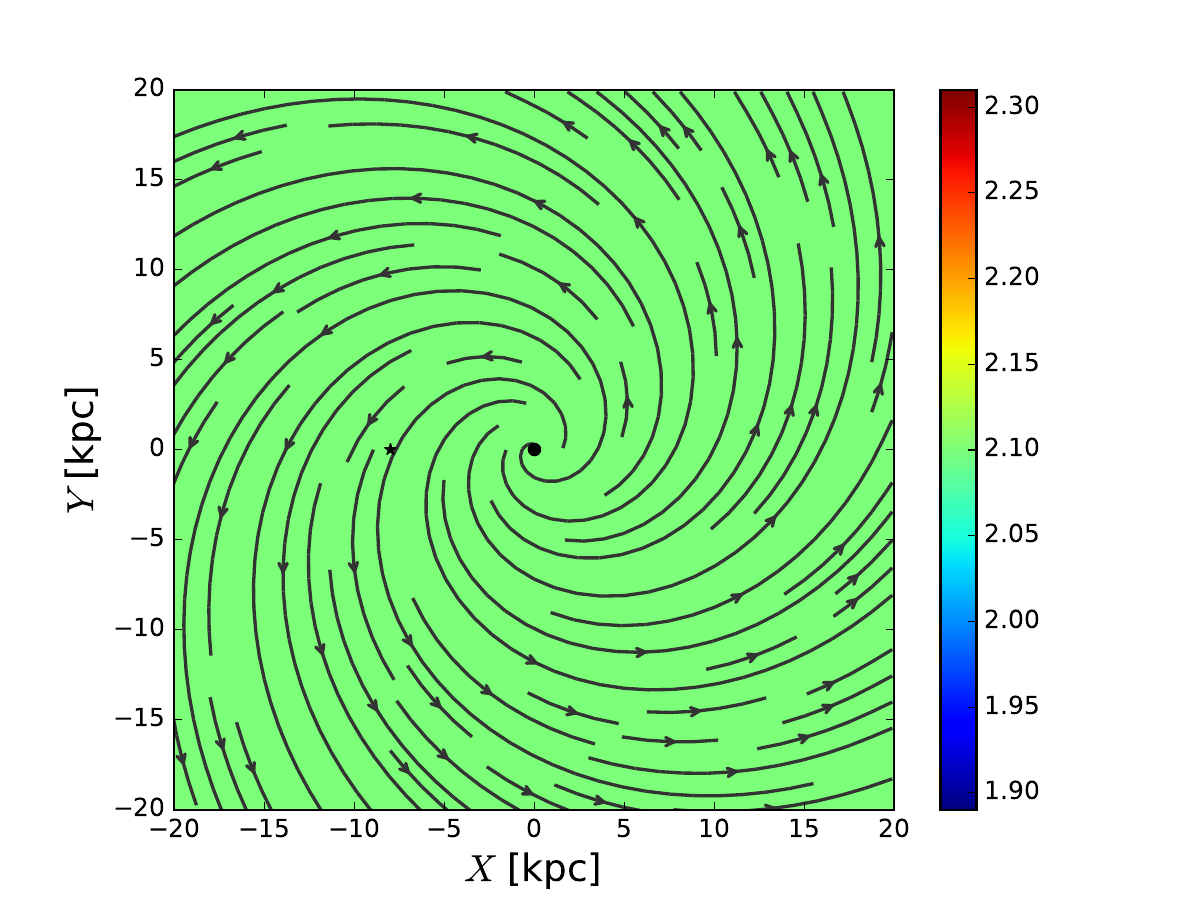} &
        \includegraphics[trim={0cm 0cm 4.5cm 0cm},clip,width=.36\linewidth]{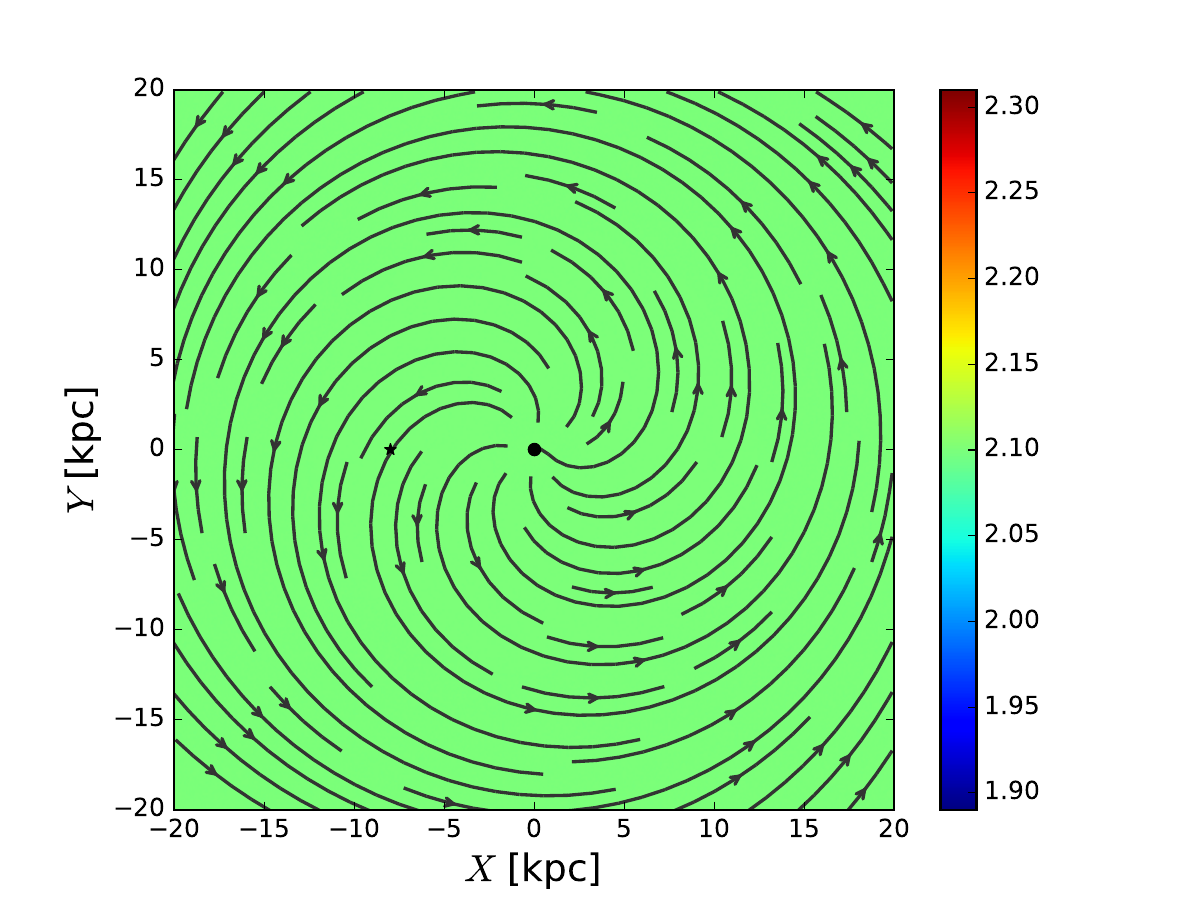} \\[-4.4ex]

\hspace{.12cm}
\includegraphics[trim={0cm 4cm 4.5cm 5.3cm},clip,width=.348\linewidth]{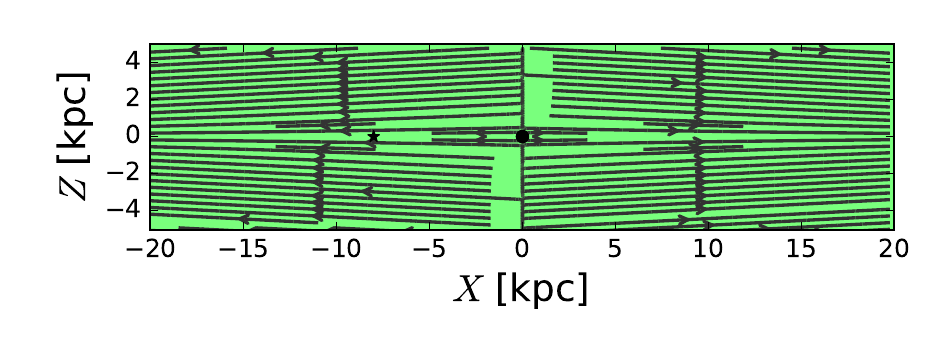} &
        \hspace{.12cm}
        \includegraphics[trim={0cm 4cm 4.5cm 5.3cm},clip,width=.348\linewidth]{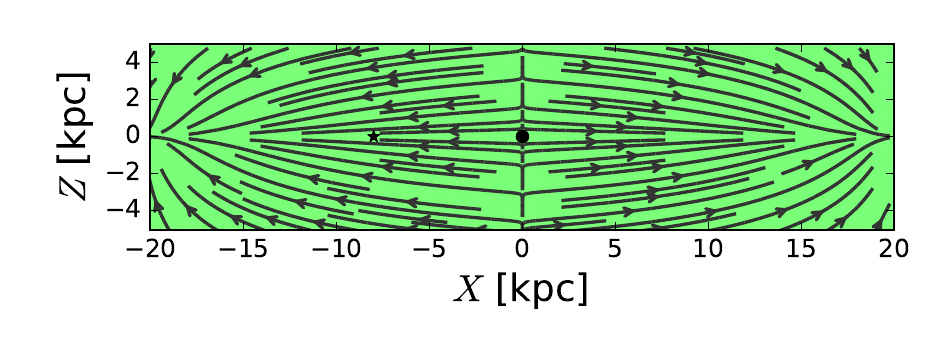} \\[-2.4ex]

\includegraphics[trim={0cm 0cm 4.5cm 0cm},clip,width=.36\linewidth]{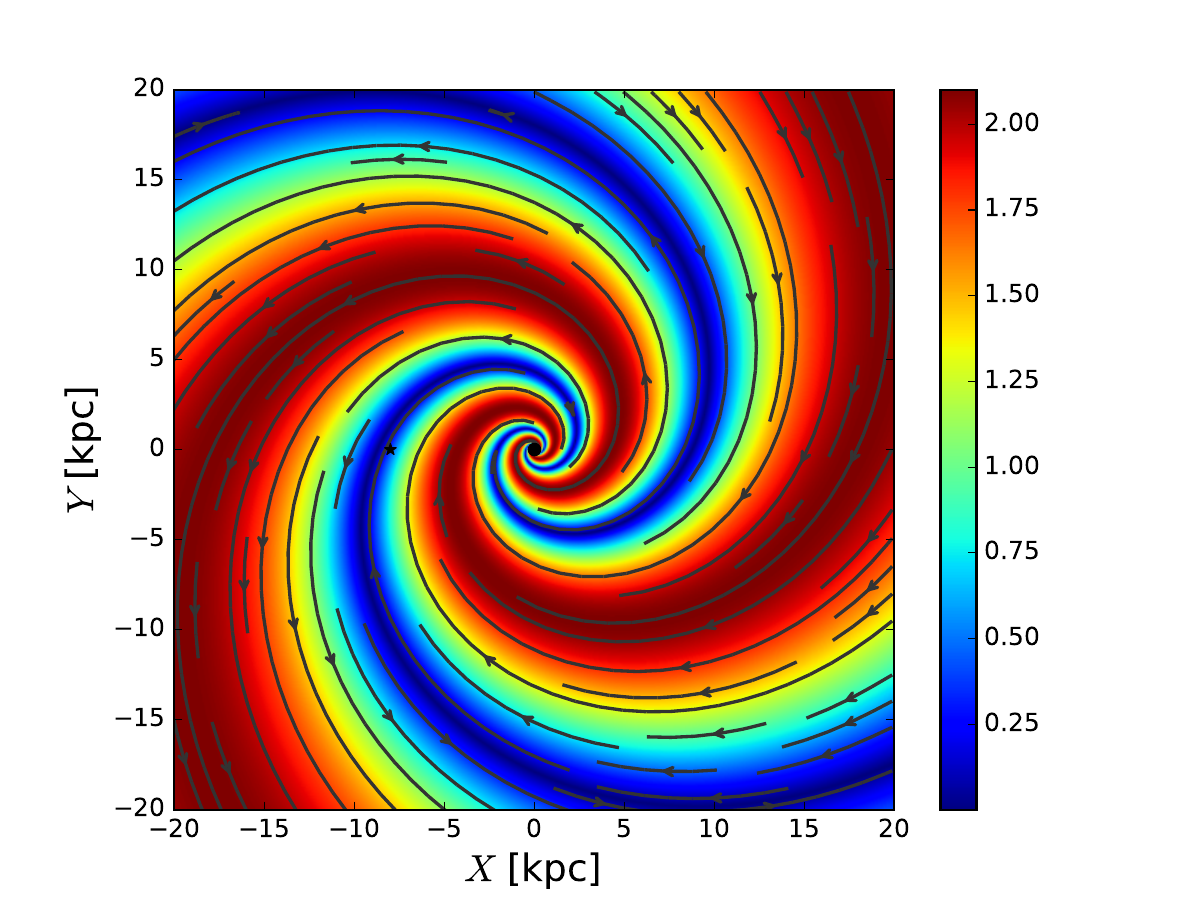} &
        \includegraphics[trim={0cm 0cm 4.5cm 0cm},clip,width=.36\linewidth]{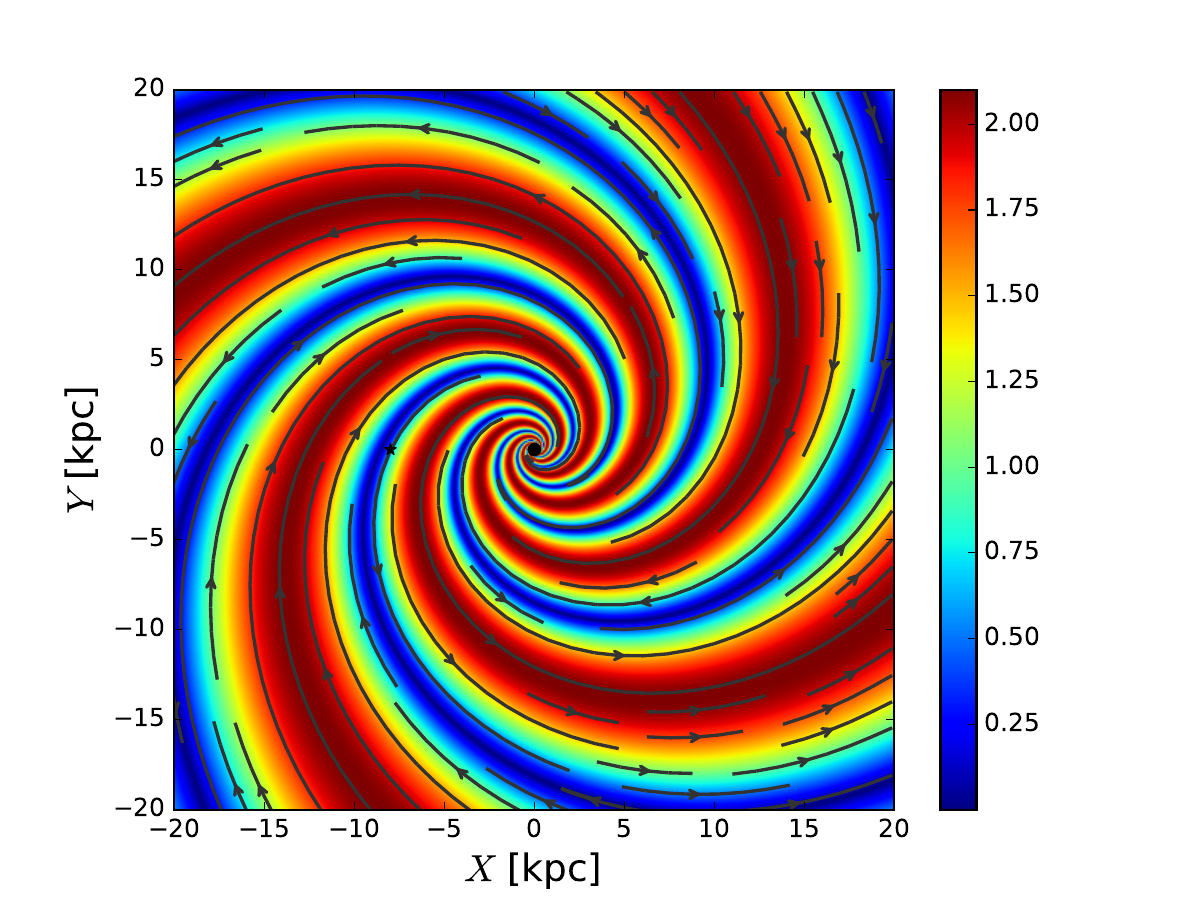} \\[-4.4ex]

\hspace{.12cm}
\includegraphics[trim={0cm 4cm 4.5cm 5.3cm},clip,width=.348\linewidth]{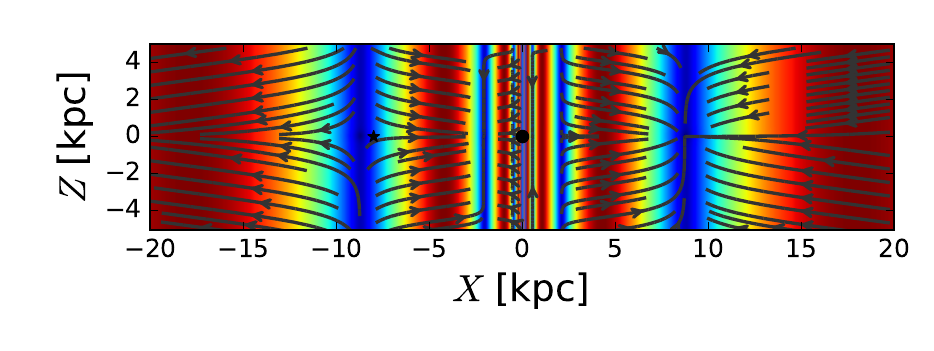} &
        \hspace{.12cm}
        \includegraphics[trim={0cm 4cm 4.5cm 5.3cm},clip,width=.348\linewidth]{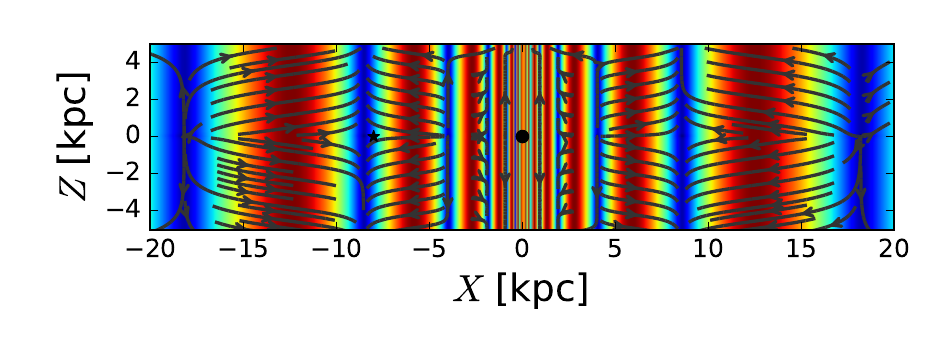} \\[-2.4ex]

\end{tabular}
\caption{Field line geometrical structures of best-fit GMF models in the $(x,\,y,\,z=0)$ and $(x,\,y=0,\,z)$ planes of the Galaxy. From top-left to bottom-right: ASS, LSA, BSS, and QSS GMF models. The GMF reconstructed models corresponds to the best-fits obtained while assuming that $n_{\rm{d}}$ is our best-fit ED model. The black dot at the center of the figure is the Galactic center and the black star is at the Sun position. The color scales offer information about the relative strength of GMF (i.e., the norm of the vector field at each location) and of its direction (red going clockwise, blue going counter-clockwise). In the case of ASS and LSA models, the (constant) norm has been fixed to 2.1.}
\label{fig:GMF_XYZplanes}
\end{figure*}
For a given GMF model (same line style but different color in Fig.~\ref{fig:GMF_pitchandtilt}), the agreement among the different choice of $n_{\rm{d}}$ model is remarkable. The values of the pitch angles and of the tilt angles agree fairly well through the different modelings. The relative difference between the parameters characterizing those angles range from few percent to about few tens of a percent when comparing very
different modelings, such as that of the model having the lowest degree of complexity (ED + ASS) and the one with the largest one (ARM4$\oplus$ED~+~QSS).
This level of uncertainty is of the same order as the one attributable to the \textit{Planck} residual systematics.
As inferred from Fig.~\ref{fig:GMF_pitchandtilt}, the agreement on the pitch angle is the strongest for a Galactic radius of about 8 kpc, with a mean pitch angle of about 27 degrees and a mean scatter among the twelve computed models of about five degrees, corresponding to a mean relative difference of 14 percent between the reconstructions.
However, we also observe that in the case of the LSA model, for which the pitch angle can vary with radius, the evolution from the inner part of the Galaxy to the outskirt is very large.
Finally, we also find an inversion of the sign of the tilt angle for the QSS model with respect to the other ones.

\smallskip

In Fig.~\ref{fig:GMF_XYZplanes} we show the direction of the large-scale GMF lines in horizontal and vertical planes of the Galaxy, namely the $(x,\,y,\,z=0)$ and the $(x,\,y=0,\,z)$ planes. Stream plots of the field lines are shown for the best fits obtained for each GMF models assuming the $n_{\rm{d}}$ to follow the ED model.
Very similar plots are obtained with the other $n_{\rm{d}}$ models, as we discuss later.
In terms of the visuals and given that the functional forms of the models may be different,
the similarity among the GMF reconstructions is remarkable.
We note that for the LSA model, a second minimum of the $\chi^2$ is found with a distant solution in the parameter space but that is similar to the solution obtained by \cite{Ste2018} in fitting Faraday-depth all-sky
data and synchrotron data.
We present this solution in Appendix~\ref{sec:WMAP2nd}. The reason why our global minimum is not reached by these authors is likely due to the fact that they do not allow the $\psi_1$ parameter (see Eqs.~\ref{eq:lsa_pitch}) to span through negative values.
An interesting feature of our best-fit LSA model is that the field lines appear to wind up on themselves at a cylindrical radius of about 18 kpc.

\subsection{Best-fit maps: Comparison and characteristics}
Based on inspection of residual maps and on $\chi^2$ values, none of the twelve best-fit models seems to provide a significantly better fit than the others.
Each combination of dust density and GMF models results on different possible features on the best-fit maps. On the maps, the differences between models are more evident at low and intermediate Galactic latitudes, where the 3D models ($n_{\rm{d}}$ and GMF) differ more strongly and where the uncertainties are very small. At the contrary, at high Galactic latitudes, at about $|b_{\rm{gal}}| \geq 60^\circ$, every best-fit maps seem to converge towards the same solution.
This is well illustrated in Figs.~\ref{fig:I_fit-orth} and~\ref{fig:qu_fit-orth}, where we present orthographic projections of the maps presented in Figs.~\ref{fig:I_fit} and~\ref{fig:qu_fit}, respectively.

\begin{figure}[t]
\centering
\includegraphics[width=\columnwidth]{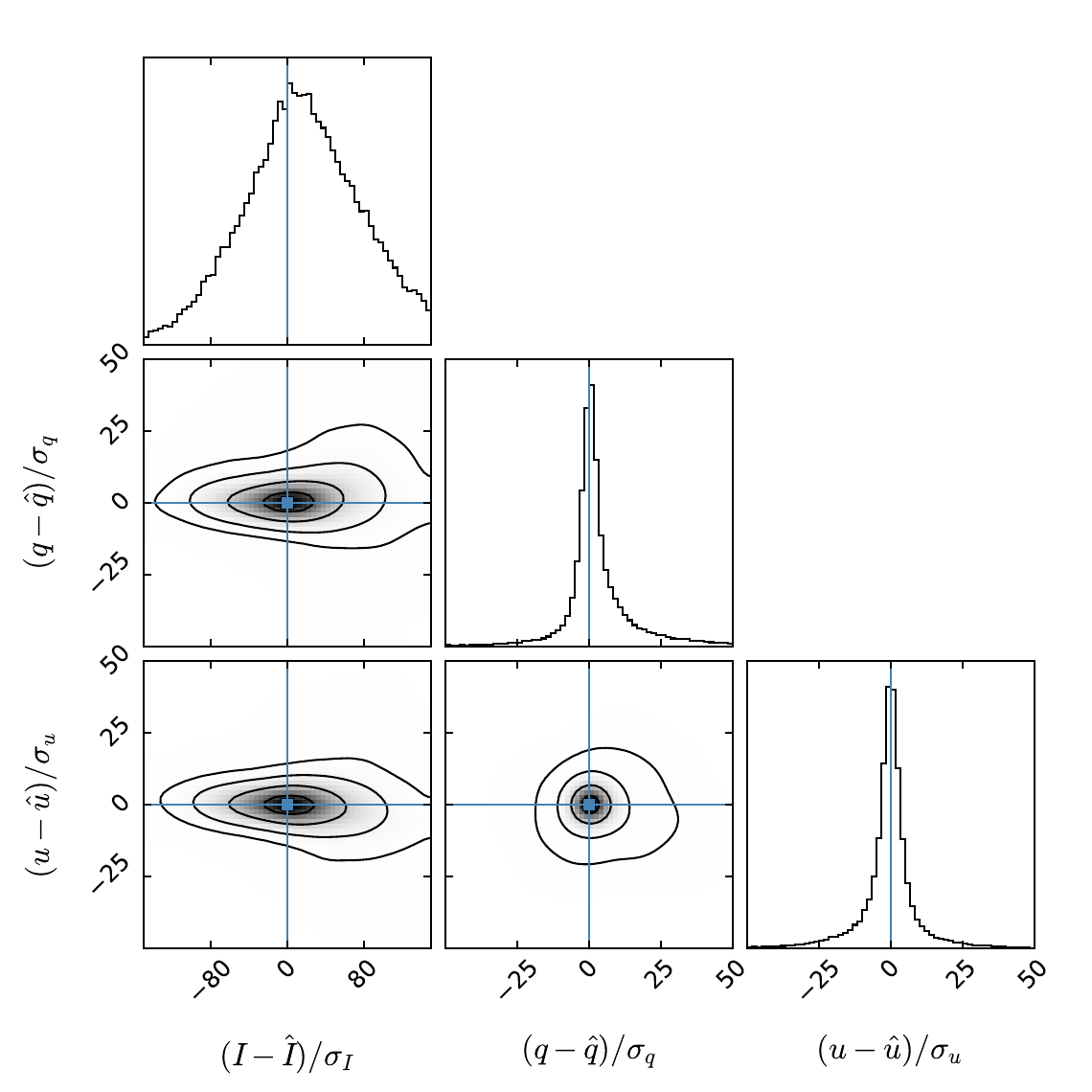}
\caption{Corner plot showing the correlations between the significance of the residuals in $I$, $q,$ and $u$. This figure corresponds to the modeling with the ARM4$\oplus$ED $n_{\rm{d}}$ model and with the QSS GMF model.
}
\label{fig:Iqufits_corner}
\end{figure}

In Fig.~\ref{fig:Iqufits_corner}, we show, a corner plot illustrating the possible correlations between the significance of the residuals in $I$, $q$ and $u$.
We show them for the modeling corresponding to $n_{\rm{d}} \equiv$ ARM4$\oplus$ED and $\rm{GMF} \equiv$ QSS; similar figures and results are obtained for all the other cases.
We observe that the residuals on the polarization maps are not strongly correlated to the residuals on the intensity map.
This reinforces our view that the reduced-Stokes parameter maps are more sensitive to the GMF than to the dust density distribution. Furthermore, we find that the polarization residuals are centered on zero and not correlated. While this is not shown on the figure, we observe that the scatter of the significance of the residuals
(in $I$, $q,$ and $u$) decreases with increasing Galactic latitude.
This is observed for all of the models and we understand it as being due to the small uncertainties on $q$ and $u$ at low $|b_{\rm{gal}}|$.

\subsection{Comparison of reconstructed GMF}
\begin{figure*}[h]
\centering
\begin{tabular}{lll}
\includegraphics[trim={0cm 0cm 4.5cm 0cm},clip,width=.3\linewidth]{_figs/qu353sig-1EDASS2-bf_xy.pdf} &
        \includegraphics[trim={0cm 0cm 4.5cm 0cm},clip,width=.3\linewidth]{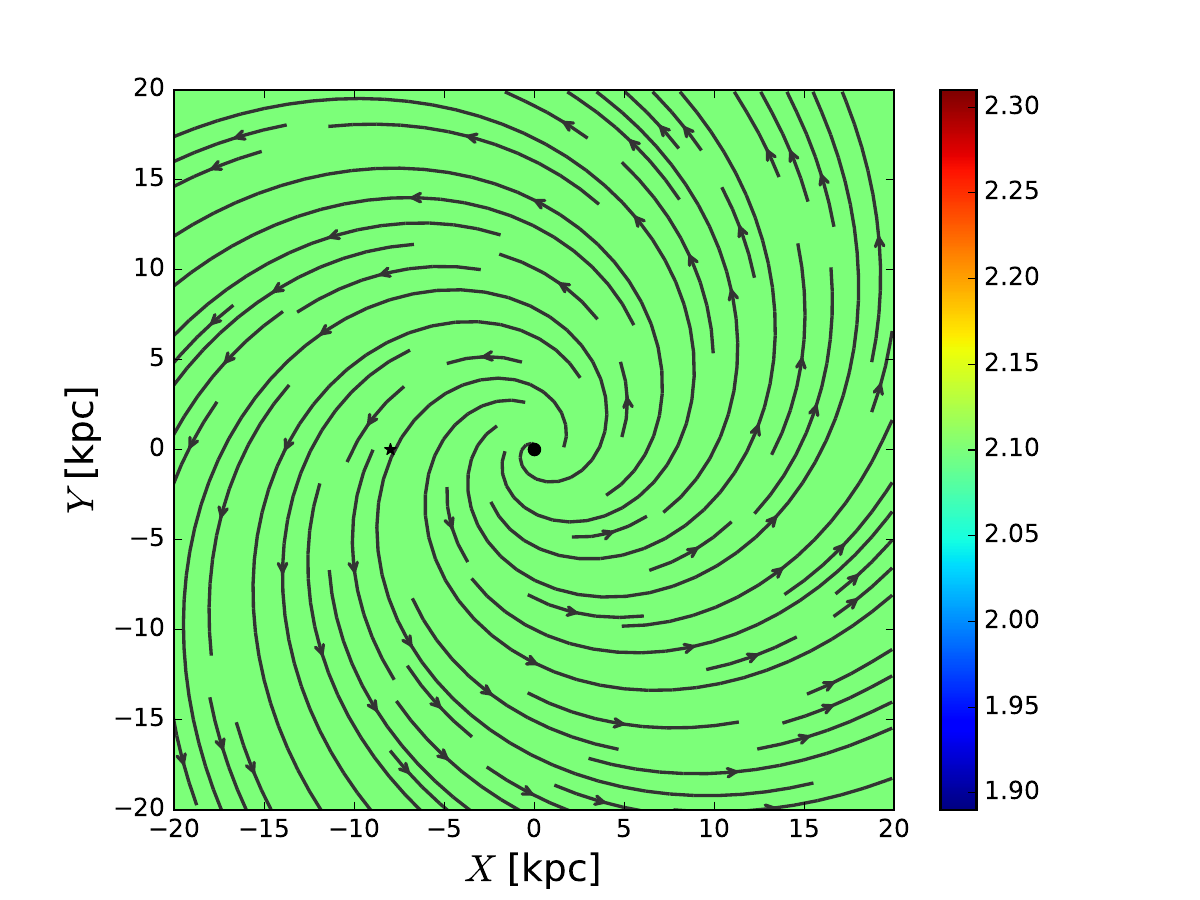} &
                \includegraphics[trim={0cm 0cm 4.5cm 0cm},clip,width=.3\linewidth]{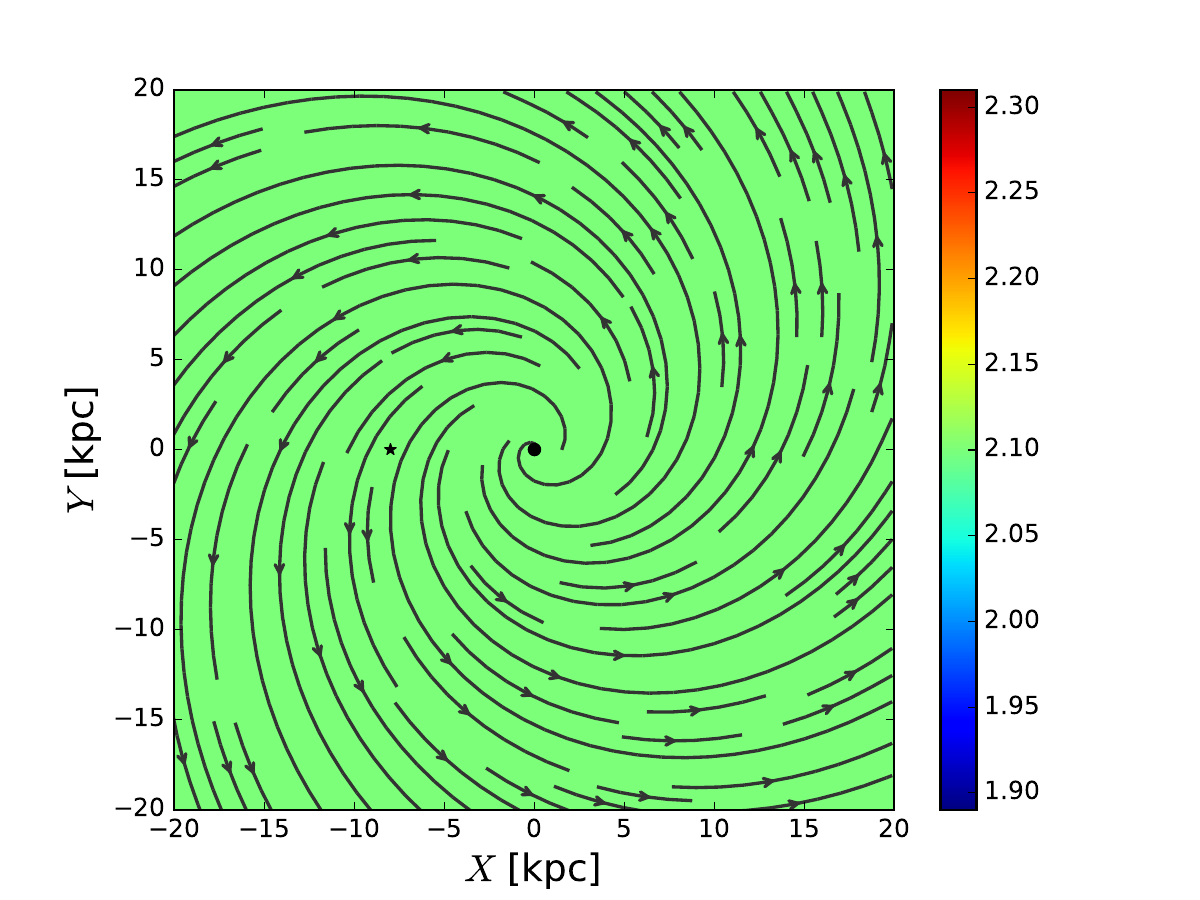} \\[-3.8ex]

\hspace{.08cm}
\includegraphics[trim={0cm 4cm 4.5cm 5.4cm},clip,width=.291\linewidth]{_figs/qu353sig-1EDASS2-bf_xz.pdf} &
        \hspace{.08cm}
        \includegraphics[trim={0cm 4cm 4.5cm 5.4cm},clip,width=.291\linewidth]{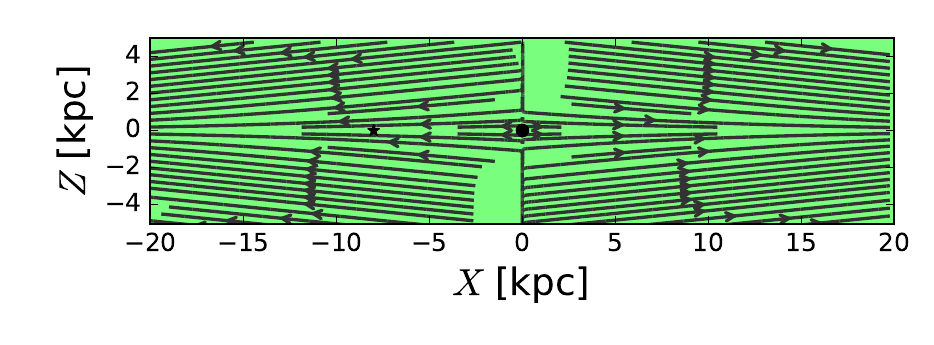} &
                \hspace{.08cm}
                \includegraphics[trim={0cm 4cm 4.5cm 5.4cm},clip,width=.291\linewidth]{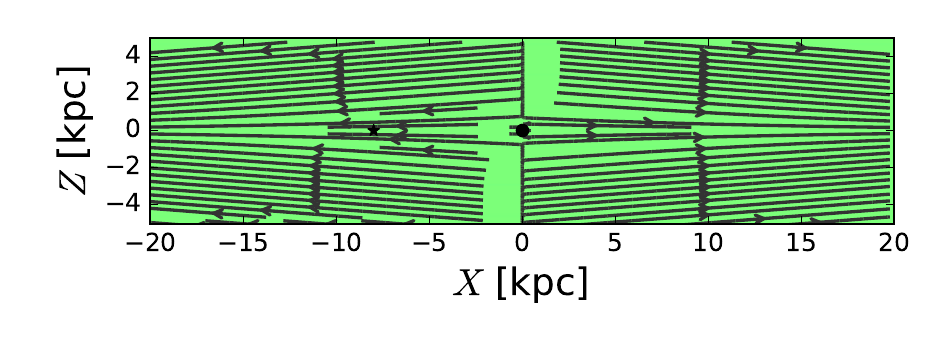} \\[-2.4ex]
\end{tabular}
\caption{Field line geometrical structures of best-fit ASS GMF models in the $(x,\,y,\,z=0)$ and $(x,\,y=0,\,z)$ planes of the Galaxy for the three best-fit $n_{\rm{d}}$ models obtained from the adjustment of the $I_{\rm{353}}$ map. Shown from left to right, with $n_{\rm{d}} \equiv$ ED, ARM4 and ARM4$\oplus$ED, respectively. The color scale indicates the strength of GMF, i.e. the norm of the vector field at each location, which is fixed to 2.1 in the case of the ASS model.}
\label{fig:GMF-vs-ndust_XYZplanes}
\end{figure*}
The fact that the best-fits GMF are very consistent with respect to the choice of the dust density distribution model can be observed, for example, in Fig.~\ref{fig:GMF-vs-ndust_XYZplanes}.
In this figure we show the geometrical structure of the GMF in the horizontal and vertical cross-cuts of the Galaxy obtained with the three best-fit models of $n_{\rm{d}}$ when inferring the GMF geometry constraining the ASS model.
The similarity of the recovered field geometrical structures is striking, especially in the $(x,\,y,\,z=0)$ plane. Equivalent figures are obtained for the other GMF models.

\begin{figure*}[h]
\centering
\begin{tabular}{ccc}
\includegraphics[width=.3\linewidth]{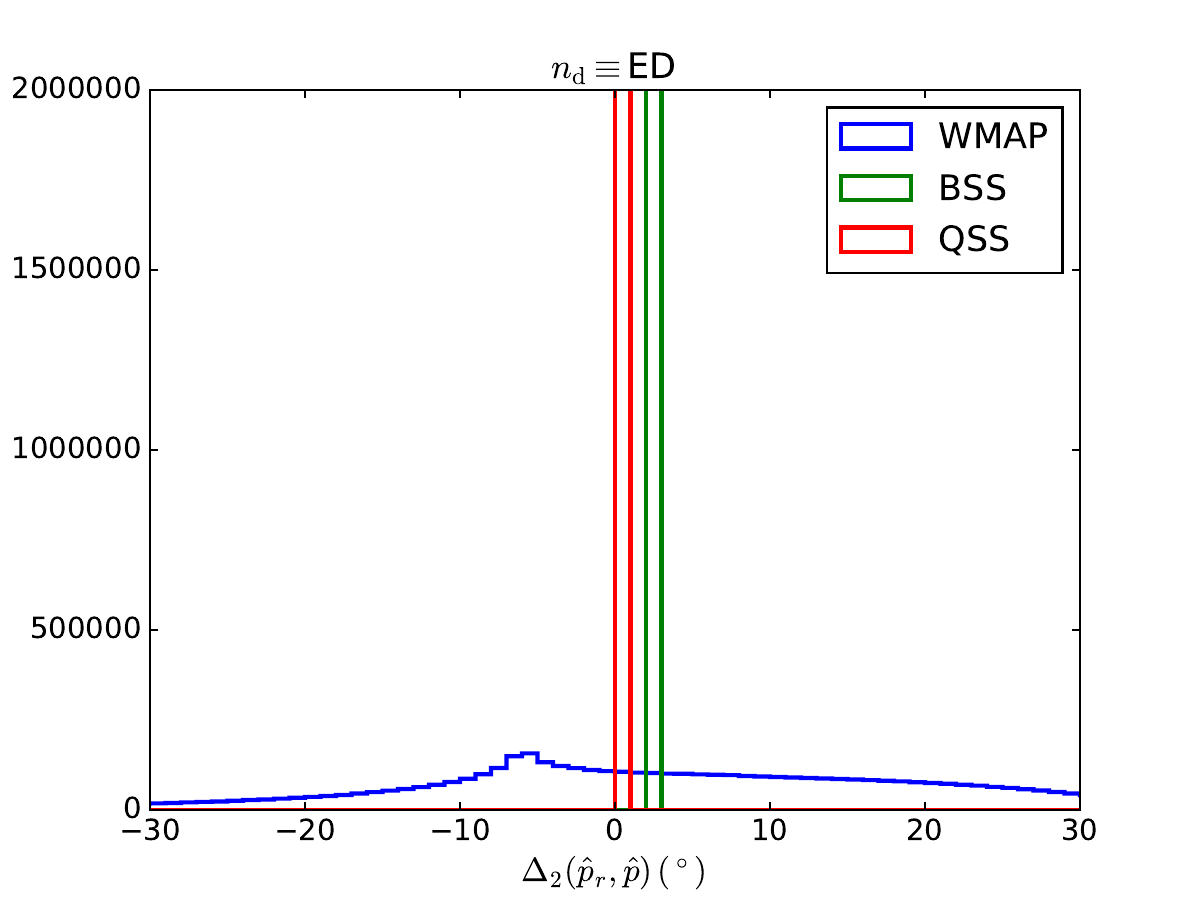} &
        \includegraphics[width=.3\linewidth]{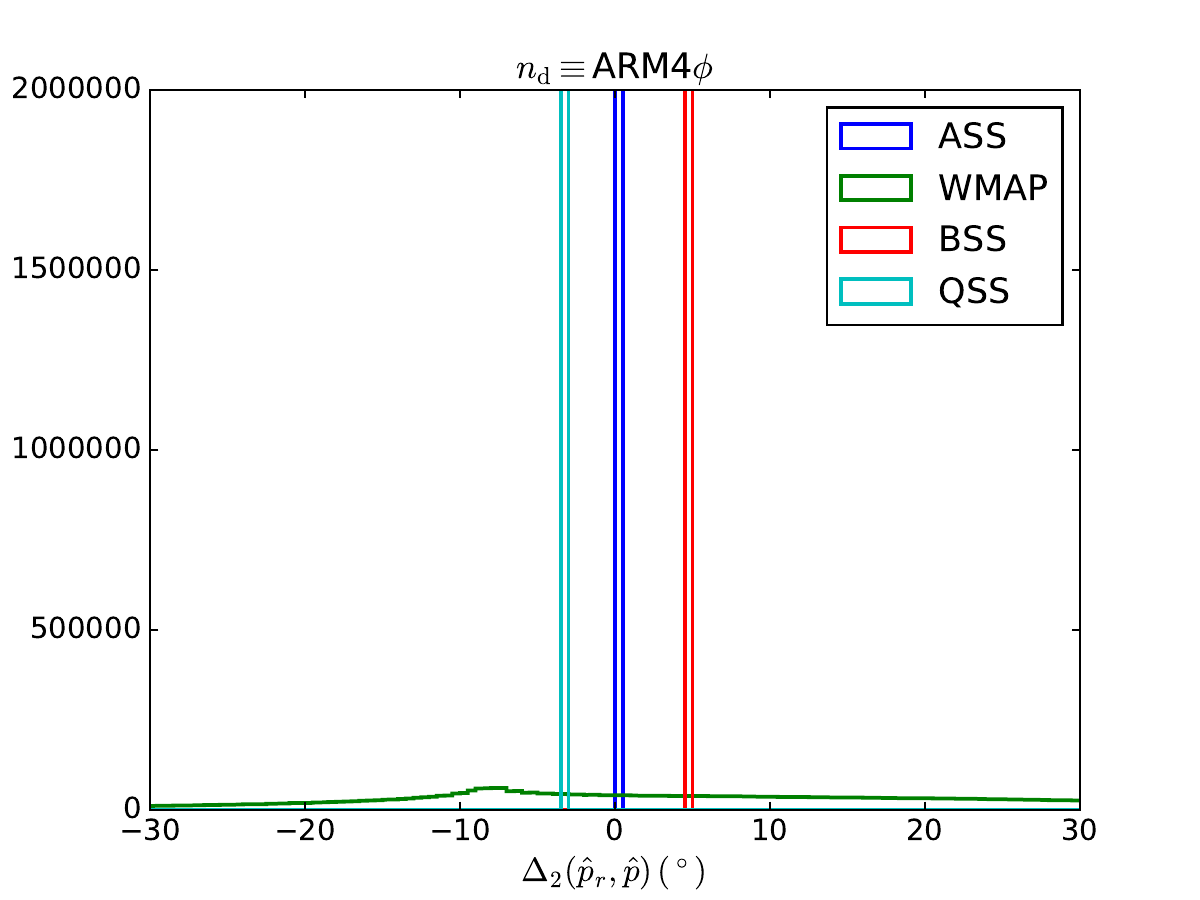} &
                \includegraphics[width=.3\linewidth]{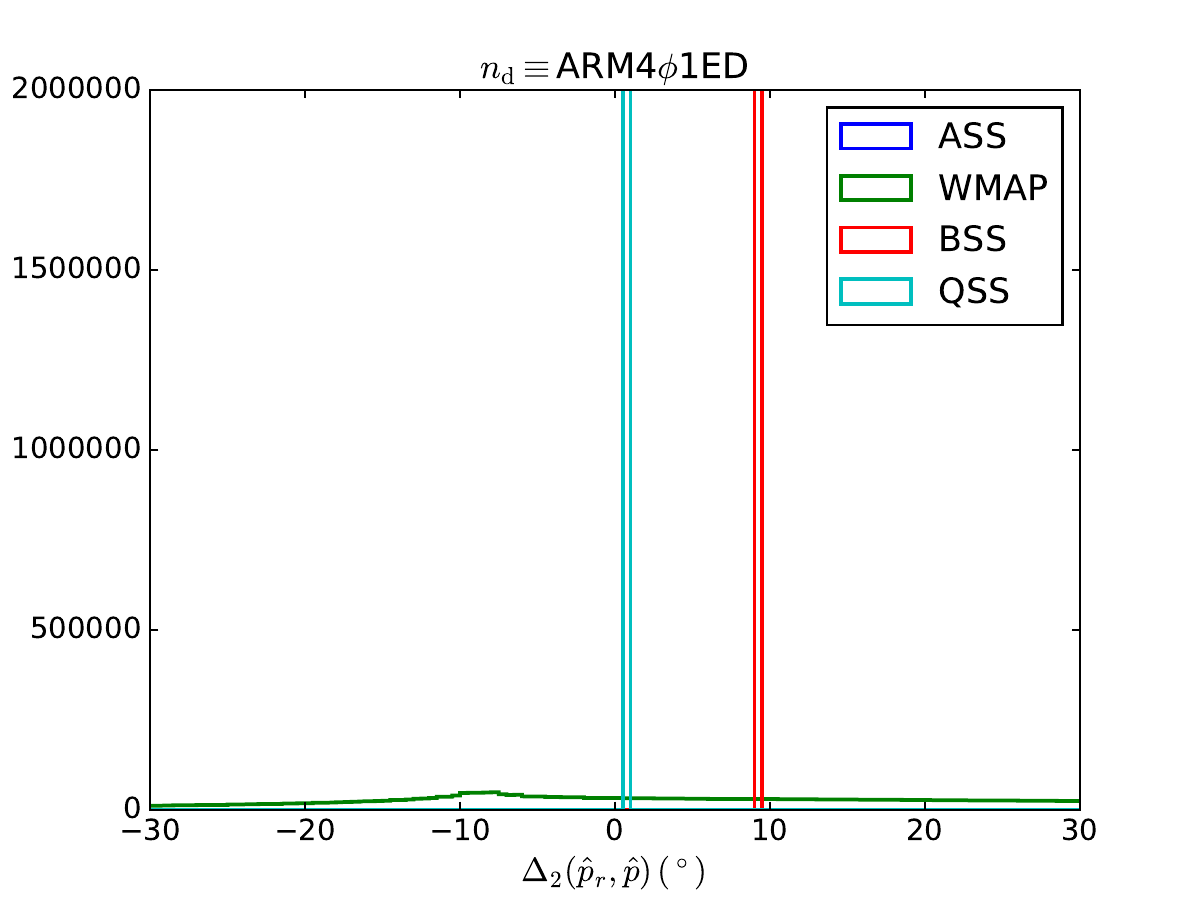} \\

\includegraphics[width=.3\linewidth]{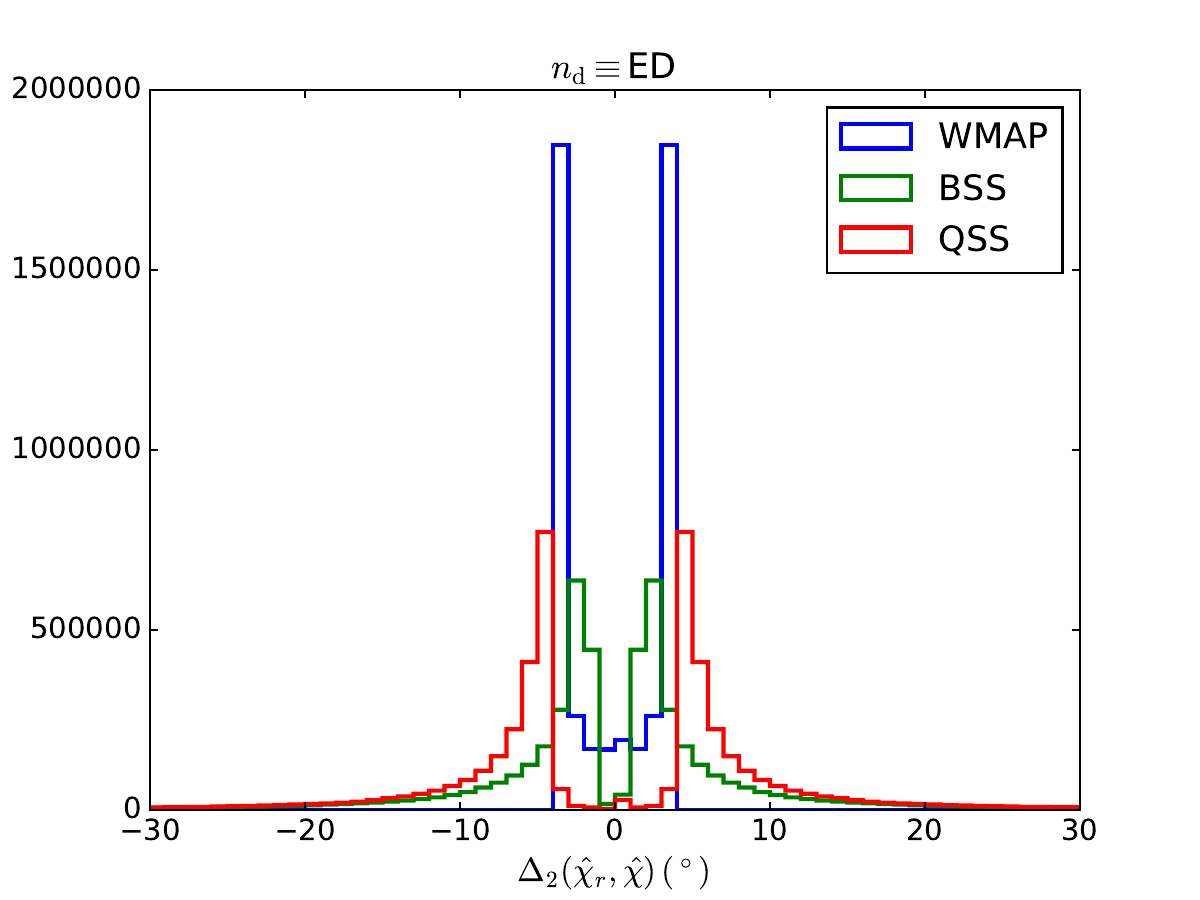} &
        \includegraphics[width=.3\linewidth]{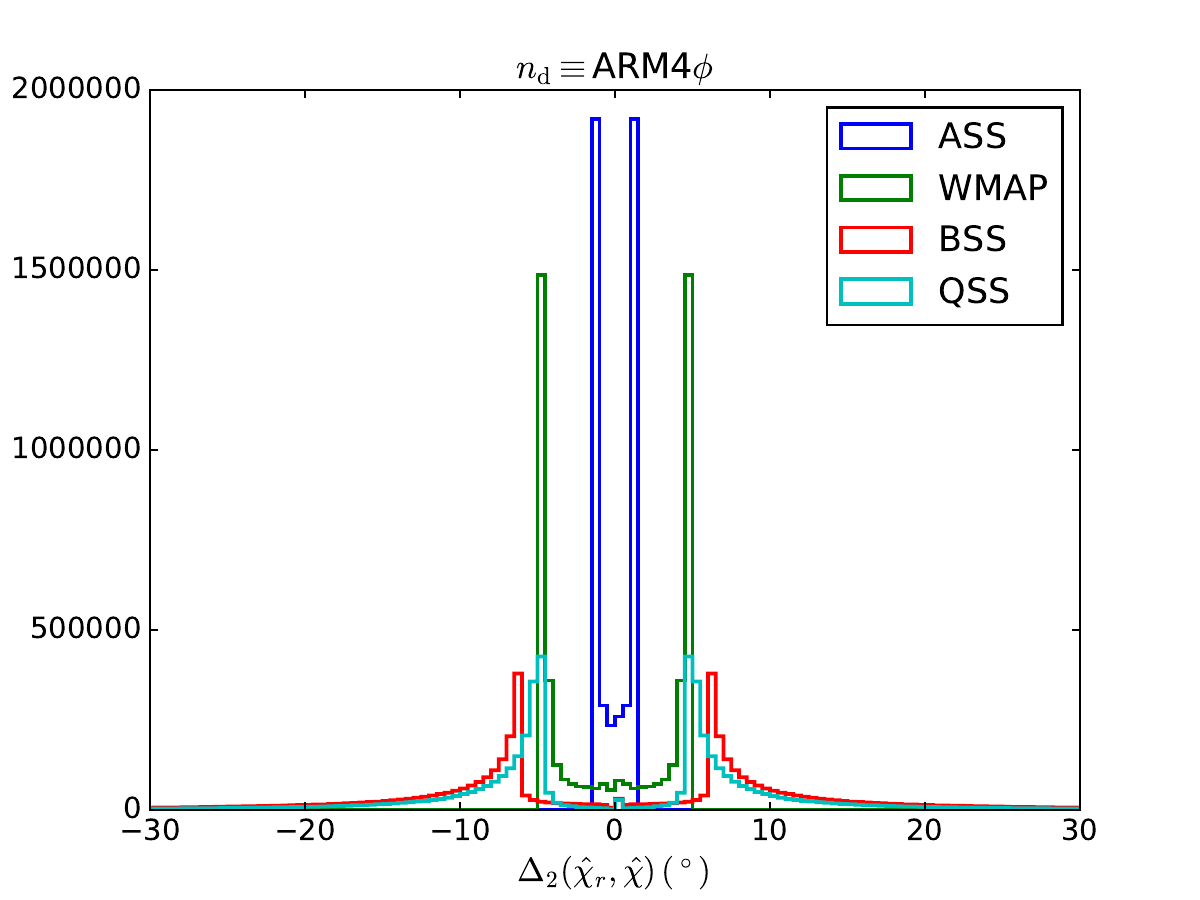} &
                \includegraphics[width=.3\linewidth]{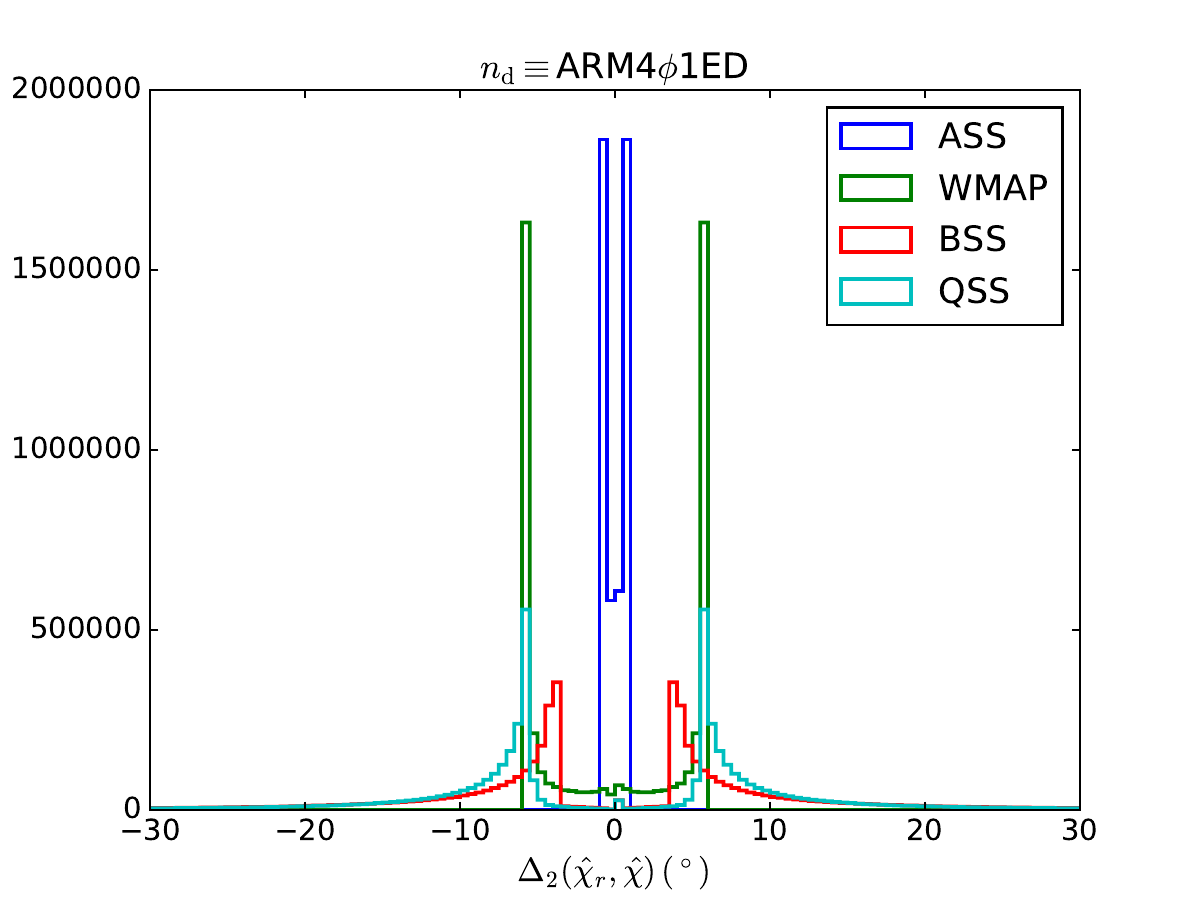} \\

\includegraphics[width=.3\linewidth]{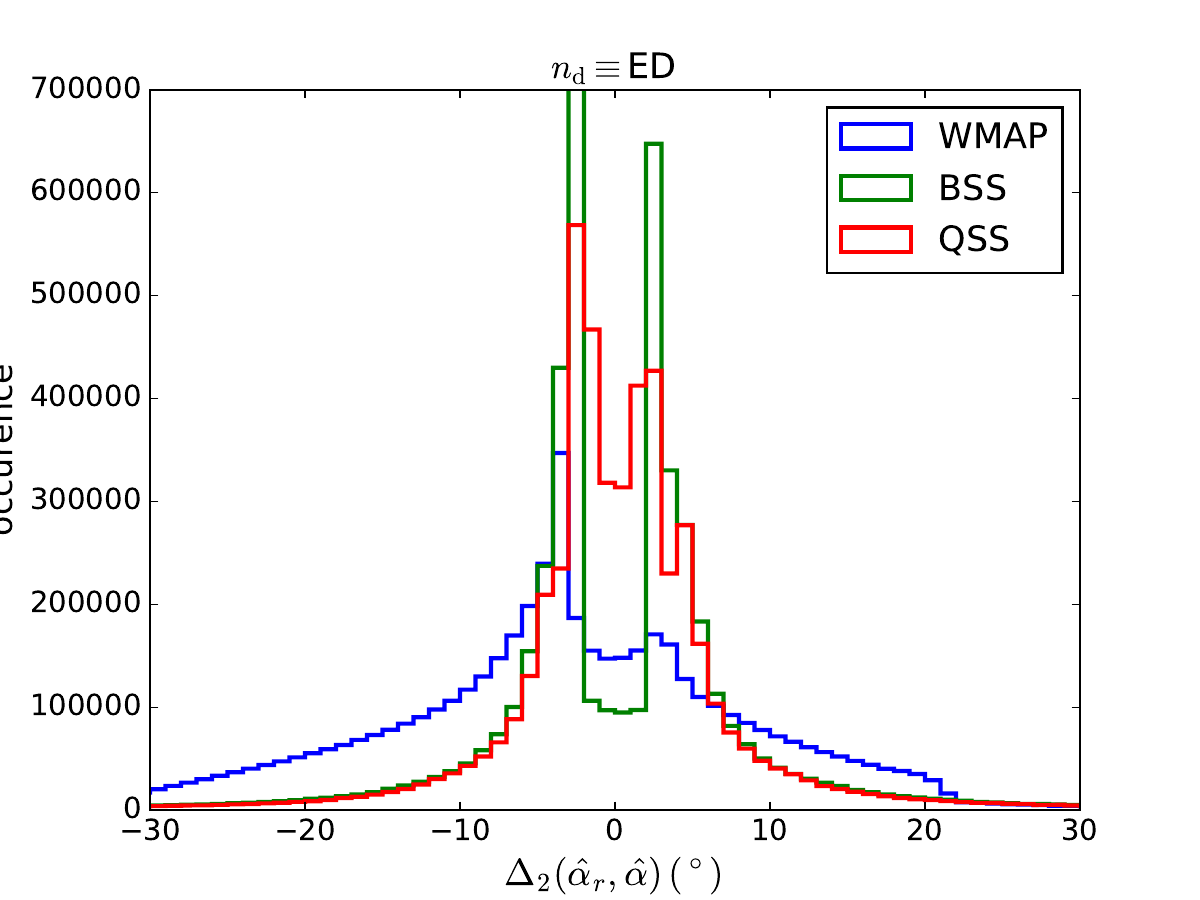} &
        \includegraphics[width=.3\linewidth]{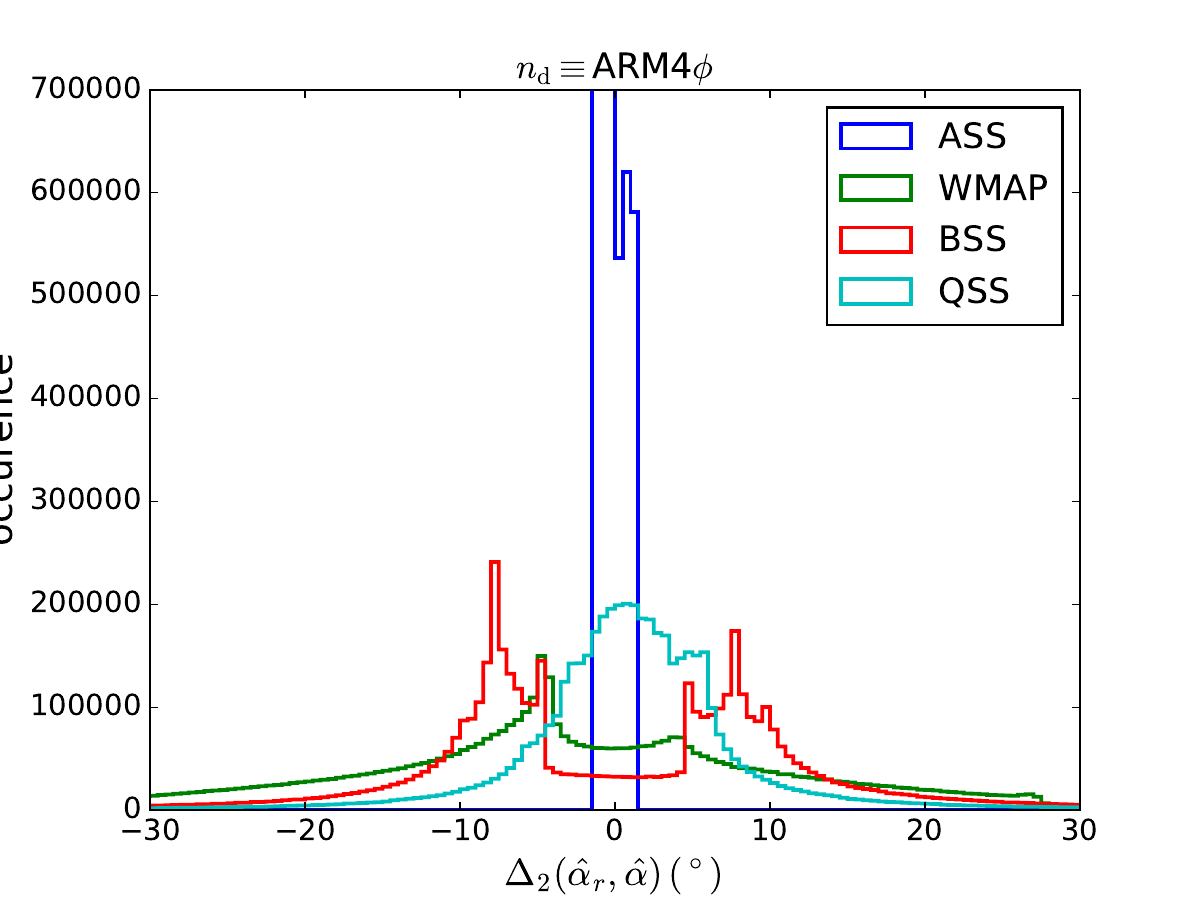} &
                \includegraphics[width=.3\linewidth]{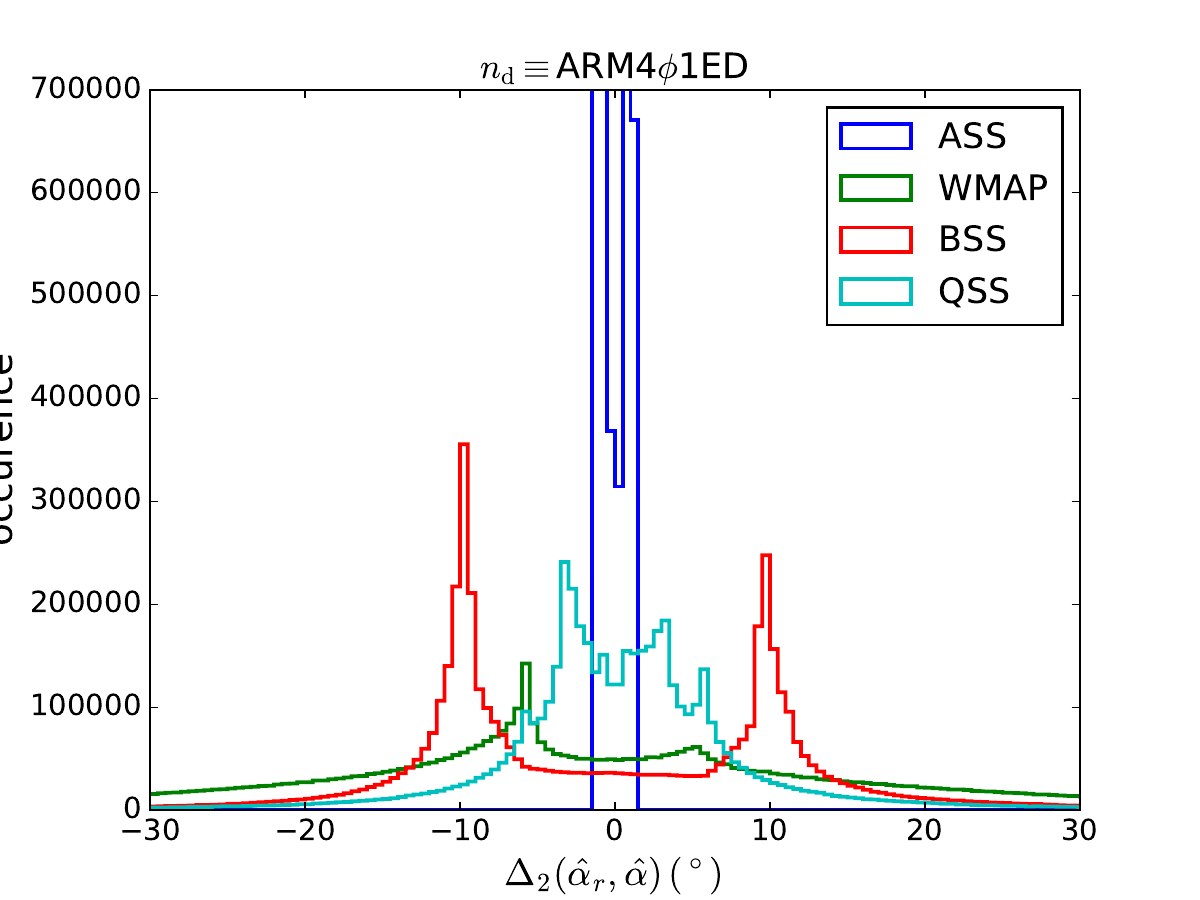} \\

\includegraphics[width=.3\linewidth]{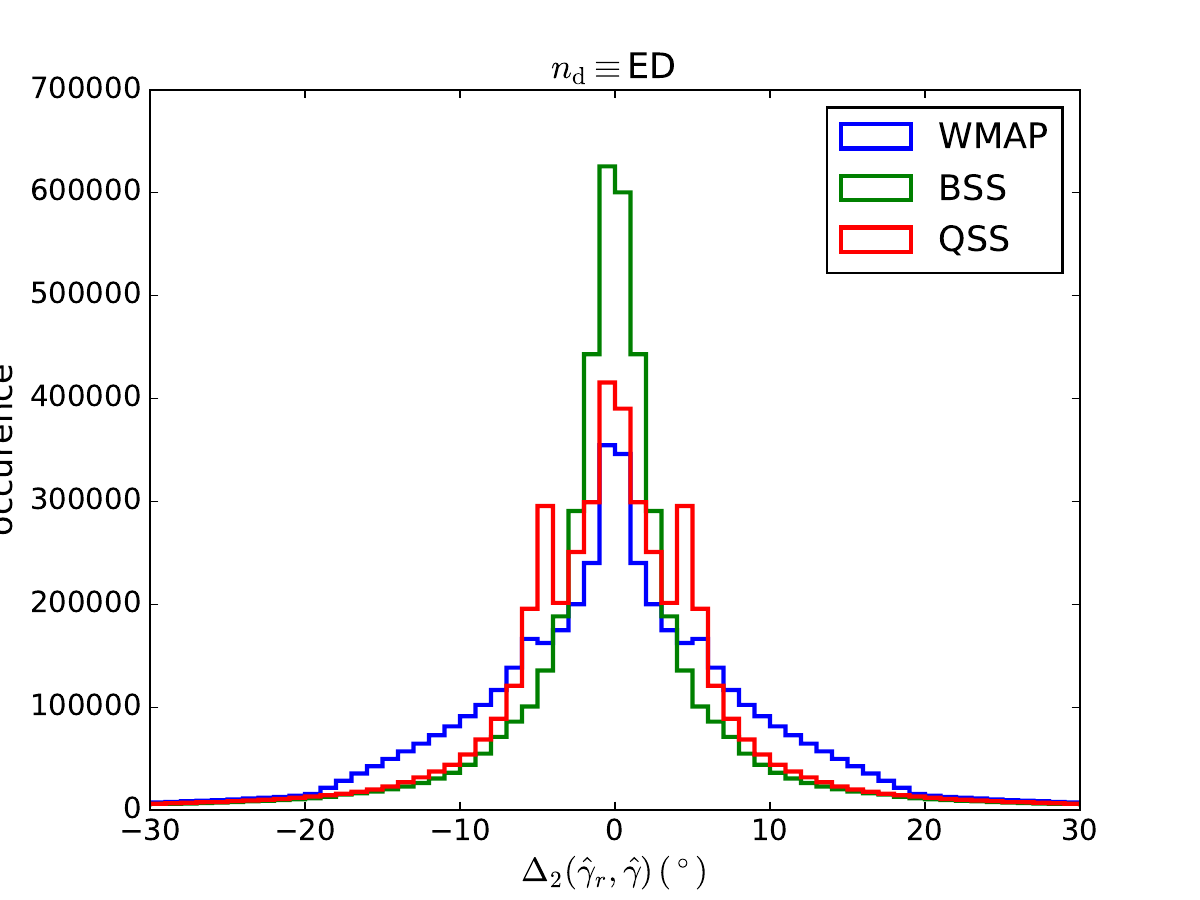} &
        \includegraphics[width=.3\linewidth]{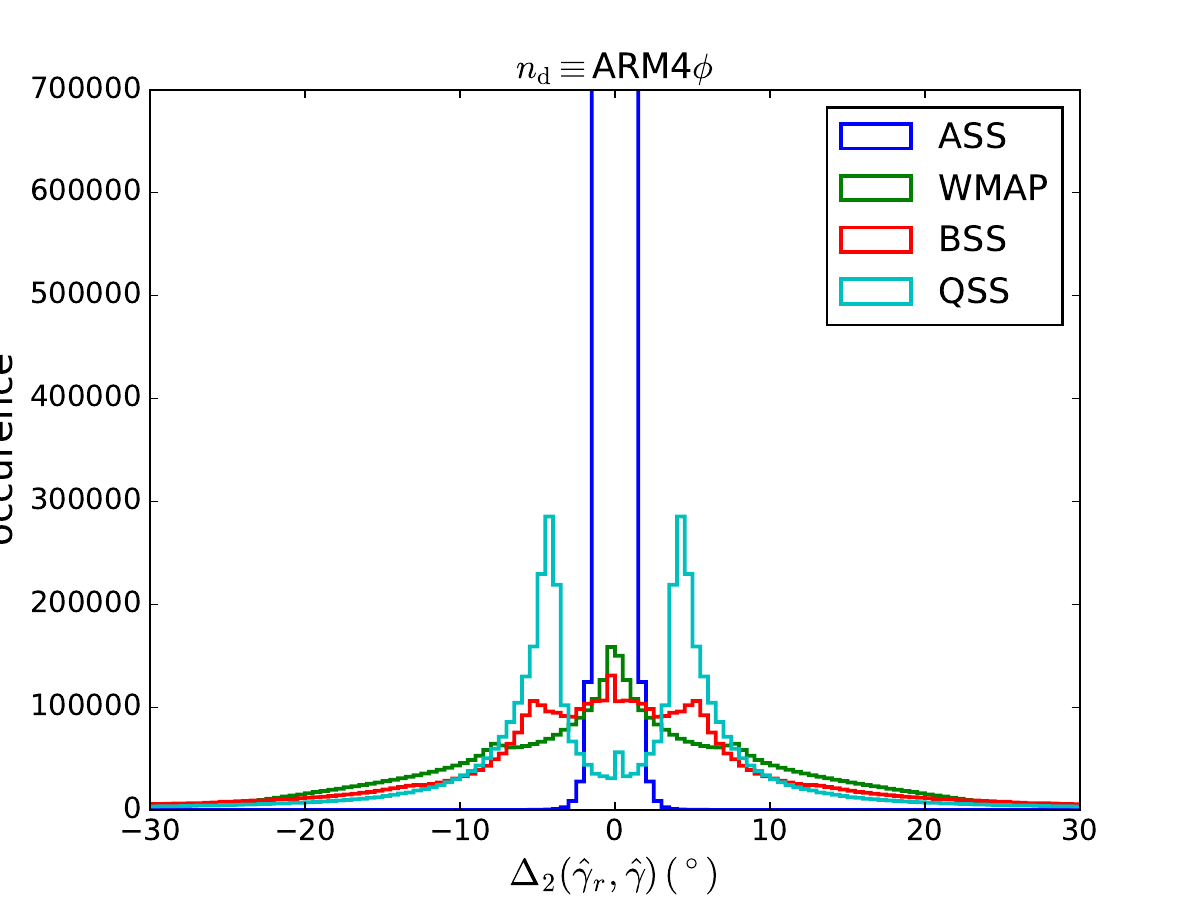} &
                \includegraphics[width=.3\linewidth]{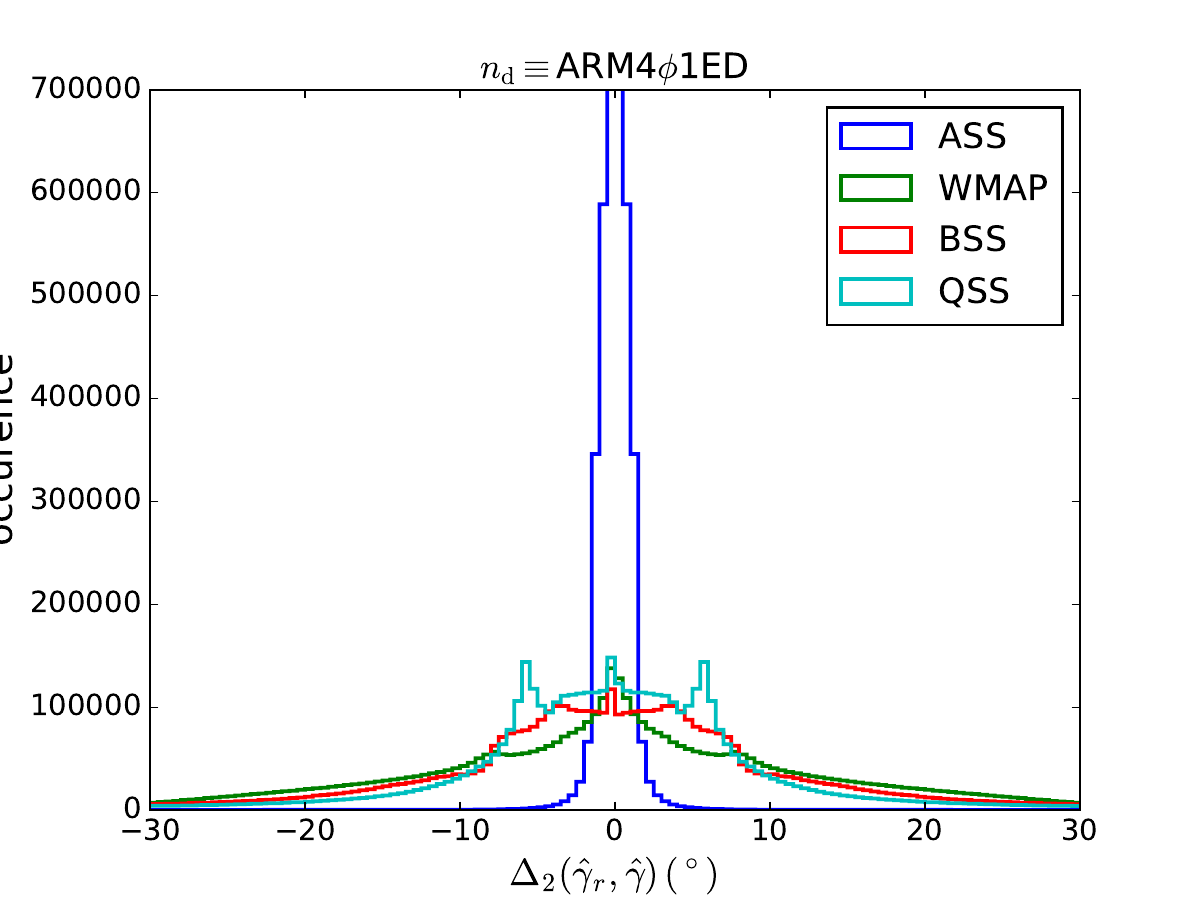} \\
\end{tabular}
\caption{Histograms of the relative (local) angles as compared to our reference model defined as being the best fit obtained with $n_{\rm{d}}$ = ED and the ASS GMF model. From top to bottom: Pitch angle, tilt angle, inclination angles, and position angle.
From left to right: $n_{\rm{d}}$ = ED, ARM$\phi$ and ARM4$\oplus$ED. The colors correspond to different GMF best-fit model.}
\label{fig:GMF_comp-angles}
\end{figure*}

In order to compare quantitatively the GMF geometrical structures that we reconstructed when fitting the polarization maps, we proceed as in Sect.~\ref{sec:qualreconstructionGMF}.
For each best-fit model, we compute the pitch ($p$), tilt ($\chi$), inclination ($\alpha$) and position ($\gamma$) angles at each position of the Galactic space used for our MC realizations.
We then compute the signed differences of each of the four angles considering the best-fit from the least evolved modeling (i.e. $n_{\rm{d}} \equiv$~ED~+~$\rm{GMF} \equiv$~ASS) as a reference to which each best-fit GMF model is compared with.
The histograms of the signed differences are presented in Fig.~\ref{fig:GMF_comp-angles}.

In terms of the pitch and tilt angles, inspection of the figure leads to the following main results:
\begin{enumerate}[(i)]
\item The pitch angles of all the best-fit GMF obtained for the ASS, BSS, and QSS agree within less than five degrees (with the exception of ARM4$\oplus$ED~+~BSS, which is about nine degrees away).
\item The histogram of the pitch angles of the LSA model (allowed to vary with Galactic radius) peaks at value near the values of the other GMF models. At the Sun radius, the pitch angle of this model is between six to eight degrees away from our reference model.
\item The tilt angles of all the models are centered around the same value. The scatter is however larger than that for the pitch angle when comparing BSS and QSS to ASS. This was expected due to the modulation of the field amplitude in planes parallel to the Galactic disk ($z$ constant) that produces a change of the tilt angle of the GMF lines.
\item Comparison of the best-fits of the same GMF model but for different $n_{\rm{d}}$ reveals a sizable coherence. We have checked that this conclusion, already informed around Fig.~\ref{fig:GMF_pitchandtilt}, equally holds for the other GMF than for the ASS, which is further discussed below.
\end{enumerate}

As an example, the good agreement observed in the spirals of Fig.~\ref{fig:GMF-vs-ndust_XYZplanes} for the ASS model is quantified by the blue histograms in the middle and right panels of the first row of Fig.~\ref{fig:GMF_comp-angles}, except that, here, we are not limited to $(x,\,y,\,z=0)$
or $(x,\,y=0,\,z)$ planes but account for the whole sampled space.
The histograms show that through the whole sampled space the differences of the measured pitch angle are below one degree.
The blue histograms in the middle and right panels of the second row quantify the small differences in the out-of-plane components among the ASS best-fits observed in lower panels of Fig.~\ref{fig:GMF-vs-ndust_XYZplanes}. The measured tilt angle differences (compared to the reference GMF) are less than about two degrees throughout the whole space.

\smallskip

In terms of the inclination ($\alpha$) and position ($\gamma$) angles, which are the angles directly linked to the observable through line-of-sight integration (see Eq.~\ref{eq:DUSTEMISSION}), the distributions of the relative angles shown in Fig.~\ref{fig:GMF_comp-angles} (last two rows) also show an overall agreement between the model reconstructions.
However, the wings of the distributions are more important for the angles $\alpha$ and $\gamma$ than those of $p$ and $\chi$.

These results illustrate the non trivial mapping between the two set of angles, $p$ and $\chi$ on one side and $\alpha$ and $\gamma$ on the other side.
Despite their similarities, the best-fit GMF models look different from the observer's  view, whereas when combined with the dust density and integrated along the lines of sight, all sets of polarization maps are equivalently a good (bad) fit to the data. This also illustrates the degree of degeneracy of the inclination and position angle of the GMF lines along the lines of sight.

\smallskip

In the above, it is the comparison of model-parameter values (and corresponding geometrical features) that is relevant, not the absolute values of the parameters since they are expected to be biased, despite the fact that we ignore the turbulent component of the GMF in the fit (see Sect.~\ref{sec:turbinthegame}).
However, we have shown that reliable constraints can be obtained for the pitch angle and therefore we limit our final conclusions to this parameter. 
We find with our analysis that the spiral pattern of the GMF has a mean pitch angle (at the Sun radius for the LSA model) of about 27 degrees, with a 14 percent scatter, and takes 17.5 and 35.4 degree as the two extreme values. Our pitch angle values are rather large but are still compatible with what can be expected from dynamo theory \cite[e.g.,][]{Cha2016}. We also find that, when allowed, the pitch angle vary dramatically from the inner part of the Galaxy to the outside region.
A value of 27 degree for the pitch angle was already reported in \cite{Pag2007}.
The out-of-plane pattern of the GMF is poorly constrained in our analysis. This is likely because it depends more on the particular GMF model under study and, at the same time, because its influence on maps is more noticeable at low and intermediate Galactic latitudes where our fits are not satisfying.
Let us emphasize, though, that none of our best-fits exhibits the expected X-shape of the GMF lines that has been found in radio observations of external galaxies (see e.g., \citealt{Jan2012a} for a discussion).

\subsection{Alternative comparison of polarization maps}
\label{sec:validation}
\begin{figure}[h]
\centering
\includegraphics[trim={2.5cm 3.cm 2.5cm 2.7cm},clip,width=\columnwidth]{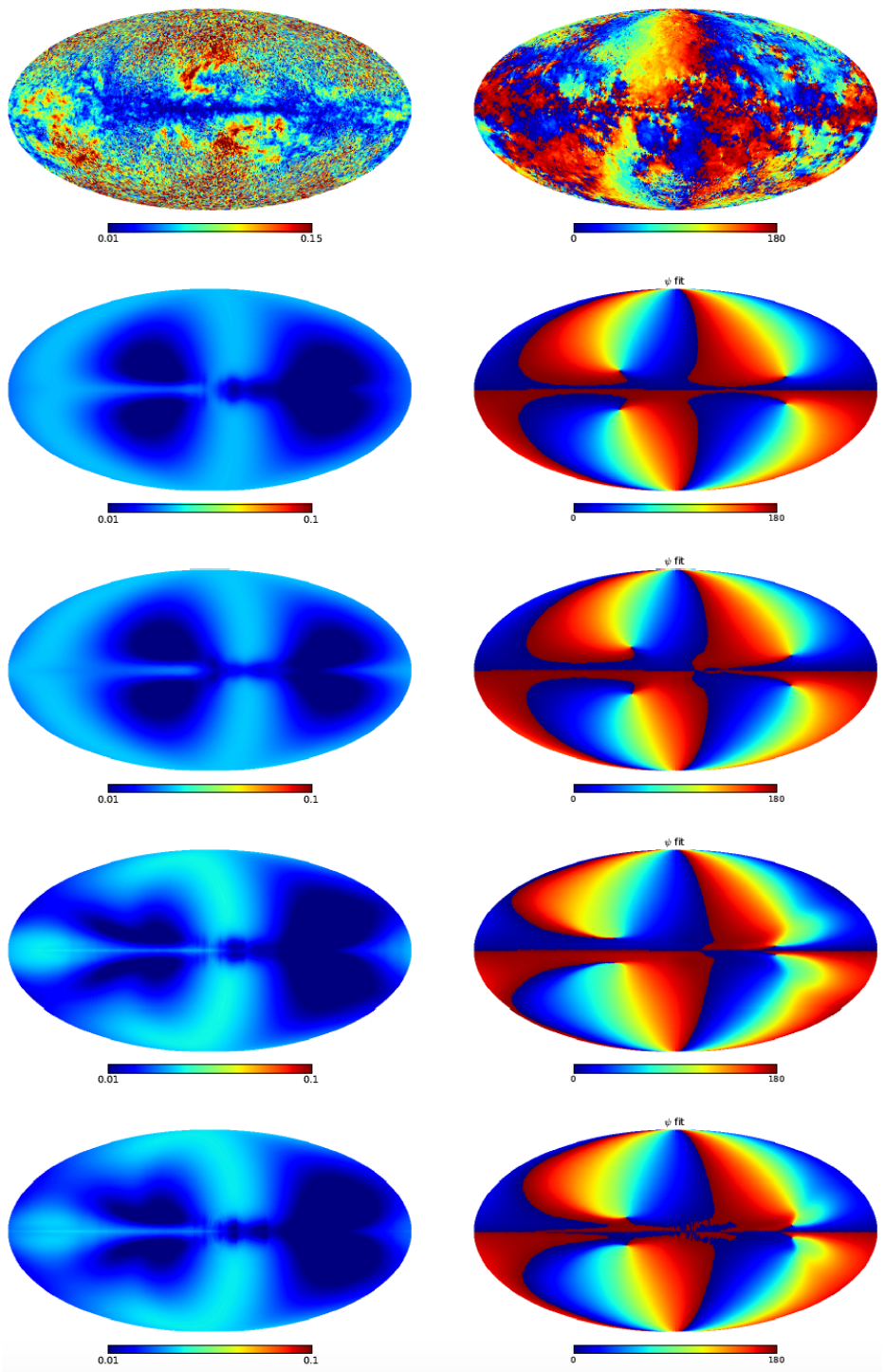}
\caption{Degree of linear polarization (left) and polarization position angle (right), as deduced from the \textit{Planck} data at 353 GHz (first row) and as deduced from our best-fits of the GMF models, while assuming the best-fit model from $n_{\rm{d}} \equiv$~ARM4. Rows two to five correspond to ASS, LSA, BSS, and QSS models for the GMF, respectively.}
\label{fig:GMF_plin-PPA}
\end{figure}

The degree of linear polarization ($p_{\rm{lin}} = \sqrt{q^2 + u^2}$) and the polarization position angle ($\psi = 1/2\, {\rm{arctan}}2(-u,q)$)\footnote{
The polarization position angles are given in the IAU convention.}
are commonly used to study and characterize the magnetized interstellar medium \cite[e.g.,][]{PlanckXIX2015}.
In Fig.~\ref{fig:GMF_plin-PPA} we show the noise debiased\footnote{Noise debiasing procedure is performed in the same way as in Appendix B.2 of \cite{PlanckXIX2015}.} $p_{\rm{lin}}$ and $\psi$ maps from the \textit{Planck} data at 353 GHz (top row) and those deduced from the polarization maps ($q$ and $u$) corresponding to our best-fits GMF reconstructions obtained with the $n_{\rm{d}} \equiv$~ARM4 model fitted to $I_{\rm{353}}$.
These maps are alternative ways at looking at the $q$ and $u$ maps presented in Fig.~\ref{fig:qu_fit} (but for another $n_{\rm{d}}$ model, which does not change the maps substantially). The joint consideration of all of the maps ($q$, $u$, $p_{\rm{lin}}$ and $\psi$) can be of interest, for instance, in diagnosing the limitation of our modeling and proposing further developments.

\smallskip

Inspection of the left column of Fig.~\ref{fig:GMF_plin-PPA} clearly illustrates qualitatively the limitation of the models at reproducing the data.
Despite the noise dominated map of the data, it is nevertheless interesting to note how the modeling can qualitatively reproduce low $p_{\rm{lin}}$ values in some region of the Galactic equator where we know the intensity to be high. These low degrees of linear polarization results from line-of-sight depolarization induced by the varying GMF orientation.
This can be inferred from the comparison, at low $|b_{\rm{gal}}|$, of the BSS and QSS models (two last rows) to the ASS model (second row), for instance. Let us stress that this depolarization is obtained using the regular GMF only.

Furthermore, for lines of sight with $p_{\rm{lin}} \lesssim 5\%$, the relatively simple models of the large-scale regular GMF that we consider are able to reproduce qualitatively the low and high values of the degree of linear polarization for low and high values of the intensity as suggested by the data \citep{PlanckXIX2015}.
However, these models fail at reproducing very high degree of linear polarization ($p_{\rm{lin}} > 5\%$), where large difference are observed between model maps and data. These difference are well localized in the sky towards regions of nearby known structures (spurs, arms, and clouds) mainly attributable to the nearby structures such as the Sco-Cen association (e.g., \cite{Das2020}).
We used the \texttt{gpempy} software to investigate this further and we reached the conclusion that, in general, $p_{\rm{lin}}$ is not boosted by the simple addition of a clump of dust towards such a maximum to solve the problem. The addition of a clump actually produces the opposite.
To reproduce the observed $p_{\rm{lin}}$ in these objects, it is likely that the effective degree of polarization of the dust population in these structures will have to be enhanced.
This can be achieved by having locally either
(\textit{i}) a more efficient alignment of the dust grains along the field line or
(\textit{ii}) a higher intrinsic degree of linear polarization of the grain population, or
(\textit{iii}) a GMF particularly well aligned and coherent with respect to the line of sight (see Eq.~\ref{eq:DUSTEMISSION}).

\smallskip

As opposed to $p_{\rm{lin}}$, it is remarkable how the studied GMF models are able to capture qualitatively the broad tendency in the polarization position angle data (right column of Fig.~\ref{fig:GMF_plin-PPA}). Indeed, the agreement between the models and the data is fair in terms of $\psi$, namely, in terms of the projected (and weighted) orientation of the GMF lines integrated along the lines of sight.
Beside the fair resemblance, the modeled $\psi$ maps show more coherence and smooth spatial variations than observed in the data. Also, inherently to the large-scale regular GMF considered in this study, the modeled maps have a mirror symmetry about the Galactic equator that is accompanied
by a change of sign. In the data, that symmetry is disturbed on small scale and twisted and skewed on global scale. This is well illustrated in the upper right quadrant of the maps and comparing the `lines' for which $\psi = 0^\circ = 180^\circ$, respectively.
The inclusion of magnetic fields attached to local structures, such as depicted by \cite{Alv2018} and \cite{Pel2020} in the case of the Local Bubble, addresses the latter issue in breaking the North-South symmetry.
The modeling of (relatively) small scale perturbations would require the modeling of magnetized structures well localized in the Galaxy and will necessitate the modeling of the turbulent component, perhaps in the form of parameterized magneto-hydrodynamic turbulence, such as that depicted in \cite{Wan2020}.
Any such investigation is beyond the scope of this paper and will be addressed somewhere else.

\subsection{Prospects}
Improvements in the modelings of the
regular part of the GMF with respect to this paper will require (\textit{i}) the implementation of local magnetic field structure (e.g., \citealt{Alv2018}; \citealt{Pel2020});
(\textit{ii}) the interconnection of the latter with the large-scale regular GMF; and (\textit{iii}) the generalization of the large-scale regular GMF models.
As a straightforward generalized model, we would propose a model that contains both the radial variation of the pitch angle, such as encoded in the LSA model, and the strength field modulation, such as in the BSS or QSS model. Independently, both features appear indeed to be tentatively favored by dust data, as compared to the ASS model.
The consideration of the other observables such as synchrotron emission and Faraday rotation measurement data promise to help significantly at constraining and modeling further the large-scale regular part of GMF and to solve possible ambiguities. In this respect, it might be worth undertaking physical models of the GMF such as those derived by \cite{Fer2014}.

A second stage of improvement in the GMF modeling will be to include the turbulent part of the GMF in the reconstruction. This step is of interest for CMB foreground studies as it allows for the statistical characterization of the polarized Galactic emission from thermal dust (e.g., see \citealt{Van2017}).
However, in order to avoid overestimating the amplitude of the turbulent component compared to the regular one, we would stress the importance to progress as far as possible in the modeling of the regular component of the field without the introduction of random components. Indeed, given its effect at the map level (also discussed in \citealt{Ste2018}), a random (turbulent) component can effectively help reconcile the model with the data. Therefore, we have to be cautious to not overestimate the turbulent component to compensate for a poor modeling of the regular part.
The turbulence is certainly present in the Galaxy and must leave its imprint in the data, but perhaps, with a lower relative amplitude compared to the regular component than previously reported (e.g., \citealt{Jan2012a}; \citealt{Jan2012b}; \citealt{PlanckXLII2016}), especially at the physical scales probed by the Galactic diffuse emission considered at low resolution.

At the map level, the inclusion of a turbulent component also results in the two following effects:
(\textit{i}) an angular spatial de-coherence of the polarization position angles and
(\textit{ii}) an overall lowering of the degree of linear polarization, due to line of sight depolarization. The latter would call for an overall increase of the global amplitude of the modeled polarization maps and therefore would tend to enlarge the range of values span by $Q$ and $U$. Such an increase might help at recovering the missing amplitude of the field strength, that can be visually inferred from comparing the data with the modeled polarization maps (see Fig.~\ref{fig:qu_fit} or Fig.~\ref{fig:GMF_plin-PPA}).
We found that this apparent missing amplitude decrease the larger the pixels (alternatively the smaller the $N_{\rm{side}}$) with which the fits are performed.
This points to the fact that turbulence in the GMF is not the dominant factor when considering sufficiently large pixels in the sky.
We found that the factor by which we should multiply the $q$ and $u$ maps for a better visual fit is, averaged on the 12 GMF reconstruction, of about 1.08 at $N_{\rm{side}} = 32$ and 1.13 at $N_{\rm{side}} = 64$.
This observational fact opens new possible ways of tackling the modeling of the GMF and its different components.
We have not investigated this possibility further and we leave this issue for a future work.

\section{Summary \& conclusions}
\label{sec:conclu}

In this paper, we present a methodology to infer the large-scale geometrical features of the GMF by constraining models of the regular components of the GMF relying on a dedicated MCMC method and a maximum-likelihood analysis of polarization data of the diffuse Galactic emission at microwave and millimeter frequencies, and more specifically, from the thermal dust emission.

Based on the current understanding of the modeling of the Galactic synchrotron and the thermal dust emission mechanisms in intensity and polarization, we built a step-by-step methodology to optimally divide and to simplify the fitting procedure.
First, assuming the Galactic thermal dust intensity to be independent of the GMF, we can obtain a fair representation of the dust density distribution across the Galaxy. Second, we use the recovered thermal dust density distribution to fit the Galactic thermal dust polarization and so, to constrain the geometrical structure of the regular GMF component,
assuming only this component is present in the data. 
Finally, the latter could be used as an input to constrain simultaneously the GMF amplitude and the Galactic relativistic electron density distribution by fitting synchrotron maps in intensity and polarization.

\smallskip

Relying on two sets of realistic simulations of the thermal polarized Galactic thermal dust emission, we demonstrate that it is possible to provide constraints on models of the regular component of the GMF and thus that it is possible to infer large-scale geometrical features of the GMF.

Assuming a non-turbulent contribution to the GMF in the simulations, we were able to provide excellent to
fair reconstruction of the regular GMF geometrical structure, even in the case where we adopt an overly simplistic model (different and simpler parameterization with respect to the input one) for the Galactic dust density distribution. The latter case is likely to be representative of what happens when dealing with real data sets given the convoluted nature of the Galactic thermal dust emission in intensity, in particular.
These relatively good results are possible thanks to the fitting of the reduced Stokes parameters (i.e., normalized to the intensity) that reduces the variations in the thermal dust polarization data induced by variations of the dust density along the lines of sight. 
In the case of simulations including a turbulent component of the GMF, we found that it is possible to satisfactorily recover the regular GMF component in the case of moderate turbulent contribution.
When the turbulence component is on the same order of the regular one, as suggested by observation of the synchrotron sky, the reconstruction of the regular GMF degrades significantly, requiring the inclusion of the turbulent component in the fitting procedure. Nevertheless, reliable constraints can be obtained for the pitch angle of the regular GMF.

As a general result, we found that the uncertainties, in terms of model parameters and GMF geometry, arising from the lack of a consideration of the turbulence component  are roughly of the same order of the ones induced by the mis-modeling of the density distribution or of the mis-modeling of the regular part of the GMF. This shows that improvements in the modelings are mandatory both for the deterministic (regular) and stochastic (turbulent) parts of the models.
We also find that the maximum-likelihood uncertainties on the best-fit parameters are underestimated as they do not account for any mismodeling effects.
Then, we argue that the scatter resulting from different modelings may be more representative of the true uncertainties.

Additionally, in the case of simulations without turbulence, we show that it appears possible to tackle the reconstruction of the distribution of the Galactic relativistic electrons by fitting the Galactic synchrotron emission using the best-fit of the GMF geometry obtained from the dust analysis as a prior. \\

Finally, we applied our methodology to a set of \textit{Planck} maps of the polarized thermal dust emission at 353 GHz.
We have been able to infer large-scale geometrical features of the GMF by constraining different models for the dust density distributions and for the regular component of the GMF.
We find that we are able to infer large-scale geometrical properties of the GMF and obtained coherent 3D views of the GMF, which, in terms of absolute parameter values, are limited by turbulence and the mismodeling of the dust density
The most robust geometrical features that we infer is that at the Sun radius, the spiral of the GMF has a mean pitch angle of about 27 degrees with a 14 percent scatter depending on the precise modeling.

\medskip

Despite some caveats, our analysis demonstrates that the Galactic thermal dust polarized emission is a powerful probe of the large-scale GMF, which is complementary to the more conventional ones (Galactic synchrotron data, Faraday rotation measures, and Faraday dispersion measures) and that it promises to play an important role in the overall efforts undertaken to reconstruct the actual 3D structure of the GMF (see e.g., \citealt{Bou2018})

\begin{acknowledgements}
Special thanks to C{\'e}line Combet and David Maurin for the numerous discussions and encouragements during the realization of this work.
This work has been funded by the European Union’s Horizon 2020 research and innovation program under grant agreement number 687312.
V.P. also acknowledges support from the European Research
Council under the European Union’s Horizon 2020 research and innovation program, under grant agreement No 771282.
We acknowledge the use of data from the Planck/ESA mission, downloaded from the Planck Legacy Archive, and of the Legacy Archive for Microwave Background Data Analysis (LAMBDA).
Support for LAMBDA is provided by the NASA Office of Space Science. Some of the results in this paper have been derived using the HEALPix (G{\'o}rski et al. 2005) package.
\end{acknowledgements}

%
\bibliographystyle{aa} 
\bibliography{PMR20bib} 
%

\begin{appendix}  
\section{3D parametric models}
\label{sec:AppendixA}
In this paper, we consider a particular set of models for the dust density distribution and for the regular and random components of the GMF. These parametric models are described below.

\subsection{Dust density distribution model}
\label{sec:ndmodel}

For the dust density distribution, we adopt two parametric models and their superposition.
The first, simple, model reproduces the bell-shape profile of the Galaxy. The second, more complex model draws spiral-arm pattern similar as the model studied in \cite{Jaff2013}. 
The two models are parameterized in cylindrical coordinate ($\rho,\, \phi,\, z$) centered on the Galactic center.

\subsubsection{Exponential Disk (ED)}

According to this model, the dust density distribution takes the parametric form:
\begin{equation}
n_{\rm{d}}(\rho,\, z) = A_0 \, \frac{\exp(- \rho / \rho_0)}{(\cosh(z / z_0))^2}\,.
\end{equation}
This model has two free parameters, $\rho_0,\, z_0$, to be fitted to the data. The global amplitude is absorbed in the maximum-profile likelihood computation and is thus irrelevant.

\subsubsection{4 Spiral Arms (ARM4)}
In this case, the dust density distribution is obtained by summing the contributions of four logarithmic spiral arms. In this implementation, and up to a relative amplitude, the arms are assumed to be identical with a rotation of ninety degrees.
Our implementation is given as:
\begin{equation}
n_{\rm{d}}(\rho,\, \phi,\, z) = A_0 \, \frac{\exp(- (\rho - \rho_c)^2 / (2 \rho_0^2))}{\left(\cosh(z / z_0)\right)^2} \, \mathcal{S}(\rho,\,\phi)
\end{equation}
where the function $\mathcal{S}$ encodes the logarithmic spiral pattern as:
\begin{equation}
\mathcal{S}(\rho,\, \phi) = \sum_i \left( A_i \, \exp \left( \frac{- (\phi - \phi_{s,i} )^2}{2 \phi_0^2} \right) \right)
\end{equation}
and
\begin{align}
\phi_{s,i} = & \frac{1}{\tan(p)} \log (\rho / \rho_{0,i})       \, , \\
\rho_{0,i} = & \exp  \left( \phi_{0,i} \, \tan(p) \right)       \, , \\
\phi_{0,i} = & \phi_{00} + i \, \pi/2           \, ,
\end{align}
with $i = \left\lbrace 1,\,2,\,3,\,4 \right\rbrace$, $\phi_{00}$ the angular
starting point of the fourth arm and $p$ the pitch angle of the logarithmic spirals. Particular attention has to be made regarding the $2\pi$ ambiguity of polar angle during the evaluation of the function $\mathcal{S}$.

This parametric model has nine free parameters to be fitted to the data $\rho_0,\, \rho_c,\, z_0,\, \phi_0,\, \phi_{00},\, p, $ and three relative amplitudes for the spiral arms. As for the ED model, the global amplitude is irrelevant.

\subsubsection{4 Spiral Arms plus 1 Exponential Disk (ARM4$\oplus$ED)}
We further consider a dust density model that results from the superposition of the ED and ARM4 models. This model has twelve free parameters to be fitted to the data. The radial and height scale lengths for the disk part and for the arm part are allowed to take different values.

\subsection{Regular Galactic Magnetic Field model}
\label{sec:GMFmodel}
Throughout this paper, we make use of four parametric models for the regular component of the GMF. These are the
Axi-Symmetric Spiral, the Logarithmic Spiral Arm, the Bi-symmetric Spiral, and the Quadri-Symmetric Spiral models.
Despite their simplicity, these four mathematical toy models are expected to capture, at least to some degree, the large-scale geometrical features of the regular part of the magnetic field of our Galaxy with two to four free-parameters.
The four GMF parametric models share the following main features:
\begin{enumerate}[(i)]
\item field lines have spiral geometries in planes parallel to the Galactic disk (at $z$ constant)
 \item field lines have out-of-the-plane component, intended to produce the so-called X-shape of the GMF
\item the field lines have some degree of symmetry with respect to the $z$-axis of cylindrical Galactic coordinate system\end{enumerate}
The first three models have already been discussed in the literature. We briefly present the models below but refer the reader to e.g. \cite{Rui2010} for detailed discussion.

\begin{figure}[h]
\centering
\includegraphics[trim={0cm 0cm 4.5cm 0cm},clip,width=.72\linewidth]{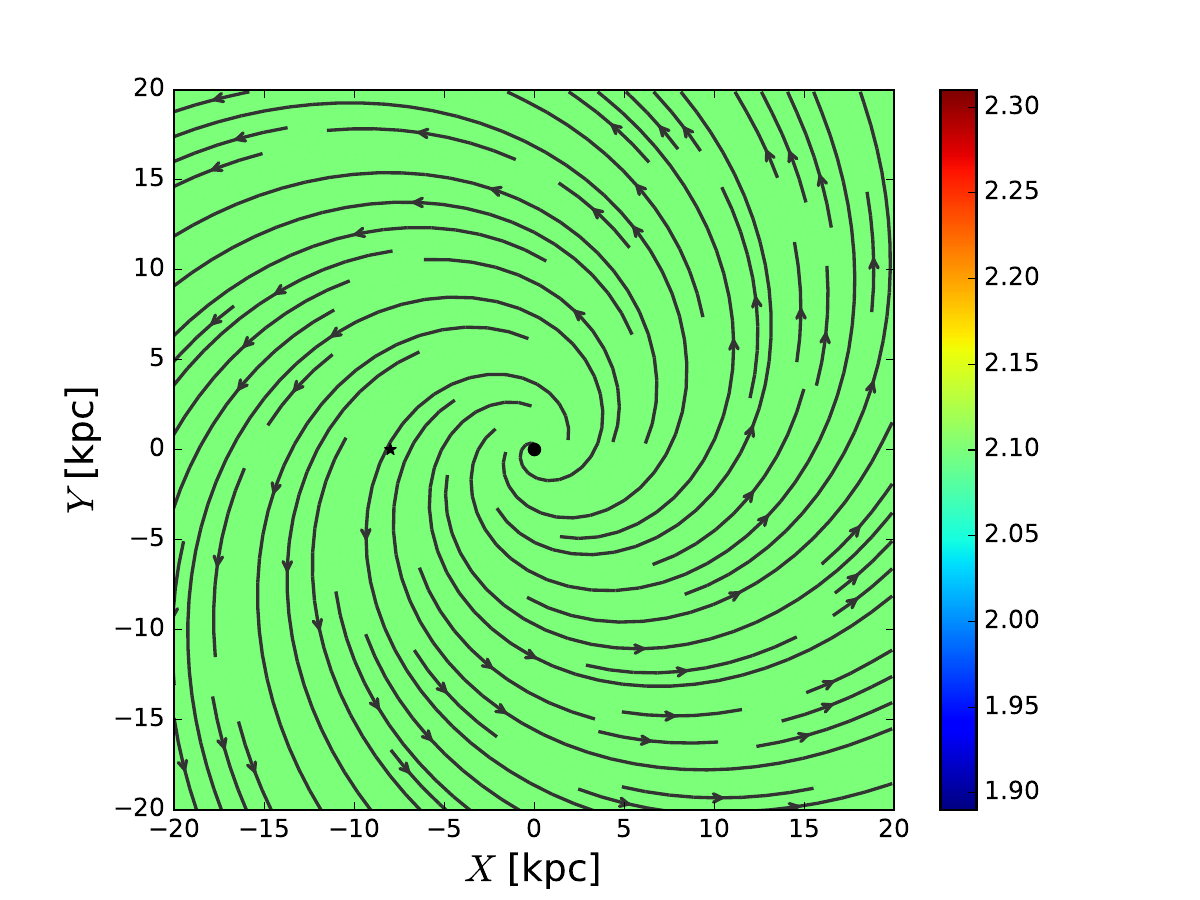} \\[-3.9ex]
        \hspace{.1cm}
        \includegraphics[trim={0cm 4cm 4.5cm 5.5cm},clip, width=.696\linewidth]{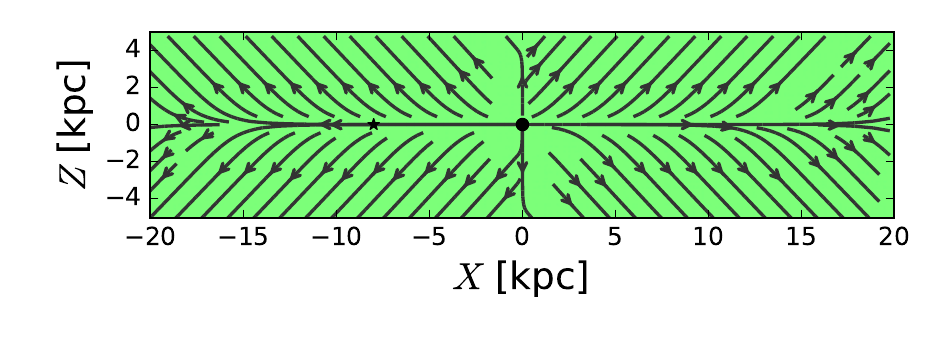} 
\\
\caption{GMF structures in the XY plane (top) and XZ plane (bottom)
for the input LSA model used as an input to build the \texttt{S1} and \texttt{S2} realistic simulations.
\label{fig:inputWMAP_xyz}}
\end{figure}
The four models take simple analytic expressions in the cylindrical coordinate
system center on the Galactic center:
\begin{equation}
{\mathbf{B}} = B_\rho {\mathbf{e}}_\rho + B_\phi {\mathbf{e}}_\phi +B_z {\mathbf{e}}_z,
\label{eq:Bfield_cyl-general}
\end{equation}
where we introduced the orthonormal basis $({\mathbf{e}}_\rho,\, {\mathbf{e}}_\phi,\, {\mathbf{e}}_z)$ with ${\mathbf{e}}_z = {\mathbf{e}}_\rho \times {\mathbf{e}}_\phi,$ where the polar angle $\phi$ increases counter-clockwise in the $(x,\,y,\,z=0)$ plane of the Galaxy. The four parametric models of the regular GMF component differ by their functional forms of the cylindrical components of the field.

\subsubsection{Axi Symmetric Spiral model: ASS}
The ASS model (see e.g., \citealt{Val1991}; \citealt{Poe1993}) is one of the simplest descriptions of the Galactic magnetic field.
It is compatible with a non-primordial origin of the Galactic magnetism, based on the dynamo theory. The field lines follow logarithmic spirals with constant pitch angle. The cylindrical components of this model are:
\begin{align}
B_\rho  = &     \,      B_0 \sin(p) \cos(\chi(z))                \nonumber \\
B_\phi  = &     \,      B_0 \cos(p) \cos(\chi(z))                \nonumber \\
B_z             = &     \,      B_0 \sin(\chi(z)),
\label{eq:ASSmodel}
\end{align}
where $B_0$ is the field strength that might, in general, depend on the distance to the Galactic center, $p$ is the (constant) pitch angle and $\chi(z)$ is a "tilt angle" that allows the field lines to have a non-zero $z$ component.
Different modeling of the field strength dependence exist.  We set $B_0$ to 2.1 $\mu$G throughout space because the polarized thermal dust emission is not sensitive to the field strength.

The pitch angle is defined as the angle between the azimuthal direction, ${\mathbf{e}}_\phi$, and the magnetic field. $p = 0^\circ$ corresponds to circle and $p = 90^\circ$ produces (cylindrical) radial lines.

The tilt angle is defined as the angle between the field line and the plane parallel to the Galactic plane. This angle usually takes the parametric form:
\begin{align}
\chi(z) = & \, \chi_0 \tanh \left( \frac{z}{z_0} \right)\,.
\label{eq:chi_tilt}
\end{align}
For the ASS model, we adopted the usual choice which is to fix the height scale at $z_0 = 1$ kpc. However, it is to be noted that this choice impact the value of $\chi_0$ as these two parameters are not strictly independent.

According to the above implementation, the ASS model has two free-parameters (the pitch angle ($p$) and the tilt angle at large Galactic height ($\chi_0$)) that can be adjusted to fit the polarization maps to the data.

\subsubsection{LSA model}
This parametric model, originally named the \textit{Logarithmic Spiral Arm}, was introduced in \cite{Pag2007} to describe the three-year \textit{WMAP} data at 22~GHz for the synchrotron emission and at 94~GHz for the dust emission.
Basically, this model implements an extension of the axi-symmetric class of models where the pitch angle defining the spiral structure depends on the cylindrical radius ($\rho$). Consequently, and despite the name of the model, the spirals are not logarithmic.
As for the ASS model, there is no arm structure in the field amplitude.

The parametric form of the field components read
\begin{align}
B_\rho = & \,   B_0 \sin( \psi(\rho)) \cos(\chi(z))     \nonumber \\
B_\phi = & \,   B_0 \cos(\psi(\rho)) \cos(\chi(z))      \nonumber \\
B_z     = & \, B_0 \sin(\chi(z)),
\label{eq:LSAmodel}
\end{align}
where the function 
\begin{equation}
\psi(\rho) = \psi_0 + \psi_1 \, \ln \left( \frac{\rho}{R_\odot} \right)
\label{eq:lsa_pitch}
\end{equation}
forces the magnetic field lines to follow a spiral pattern with varying pitch angle; $\psi_0$ is the pitch angle at Sun radius ($R_\odot = 8.0$ kpc); and $\psi_1$ the amplitude of the logarithmic radial modulation of the pitch angle.
The tilt angle $\chi(z)$ is taken as in Eq.~\ref{eq:chi_tilt}  with $z_0 = 1$ kpc.
Having fixed $B_0$ to a constant, the LSA model has three free
parameters to be fixed by the data: ($\psi_0$, $\psi_1$ and $\chi_0$).
By construction, there is no reason to constrain the $\psi_1$ parameter to only positive or negative values. Fixing $\psi_1 = 0$ reduces the LSA model to the ASS model.

\subsubsection{Bi Symmetric Spiral model: BSS}
The class of bi-symmetric spiral models produces GMF that could be compatible with a primordial origin. This model includes magnetic field line reversion as suggested from pulsar rotation measurements \citep[e.g.,][]{Han1994, Han2006}.
This model was also referred to as the \textit{Modified Logarithmic Spiral} model in
\cite{Fau2011}.
The field lines and the spirals drawn by the surfaces of iso-amplitude of
the field have a constant and same pitch angle.
The cylindrical components of this field model read
\begin{align}
B_\rho  = & \, B_0 \cos \left(
                                \phi \pm \beta \ln\left(\frac{\rho}{\rho_0}
                                \right)\right) \sin(p) \cos(\chi(z))    \nonumber \\
B_\phi = & \, B_0  \cos \left(
                                \phi \pm \beta \ln\left(\frac{\rho}{\rho_0}
                                \right)\right) \cos(p) \cos(\chi(z))    \nonumber \\
B_z = & \, B_0 \sin(\chi(z)),
\label{eq:BSSmodel}
\end{align}
where $\beta = 1/\tan(p)$, with $p$ the pitch angle, $\chi(z)$ the tilt angle defined as in Eq.~\ref{eq:chi_tilt}, $B_0$ the global amplitude of the field strength, which we assume to be a constant, and $\rho_0$ a radial scaling factor dictating the position of the spiral arms in the $(x,\,y,\,z=0)$ plane.
In this model, the amplitude of the GMF is shaped in spirals by the term $\cos \left(\phi \pm \beta \ln\left({\rho}/{\rho_0} \right)\right)$. Because this modulation appears only in the components $B_\rho$ and $B_\phi$, it is not a simple scaling of the GMF strength. It produces changes in field line directions and thus produces the reversal of the field.
The $\pm$ in the parentheses was introduced by \citet{Rui2010} to account for the different conventions met in the literature.
Given our working scheme, we adopt the "$-$" sign so that the spiral pattern of the GMF amplitude and of the GMF lines coincide, which is reasonable to postulate.

Unlike for the ASS and LSA models, we let the parameter $z_0$, involved in the tilt angle modeling, to be free.
The BSS model has therefore four free parameters that can be fitted by the observations.

\subsubsection{Quadri Symmetric Spiral model: QSS}
This is a class of GMF models is a slight modification of the BSS in which we multiply the argument responsible for the amplitude modulation by a factor of 2. This small change results in a class of GMF models having four spiral arms located ninety degrees away from one another.
This four-parameter model therefore shares the four spiral arm features of the model presented in \cite{Jaff2010} and \cite{Jaff2013} but with a much lower number of free-parameters, with field reversal and with out-of-plane component.
This model has twice the number of field reversals than the BSS does. Such regular GMF models produce more efficient line-of-sight depolarization of the dust emission than the other models, a feature that might be called by the data.

\subsection{Turbulent Galactic Magnetic Field model}
\label{sec:turbModel}
We considered modeling the 3D turbulent part of the GMF at each position in the Galaxy ($\mathbf{B}^{\rm{turb}}(\mathbf{r})$), as derived from an isotropic random component that follows a two-slopes power-law magnetic energy spectrum.

To simulate the turbulent part of the GMF, each component of the turbulent GMF vector is obtained from a 3D Gaussian realization that follows a power-law power spectrum with two spectral indices, as suggested by \citep{Han2004,Han2017}.
The composite spatial magnetic-energy spectrum takes the form
$P(k) = C \, k^\alpha$ with $\alpha = - 0.37$ for $k < k_{\rm{out}}$ and $\alpha = - 5/3$ for $k \geq k_{\rm{out}}$ with the turnover at  $k_{\rm{out}} = 10$ corresponding to the outer-scale of the turbulence  estimated at 0.1 kpc \citep{Hav2006,Hav2008}.

Following \citet{Fau2011,Wae2009,PlanckXLII2016}, the turbulent Gaussian random field, is computed in Fourier space leading to a 3D cube that contains the sampled Galaxy. In order to compute the turbulent component with the highest possible resolution in reasonable computing time, we perform both a low and a high spatial resolution simulation of the turbulent GMF in a cubic Cartesian grid of 1024 cells on a side. The physical resolution of the grids are 0.39 and 3.9 kpc, respectively. In the Fourier space the high-resolution and low-resolution simulations span the ranges of $k$-values of $\left[ 0.25,\, 256 \right] $ and $\left[0.025,\, 25.6\right]$ kpc$^{-1}$, respectively. The vector components of the turbulent magnetic field are then extracted at each location of our spherical sampling of the Galactic space (with $N_{\rm{side}} = 1024$ as in Sect.~\ref{sec:simu}), assuming a nearest neighbor scheme to map the two space samplings.
The high- and low-resolution simulations are normalized in the overlapping range of $k$ values (i.e., $k \in \left[0.25,\, 25.6\right]$ kpc$^{-1}$). Inside the heliocentric radius of 2 kpc, the two turbulent components (high and low resolution) are considered in order to preserve continuity of large-scale fluctuations in the neighborhood of the Sun. Outside the heliocentric radius of 2 kpc, only the low resolution box is used, we add random white noise with an $rms$ that corresponds to the spatial frequencies not sampled in the low-resolution box but that are included in the high-resolution one (namely, for $k \in \left[25.6,\, 256 \right]$). This procedure ensures an identical $rms$ of the turbulent GMF vectors throughout the sampled space. Unlike \cite{PlanckXLII2016,Ste2018}, we do not re-scale the turbulent GMF component as a function of position in the Galaxy. This is because the regular GMF model of \cite{Pag2007}, the only one to which the turbulent component is added in this study, has a constant amplitude.

\section{Resolution bias}
\label{sec:resolutionBias}

\begin{figure*}
\begin{center}
\includegraphics[trim={1cm 0cm 1cm 0cm},clip,width=.31\linewidth]{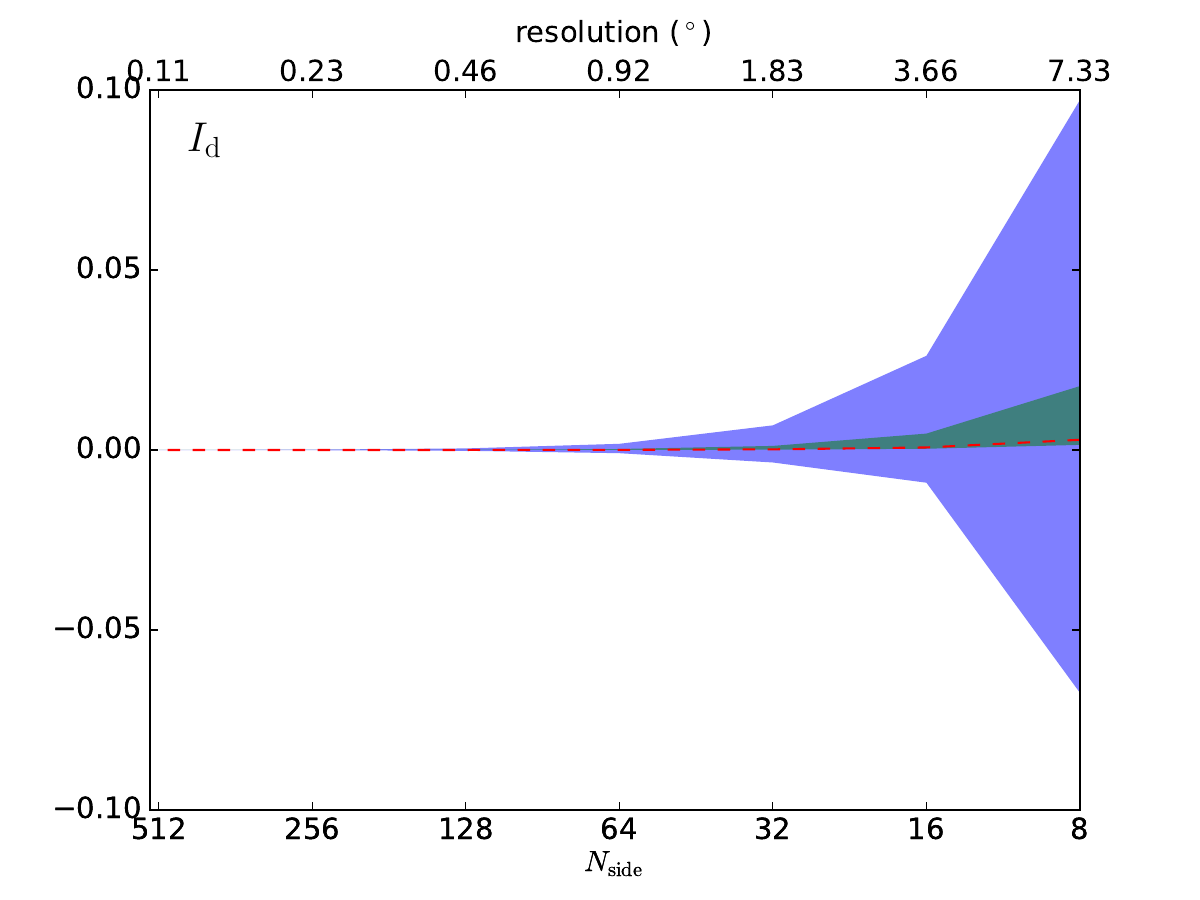} 
\includegraphics[trim={1cm 0cm 1cm 0cm},clip,width=.31\linewidth]{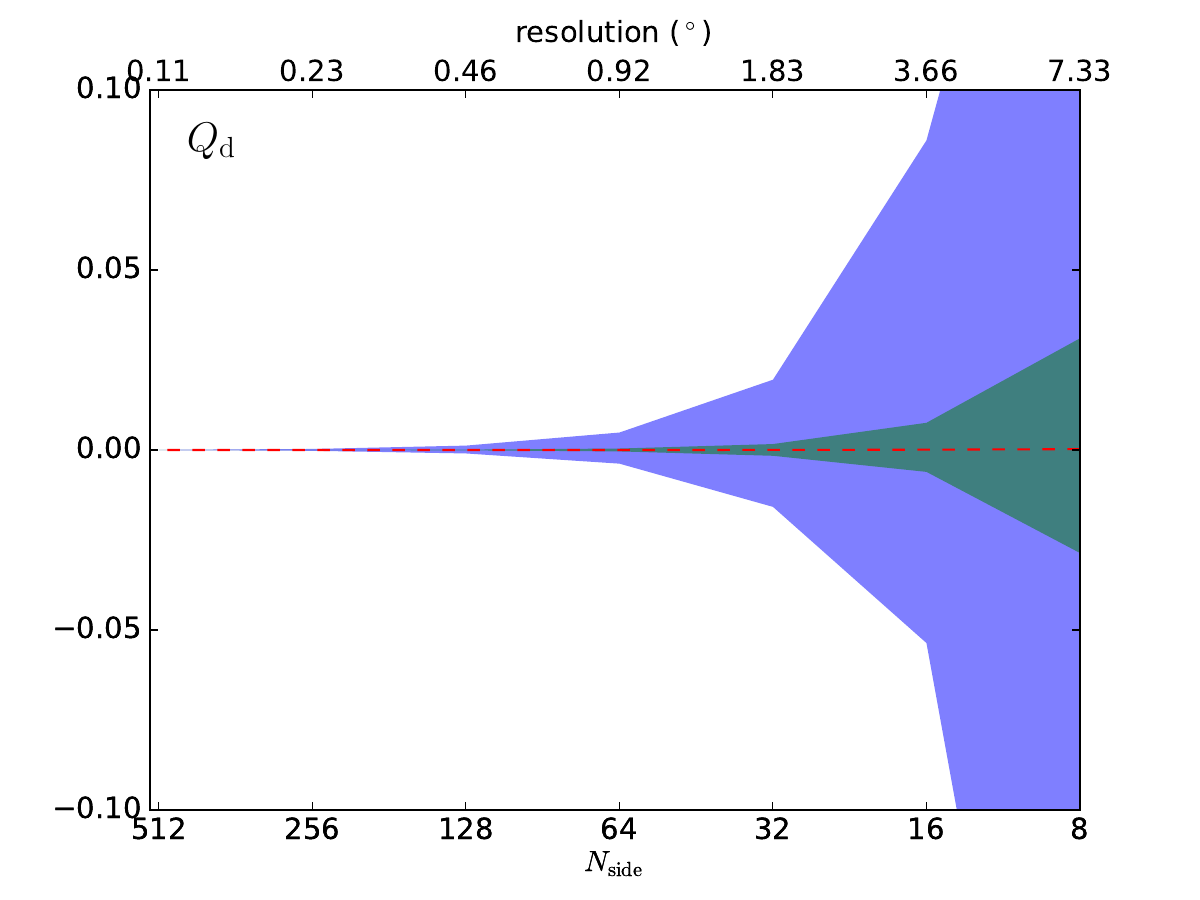}
\includegraphics[trim={1cm 0cm 1cm 0cm},clip,width=.31\linewidth]{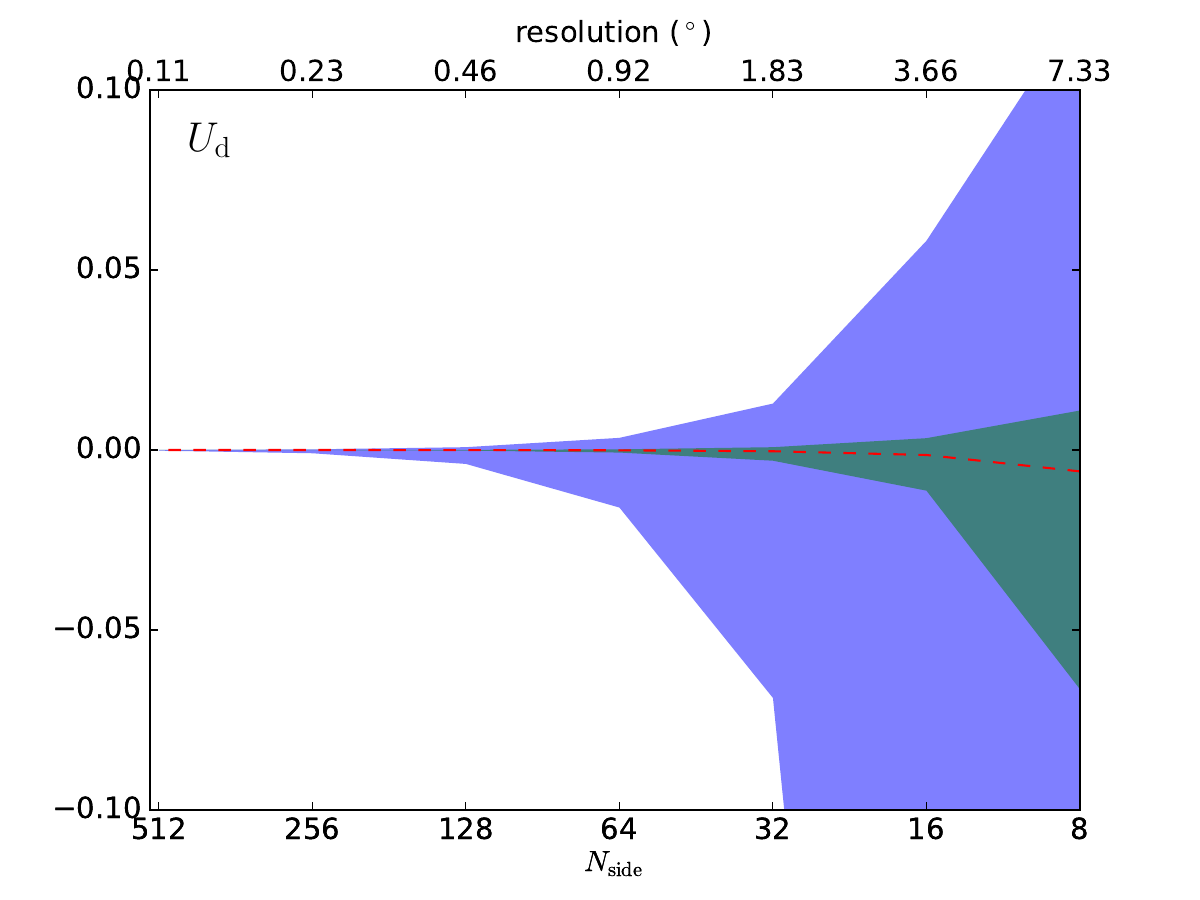}
\caption{Distribution of the relative difference per pixel between low
resolution maps produced at several HEALPix $N_{\rm{side}}$ (shown
in the x-axis) and the high resolution map at $N_{\rm{side}}=1024$
downgraded to equivalent resolution for the signal term of the \texttt{S2}
simulated maps. From left to right, we present this distribution for Stokes parameter $I$, $Q$
and $U$. The shaded-blue (-green) regions contains the 95\% (68\%) of the
total number of pixels. 
\label{fig:IQU_vs_NSIDE}}
\end{center}
\end{figure*}

To efficiently explore the model parameter spaces while running MCMC method, large number of simulations are necessary.  It is therefore often necessary to work at resolution lower than that of the original data to keep the computing time reasonable. Here, we explore possible consequences of this choice in the final results. 

First, we have tested the impact of the choice of the angular resolution of the HEALPix tessellation on the simulated observables for the models presented in Sect.~\ref{sec:simu}. This comparison study is illustrated in Fig.~\ref{fig:IQU_vs_NSIDE}, where we represent the distribution of the relative differences of the models as computed for $N_{\rm{side}} = 8, 16, 32, 64, 128, 256, 512$ with respect to the ones computed at $N_{\rm{side}} = 1024$  and downgraded to the respective lower $N_{\rm{side}}$ values.
For each pixel, the relative differences is computed as $2(S_{\rm{HR}} - S_{\rm{LR}})/(S_{\rm{HR}} + S_{\rm{LR}})$, where the subscripts HR and LR stand for high resolution and low resolution, respectively; and $S$ represents any of the linear polarization Stokes parameters.

In Fig.~\ref{fig:IQU_vs_NSIDE}, the dashed-red line represents the median of the distribution of the $N_{\rm{pix}}$ measurements of the relative difference, the green-shaded region contains 68\% of the pixels and the blue-shaded region 95\%.
In that figure we see that, at the pixel level, a scatter arises due to the resolution at which the simulations are computed. This is a sampling issue.
The lower the map resolution, the coarser the sampling of the 3D space. At the map level this leads to features that are either missed or grossly represented.
As a consequence, other values of the model parameters may lead to a better fit to the observed downgraded features, thus producing an overall shift (or bias) in the parameter space.

To remedy this issue, \cite{Wan2020} suggested to interpolate high-resolution maps at the sky coordinates of the centers of the low-resolution-map pixels.
Interpolation can indeed reduce the magnitude of this sampling issue but will not make it disappear in general. Moreover, interpolation has  its drawbacks.  The resulting maps depend on the interpolation scheme adopted and that the propagation of uncertainties is more complex and in any case not analytic. In this work, we do not consider  this possibility any further.
We note that, as shown in Fig.~\ref{fig:IQU_vs_NSIDE}, the high and low angular-resolution simulations agree on pixel values at the percent level starting from $N_{\rm{side}} = 64$ and the agreement increases as $N_{\rm{side}}$ increases.

Second, we investigate the bias introduced in the best-fit parameters when 
working at resolution lower than that of the native data. For this purpose we consider the case of fitting the simulation \texttt{S1} because the underlying dust density distribution model is the simplest and the loss of information due to line-of-sight integration is expected to be negligible, if any. That is, for this simple case the source of bias in recovered parameter values mainly comes from the effect of the simulation resolution.
To demonstrate the significance of the resolution bias, we performed the MCMC fits of the intensity map and of the polarization maps at the $N_{\rm{side}}$ values of 8, 16, and 32, in addition to the fits in $N_{\rm{side}}$ = 64 presented in the core of the paper.
For the two parameters of the dust density distribution and for the four parameters of the LSA GMF model, we then evaluate the difference between the best-fit parameters and the input ones, in sigma units,:
$|p_0 - \hat{p}|/\sigma_{\hat{p}}$. Here, $p_0$ is the input-parameter value, $\hat{p}$ is the best-fit parameter value corresponding to the minimum $\chi^2$ and $\sigma_{\hat{p}}$ is computed as being the standard deviation of the marginalized distribution of the considered parameter as sampled by the MCMC algorithm.

\begin{figure}[t]
\centering
\includegraphics[trim={0cm 0cm 0cm 0cm},clip,width=\columnwidth]{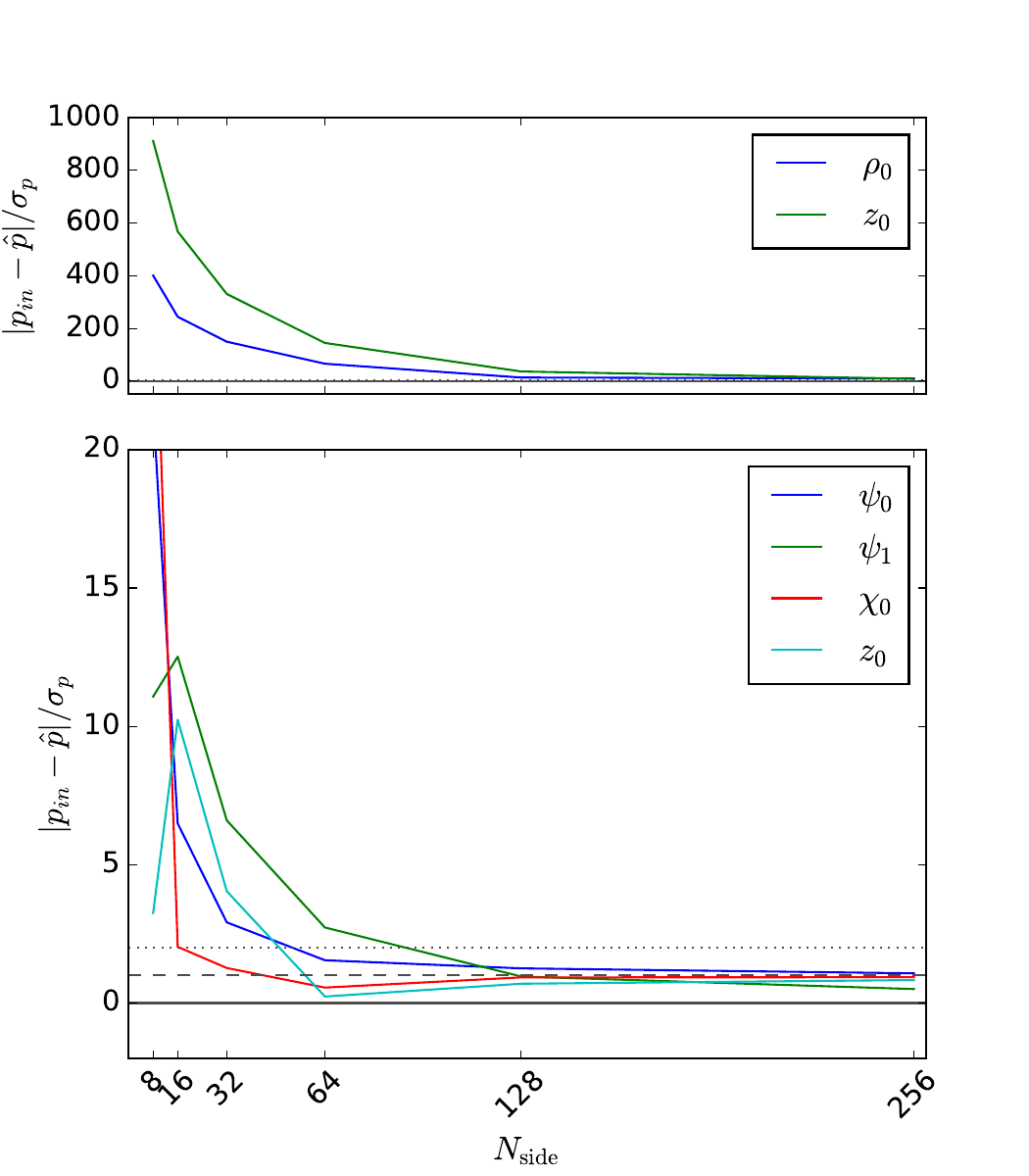} \\
\caption{Significance of the resolution bias of the best-fit parameter
values as compared to the input parameter values as a function of the
$N_{\rm{side}}$ value used to generate the models in the MCMC algorithm.
for simulation \texttt{S1}.
Top panel shows the two parameters of the ED dust density distribution and
bottom panel shows that behavior for the four GMF parameters.
\label{fig:resolbias_bfits}}
\end{figure}

The results of this analysis are shown in Fig.~\ref{fig:resolbias_bfits}. It is seen that the best-fit parameters converge toward the input values as the resolution of the model used in the MCMC analysis increases. The convergence is however not trivial and depends upon the considered parameter and also on its input values.
To this respect, we note that for some parameters, the best-fit values may oscillate around the input across the $N_{\rm{side}}$ values.
The value of the reduced $\chi^2$ obtained at the different resolutions go from 984 to 1.02 for the intensity fits and from
24.2 to 1.935 for the fits of the $(q,\,u)$ maps.

We conclude that due to the resolution bias the values of the best-fit parameters for dust density distribution and the GMF, these models need to be taken with caution.
Furthermore, the best-fit parameters obtained at a given resolution should not be used at other resolutions.
Formally speaking, it should be possible to account for such a bias in the definition of the likelihood function. However, we have observed that the bias depends strongly both on the parametric form of the models and on the parameter values themselves.
Therefore, accounting for it is cumbersome and goes beyond the scope of this paper. Overcoming the resolution bias will be mandatory once the models to be fitted to the real data sets will be sufficiently evolved to account for all the complexity and the richness contained in them.
Indeed, the resolution bias can be significant only if it turns out to be the leading source of uncertainties.
However, this can happen in the case signal dominated data, as it is already the case for the  353 GHz {\it Planck} data, and when the model  accurately represents the data, what is clearly not the case as of today  (see also Sect.~\ref{sec:fit2Planck}).
Therefore, for now, we notice that this bias exists and that the larger the $N_{\rm{side}}$ value of the working resolution, the smaller the bias is.

\smallskip

Third, we tested the impact of the adopted value for the radial sampling (sampling along the line-of-sight). As far as we do not pretend to model the very nearby structures, we found that the radial step can be as large as few hundreds of parsecs.
The difference in pixel space is more pronounced due to the adopted angular sampling than due to the radial sampling. However, a fine radial sampling would be essential as soon as nearby and physically small features are introduced in the modeling.

\smallskip

In summary,
it turns out to be important to work at the highest resolution allowed by the computing facilities.
As a result, in this paper we choose to simulate the maps at $N_{\rm{side}} = 64$ and with a radial bin of 0.2 kpc.
For the models considered the computation time of a full set of Stokes parameters maps is always of the order of few tenths of a second. 
A greater computing time would limit significantly our ability to run the MCMC algorithm.

\section{Orthographic view of the Planck data and of the best-fit models}
\begin{figure}[h]
\centering
\begin{tabular}{cc}
\includegraphics[width=.47\linewidth]{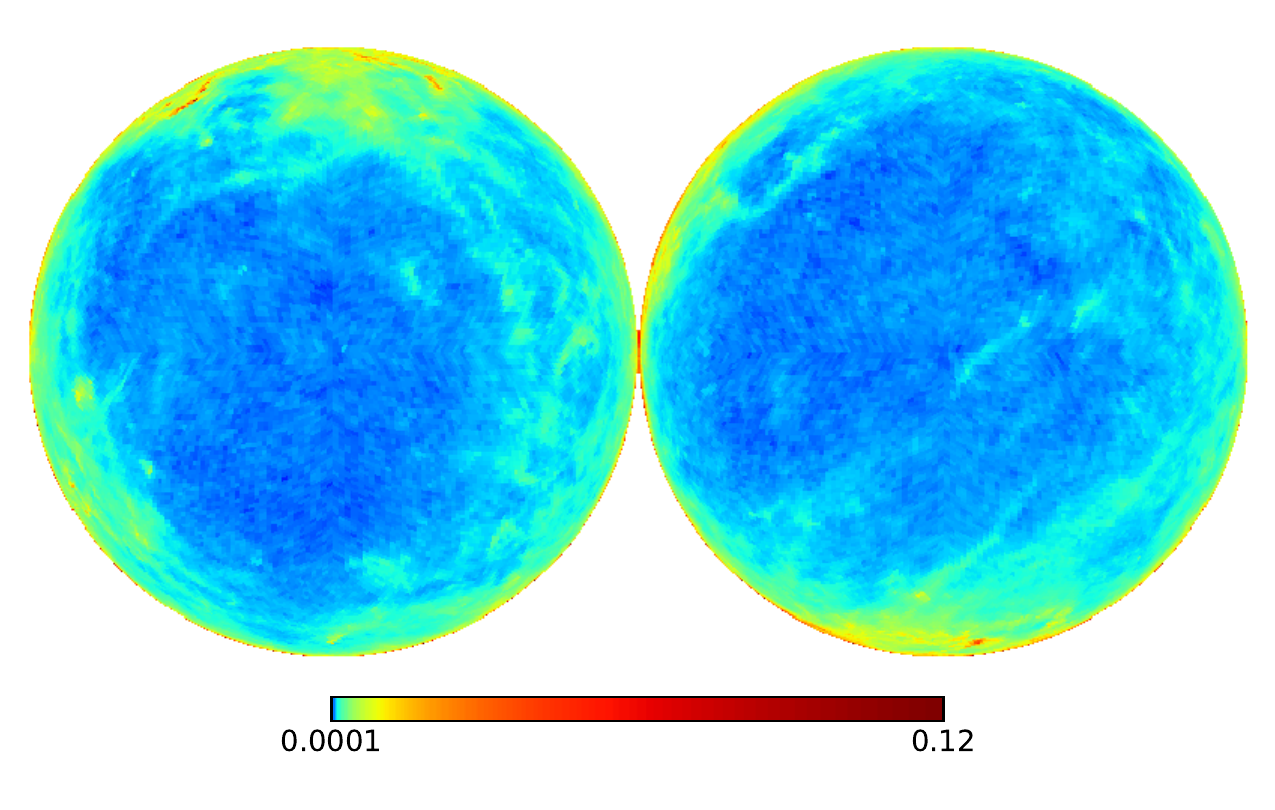} &
        \includegraphics[width=.47\linewidth]{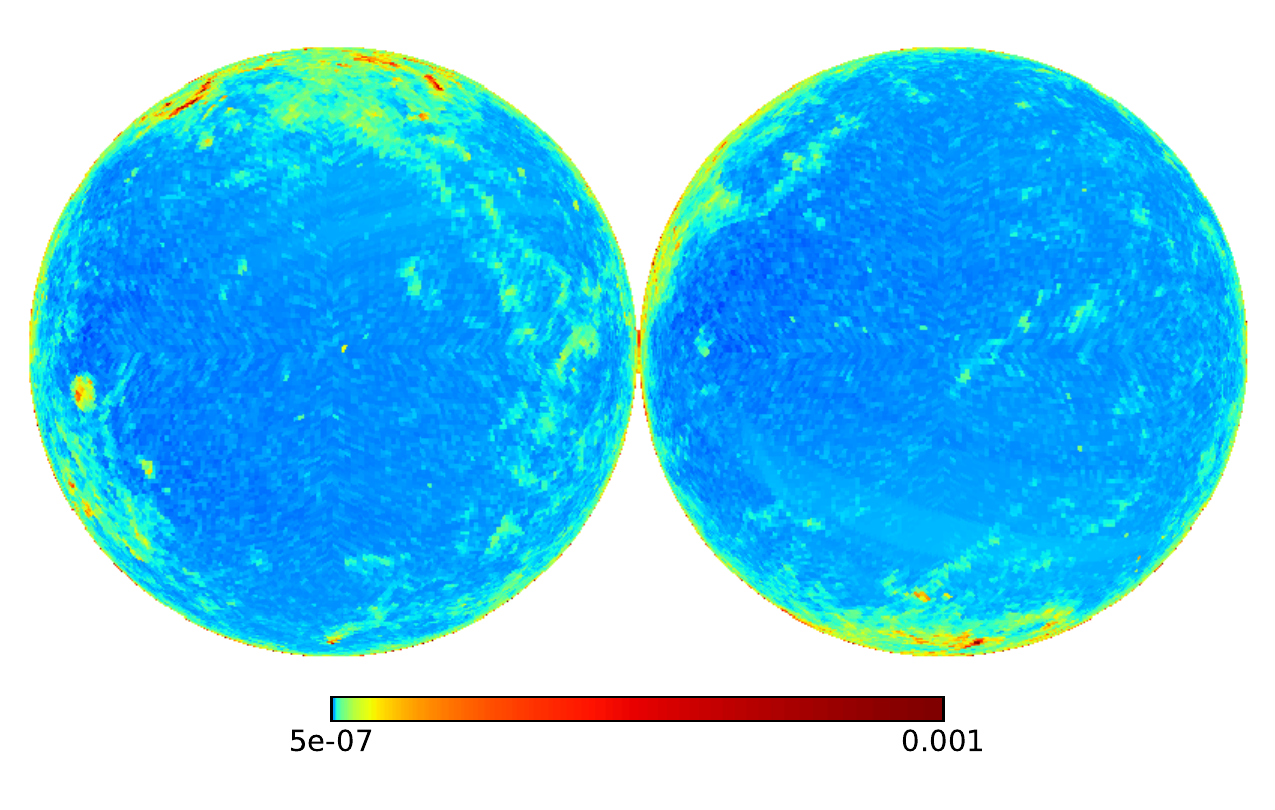} \\

\includegraphics[width=.47\linewidth]{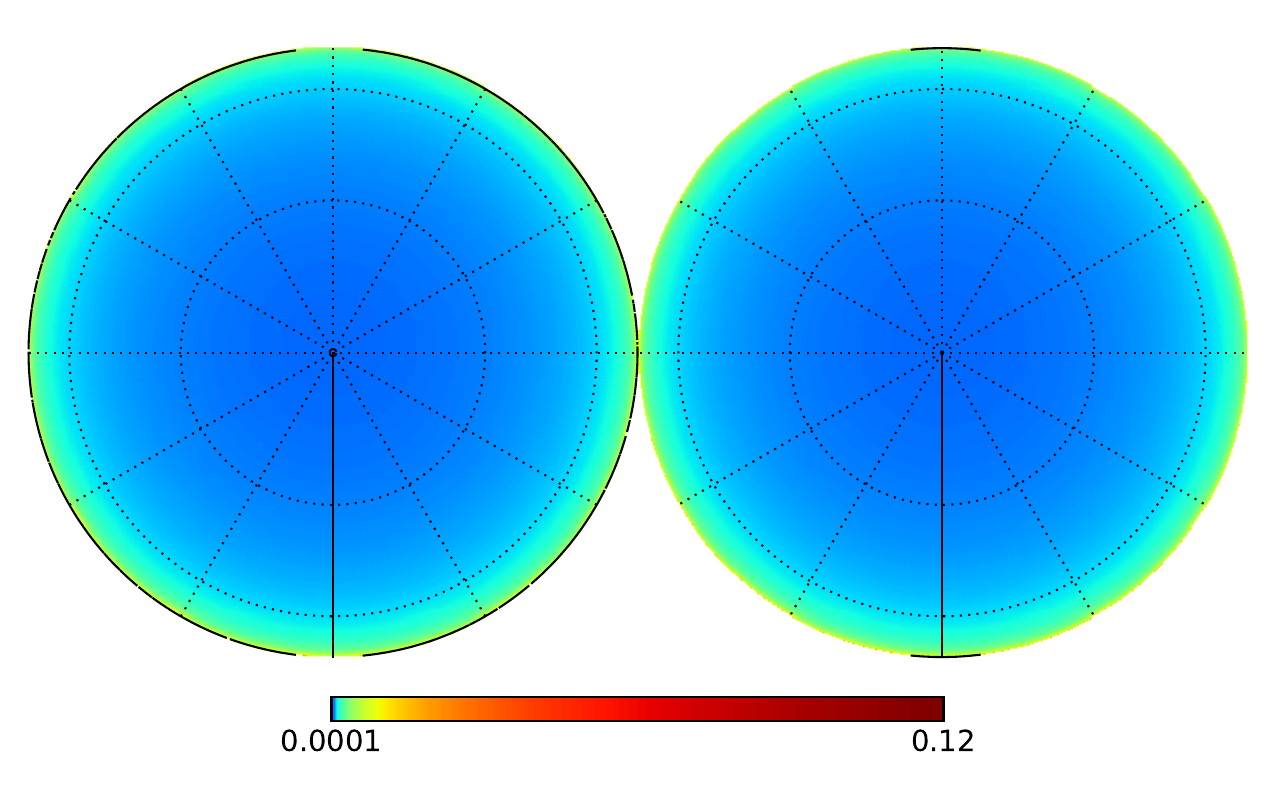} &
        \includegraphics[width=.47\linewidth]{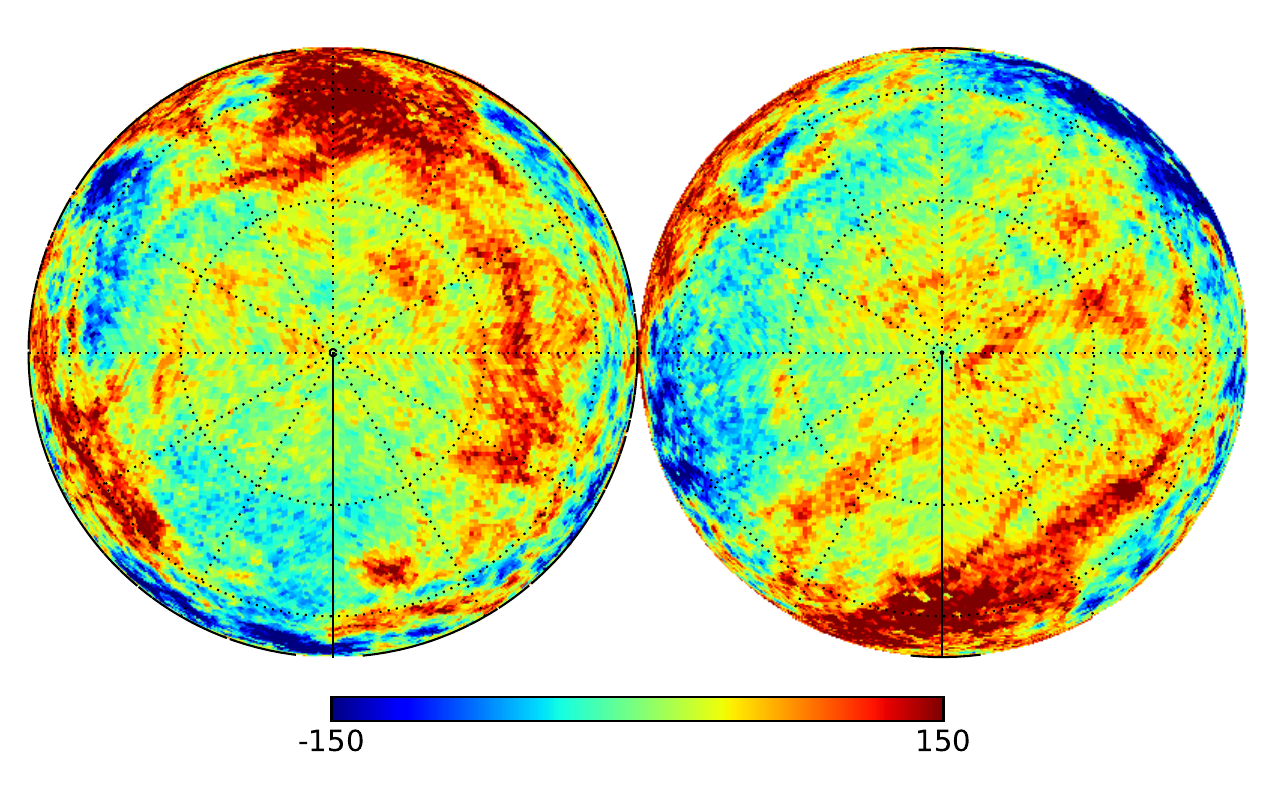} \\

\includegraphics[width=.47\linewidth]{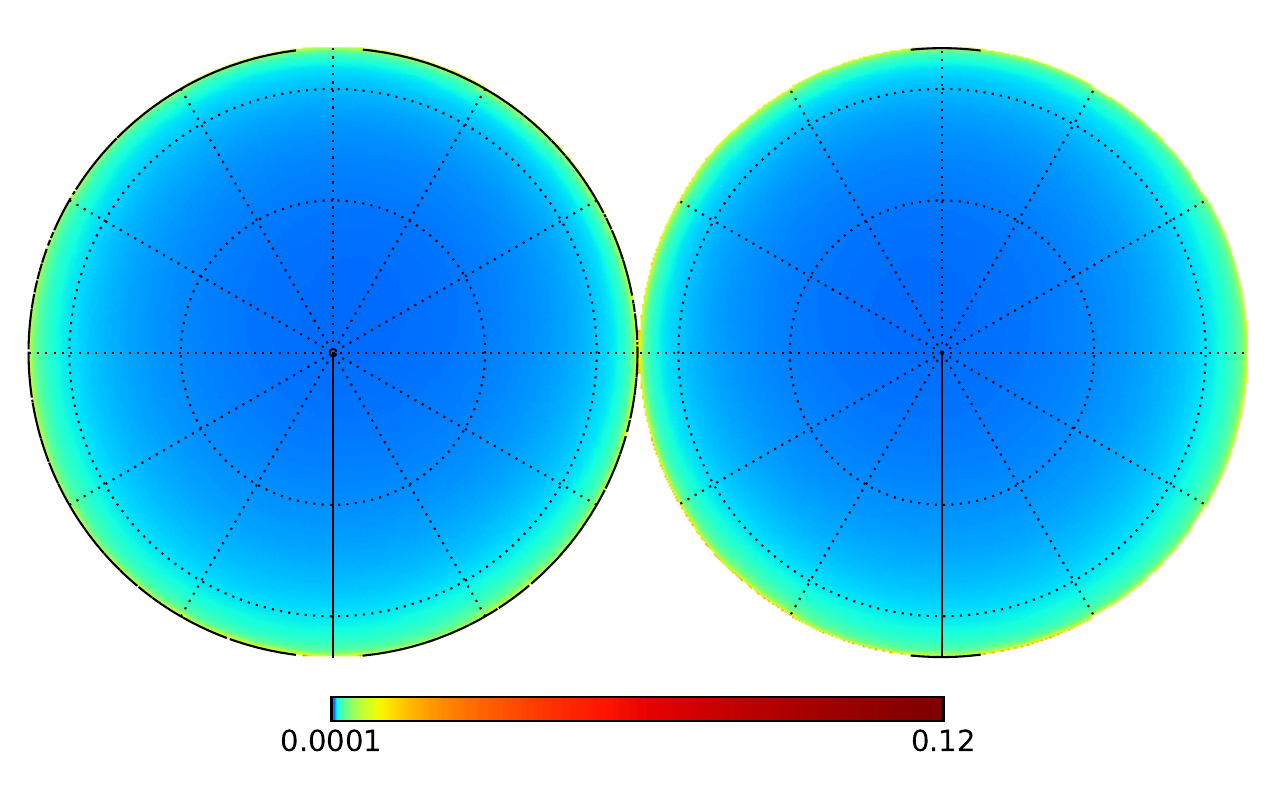} &
        \includegraphics[width=.47\linewidth]{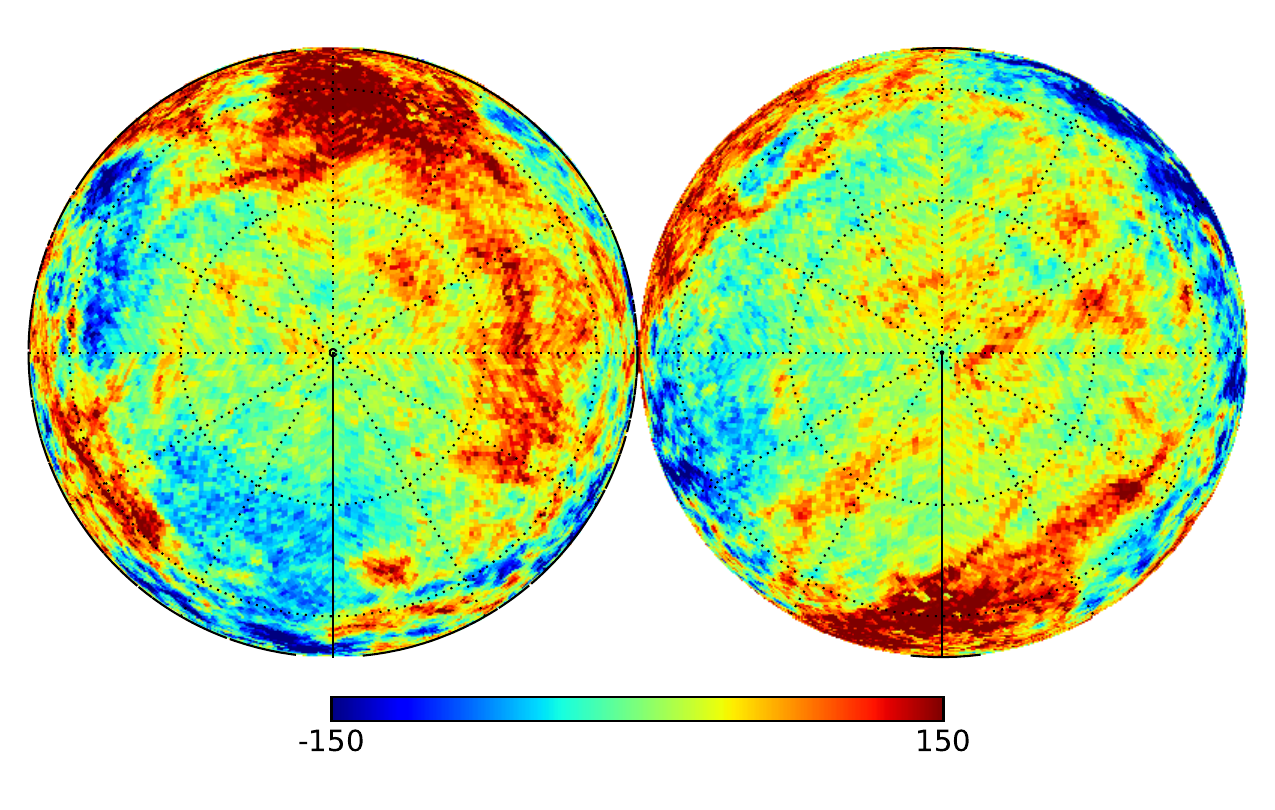} \\

\includegraphics[width=.47\linewidth]{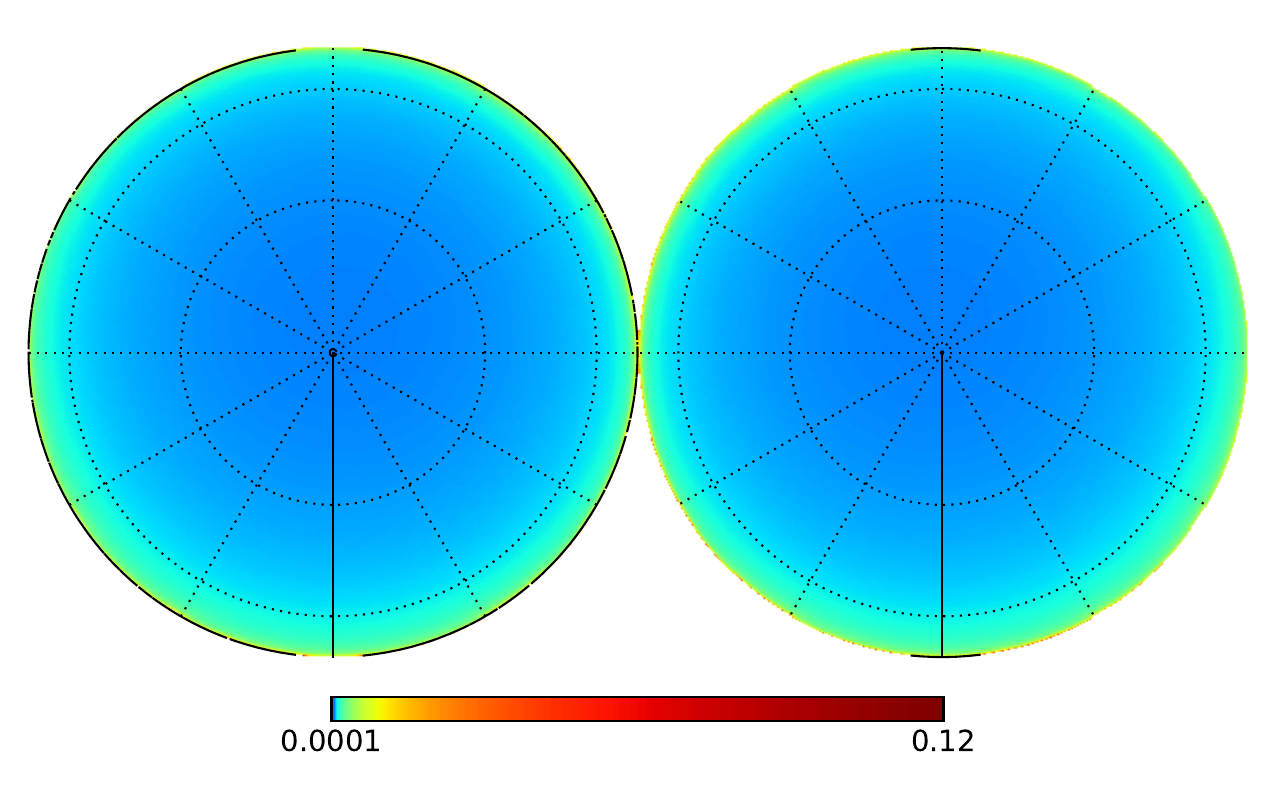} &
        \includegraphics[width=.47\linewidth]{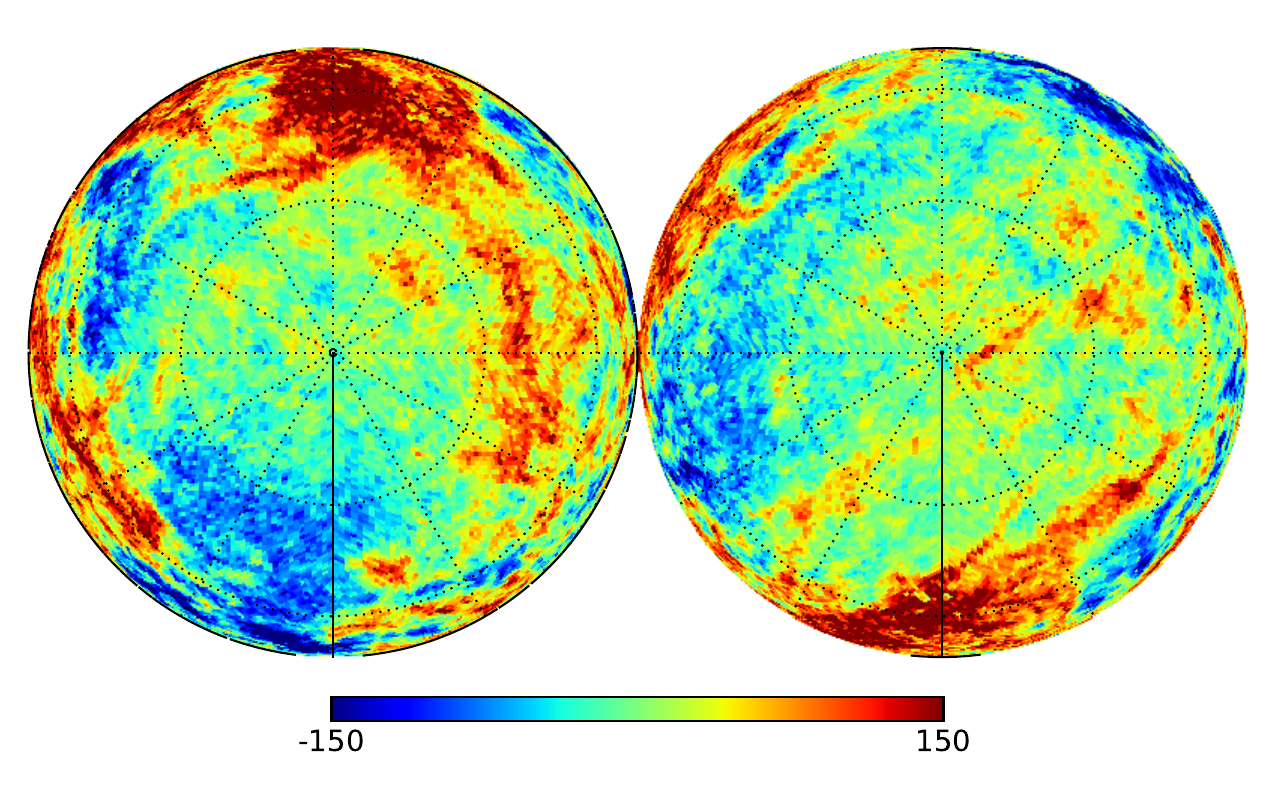} \\
\end{tabular}
\caption{Orthographic view of the intensity maps. First row show the 353-GHz map from \textit{Planck} downgraded at $N_{\rm{side}} = 64$ and the corresponding map of uncertainties that we use to compute the $\chi^2$. Rows two to four correspond to dust density distribution models labeled ED, ARM4 and ARM4$\oplus$ED, respectively. The obtained best fits are shown in the first column and the statistical significance of the residual, per-pixel, are shown in the second column. The north Galactic pole is on the right-hand side panel and the vertical solid line is for Galactic longitude zero. Galactic latitude decrease counter-clockwise on the right panel and clockwise in the left panel.}
\label{fig:I_fit-orth}
\end{figure}

\begin{figure*}[h]
\centering
\begin{tabular}{cccc}
\includegraphics[width=.23\linewidth]{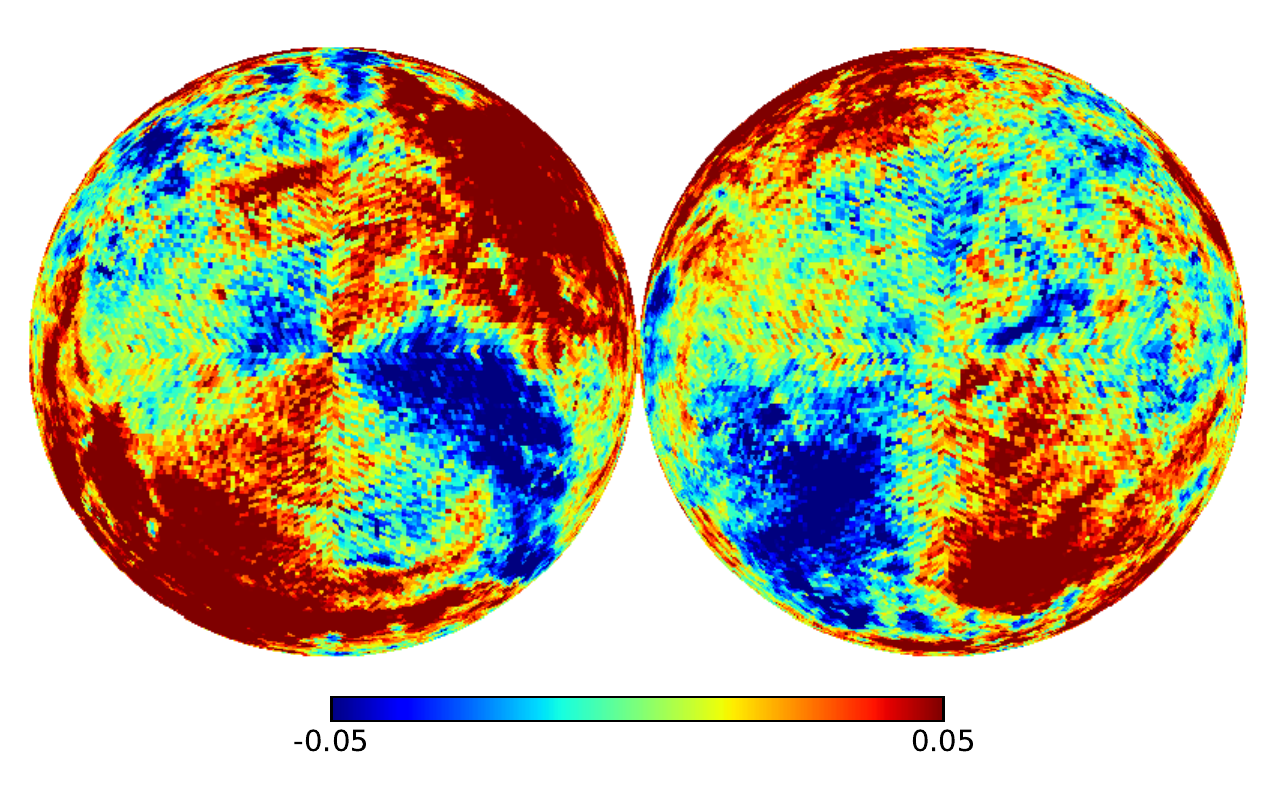} &
        \includegraphics[width=.23\linewidth]{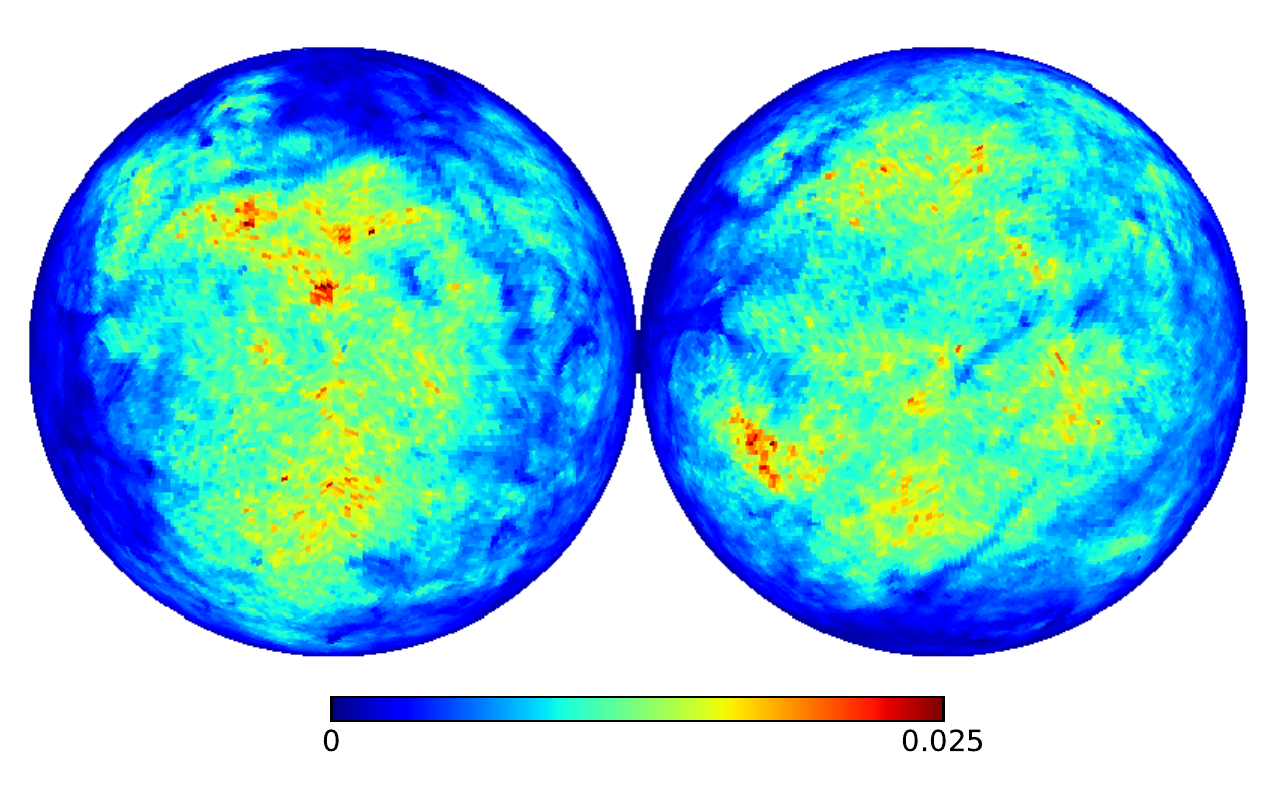} &
                \includegraphics[width=.23\linewidth]{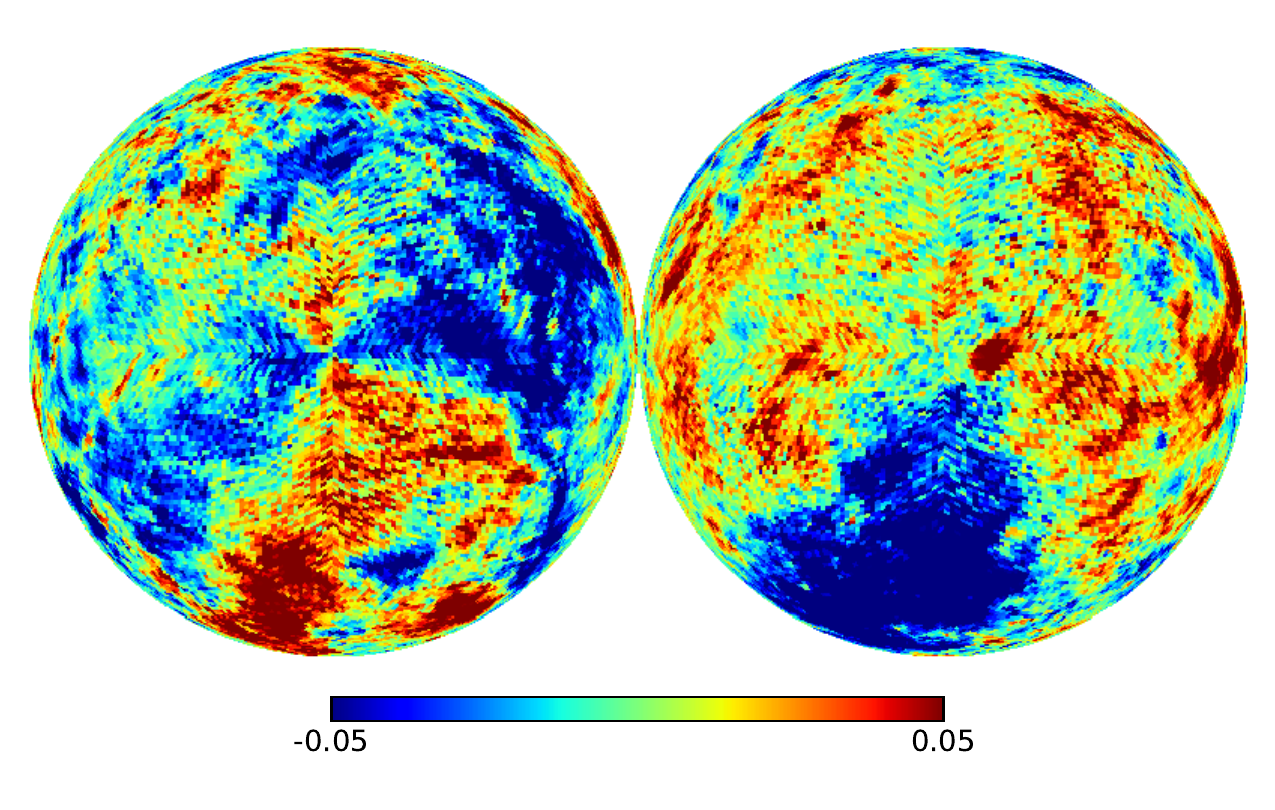} &
                        \includegraphics[width=.23\linewidth]{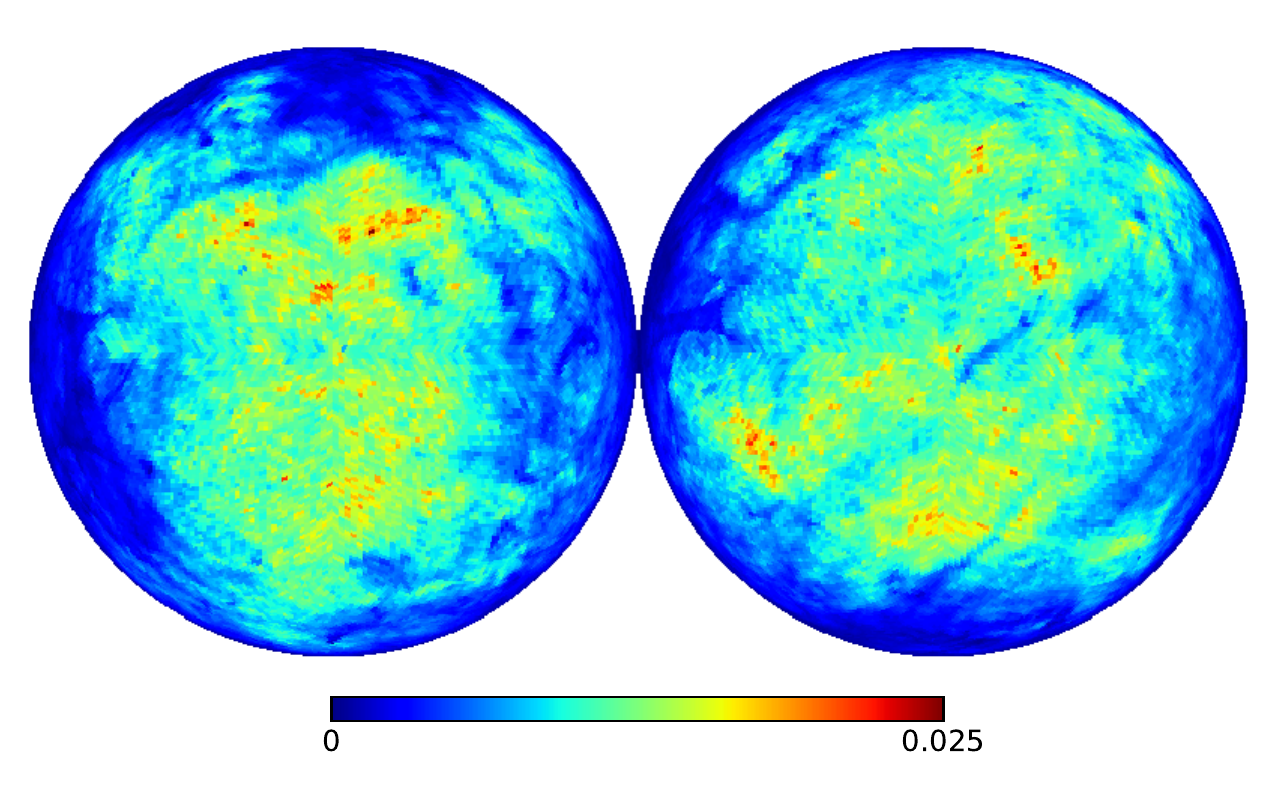} \\

\includegraphics[width=.23\linewidth]{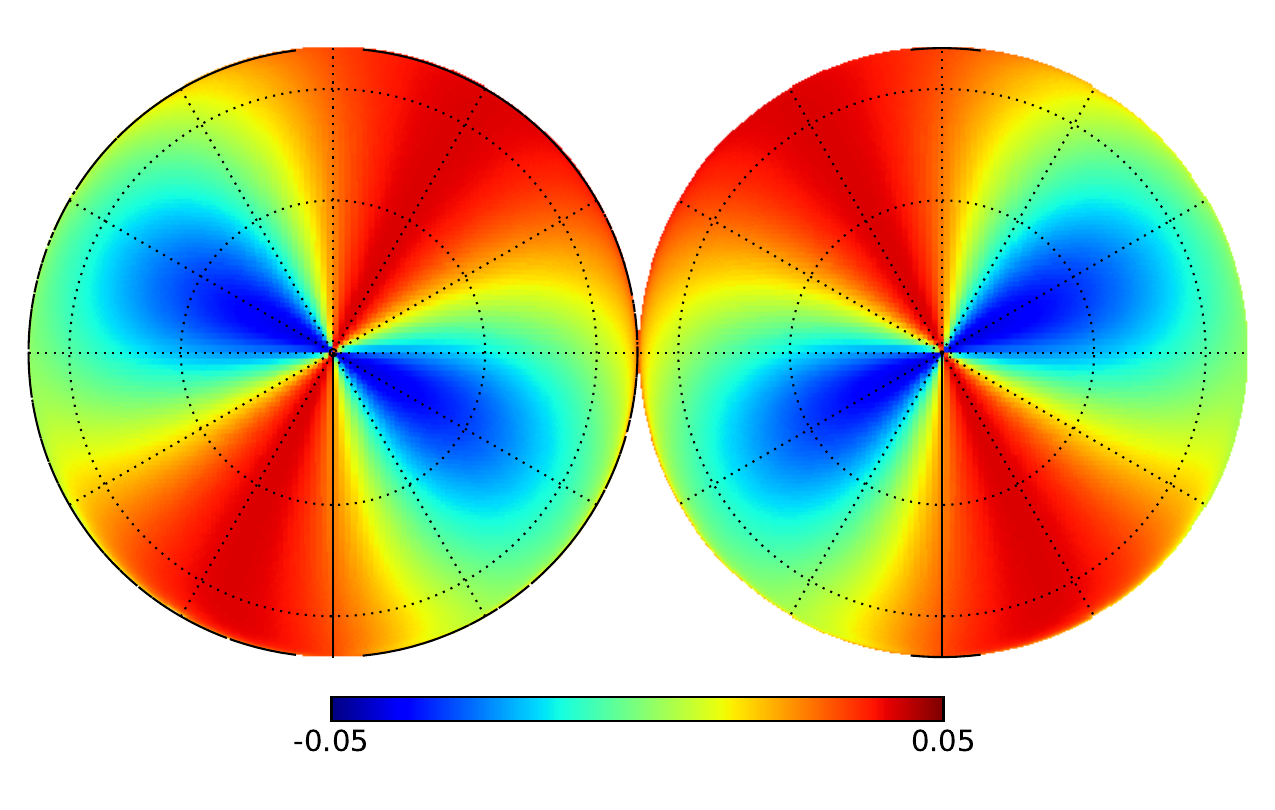} &
        \includegraphics[width=.23\linewidth]{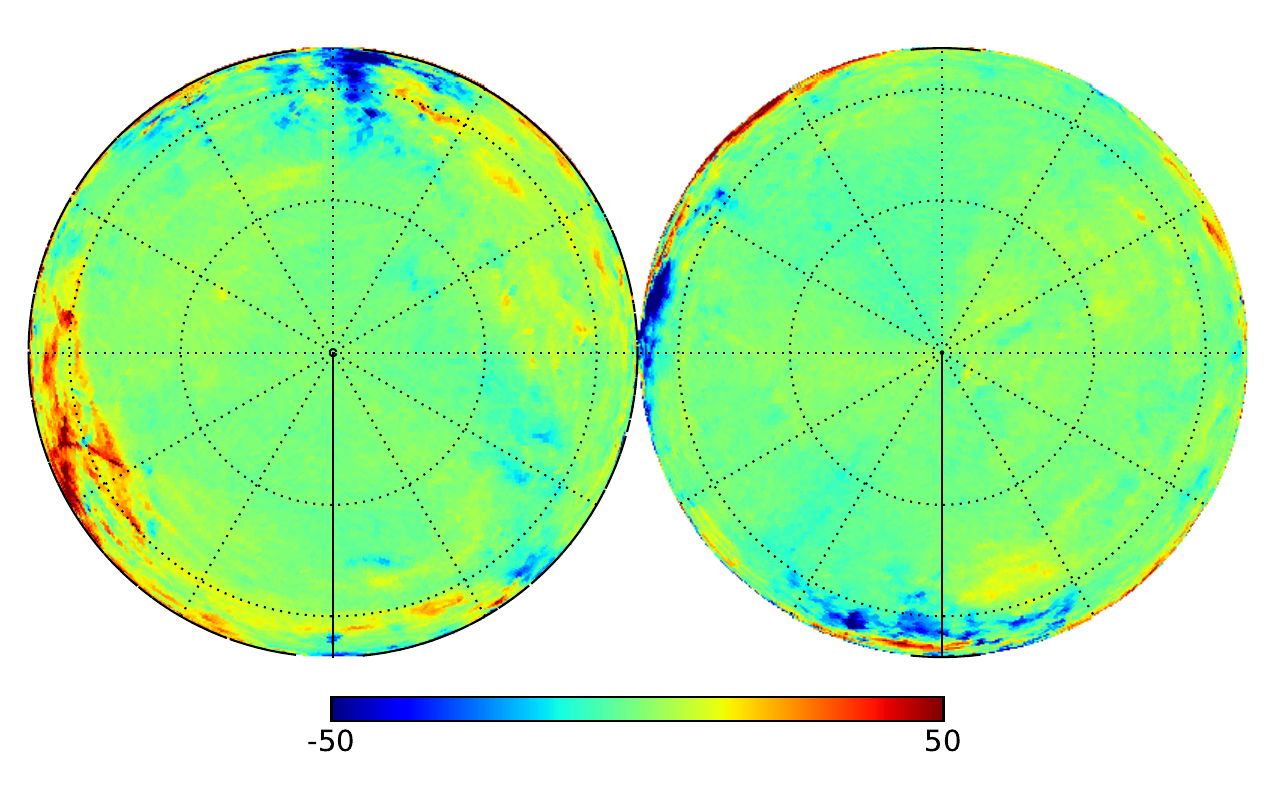} &
                \includegraphics[width=.23\linewidth]{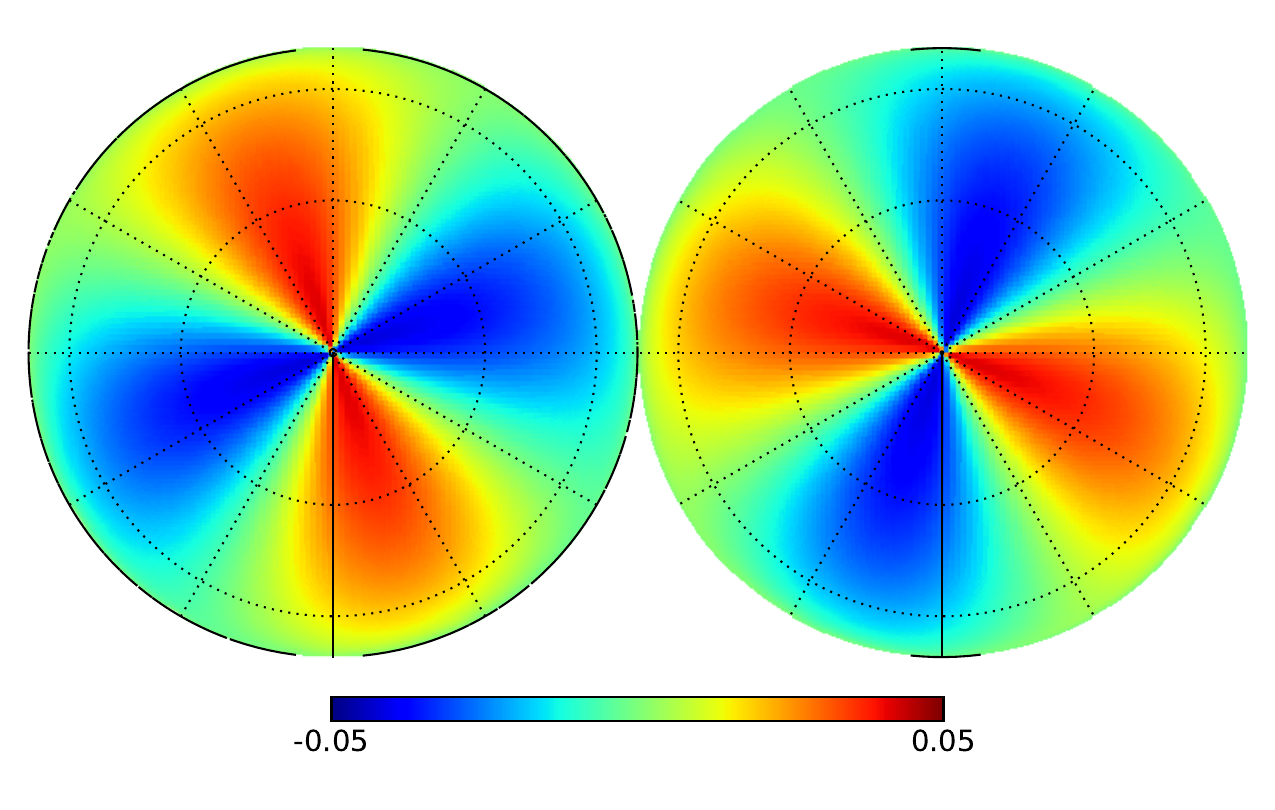} &
                        \includegraphics[width=.23\linewidth]{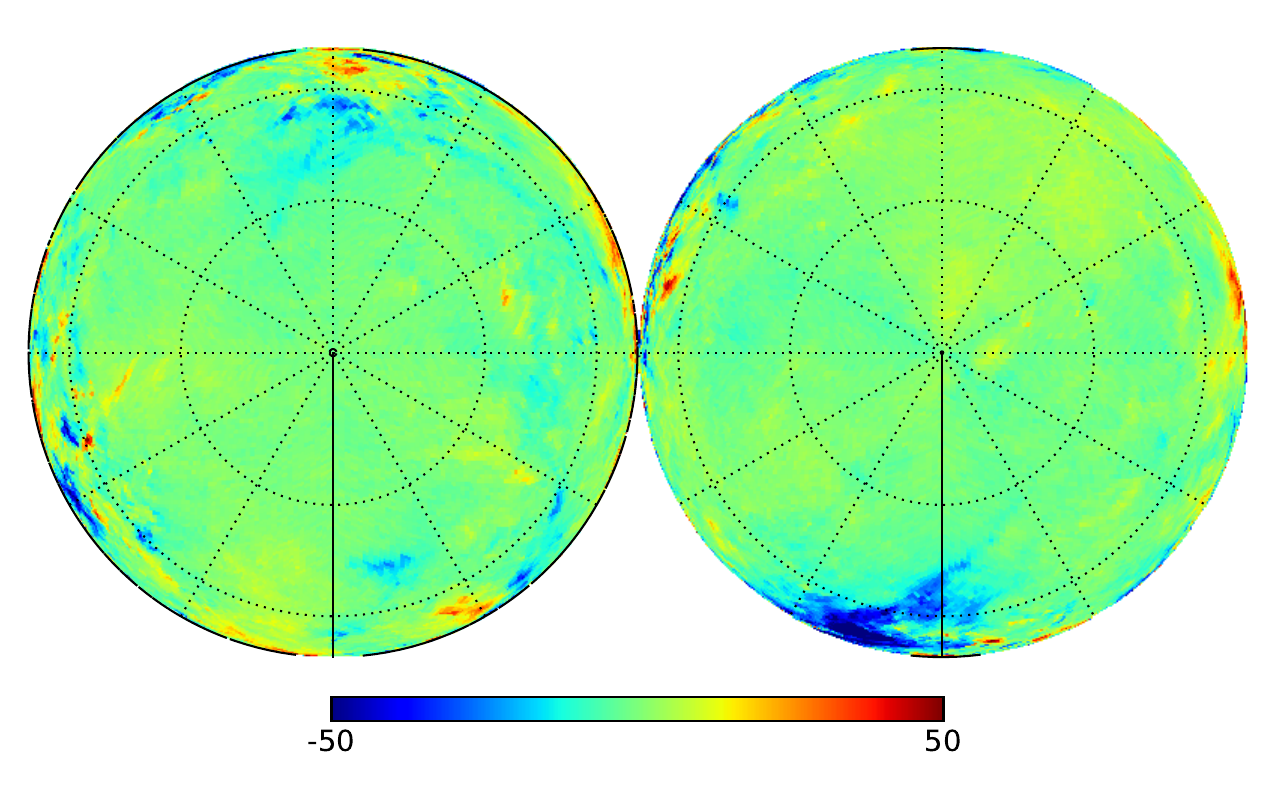} \\

\includegraphics[width=.23\linewidth]{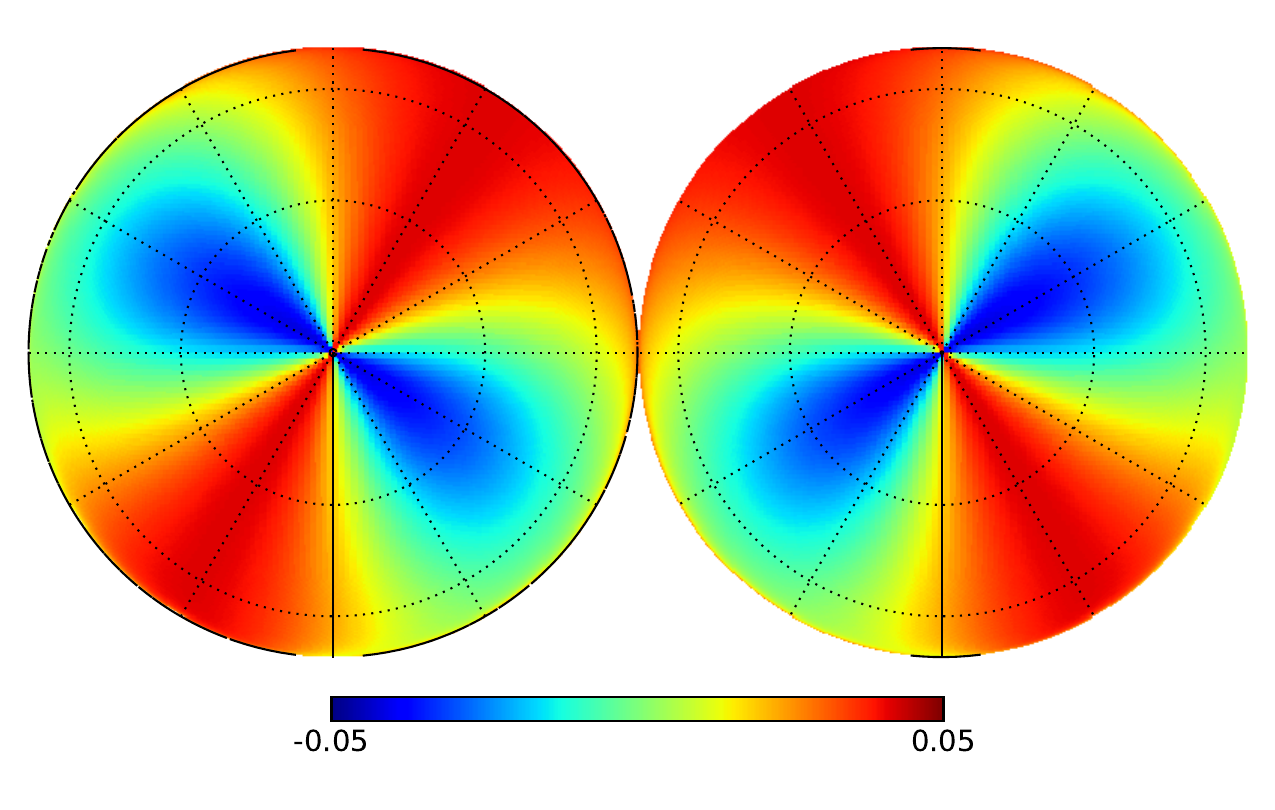} &
        \includegraphics[width=.23\linewidth]{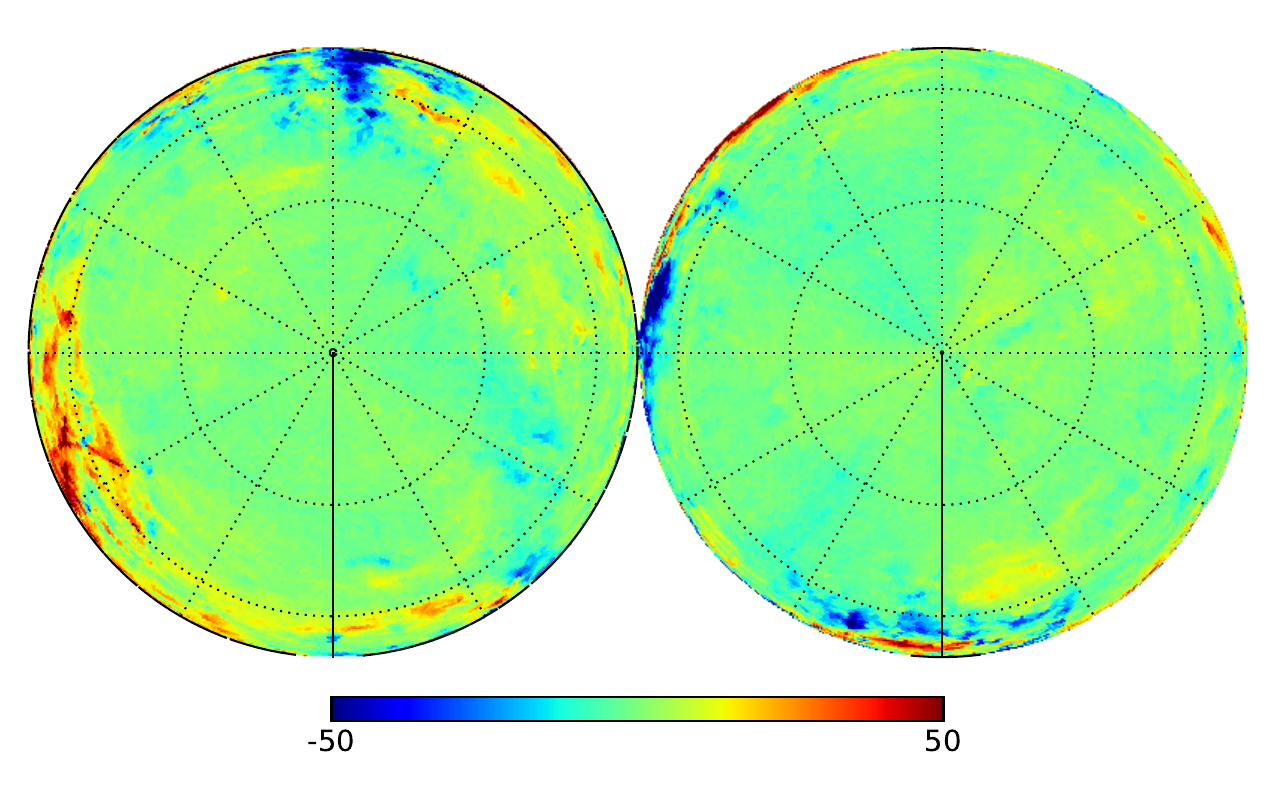} &
                \includegraphics[width=.23\linewidth]{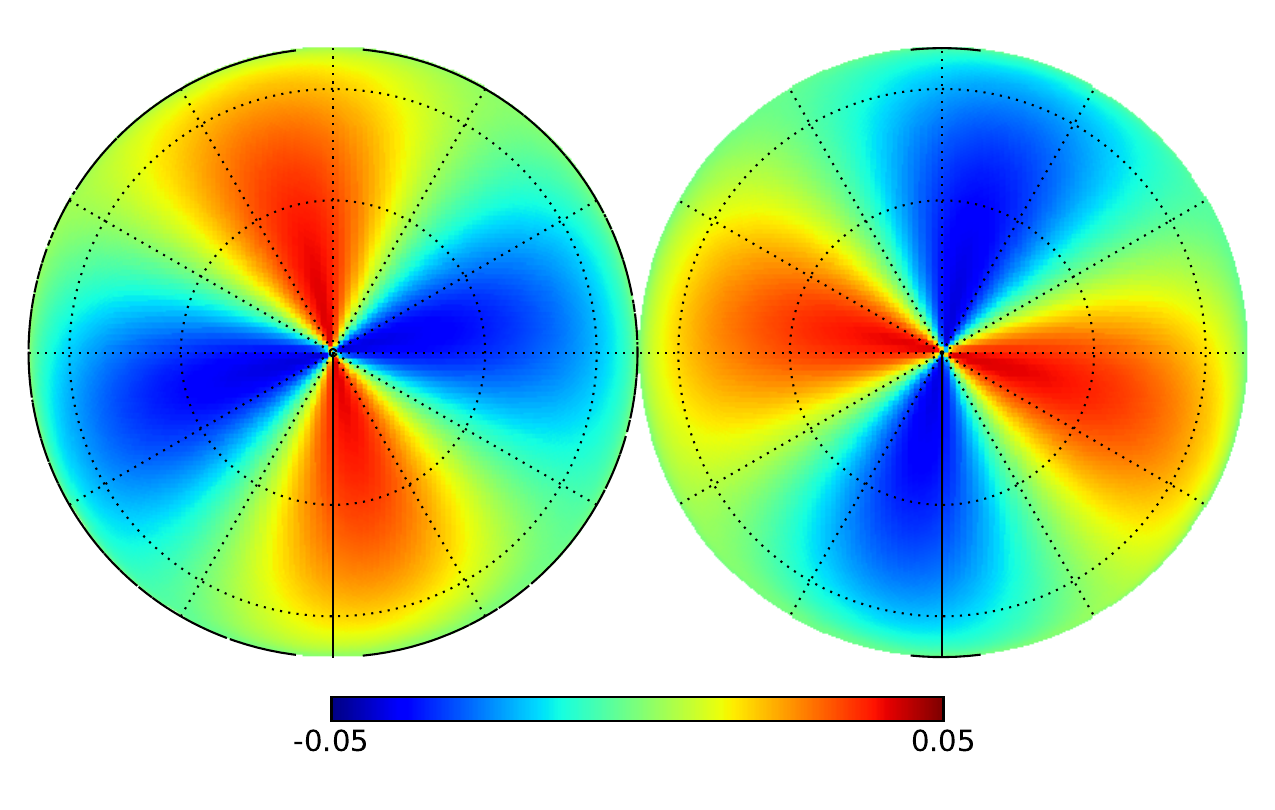} &
                        \includegraphics[width=.23\linewidth]{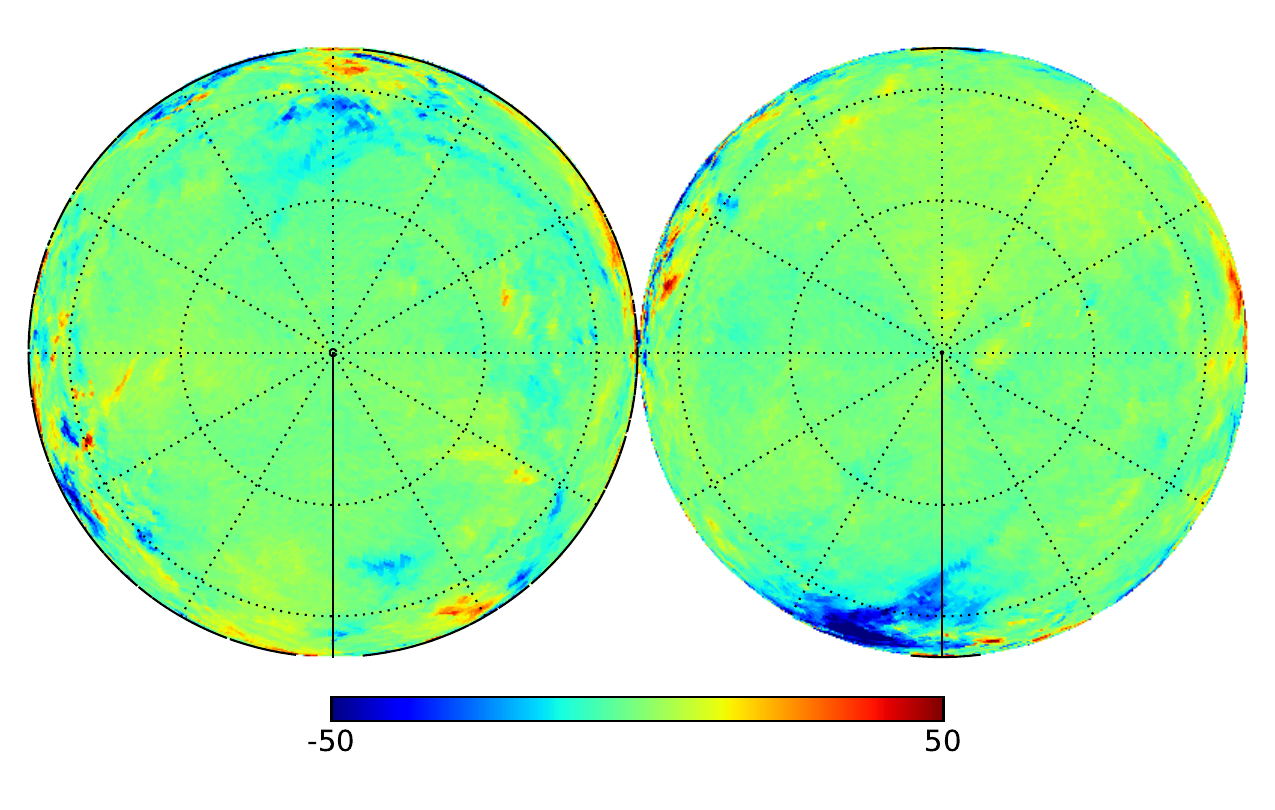} \\

\includegraphics[width=.23\linewidth]{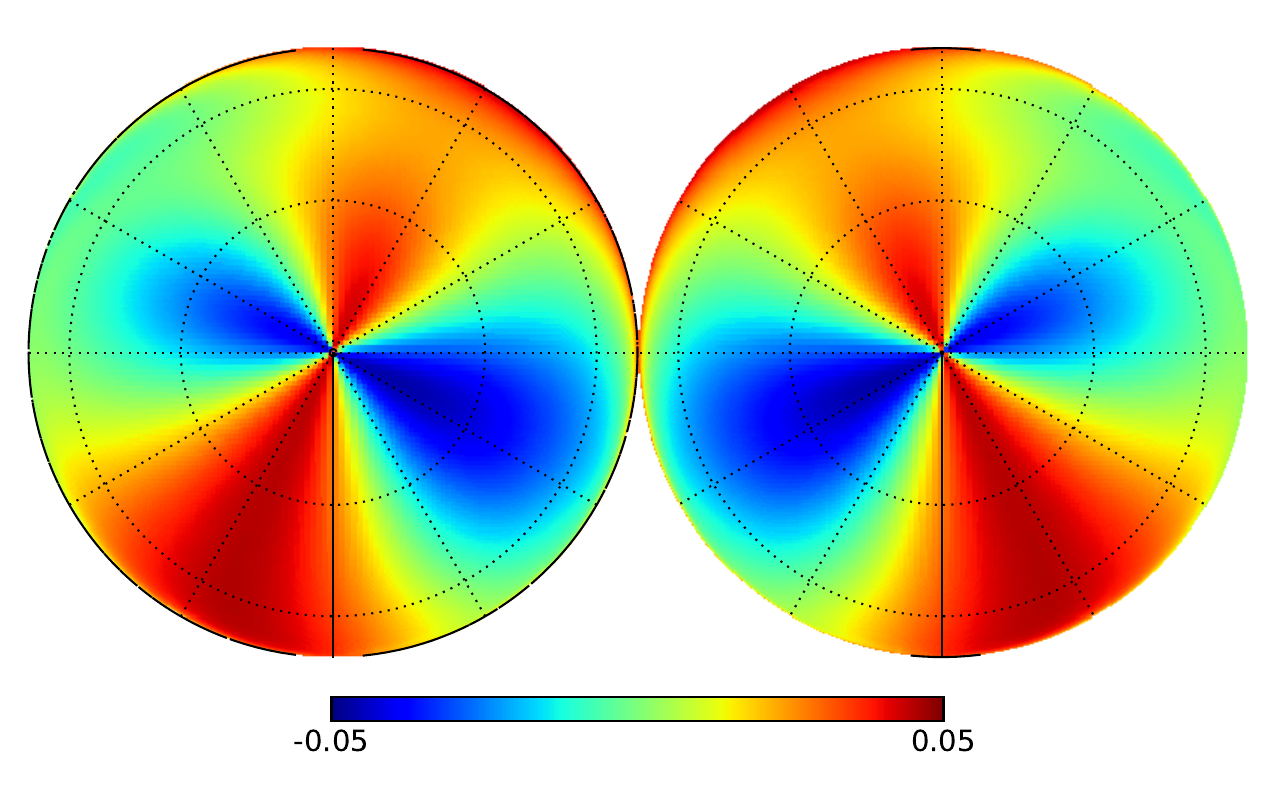} &
        \includegraphics[width=.23\linewidth]{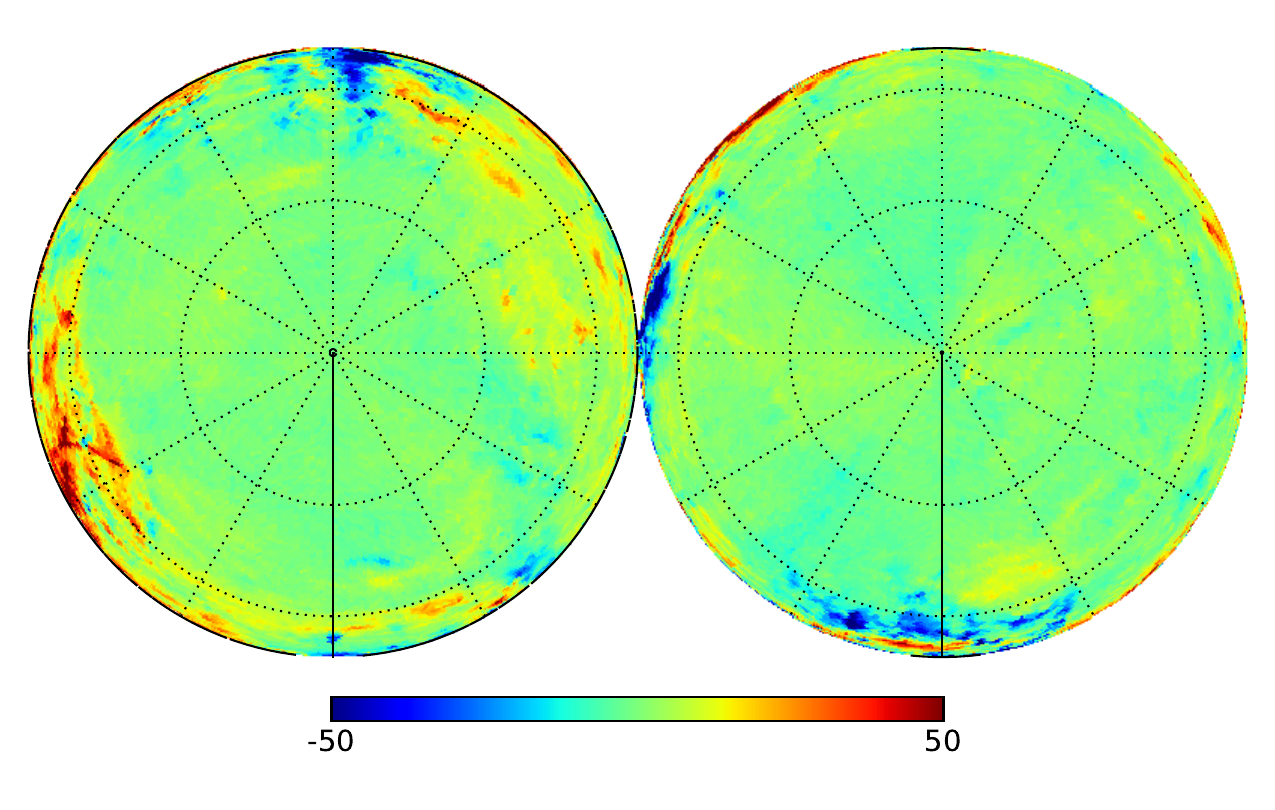} &
                \includegraphics[width=.23\linewidth]{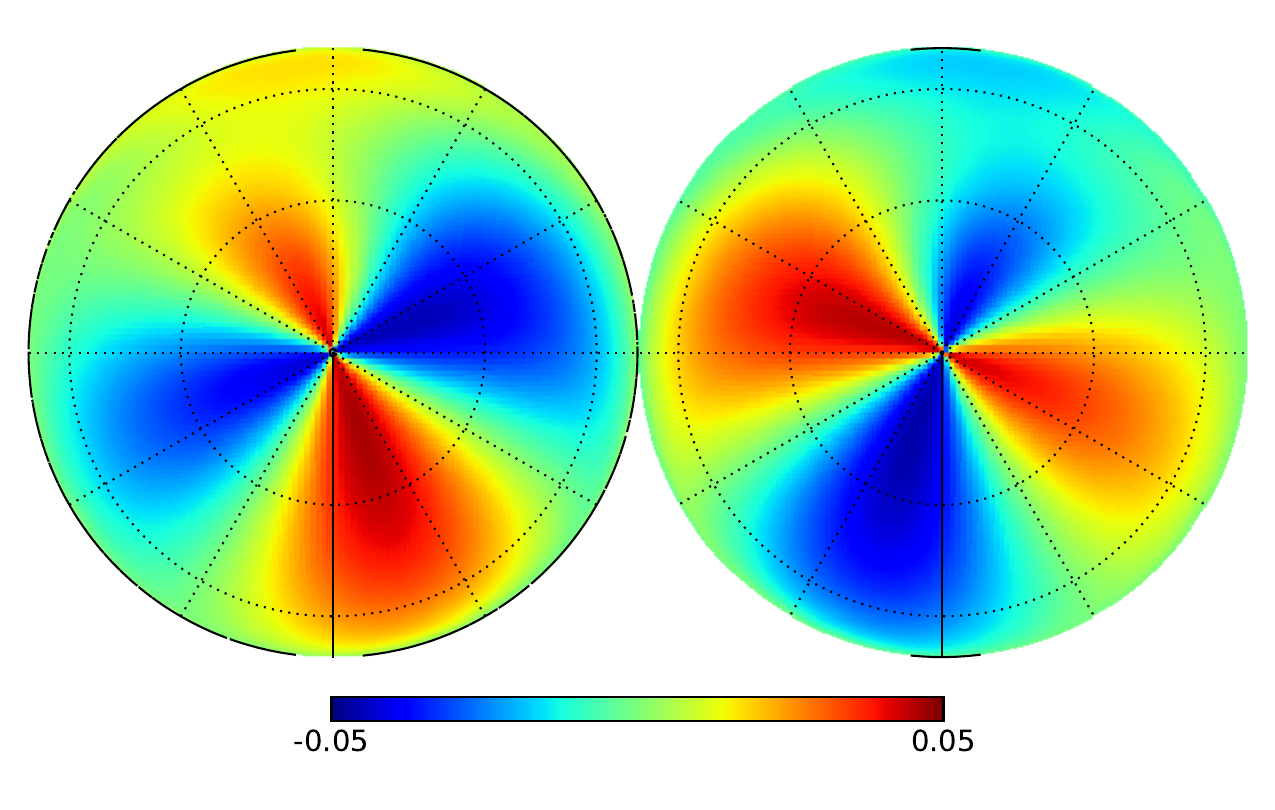} &
                        \includegraphics[width=.23\linewidth]{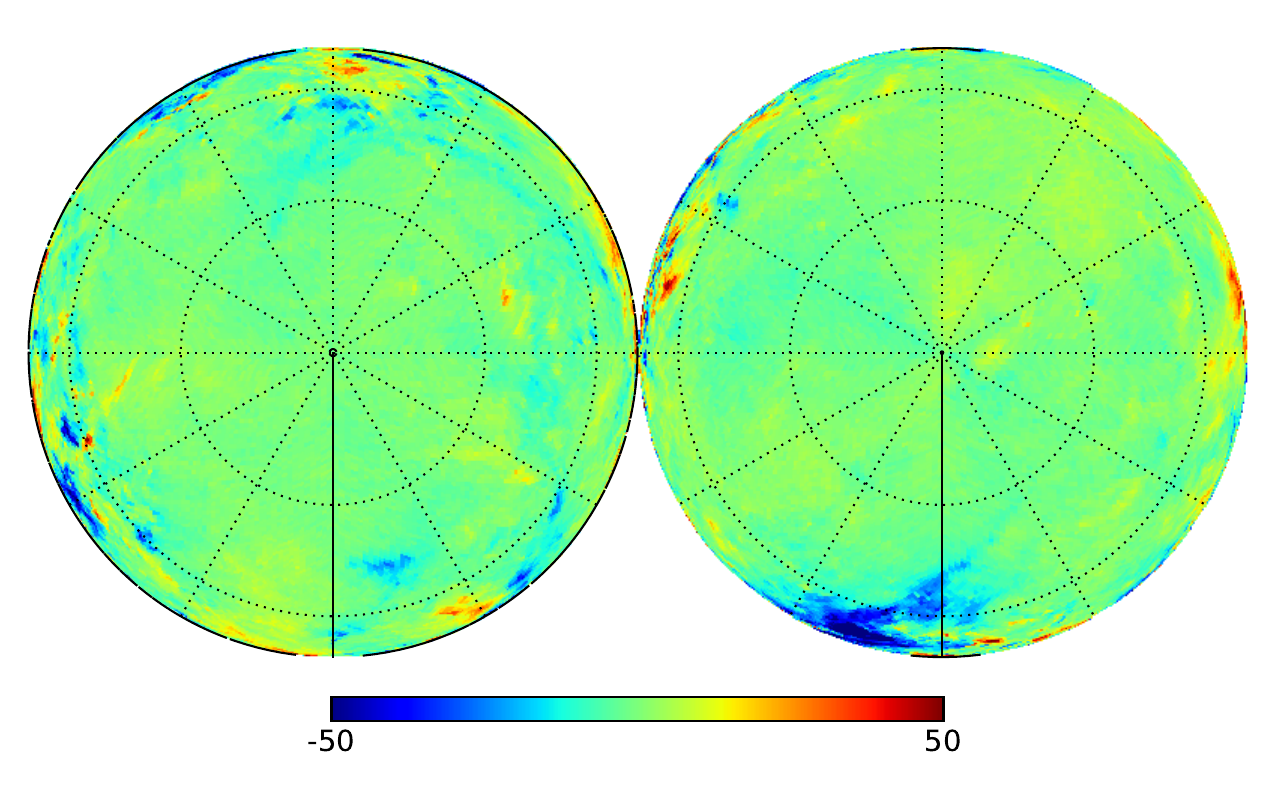} \\

\includegraphics[width=.23\linewidth]{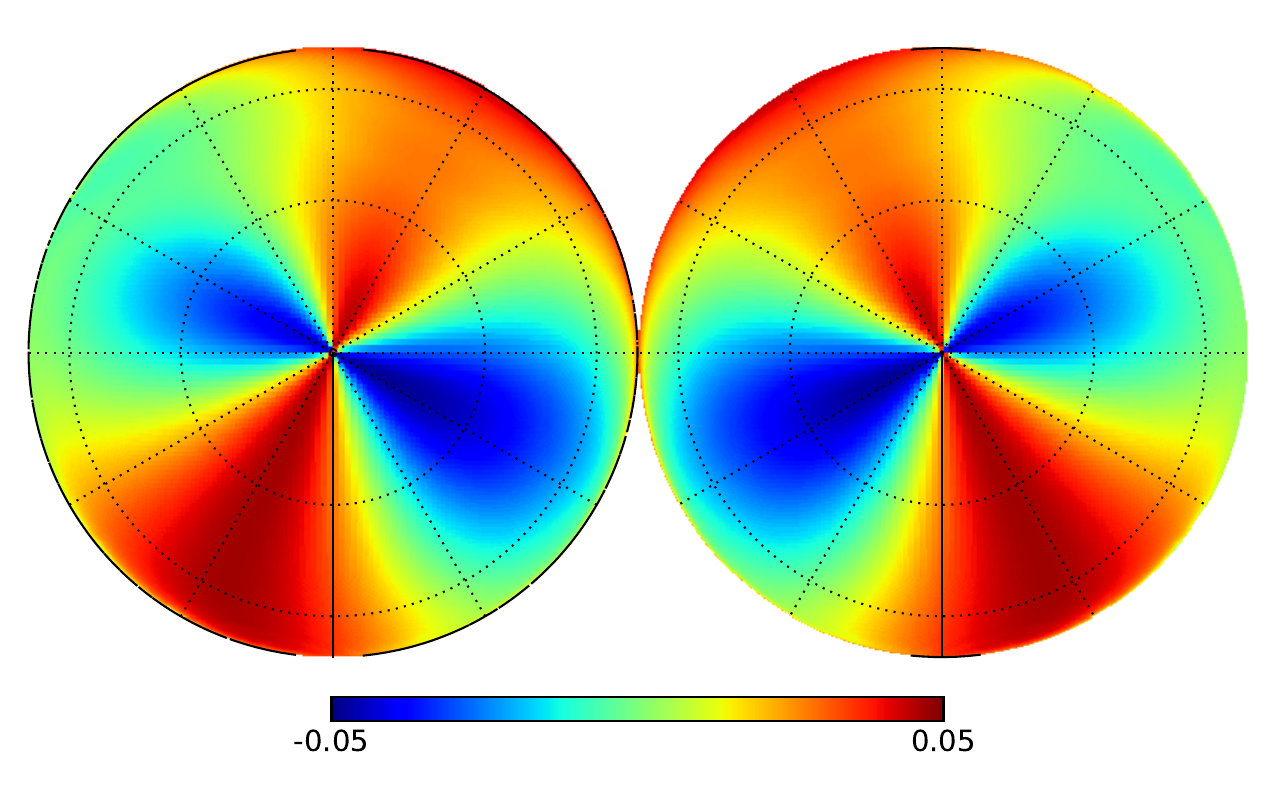} &
        \includegraphics[width=.23\linewidth]{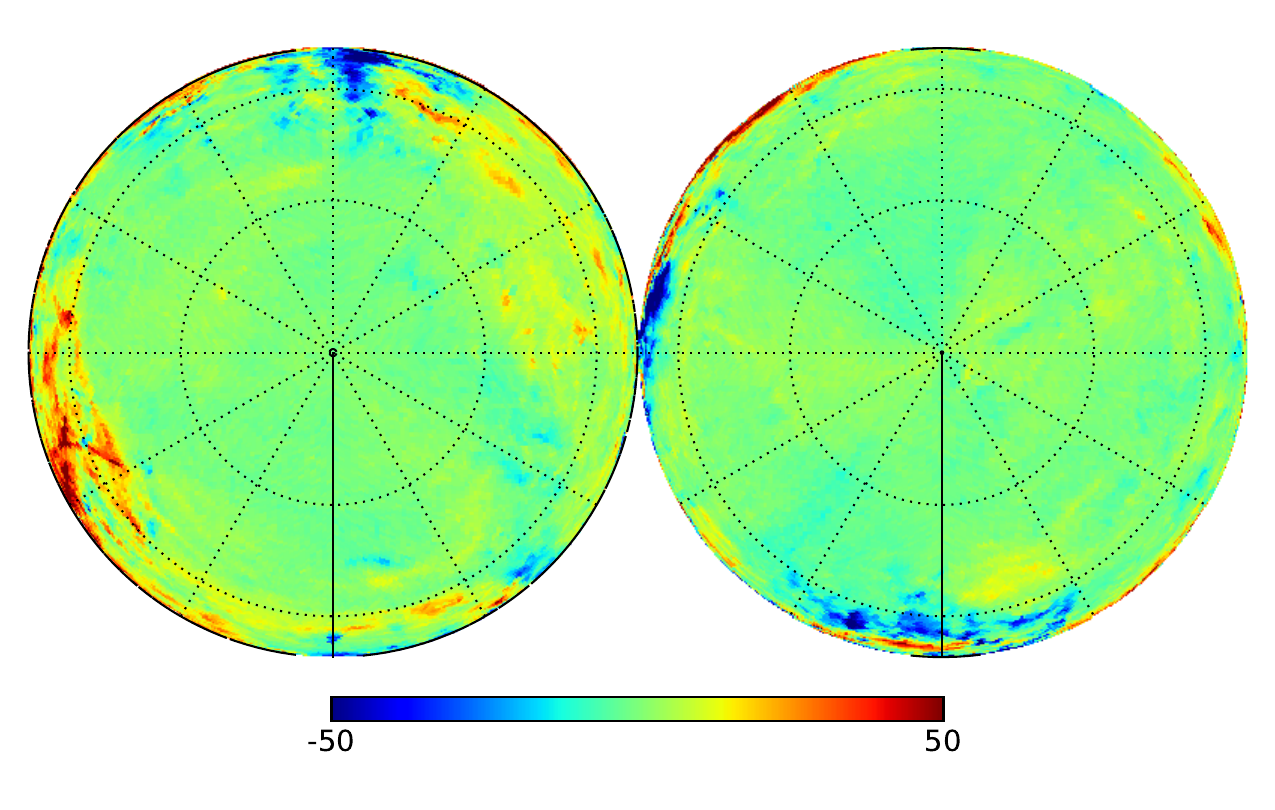} &
                \includegraphics[width=.23\linewidth]{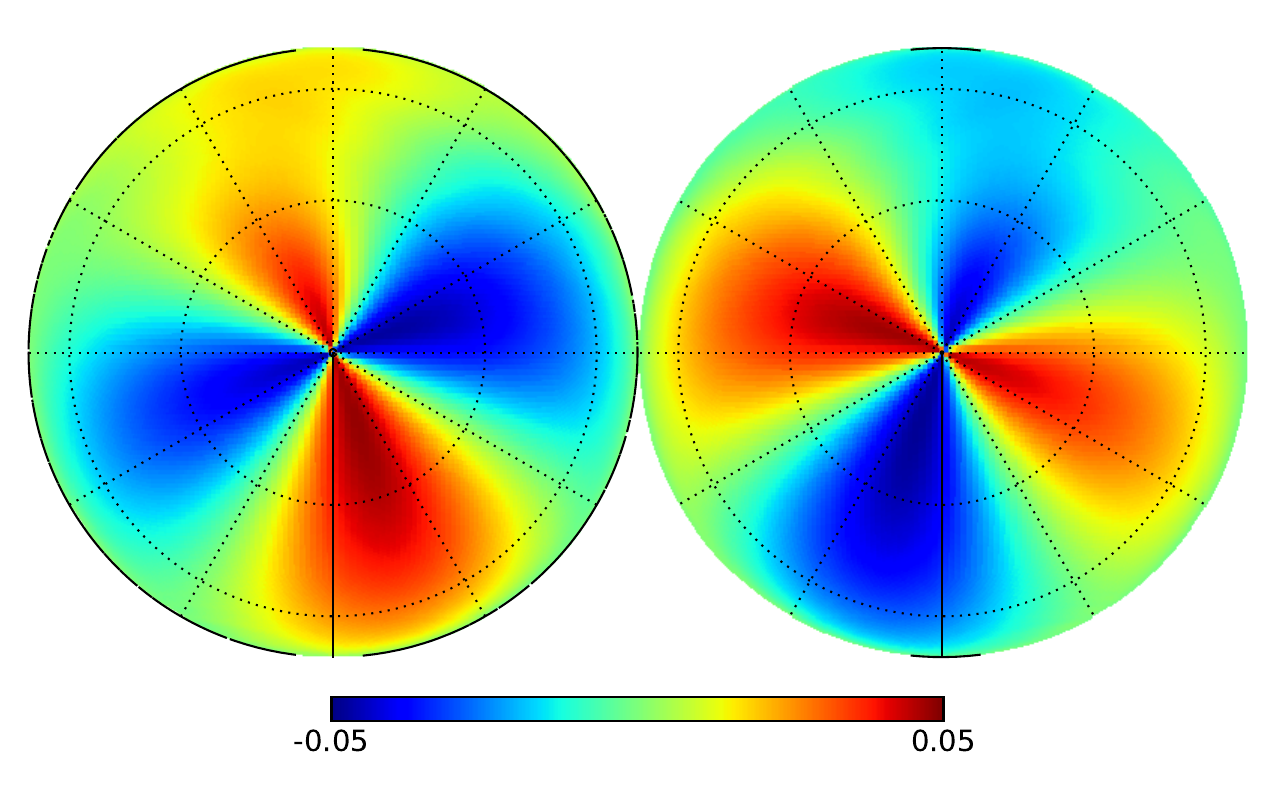} &
                        \includegraphics[width=.23\linewidth]{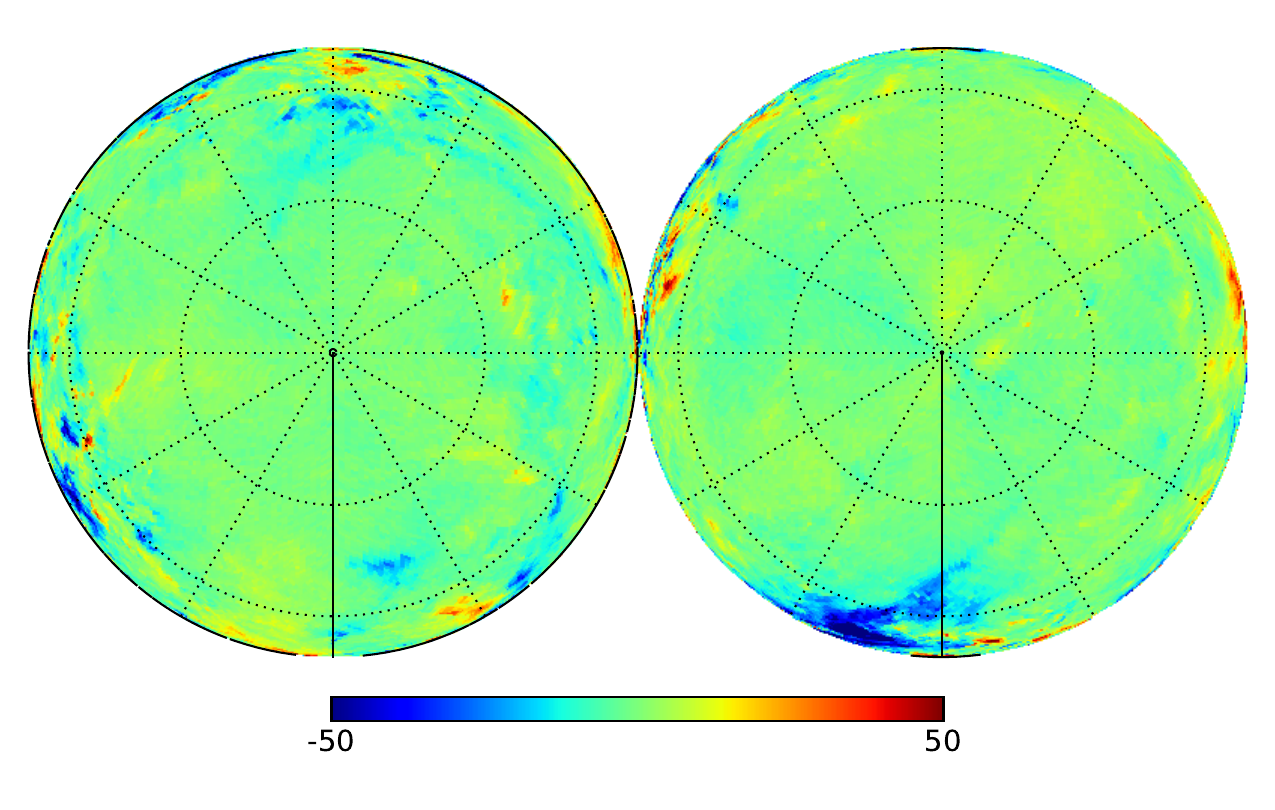} \\

\end{tabular}
\caption{Orthographic view of the polarization maps. First row shows the 353-GHz maps of the reduced Stokes parameters from \textit{Planck} downgraded at $N_{\rm{side}} = 64$ and the corresponding map of uncertainties that we use to compute the $\chi^2$ ($q$, $\sigma_{\rm{q}}$, $u$, $\sigma_{\rm{u}}$). Rows 2-5 correspond to GMF models labeled ASS, LSA, BSS, and QSS using the best-fit of the ED $n_{\rm{d}}$ model. The obtained best fits are shown in the first and third columns and the statistical significance of their residuals, per-pixel, are shown in the second and fourth columns. The same convention is used as for Fig.~\ref{fig:I_fit-orth}}
\label{fig:qu_fit-orth}
\end{figure*}

In Fig.~\ref{fig:I_fit-orth}, we provide an alternative view of the intensity
data and obtained best fits than the one presented in the core of the
paper. The agreement between the best-fit models at high Galactic
latitudes is salient in these orthographic views. We note that here we maintain
the same color scale as in Fig.~\ref{fig:I_fit}.

In Fig.~\ref{fig:qu_fit-orth} we provide an alternative view of the fitted
polarization data and of the obtained best-fits than the one presented
in the core of the paper. Inspection of the maps of the significance of
the residuals teaches us that our best-fit models agree well at high Galactic latitudes.
The same color scale as in Fig.~\ref{fig:qu_fit-orth} is adopted.

\section{Second solution for the LSA GMF model}
\label{sec:WMAP2nd}
\begin{figure*}[h]
\centering
\begin{tabular}{lll}

\includegraphics[width=.3\linewidth]{_figs/qu353sig-1EDWMAP3-bf_xy.pdf} &
        \includegraphics[width=.3\linewidth]{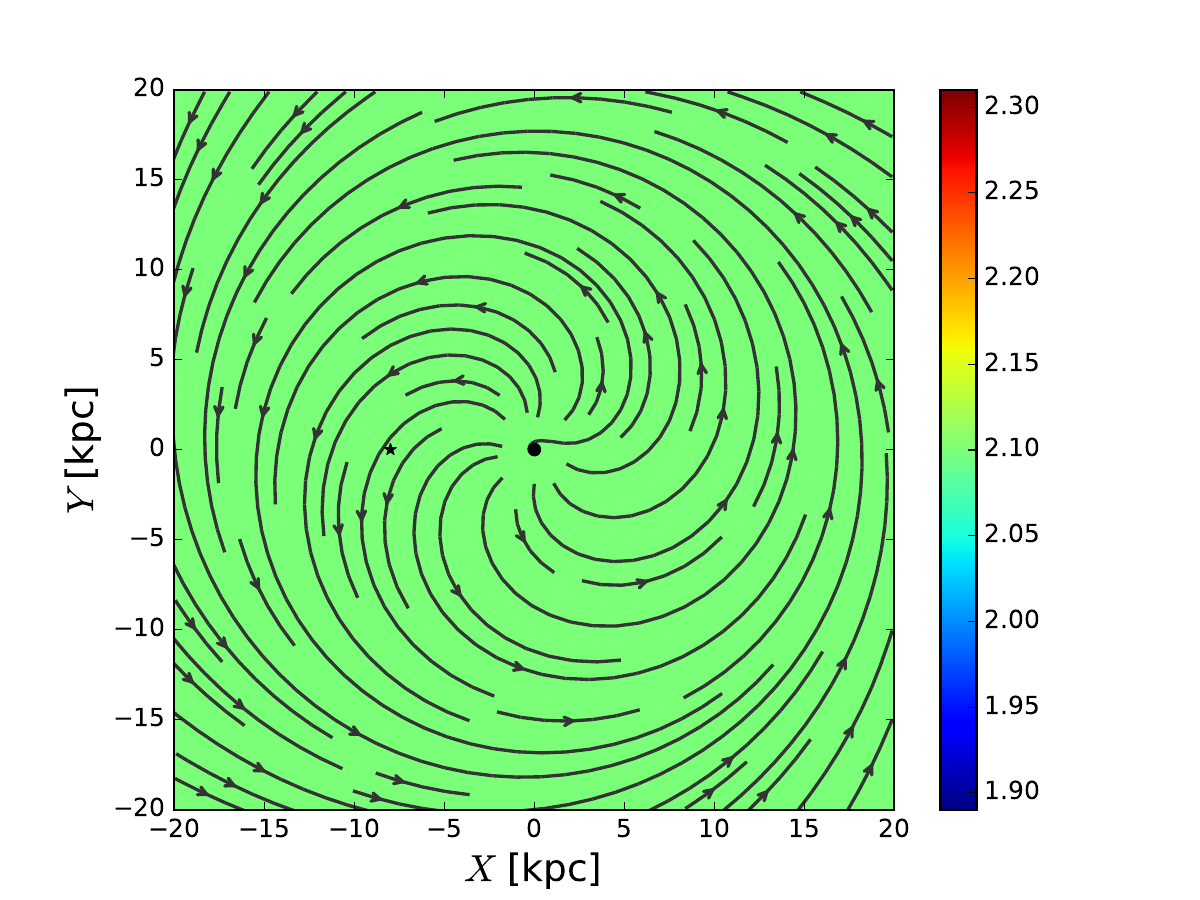} &
                \includegraphics[width=.3\linewidth]{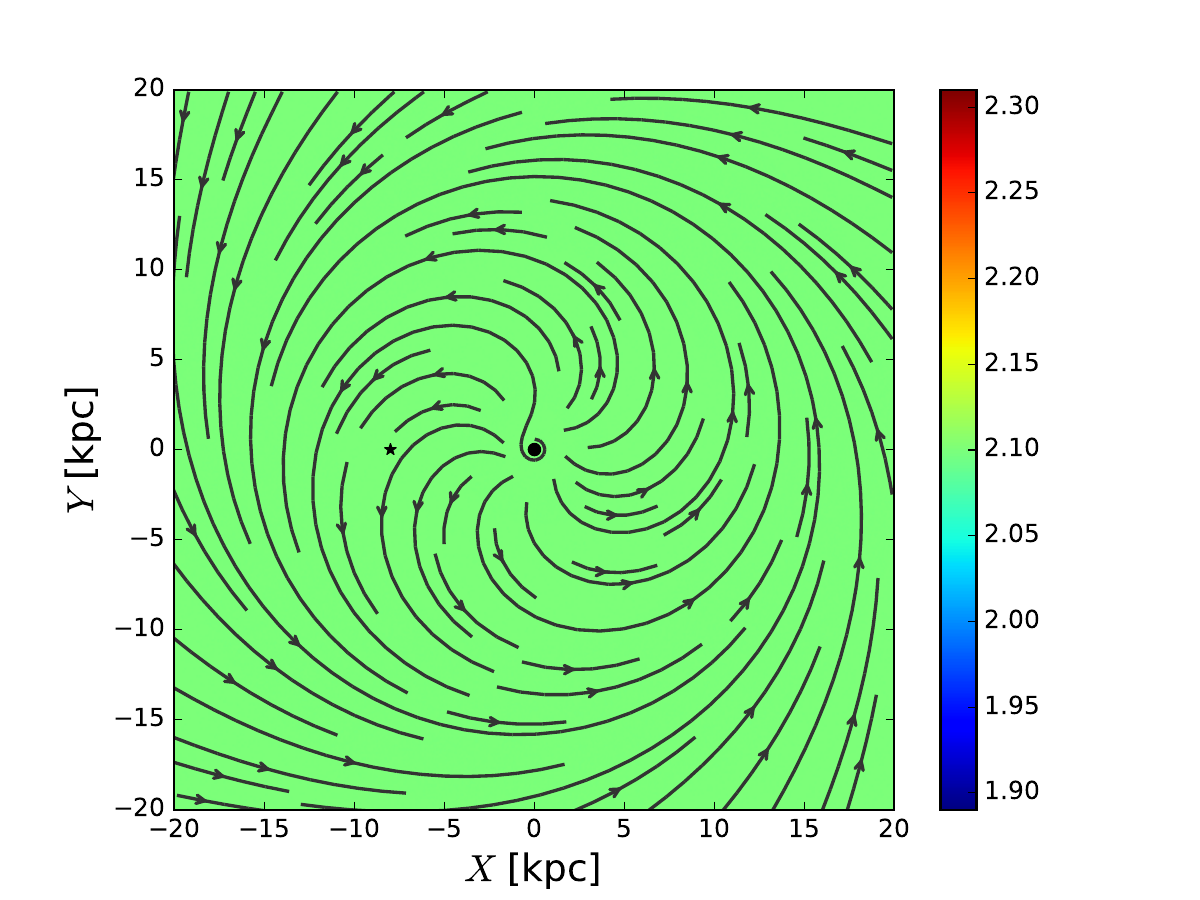} \\

\hspace{.05cm}
\includegraphics[trim={0cm 4cm 0cm 6cm},clip, width=.29\linewidth]{_figs/qu353sig-1EDWMAP3-bf_xz.pdf} &
        \hspace{.05cm}
        \includegraphics[trim={0cm 4cm 0cm 6cm},clip, width=.29\linewidth]{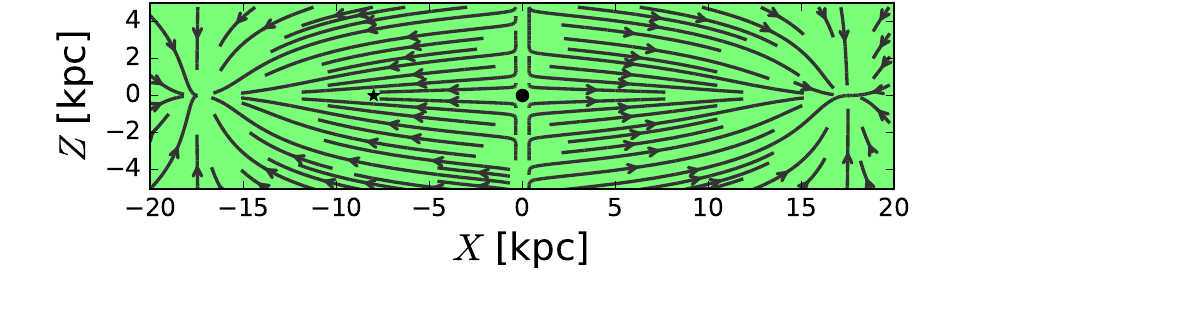} &
                \hspace{.05cm}
                \includegraphics[trim={0cm 4cm 0cm 6cm},clip, width=.29\linewidth]{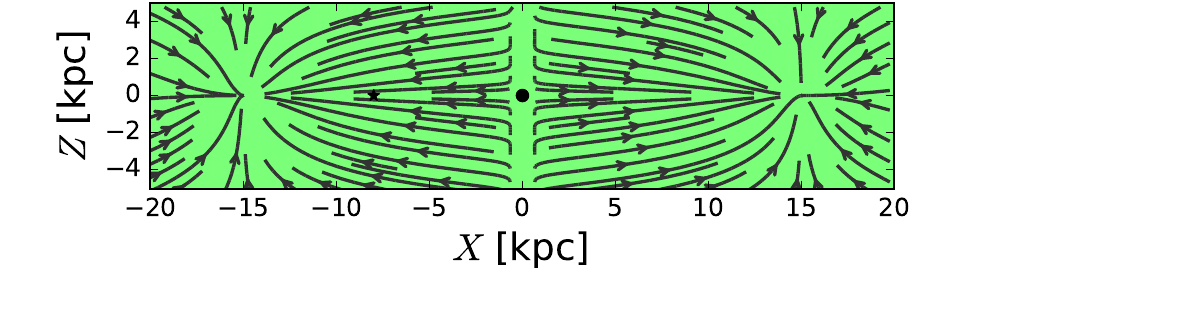} \\

\\

\includegraphics[width=.3\linewidth]{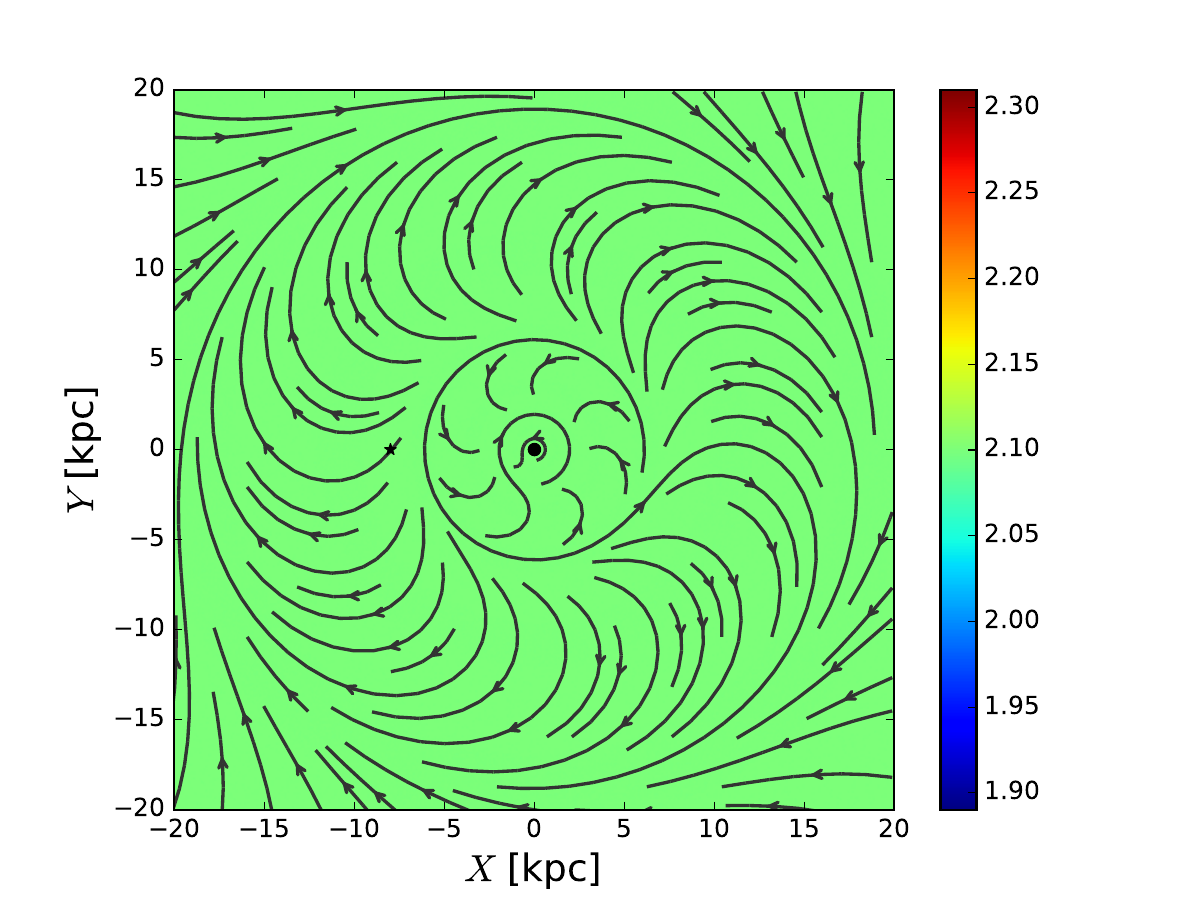} &
        \includegraphics[width=.3\linewidth]{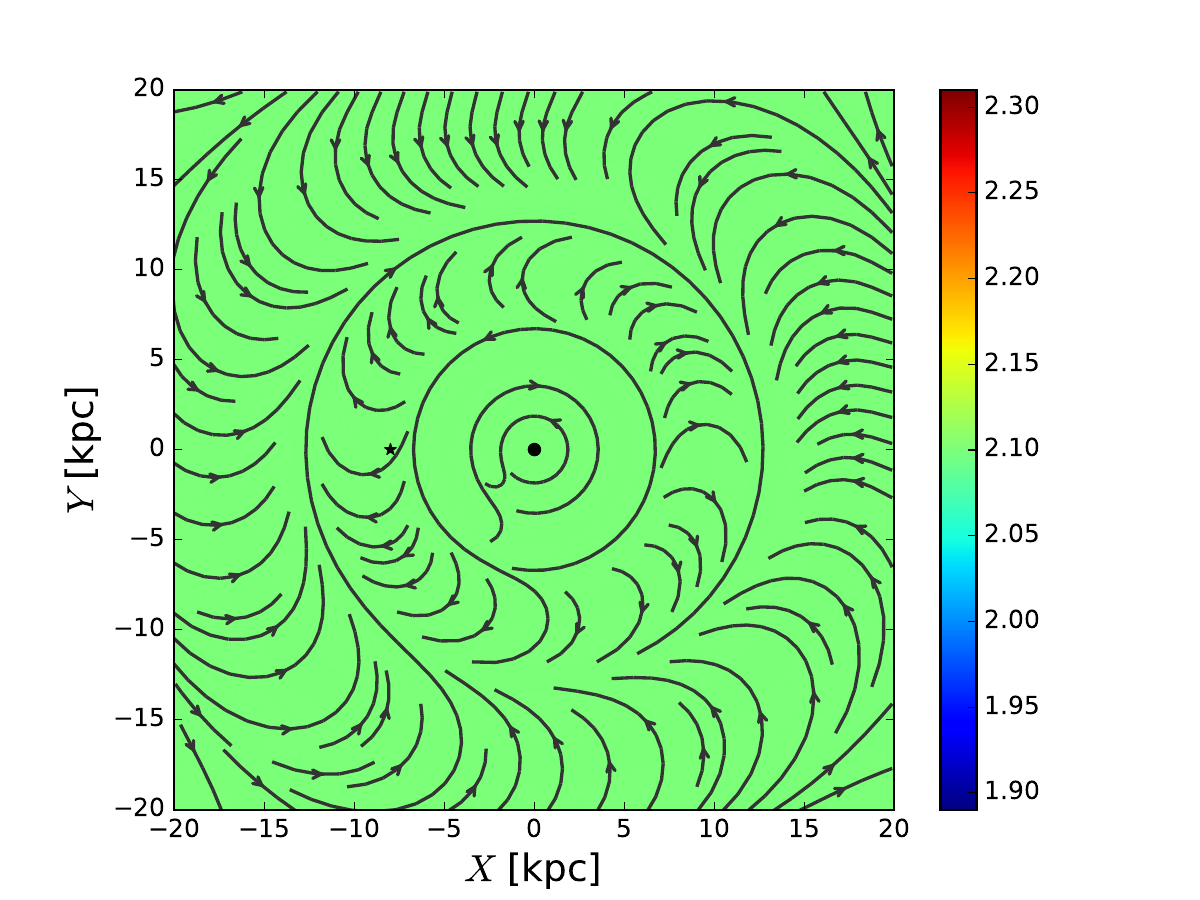} &
                \includegraphics[width=.3\linewidth]{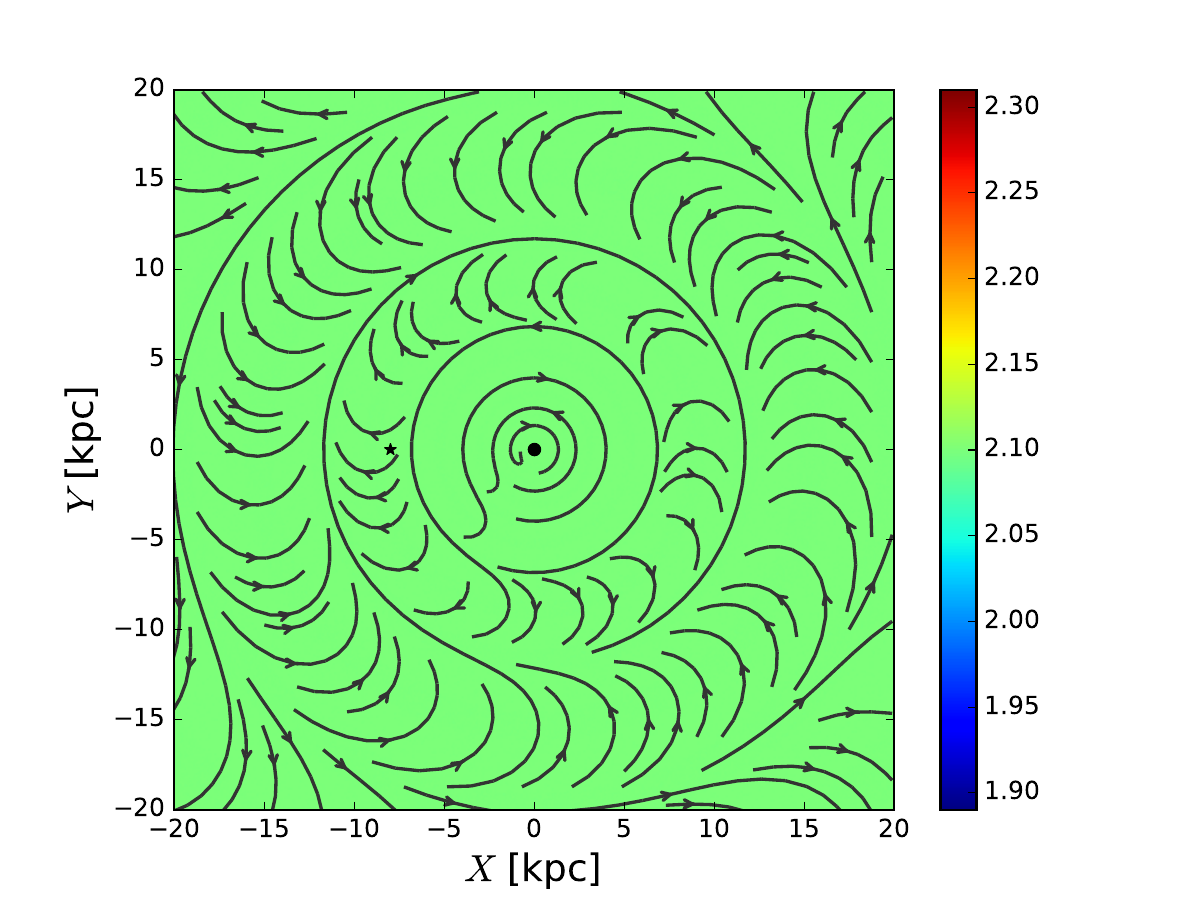} \\

\hspace{.05cm}
\includegraphics[trim={0cm 4cm 0cm 6cm},clip, width=.29\linewidth]{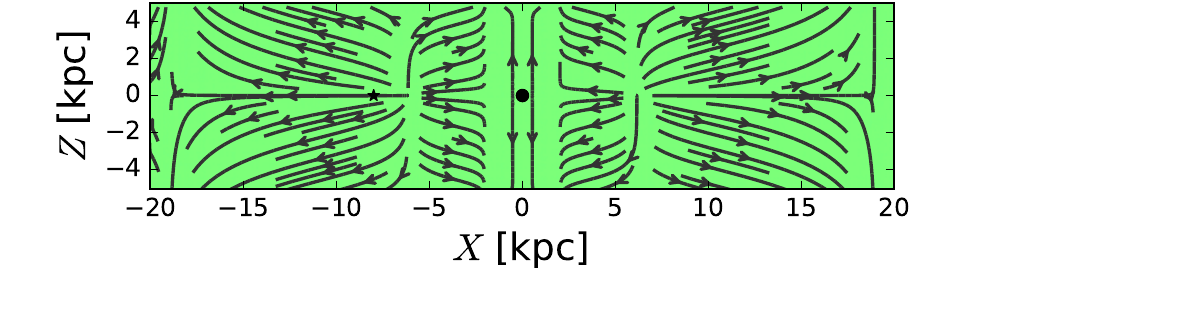} &
        \hspace{.05cm}
        \includegraphics[trim={0cm 4cm 0cm 6cm},clip, width=.29\linewidth]{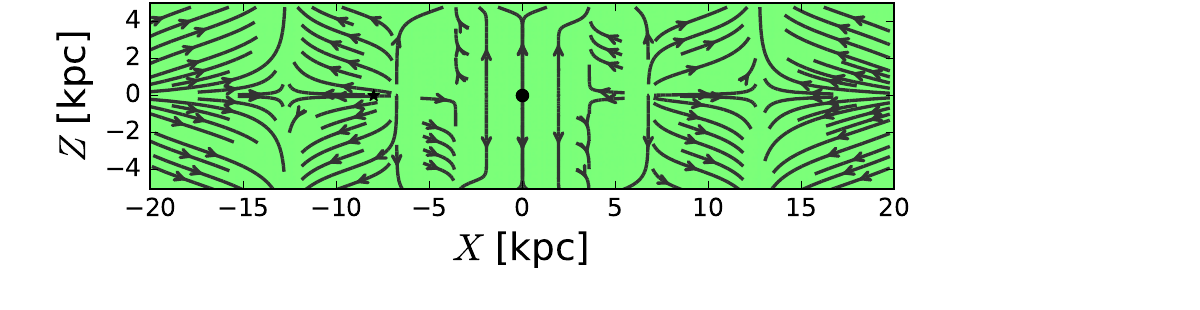} &
                \hspace{.05cm}
                \includegraphics[trim={0cm 4cm 0cm 6cm},clip, width=.29\linewidth]{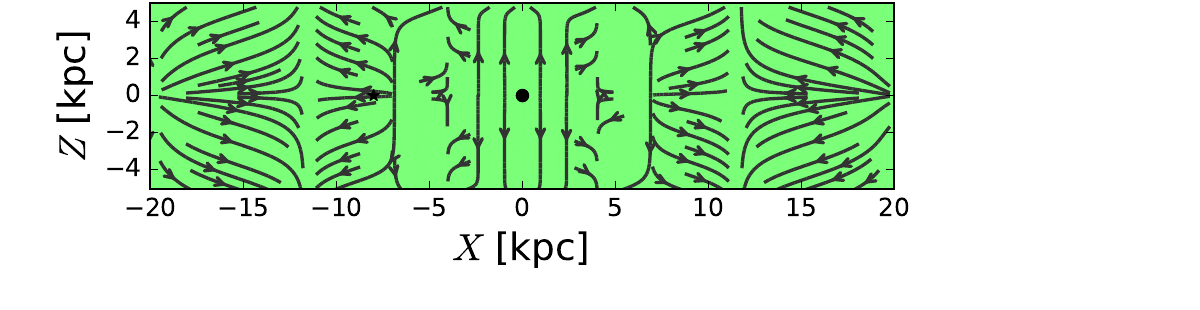} \\
\end{tabular}
\caption{GMF geometrical structure for the best-fit presented in the core of the paper (top row) and for the one obtained from the second local minimum (bottom row).
We show the best-fit ASS GMF models in the $(x,\,y,\,z=0)$ and $(x,\,y=0,\,z)$ planes of the Galaxy for the three best-fit $n_{\rm{d}}$ models obtained from the adjustment of the $I_{\rm{353}}$ map. 
From left to right, with $n_{\rm{d}} \equiv$ ED, ARM4, and ARM4$\oplus$ED, respectively.  }
\label{fig:WMAP2nd}
\end{figure*}

We illustrate here the existence of local minima in the hyper surface of the $\chi^2$ that we minimize to determine the best-fit to the {\it Planck} data.
For this purpose, we present the second solution that we obtained when fitting the LSA GMF model in the second row of Fig.~\ref{fig:WMAP2nd}.
We also present this solution because it illustrates the importance of considering large and as much un-informative prior as possible when fitting models to the existing dataset. Indeed, the LSA GMF model has recently been fitted to Faraday rotation measures and synchrotron data by \cite{Ste2018}, but while exploring a reduced parameter space compared to the one we considered in this paper.
It is interesting to notice that the second solution of the GMF, which we present here as a local minimum, appears to be consistent with their solution.

\smallskip

The best-fit model for this second solution depends more strongly on the underlying dust density distribution than the best-fit models that we discussed in the core of the paper and that we present (for the three dust density model) in the first row of Fig.~\ref{fig:WMAP2nd}.

\end{appendix}
\end{document}